\documentclass[useAMS,usenatbib,fleqn]{mn2e}
\usepackage{mathptmx}
\usepackage{deluxetable}
\usepackage{natbib,float,longtable,subfigure,graphicx,amsmath,amssymb}
\usepackage[T1]{fontenc}
\usepackage{ae,aecompl}
\usepackage{lscape}
\usepackage{color}

\newcommand{\todo}{\ifmmode \text{\Huge{\(\bullet\)}} \else {\Huge$\bullet$}\fi}
\newcommand{\tido}{\ifmmode {\bullet} \else $\bullet$\fi}

\newcommand{\E        }[1]{\ifmmode 10^{#1} \else $10^{#1}$\fi}
\newcommand{\tE        }[1]{\ifmmode \times10^{#1} \else $\times10^{#1}$\fi}
\newcommand{\til}{\ifmmode \sim \else $\sim$\fi}
\renewcommand{\~} {\ifmmode \sim \else $\sim$\fi}

\newcommand{\pc}	{\ifmmode {\rm pc} \else pc\fi}
\newcommand{\ld}	{\ifmmode {\rm l.d.} \else l.d.\fi}
\newcommand{\kms}	{\ifmmode {\rm km\,s}^{-1} \else km\,s$^{-1}$\fi}
\newcommand{\cc}	{\ifmmode {\rm cm}^{-3}    \else cm$^{-3}$\fi}
\newcommand{\cmii}	{\ifmmode {\rm cm}^{-2}    \else cm$^{-2}$\fi}
\newcommand{\ergs}	{\ifmmode {\rm erg\,s}^{-1} \else erg s$^{-1}$\fi}
\newcommand{\ergcms}	{\ifmmode {\rm erg\,cm}^{-2}\,{\rm s}^{-1} \else erg\,cm$^{-2}$\,s$^{-1}$\fi}
\newcommand{\ergscm}	{\ifmmode {\rm erg\,s}^{-1}\,{\rm cm}^{-2} \else erg\,s$^{-1}$\,cm$^{-2}$\fi}
\newcommand{\ergcmsA}	{\ifmmode {\rm erg\,cm}^{-2}\,{\rm s}^{-1}\,{\rm\AA}^{-1}
\else erg\,cm$^{-2}$\,s$^{-1}$\,\AA$^{-1}$\fi}
\newcommand{  \ergcmsHz  }{\ifmmode{\rm erg\,cm}^{-2}\,{\rm s}^{-1}\,{\rm Hz}^{-1}
                       \else ergs\,cm$^{-2}$\,s$^{-1}$\,Hz$^{-1}$\fi}
\newcommand{\kev}	{\ifmmode {\rm keV} \else keV\fi}

\newcommand{\mic}	{\ifmmode {\rm \mu m} \else $\mu$m\fi}
\newcommand{\vFWHM}	{\ifmmode v_{\mbox{\tiny FWHM}} \else $v_{\mbox{\tiny FWHM}}$\fi}
\newcommand{\vBLR}	{\ifmmode v_{\mbox{\tiny BLR}} \else $v_{\mbox{\tiny BLR}}$\fi}
\newcommand{\sigBLR}	{\ifmmode \sigma_{\mbox{\tiny BLR}} \else $\sigma_{\mbox{\tiny BLR}}$\fi}
\newcommand{\vNLR}	{\ifmmode v_{\mbox{\tiny NLR}} \else $v_{\mbox{\tiny NLR}}$\fi}
\newcommand{\tauBLR}	{\ifmmode \tau_{\mbox{\tiny BLR}} \else $\tau_{\mbox{\tiny BLR}}$\fi}

\newcommand{\Hubble}	{\ifmmode {\rm km\,s}^{-1}\,{\rm Mpc}^{-1} \else km\,s$^{-1}$\,Mpc$^{-1}$\fi}
\newcommand{\NDunit}	{\ifmmode {\rm Mpc}^{-3} \else Mpc$^{-3}$\fi}
\newcommand{\LFunit}	{\ifmmode {\rm Mpc}^{-3}\,{\rm mag}^{-1} \else Mpc$^{-3}$\,mag$^{-1}$\fi}
\newcommand{\MFunit}	{\ifmmode {\rm Mpc}^{-3}\,{\rm dex}^{-1} \else Mpc$^{-3}$\,dex$^{-1}$\fi}

\newcommand{\Msun}{\ifmmode M_{\odot} \else $M_{\odot}$\fi}
\newcommand{\Lsun}{\ifmmode L_{\odot} \else $L_{\odot}$\fi}
\newcommand{\Zsun}{\ifmmode Z_{\odot} \else $Z_{\odot}$\fi}
\newcommand{\mpyr}{\ifmmode \Msun\,{\rm yr}^{-1} \else $\Msun\,{\rm yr}^{-1}$\fi}

\newcommand{\qnote}{\ifmmode q_{0} \else $q_{0}$\fi}
\newcommand{\Hnote}{\ifmmode H_{0} \else $H_{0}$\fi}
\newcommand{\hnote}{\ifmmode h_{0} \else $h_{0}$\fi}
\newcommand{\anote}{\ifmmode a_{0} \else $a_{0}$\fi}
\newcommand{\tnote}{\ifmmode t_{0} \else $t_{0}$\fi}

\newcommand{  \Halpha   }{\ifmmode {\rm H}\alpha \else H$\alpha$\fi}
\newcommand{  \ha   	}{\ifmmode {\rm H}\alpha \else H$\alpha$\fi}
\newcommand{  \Hbeta    }{\ifmmode {\rm H}\beta \else H$\beta$\fi}
\newcommand{  \hb    	}{\ifmmode {\rm H}\beta \else H$\beta$\fi}
\newcommand{  \Hgamma   }{\ifmmode {\rm H}\gamma \else H$\gamma$\fi}
\newcommand{  \Hdelta   }{\ifmmode {\rm H}\delta \else H$\delta$\fi}
\newcommand{  \Lya      }{\ifmmode {\rm Ly}\alpha \else Ly$\alpha$\fi}
\newcommand{  \Lyb      }{\ifmmode {\rm Ly}\beta \else Ly$\beta$\fi}
\newcommand{  \Pa       }{\ifmmode {\rm P}\alpha \else P$\alpha$\fi}
\newcommand{  \Pb       }{\ifmmode {\rm P}\beta \else P$\beta$\fi}
\newcommand{  \Bra      }{\ifmmode {\rm Br}\alpha \else Br$\alpha$\fi}
\newcommand{  \Brg      }{\ifmmode {\rm Br}\gamma \else Br$\gamma$\fi}
\newcommand{  \hii      }{\ifmmode {\rm H}\,\textsc{ii} \else H\,\textsc{ii}\fi}
\newcommand{  \hei      }{\ifmmode {\rm He}\,\textsc{i} \else He\,\textsc{i}\fi}
\newcommand{  \heii     }{\ifmmode {\rm He}\,\textsc{ii} \else He\,\textsc{ii}\fi}
\newcommand{  \HeIIuv   }{\ifmmode {\rm He}\,\textsc{ii}\,\lambda1640 \else He\,\textsc{ii}\,$\lambda1640$\fi}
\newcommand{  \HeIIop   }{\ifmmode {\rm He}\,\textsc{ii}\,\lambda4686 \else He\,\textsc{ii}\,$\lambda4686$\fi}
\newcommand{  \cii      }{\ifmmode {\rm C}\,\textsc{ii}  \else C\,\textsc{ii}\fi}
\newcommand{  \ciii     }{\ifmmode {\rm C}\,\textsc{iii}\right] \else C\,\textsc{iii}]\fi}
\newcommand{  \CIII     }{\ifmmode {\rm C}\,\textsc{iii}\right]\,\lambda1909 \else C\,\textsc{iii}]\,$\lambda1909$\fi}
\newcommand{  \civ      }{\ifmmode {\rm C}\,\textsc{iv}  \else C\,\textsc{iv}\fi}
\newcommand{  \CIV      }{\ifmmode {\rm C}\,\textsc{iv}\,\lambda1549 \else C\,\textsc{iv}\,$\lambda1549$\fi}
\newcommand{  \nii      }{\ifmmode [{\rm N}\,\textsc{ii}]  \else [N\,\textsc{ii}]\fi}
\newcommand{  \niii     }{\ifmmode {\rm N}\,\textsc{iii} \else N\,\textsc{iii}\fi}
\newcommand{  \niv      }{\ifmmode {\rm N}\,\textsc{iv}  \else N\,\textsc{iv}\fi}
\newcommand{  \NIVuv    }{\ifmmode {\rm N}\,\textsc{iv}\,\lambda1486 \else N\,\textsc{iv}\,$\lambda1486$\fi}
\newcommand{  \nv       }{\ifmmode {\rm N}\,\textsc{v}   \else N\,\textsc{v}\fi}
\newcommand{\oi}{\ifmmode \left[{\rm O}\,\textsc{i}\right] \else [O\,{\sc i}]\fi}
\newcommand{\OI}{\ifmmode \left[{\rm O}\,\textsc{i}\right]\,\lambda6300 \else [O\,{\sc i}]$\,\lambda6300$\fi}

\newcommand{\oii}{\ifmmode \left[{\rm O}\,\textsc{ii}\right] \else [O\,{\sc ii}]\fi}
\newcommand{\OII}{\ifmmode \left[{\rm O}\,\textsc{ii}\right]\,\lambda3727 \else [O\,{\sc ii}]\,$\lambda3727$\fi}
\newcommand{\oiii}{\ifmmode \left[{\rm O}\,\textsc{iii}\right] \else [O\,{\sc iii}]\fi}
\newcommand{\OIII}{\ifmmode \left[{\rm O}\,\textsc{iii}\right]\,\lambda5007 \else [O\,{\sc iii}]\,$\lambda5007$\fi}

\newcommand{\NII}{\ifmmode \left[{\rm N}\,\textsc{ii}\right]\,\lambda6583 \else [N\,{\sc ii}]$\,\lambda6583$\fi}

\newcommand{\NeIII}{\ifmmode \left[{\rm Ne}\,\textsc{iii}\right]\,\lambda3968 \else [Ne\,{\sc iii}]$\,\lambda3968$\fi}

\newcommand{\NeV}{\ifmmode \left[{\rm Ne}\,\textsc{v}\right]\,\lambda3426 \else [Ne\,{\sc v}]$\,\lambda3426$\fi}

\newcommand{\HeII}{\ifmmode {\rm He}\,\textsc{ii}\,\lambda4686 \else He\,{\sc ii}$\,\lambda4686$\fi}

\newcommand{\sii}{\ifmmode \left[{\rm S}\,\textsc{ii}\right] \else [S\,{\sc ii}]\fi}

\newcommand{\SII}{\ifmmode \left[{\rm S}\,\textsc{ii}\right]\,\lambda6717,6731 \else [S\,{\sc ii}]$\,\lambda6717,6731$\fi}

\newcommand{  \OIIIuv   }{\ifmmode {\rm O}\,\textsc{iii}\,\lambda1663 \else O\,\textsc{iii}\,$\lambda1663$\fi}
\newcommand{  \oiv      }{\ifmmode {\rm O}\,\textsc{iv}  \else O\,\textsc{iv}\fi}
\newcommand{  \OIVuv    }{\ifmmode {\rm O}\,\textsc{iv}\,\lambda1402  \else O\,\textsc{iv}\,$\lambda1402$\fi}
\newcommand{  \OIVIR    }{\ifmmode {\rm O}\,\textsc{iv}\,25.9\,\mu {\rm m} \else O\,\textsc{iv}\,$25.9\,\mu$m\fi}
\newcommand{  \ovi      }{\ifmmode {\rm O}\,\textsc{vi}   \else O\,\textsc{vi}\fi}
\newcommand{  \Ovi      }{\ifmmode {\rm O}\,\textsc{vi}\,\lambda1035 \else O\,\textsc{vi}\,$\lambda1035$\fi}
\newcommand{  \nei      }{\ifmmode {\rm Ne}\,\textsc{i}   \else Ne\,\textsc{i}\fi}
\newcommand{  \neii     }{\ifmmode {\rm Ne}\,\textsc{ii}  \else Ne\,\textsc{ii}\fi}
\newcommand{  \NeiiIR   }{\ifmmode {\rm Ne}\,\textsc{ii}\,12.8\,\mu {\rm m} \else Ne\,\textsc{ii}\,$12.8\,\mu$m\fi}
\newcommand{  \neiii    }{\ifmmode {\rm Ne}\,\textsc{iii} \else Ne\,\textsc{iii}\fi}
\newcommand{  \neiv     }{\ifmmode {\rm Ne}\,\textsc{iv}  \else Ne\,\textsc{iv}\fi}
\newcommand{  \nev      }{\ifmmode {\rm Ne}\,\textsc{v}   \else Ne\,\textsc{v}\fi}
\newcommand{  \NevIR    }{\ifmmode {\rm Ne}\,\textsc{v}\,24.3\,\mu {\rm m} \else Ne\,\textsc{v}\,$24.3\,\mu$m\fi}
\newcommand{  \nevi     }{\ifmmode {\rm Ne}\,\textsc{vi}  \else Ne\,\textsc{vi}\fi}
\newcommand{  \mgi      }{\ifmmode {\rm Mg}\,\textsc{i}   \else Mg\,\textsc{i}\fi}
\newcommand{  \mgii     }{\ifmmode {\rm Mg}\,\textsc{ii}  \else Mg\,\textsc{ii}\fi}
\newcommand{  \MgII     }{\ifmmode {\rm Mg}\,\textsc{ii}\,\lambda2798 \else Mg\,\textsc{ii}\,$\lambda2798$\fi}
\newcommand{  \siii     }{\ifmmode {\rm S}\,\textsc{iii} \else S\,\textsc{iii}\fi}
\newcommand{  \siv      }{\ifmmode {\rm S}\,\textsc{iv}  \else S\,\textsc{iv}\fi}
\newcommand{  \sili     }{\ifmmode {\rm Si}\,\textsc{i}   \else Si\,\textsc{i}\fi}
\newcommand{  \silii    }{\ifmmode {\rm Si}\,\textsc{ii}  \else Si\,\textsc{ii}\fi}
\newcommand{  \Siliv    }{\ifmmode {\rm Si}\,\textsc{iv}  \else Si\,\textsc{iv}\fi}
\newcommand{  \SilIVuv  }{\ifmmode {\rm Si}\,\textsc{iv}\,\lambda1400  \else Si\,\textsc{iv}\,$\lambda1400$\fi}

\newcommand{  \caii     }{\ifmmode {\rm Ca}\,\textsc{ii}   \else Ca\,\textsc{ii}\fi}
\newcommand{  \feii     }{\ifmmode {\rm Fe}\,\textsc{ii}  \else Fe\,\textsc{ii}\fi}
\newcommand{  \feiii    }{\ifmmode {\rm Fe}\,\textsc{iii} \else Fe\,\textsc{iii}\fi}

\newcommand{ \Lhb   }{\ifmmode L\left(\hb\right) \else $L\left(\hb\right)$\fi}
\newcommand{ \fwhb  }{\ifmmode {\rm FWHM}\left(\hb\right) \else FWHM(\hb)\fi}
\newcommand{ \Lha   }{\ifmmode L\left(\ha\right) \else $L\left(\ha\right)$\fi}
\newcommand{ \fwha  }{\ifmmode {\rm FWHM}\left(\ha\right) \else FWHM(\ha)\fi}
\newcommand{ \Lmg   }{\ifmmode L\left(\mgii\right) \else $L\left(\mgii\right)$\fi}
\newcommand{ \fwmg  }{\ifmmode {\rm FWHM}\left(\mgii\right) \else FWHM(\mgii)\fi}
\newcommand{ \Lciv  }{\ifmmode L\left(\civ\right) \else $L\left(\civ\right)$\fi}
\newcommand{ \fwciv }{\ifmmode {\rm FWHM}\left(\civ\right) \else FWHM(\civ)\fi}
\newcommand{ \fwhm  }{\ifmmode {\rm FWHM} \else FWHM\fi} 
\newcommand{ \voff  }{\ifmmode v_{\rm off} \else $v_{\rm off}$\fi} 

\newcommand{ \mumg  }{\ifmmode \mu\left(\mgii\right) \else $\mu\left(\mgii\right)$\fi}
\newcommand{ \fmg   }{\ifmmode f\left(\mgii\right) \else $f\left(\mgii\right)$\fi}
\newcommand{ \muciv }{\ifmmode \mu\left(\civ\right) \else $\mu\left(\civ\right)$\fi}
\newcommand{ \fciv  }{\ifmmode f\left(\civ\right) \else $f\left(\civ\right)$\fi}
\newcommand{  \auvo     }{\ifmmode \alpha_{\nu,{\rm UVO}} \else $\alpha_{\nu,{\rm UVO}}$\fi}
\newcommand{  \Ledd     }{\ifmmode L_{\rm Edd} \else $L_{\rm Edd}$\fi}
\newcommand{  \lamLlam  }{\ifmmode \lambda L_{\lambda} \else $\lambda L_{\lambda}$\fi}
\newcommand{  \lLl      }{\ifmmode \lambda L_{\lambda} \else $\lambda L_{\lambda}$\fi}
\newcommand{  \nuLnu    }{\ifmmode \nu L_{\nu} \else $\nu L_{\nu}$\fi}
\newcommand{  \nLn      }{\ifmmode \nu L_{\nu} \else $\nu L_{\nu}$\fi}
\newcommand{  \Luv      }{\ifmmode L_{1450} \else $L_{1450}$\fi}
\newcommand{  \Lop      }{\ifmmode L_{5100} \else $L_{5100}$\fi}
\newcommand{  \lLop     }{\ifmmode \log\left(\Lop/\ergs\right) \else $\log\left(\Lop/\ergs\right)$\fi}
\newcommand{  \Lthree   }{\ifmmode L_{3000} \else $L_{3000}$\fi}
\newcommand{  \lLthree  }{\ifmmode \log\left(\Lthree/\ergs\right) \else $\log\left(\Lthree/\ergs\right)$\fi}

\newcommand{\Fthree}{\ifmmode F_{3000} \else $F_{3000}$\fi}
\newcommand{\fuv}{\ifmmode f_{\lambda}\left(1450{\rm \AA}\right) \else $f_{\lambda}\left(1450 {\rm \AA}\right)$\fi}
\newcommand{\fthree}{\ifmmode f_{\lambda}\left(3000{\rm \AA}\right) \else $f_{\lambda}\left(3000{\rm \AA}\right)$\fi}
\newcommand{\fH}{\ifmmode f_{\lambda}\left(1.65\micron\right) \else
$f_{\lambda}\left(1.65\micron\right)$\fi}

\newcommand{\fbol}{\ifmmode f_{\rm bol} \else $f_{\rm bol}$\fi}
\newcommand{\fbolwv}{\ifmmode f_{\rm bol}\left(\lambda\right) \else $f_{\rm bol}\left(\lambda\right)$\fi}
\newcommand{\fbolopt}{\ifmmode f_{\rm bol}\left(5100{\rm \AA}\right) \else $f_{\rm bol}\left(5100{\rm \AA}\right)$\fi}
\newcommand{\fbolthree}{\ifmmode f_{\rm bol}\left(3000{\rm \AA}\right) \else $f_{\rm bol}\left(3000{\rm \AA}\right)$\fi}
\newcommand{\fboluv}{\ifmmode f_{\rm bol}\left(1450{\rm \AA}\right) \else $f_{\rm bol}\left(1450{\rm \AA}\right)$\fi}

\newcommand{  \mbh      }{\ifmmode M_{\rm BH} \else $M_{\rm BH}$\fi}
\newcommand{  \lmbh     }{\ifmmode \log\left(\mbh/\Msun\right) \else $\log\left(\mbh/\Msun\right)$\fi} 
\newcommand{  \lledd    }{\ifmmode L/L_{\rm Edd} \else $L/L_{\rm Edd}$\fi}
\newcommand{  \Lbol     }{\ifmmode L_{\rm bol} \else $L_{\rm bol}$\fi}
\newcommand{  \lbol     }{\ifmmode L_{\rm bol} \else $L_{\rm bol}$\fi}
\newcommand{  \lLbol    }{\ifmmode \log\left(\Lbol/\ergs\right) \else $\log\left(\Lbol/\ergs\right)$\fi} 
\newcommand{  \Lagn     }{\ifmmode L_{\rm AGN} \else $L_{\rm AGN}$\fi}
\newcommand{  \lagn     }{\ifmmode L_{\rm AGN} \else $L_{\rm AGN}$\fi}

\newcommand{  \tgrow     }{\ifmmode t_{\rm growth} \else $t_{\rm growth}$\fi}
\newcommand{  \tUni      }{\ifmmode t_{\rm Universe} \else $t_{\rm Universe}$\fi}

\newcommand{  \Mindot	}{\ifmmode \dot{M}_{\rm infall} \else $\dot{M}_{\rm infall}$\fi}
\newcommand{  \Mbhdot	}{\ifmmode \dot{M}_{\rm BH} \else $\dot{M}_{\rm BH}$\fi}
\newcommand{  \Maddot	}{\ifmmode \dot{M}_{\rm AD} \else $\dot{M}_{\rm AD}$\fi}

\newcommand{  \as	}{\ifmmode a_{\rm *} 		\else $a_{\rm *}$\fi}
\newcommand{  \avec	}{\ifmmode \vec{a}_{\rm *} 	\else $\vec{a}_{\rm *}$\fi}
\newcommand{  \re	}{\ifmmode \eta      	\else $\eta$\fi}

\newcommand{  \mseed    }{\ifmmode M_{\rm seed} \else $M_{\rm seed}$\fi}
\newcommand{  \mbul     }{\ifmmode M_{\rm Bulge} \else $M_{\rm Bulge}$\fi} 
\newcommand{  \mstar    }{\ifmmode M_{*} \else $M_{*}$\fi} 
\newcommand{  \mgal     }{\ifmmode M_{*} \else $M_{*}$\fi} 
\newcommand{  \mhost    }{\ifmmode M_{\rm Host} \else $M_{\rm Host}$\fi}
\newcommand{  \mm       }{\ifmmode M_{*}/M_{\rm BH} \else $M_{*}/M_{\rm BH}$\fi}
\newcommand{  \mmsmall  }{\ifmmode M_{\rm BH}/M_{*} \else $M_{\rm BH}/M_{*}$\fi}
\newcommand{  \mmlarge  }{\ifmmode M_{*}/M_{\rm BH} \else $M_{*}/M_{\rm BH}$\fi}
\newcommand{  \mmwp     }{\ifmmode \left(M_{*}/M_{\rm BH}\right) \else $\left(M_{*}/M_{\rm BH}\right)$\fi}
\newcommand{  \ml       }{\ifmmode M_{*}/L_{*} \else $M_{*}/L_{*}$\fi}
\newcommand{  \mlwp     }{\ifmmode \left(M_{*}/L\right) \else $\left(M_{*}/L\right)$\fi}
\newcommand{  \mlk      }{\ifmmode \left(M_{*}/L_{K}\right) \else $\left(M_{*}/L_{K}\right)$\fi}
\newcommand{  \sigs     }{\ifmmode \sigma_{*} \else $\sigma_{*}$\fi}
\newcommand{  \Reff     }{\ifmmode R_{\rm e} \else $R_{\rm e}$\fi}

\newcommand  {\RBLR}        {\hbox{$ {R_{\rm BLR}} $}}

\def \swiftxrt {{\em Swift}/XRT\ }

\def \swiftbat {{\em Swift}/BAT\ }
\def \swiftbatsh {{\em Swift}/BAT}

\def \chandrash {{\em Chandra}}

\def \xmmsh{{\em XMM-Newton}}

\def \suzakush{{\em Suzaku}}

\def\kmps{\hbox{$\km\s^{-1}\,$}}

\newcommand{\bj}{\ifmmode b_{\rm J} \else $b_{\rm J}$\fi}

\newcommand{\iab}{\ifmmode i_{\rm AB} \else $i_{\rm AB}$\fi}

\newcommand{\jab}{\ifmmode J_{\rm AB} \else $J_{\rm AB}$\fi}
\newcommand{\hab}{\ifmmode H_{\rm AB} \else $H_{\rm AB}$\fi}
\newcommand{\kab}{\ifmmode K_{\rm AB} \else $K_{\rm AB}$\fi}

\newcommand{\jveg}{\ifmmode J_{\rm Vega} \else $J_{\rm Vega}$\fi}
\newcommand{\hveg}{\ifmmode H_{\rm Vega} \else $H_{\rm Vega}$\fi}
\newcommand{\kveg}{\ifmmode K_{\rm Vega} \else $K_{\rm Vega}$\fi}

\def\arcsec{\hbox{$^{\prime\prime}$}}

\newcommand{  \Chisq    }{\ifmmode \chi^{2} \else $\chi^{2}$}
\newcommand{  \nelec    }{\ifmmode n_{e} \else $n_{e}$\fi}     % electron density
\newcommand{  \nh       }{\ifmmode n_{H} \else $n_{H}$\fi}     % hydrogen density
\newcommand{  \Ncol     }{\ifmmode N_{col} \else $N_{col}$\fi} % column density
\newcommand{  \NH       }{\ifmmode N_{H} \else $N_{\rm H}$\fi}     % column density

\def\arcsec{\hbox{$^{\prime\prime}$}}
\def\ion#1#2{#1$\;${\small\rm\@Roman{#2}}\relax}

\newcommand{\OIIIa}{\ifmmode \left[{\rm O}\,\textsc{iii}\right]\,\lambda4959 \else [O\,{\sc iii}]\,$\lambda4959$\fi}
\newcommand{\NIIa}{\ifmmode \left[{\rm N}\,\textsc{ii}\right]\,\lambda6548 \else [N\,{\sc ii}]\,$\lambda6548$\fi}
\newcommand{\SIIa}{\ifmmode \left[{\rm S}\,\textsc{ii}\right]\,\lambda6716 \else [S\,{\sc ii}]\,$\lambda6716$\fi}
\newcommand{\SIIb}{\ifmmode \left[{\rm S}\,\textsc{ii}\right]\,\lambda6732 \else [S\,{\sc ii}]\,$\lambda6731$\fi}
\newcommand{\NeVa}{\ifmmode \left[{\rm Ne}\,\textsc{v}\right]\,\lambda3346 \else [Ne\,{\sc v}]\,$\lambda3346$\fi}
\newcommand{\NeVb}{\ifmmode \left[{\rm Ne}\,\textsc{v}\right]\,\lambda3426 \else [Ne\,{\sc v}]\,$\lambda3426$\fi}
\newcommand{\NeIIIa}{\ifmmode \left[{\rm Ne}\,\textsc{iii}\right]\,\lambda3869 \else [Ne\,{\sc iii}]\,$\lambda3869$\fi}
\newcommand{\NeIIIb}{\ifmmode \left[{\rm Ne}\,\textsc{iii}\right]\,\lambda3968 \else [Ne\,{\sc iii}]\,$\lambda3968$\fi}
\newcommand{\Mgb}{\ifmmode \left{\rm Mg}\,\textsc{i}\right\,\lambda5175 \else Mg\,{\sc i}\,$\lambda5175$\fi}

\newcommand{\mgb}{\ifmmode \left{\rm Mg}\,\textsc{i}\right \else Mg\,{\sc i}\fi}

\newcommand{\Cahk}{\ifmmode \left[{\rm Ca H+K}\,\textsc{ii}\right\,\lambda3935,3968 \else Ca H+K$\,\lambda3935,3968$\fi}

\def\arcsec{{\mbox{$^{\prime \prime}$}}}

\def\ga{{\rm\thinspace gauss}}

\def\km{{\rm\thinspace km}}

\def\Lsun{\hbox{$\rm\thinspace L_{\odot}$}}

\def\pc{{\rm\thinspace pc}}

\def\s{{\rm\thinspace s}}

\newcommand{\Silivi}{\ifmmode \left[{\rm Si}\,\textsc{vi}\right] \else [Si\,{\sc vi}]\fi}
\newcommand{\Silix}{\ifmmode \left[{\rm Si}\,\textsc{x}\right] \else [Si\,{\sc x}]\fi}
\newcommand{\Silixi}{\ifmmode \left[{\rm Si}\,\textsc{xi}\right] \else [Si\,{\sc xi}]\fi}

\newcommand{\Caviii}{\ifmmode \left[{\rm Ca}\,\textsc{viii}\right] \else [Ca\,{\sc viii}]\fi}
\newcommand{\Sviii}{\ifmmode \left[{\rm S}\,\textsc{viii}\right] \else [S\,{\sc viii}]\fi}
\newcommand{\Six}{\ifmmode \left[{\rm S}\,\textsc{ix}\right] \else [S\,{\sc ix}]\fi}
\newcommand{\Sxi}{\ifmmode \left[{\rm S}\,\textsc{xi}\right] \else [S\,{\sc xi}]\fi}

\newcommand{\Fexiii}{\ifmmode \left[{\rm Fe}\,\textsc{xiii}\right] \else [Fe\,{\sc xiii}]\fi}
\newcommand{\Alix}{\ifmmode \left[{\rm Al}\,\textsc{ix}\right] \else [Al\,{\sc ix}]\fi}

\newcommand{\feiif}{\ifmmode \left[{\rm Fe}\,\textsc{ii}\right] \else [Fe\,{\sc ii}]\fi}

\newcommand{\siiif}{\ifmmode \left[{\rm S}\,\textsc{iii}\right] \else [S\,{\sc iii}]\fi}

\newcommand{\pii}{\ifmmode \left[{\rm P}\,\textsc{ii}\right] \else [P\,{\sc ii}]\fi}
\newcommand{\NiII}{\ifmmode \left[{\rm Ni}\,\textsc{II}\right] \else [Ni\,{\sc II}]\fi}

\newcommand{\NI}{\ifmmode \left[{\rm N}\,\textsc{I}\right] \else [N\,{\sc i}]\fi}
\newcommand{\CI}{\ifmmode \left[{\rm C}\,\textsc{I}\right] \else [C\,{\sc i}]\fi}

\def\apjl{ApJ}%
\def\apjs{ApJS}%
\def\ao{Appl.~Opt.}%
\def\aap{A\&A}%
\title[BASS IV: NIR Spectroscopy]{BAT AGN Spectroscopic Survey - IV: Near-Infrared Coronal Lines, Hidden Broad Lines, and Correlation with Hard X-ray Emission}

\author[Lamperti et al.]
{Isabella Lamperti$^{1}$\thanks{E-mail: lamperis@phys.ethz.ch},
Michael Koss$^{1,2}$\thanks{Ambizione fellow},
Benny Trakhtenbrot$^{1}$\thanks{Zwicky fellow},
Kevin Schawinski$^{1}$, 
\newauthor Claudio Ricci$^{3}$,
Kyuseok Oh$^{1}$, 
Hermine Landt$^{4}$,
Rog{\'e}rio Riffel$^{5}$,
\newauthor Alberto Rodr{\'\i}guez-Ardila$^{6}$,
Neil Gehrels$^{7}$,
Fiona Harrison$^{8}$,
Nicola Masetti$^{9,10}$,
\newauthor Richard Mushotzky$^{11}$,
Ezequiel Treister$^{3}$,
Yoshihiro Ueda$^{12}$,
Sylvain Veilleux$^{11}$\\
$^{1}$Institute for Astronomy, Department of Physics, ETH Zurich, Wolfgang-Pauli-Strasse 27, CH-8093 Z\"{u}rich, Switzerland\\
$^{2}$Institute for Astronomy, University of Hawaii, 2680 Woodlawn Drive, Honolulu, HI 96822, USA\\
$^{3}$Instituto de Astrof\'{\i}sica, Facultad de F\'{\i}sica, Pontificia Universidad Cat\'olica de Chile, Casilla 306, Santiago 22, Chile\\
$^{4}$Centre for Extragalactic Astronomy, Department of Physics, Durham University, South Road, Durham, DH1 3LE, UK \\
$^{5}$Departamento de Astronomia, Universidade Federal do Rio Grande do Sul. Av. Bento Gon\c{c}alves 9500 Porto Alegre, RS, Brazil\\
$^{6}$Laborat\'{\o}rio Nacional de Astrof\'{\i}sica
Rua dos Estados Unidos 154, Bairro das Na\c{c}$\tilde{o}$es CEP 37504-364, Itajub\'a, MG, Brazil\\
$^{7}$NASA Goddard Space Flight Center, Greenbelt, MD 20771, USA\\
$^{8}$Cahill Center for Astronomy and Astrophysics, California Institute of Technology, Pasadena, CA 91125, USA\\
$^{9}$INAF -- Istituto di Astrofisica Spaziale e Fisica Cosmica di Bologna, via Gobetti 101, 40129 Bologna, Italy\\
$^{10}$Departamento de Ciencias F\'{\i}sicas, Universidad Andr\'es Bello, Fern\'andez Concha 700, Las Condes, Santiago, Chile\\
$^{11}$Department of Astronomy and Joint Space-Science Institute, University of Maryland, College Park, MD 20742, USA\\
$^{12}$Department of Astronomy, Kyoto University, Kyoto 606-8502, Japan}

\date{Accepted to MNRAS}
\pubyear{2016}

\begin{document}
\label{firstpage}
\pagerange{\pageref{firstpage}--\pageref{lastpage}}
\maketitle

% Number of AGN 
\newcommand{  \Ntot  }{102}
\newcommand{  \Ntotwithdup	}{122}
\newcommand{  \NPBC	}{6}
\newcommand{  \NIRTF  }{48}
\newcommand{  \NFlamingos }{7}
\newcommand{  \NLandtIRTF  }{14}
\newcommand{  \NLandtGNIRS  }{3}
\newcommand{  \NRiffel  }{15}
\newcommand{  \NRodriguez  }{1}
\newcommand{  \NMason  }{13}

\newcommand{  \NBASSmatch  }{94}

\newcommand{  \NSyone  }{69}
\newcommand{  \NSytwo  }{33}

\newcommand{ \Pab } {Pa$\beta$}
\newcommand{ \Paa } {Pa$\alpha$}

\newcommand{ \MBH } {$M_{\rm BH}$}
\newcommand{ \sig } {$\sigma_*$}
\newcommand{ \spextool }{{\tt Spextool}}

\newcommand {\nhunit} {cm$^{-2}$}

%In total we could measure \MBH\ for 69 out of 102 AGN.
\begin{abstract}
We provide a comprehensive census of the near-Infrared (NIR, 0.8-2.4 $\mu$m) spectroscopic properties of \Ntot\ nearby ($z < 0.075$) active galactic nuclei (AGN), selected in the hard X-ray band (14-195 keV) from the \textit{Swift}-Burst Alert Telescope (BAT) survey.  With the launch of the James Webb Space Telescope this regime is of increasing importance for dusty and obscured AGN surveys.   We measure black hole masses in 68\% (69/102) of the sample using broad emission lines (34/102) and/or the velocity dispersion of the Ca II triplet or the CO band-heads (46/102).  We find that emission line diagnostics in the NIR are ineffective at identifying bright, nearby AGN galaxies because (\feiif\ 1.257$\mu$m/Pa$\beta$ and H$_2$ 2.12$\mu$m/Br$\gamma$) identify only 25$\%$ (25/102) as AGN with significant overlap with star forming galaxies and only 20\% of Seyfert 2 have detected coronal lines (6/30).  We measure the coronal line emission in Seyfert 2 to be weaker than in Seyfert 1 of the same bolometric luminosity suggesting obscuration by the nuclear torus. We find that the correlation between the hard X-ray and the \Silivi\ coronal line luminosity is significantly better than with the \OIII\ luminosity.   Finally, we find 3/29 galaxies (10\%) that are optically classified as Seyfert 2 show broad emission lines in the NIR.  These AGN have the lowest levels of obscuration among the Seyfert 2s in our sample ($\log N_{\rm H} < 22.43$ cm$^{-2}$), and all show signs of galaxy-scale interactions or mergers suggesting that the optical broad emission lines are obscured by host galaxy dust.

\end{abstract}


%%%%%%%%%%%%%%%%%%%%%%%%%%%%%%%%%%%%%%%%%%%%%%%%%%

%%%%%%%%%%%%%%%%% BODY OF PAPER %%%%%%%%%%%%%%%%%%

\section{Introduction}

The near-infrared (NIR) spectral regime (0.8-2.4 \mic) provides numerous emission lines for studies of active galactic nuclei (AGN) and has the advantage to be ten time less obscured than the optical \citep{Veilleux2002}.  
For example, the vibrational and rotational modes of H$_2$
can be excited by UV fluorescence  \citep{Black1987,Rodriguez2004,Rodriguez2005,Riffel2013}, shock heating \citep{Hollenbach1989}, and heating by X-rays, where hard X-ray photons penetrate into molecular clouds and heat the molecular gas \citep{Maloney1996}. 
The \feiif\ emission in AGN can be produced by shocks from the radio jets or X-ray heating, while in star-forming galaxies (SFGs) the \feiif\ emission is produced by photoionization or supernovae shocks \citep{Rodriguez2004}.

NIR emission line diagnostics \citep[e.g.][]{Larkin1998,Rodriguez2004, Riffel2013, Colina2015} are based on the close relation between the line ratios \feiif\ 1.257$\mu$m/Pa$\beta$ and H$_2$ 2.12$\mu$m/Br$\gamma$, and the nuclear black hole activity.
In principle, the low levels of dust attenuation in the NIR imply that such line diagnostics can be used even among highly reddened objects. 
However, several studies \cite[e.g.,][]{Dale2004,Martins2013} found that this diagnostic is not effective in separating SFGs from AGN.

The NIR regime also includes the Hydrogen Pa$\alpha$ and Pa$\beta$ emission lines, which by virtue of being less obscured may at times ($\sim30\%$ of sources) present a `hidden' broad line region (BLR), in galaxies with narrow optical \Hbeta\ and/or \Halpha\ lines \citep[e.g.,][]{Veilleux1997, Onori2014, Smith2014, LaFranca2015}.  \cite{Onori2016} recently studied a sample of 41 obscured (Sy2) and intermediate class AGN (Sy1.8$-$1.9) and found broad Pa$\alpha$, Pa$\beta$, or \hei\ lines in 32\% (13/41) of sources.
The study of \cite{Landt2008} compared the width of the Paschen and Balmer lines (in terms of FWHM) and found that H$\beta$ is generally broader than Pa$\beta$. 
One possible reason for this trend is the presence of the H$\beta$ ``red shelf''  \citep[e.g.,][]{DeRobertis1985,Marziani1996}, which originates from  Fe II multiplets \citep{Veron2002}. 
Another possibility is that the broad Balmer lines originate from a region closer to the black hole than the Paschen lines \citep{Kim2010}.

The NIR broad Paschen emission lines can be used to derive black hole mass (\MBH) estimates, based on their luminosities and widths, following a similar approach to that based on the optical Balmer lines \citep[e.g.][]{Kim2010,Landt2011b,Landt2013}. 
Alternatively, the size of the BLR (\RBLR), and therefore \MBH, can also be estimated using the 1 \mic\ continuum luminosity \citep{Landt2011a} or the hard-band X-ray luminosity \cite[e.g.,][]{LaFranca2015}, again mimicking similar methods to those put forward for the Balmer lines \citep{Greene2010}. 
For obscured AGN, the spectral region surrounding the \caii\ triplet of absorption features \citep[0.845 - 0.895 \mic; e.g.,][]{Rothberg2010, Rothberg2013} and CO bandheads in the $H$ and $K$ band \citep[e.g.,][]{Dasyra2006a, Dasyra2006b, Dasyra2007, Kang2013,Riffel2015} are useful regions to measure the stellar velocity dispersion (\sig) to infer \MBH\ following the $\mbh-\sigs$ relation \citep{Ferrarese2000,Tremaine2002,Mcconnell2013,Kormendy2013}.

Moreover, the NIR regime is important because it includes several high ionization ($>$100 eV) coronal lines (CLs) generated through photoionization by the hard UV and soft X-ray continuum produced by AGN or by shocks \citep[e.g.,][]{Rodriguez2002, Rodriguez2011}. CLs tend to be broader than low-ionization emission lines emitted in the NLR, but narrower than lines in the BLR (400 \kmps $<\ $ FWHM $<\ $ 1000 \kmps). This suggests that they are produced in a region between the BLR and the NLR \citep{Rodriguez2006}.  \cite{Mueller2011} investigated the spatial distribution of CLs and found that the size of the coronal line region (CLR) only extends to a distances of $<$ 200 pc from the center of the galaxy \citep{Mazzalay2010, Mueller2011}.\\   

Recently, it has been found that high-ionization optical emission lines, such as \OIII, are only weakly correlated with AGN X-ray luminosity or with bolometric luminosity \citep{Berney2015,Ueda2015}. 
The considerable intrinsic scatter in this relation ($0.6$ dex) may be related to physical properties of the NLR such as covering factor of the NLR \citep[e.g.,][]{Netzer1993}, density dependence of \oiii\ \citep{Crenshaw2005}, differing ionization parameter \citep{Baskin2005} and SED shape changes with luminosity \citep{Netzer2006}.  Other possible reasons are AGN variability \citep{Schawinski2015}, dust-obscuration, or contamination from star formation. Coronal lines in the NIR, by virtue of the higher ionization potentials and the minor sensitivity to dust obscuration, provide an additional method to study this correlation.\\

In this work, we aim to study the NIR properties of one of the largest samples of nearby AGN (\Ntot) from the \swiftbat survey selected at 14--195 keV \citep{Baumgartner2013} as part of the BAT AGN Spectroscopic Survey (BASS). This sample, by virtue of its selection, is nearly unbiased to obscuration up to Compton-thick levels \citep{Koss2016}. The first BASS paper (Koss et al., submitted) detailed the optical spectroscopic data and measurements.
The second BASS paper \citep{Berney2015} studied the large scatter in X-rays and high ionization emission lines like \oiii.  Additionally, all of the AGN have been analyzed using X-ray observations including the best available soft X-ray data in the 0.3$-$10\,keV band from \xmmsh, \chandrash, \suzakush, or \swiftxrt and the 14--195\,keV band from \swiftbatsh\ which provide measurements of the obscuring column density \citep[\NH;][Ricci et al., submitted]{Ricci2015}.    Therefore, we are able to compare the NIR properties of these AGN with their optical and X-ray characteristics, to better understand AGN variability and obscuration.  Throughout this work, we use a cosmological model with $\Omega_{\Lambda}=0.7$, $\Omega_{\rm M}=0.3$, and $H_{0}=70$\,\kms\,Mpc$^{-1}$ to determine distances.
However, for the most nearby sources in our sample ($z<0.01$), we use the mean of the redshift independent distance in Mpc from the NASA/IPAC Extragalactic Database (NED), whenever available.\\

\section{Data and Reduction}
\label{sec:samples_data}

\subsection{Sample}
The \swiftbat observatory carried out an all-sky survey in the ultra-hard X-ray range ($>10$\,keV) that, as of the first 70 months of operation, has identified 1210 objects \citep{Baumgartner2013} of which 836 are AGN.  The majority ($\approx90\%$) of BAT-detected AGN are relatively nearby ($z<0.2$).  

\begin{figure}
\centering
\subfigure{\includegraphics[width=0.49\textwidth]
{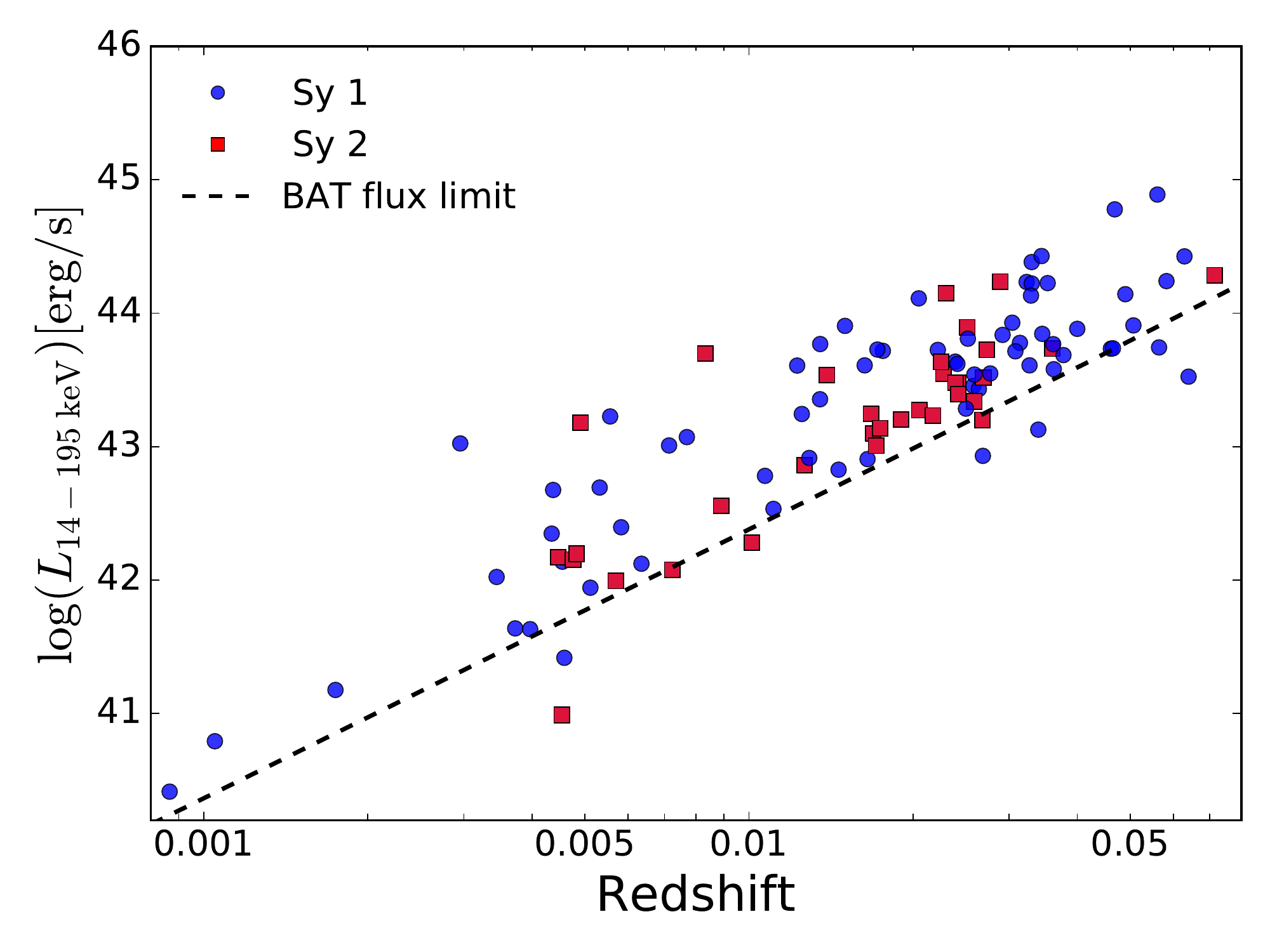}}
\caption{Distribution of the hard X-ray luminosity of the BAT AGN with respect to the redshift. The redshift is taken from the NASA/IPAC Extragalactic Database (NED). The dashed line shows the flux limit of the BAT 70 month survey over 90$\%$ of the sky ($1.34\times 10^{-11}$ erg cm$^2$ s$^{-1}$).}
\label{Lum_vs_redshift}
\end{figure}

We limited our NIR spectra sample to \Ntot\ nearby AGN with redshift $z<0.075$ to ensure high quality observations for both Seyfert 1 and Seyfert 2.  In our sample there are \NSyone\ Seyfert 1 and \NSytwo\ Seyfert 2. A full list of the AGN can be found in Tables \ref{tab:Table_IRTF}, \ref{tab:Flamingos}, and \ref{tab:arch_obs}.  Figure \ref{Lum_vs_redshift} shows the distribution of the hard X-ray luminosity of the AGN in our sample as a function of redshift.  Most of the AGN (96) are listed in the \swiftbat 70 month catalog \citep{Baumgartner2013} with an additional 6 AGN from the Palermo 100 month BAT catalog \citep{Cusumano2010} and the upcoming \swiftbat 104 month catalog (Oh et al., in prep.).  
 
The goal of our survey was to use the largest available NIR spectroscopic sample of \swiftbat sources using dedicated observations and published data. More than half (55\%, 56/102) of the AGN were observed as part of a targeted campaigns using the SpeX spectrograph \citep{Rayner2003} at the NASA 3m Infrared Telescope Facility (IRTF; 49 sources)  or with the Florida Multi-object Imaging Near-IR Grism Observational Spectrometer \citep[FLAMINGOS;][]{Elston1998} on the Kitt Peak 4m telescope (7 sources). The targets were selected at random based on visibility from the northern sky. Additionally, we included spectra from available archival observations.  We used \NRiffel\ spectra from \cite{Riffel2006}, \NLandtIRTF\ spectra from \cite{Landt2008}, two spectra from \cite{Rodriguez2002} and one spectrum from \cite{Riffel2013} observed with IRTF.  We used three spectra observed by \cite{Landt2013}  and \NMason\ publicly available spectra from \citet{Mason2015} observed with the Gemini Near-Infrared Spectrograph \citep[GNIRS;][]{Elias2006} on the 8.1 m Gemini North telescope.  We note there is some bias towards observing more Seyfert 1 in our sample compared to a blind survey of BAT AGN ($\approx50\%$ Seyfert 1) because some archival studies focused on Seyfert 1 \citep[e.g.,][]{Landt2008}.

We used the \oiii\ redshift, optical emission line measurements and classification in Seyfert 1 and Seyfert 2 from Koss et al. (submitted).  Optical counterparts of BAT AGN and the 14-195 keV measurements were based on \cite{Baumgartner2013}. We also used the \NH\ and 2-10 keV flux measurements from \cite[Ricci et al., submitted]{Ricci2015}.
The optical observations were not taken simultaneously to the NIR observations but come from separate targeted campaigns or public archives.

\begin{table*}
\centering
\caption{Spectra from the IRTF observations}
%\begin{tabular}{rlrlrrr}
\begin{tabular}{rlcccccccc}
\hline
\multicolumn{1}{c}{ID\tablenotemark{a}} & Counterpart Name  & \multicolumn{1}{c}{ Redshift } & Slit size & Slit size & \multicolumn{1}{c}{Date} & \multicolumn{1}{c}{ Airmass } & Exp. time & Type\tablenotemark{b} & H\tablenotemark{c}\\ 
 &  &  & [''] & [kpc]  & dd.mm.yy & & [s] & & (mag)\\ \hline

316 & IRAS 05589+2828 & 0.033 & 0.8 & 0.54 & 12.04.11 & 1.23 & 2160 & 1.2 & - \\ 
399 & 2MASX J07595347+2323241 & 0.0292 & 0.8 & 0.48 & 14.03.12 & 1.01 & 2160 & 1.9 & 10.3 \\ 
434 & MCG +11-11-032 & 0.036 & 0.8 & 0.59 & 14.03.12 & 1.42 & 2160 & 2 & 11.4 \\ 
439 & Mrk 18 & 0.0111 & 0.8 & 0.18 & 12.04.11 & 1.33 & 2520 & 1.9 & 10.5 \\ 
451 & IC 2461 & 0.0075 & 0.8 & 0.13 & 14.03.12 & 1.11 & 1080 & 2 & 10.1 \\ 
515 & MCG +06-24-008 & 0.0259 & 0.8 & 0.43 & 19.03.12 & 1.12 & 2520 & 2 & 10.8 \\ 
517 & UGC 05881 & 0.0205 & 0.8 & 0.34 & 12.04.11 & 1.01 & 1800 & 2 & 10.9 \\ 
533 & NGC 3588 NED01 & 0.0268 & 0.8 & 0.44 & 12.04.11 & 1.02 & 1620 & 2 & 10.6 \\ 
548 & NGC 3718 & 0.0033 & 0.8 & 0.05 & 11.06.10 & 1.24 & 4140 & 1.9 & 7.9 \\ 
586 & ARK 347 & 0.0224 & 0.8 & 0.37 & 19.03.12 & 1.17 & 2520 & 2 & 10.7 \\ 
588 & UGC 7064 & 0.025 & 0.8 & 0.41 & 19.03.12 & 1.58 & 2340 & 1.9 & 10.3 \\ 
590 & NGC 4102 & 0.0028 & 0.8 & 0.05 & 11.06.10 & 1.26 & 1200 & 2 & 7.8 \\ 
592 & Mrk 198 & 0.0242 & 0.8 & 0.40 & 19.03.12 & 1.16 & 2520 & 2 & 11.2 \\ 
593 & NGC 4138 & 0.003 & 0.8 & 0.05 & 11.06.10 & 1.24 & 3240 & 2 & 8.3 \\ 
635 & KUG 1238+278A & 0.0565 & 0.8 & 0.93 & 02.03.12 & 1.07 & 3600 & 1.9 & 11.8 \\ 
638 & NGC 4686 & 0.0167 & 0.8 & 0.28 & 12.04.11 & 1.22 & 3240 & 2 & 9.1 \\ 
659 & NGC 4992 & 0.0251 & 0.8 & 0.42 & 12.04.11 & 1.04 & 3960 & 2 & 10.3 \\ 
679 & NGC 5231 & 0.0218 & 0.8 & 0.36 & 12.06.10 & 1.07 & 1800 & 2 & 10.4 \\ 
682 & NGC 5252 & 0.023 & 0.8 & 0.38 & 12.06.10 & 1.12 & 1800 & 2 & 9.9 \\ 
686 & NGC 5273 & 0.0036 & 0.8 & 0.06 & 12.06.10 & 1.19 & 1800 & 1.5 & 8.8 \\ 
687 & CGCG 102-048 & 0.0269 & 0.8 & 0.44 & 19.03.12 & 1.35 & 2160 & 2 & 11.3 \\ 
695 & UM614 & 0.0327 & 0.8 & 0.54 & 12.04.11 & 1.53 & 1200 & 1.5 & 11.6 \\ 
712 & NGC 5506 & 0.0062 & 0.8 & 0.10 & 11.06.10 & 1.21 & 1200 & 1.9 & 8.2 \\ 
723 & NGC 5610 & 0.0169 & 0.8 & 0.28 & 12.04.11 & 1.02 & 2520 & 2 & 10.1 \\ 
734 & NGC 5683 & 0.0362 & 0.8 & 0.60 & 13.03.12 & 1.19 & 2160 & 1.2 & 11.8 \\ 
737 & 2MASX J14391186+1415215 & 0.0714 & 0.8 & 1.17 & 12.04.11 & 1.08 & 3420 & 2 & 13.1 \\ 
738 & Mrk 477 & 0.0377 & 0.8 & 0.62 & 12.04.11 & 1.49 & 1440 & 1.9 & 12.1 \\ 
748 & 2MASX J14530794+2554327 & 0.049 & 0.8 & 0.80 & 11.06.10 & 1.10 & 1800 & 1 & 12.5 \\ 
754 & Mrk 1392 & 0.0361 & 0.8 & 0.60 & 12.04.11 & 1.56 & 1200 & 1.5 & 11.1 \\ 
783 & NGC 5995 & 0.0252 & 0.8 & 0.42 & 16.06.11 & 1.24 & 1800& 1.9 & 9.5 \\ 
883 & 2MASX J17232321+3630097 & 0.04 & 0.8 & 0.66 & 16.06.11 & 1.04 & 960 & 1.5 & 11.4 \\ 
1041 & 2MASS J19334715+3254259 & 0.0583 & 0.8 & 0.96 & 11.06.10 & 1.08 & 1680 & 1.2 & - \\ 
1042 & 2MASX J19373299-0613046 & 0.0107 & 0.8 & 0.18 & 14.09.11 & 1.16 & 720 & 1.5 & 9.7 \\ 
1046 & NGC 6814 & 0.0052 & 0.8 & 0.09 & 14.09.11 & 1.19 & 1200 & 1.5 & 7.8 \\ 
1077 & MCG +04-48-002 & 0.0139 & 0.8 & 0.23 & 12.06.10 & 1.02 & 1800 & 2 & 9.9 \\ 
1099 & 2MASX J21090996-0940147 & 0.0277 & 0.8 & 0.46 & 14.09.11 & 1.23 & 1800 & 1.2 & 10.6 \\ 
1106 & 2MASX J21192912+3332566 & 0.0507 & 0.8 & 0.83 & 15.09.11 & 1.04 & 1800 & 1.5 & 11.5 \\ 
1117 & 2MASX J21355399+4728217 & 0.0259 & 0.8 & 0.43 & 11.06.10 & 1.32 & 1800 & 1.5 & 11.5 \\ 
1133 & Mrk 520 & 0.0273 & 0.8 & 0.45 & 12.06.10 & 1.08 & 1800 & 2 & 10.8 \\ 
1157 & NGC 7314 & 0.0048 & 0.8 & 0.08 & 15.09.11 & 1.43 & 1440 & 1.9 & 8.3 \\ 
1158 & NGC 7319 & 0.0225 & 0.8 & 0.37 & 11.06.10 & 1.28 & 1800 & 2 & 10.4 \\ 
1161 & Mrk 915 & 0.0241 & 0.8 & 0.40 & 15.09.11 & 1.2 & 1440 & 1.9 & 10.4 \\ 
1178 & KAZ 320 & 0.0345 & 0.8 & 0.57 & 12.06.10 & 1.08 & 1800 & 1.5 & 12.1 \\ 
1183 & Mrk 926 & 0.0469 & 0.8 & 0.77 & 11.06.10 & 1.42 & 1800 & 1.5 & 10.8 \\ 
1196 & 2MASX J23272195+1524375 & 0.0465 & 0.8 & 0.76 & 15.09.11 & 1 & 1440 & 1.9 & 11.1 \\ 
1202 & UGC 12741 & 0.0174 & 0.8 & 0.29 & 11.06.10 & 1.15 & 1200 & 2 & 10.8 \\ 
1472 & CGCG 198-020 & 0.0269 & 0.8 & 0.44 & 12.06.10 & 1.13 & 1800 & 1.5 & - \\ 
1570 & NGC 5940 & 0.0339 & 0.8 & 0.56 & 12.06.10 & 1.31 & 1800 & 1 & - \\ 
\hline
\end{tabular}
\label{tab:Table_IRTF}

\tablenotetext{a}{\swiftbat 70-month hard X-ray survey ID (http://swift.gsfc.nasa.gov/results/bs70mon/).}
\tablenotetext{b}{AGN classification following \citet{Osterbrock1981}.}
\tablenotetext{c}{H 20mag/sq." isophotal fiducial elliptical aperture magnitudes from the 2MASS extended source catalog.}

\end{table*}

\begin{table*}
\centering
\caption{Spectra from the Flamingos observations}
\begin{tabular}{rlcccccccc}
\hline

\multicolumn{1}{l}{ID\tablenotemark{a}} & Counterpart Name  & \multicolumn{1}{l}{Redshift } & Slit size  & Slit size & \multicolumn{1}{l}{Date} & \multicolumn{1}{l}{ Airmass } & Exp. time & Type\tablenotemark{b} & H\tablenotemark{c}\\ 
 &  &  & [''] & \multicolumn{1}{c}{[kpc]}  & \multicolumn{1}{c}{dd.mm.yy} & & [s] & & (mag) \\ \hline

310 & MCG +08-11-011 & 0.0205 & 1.5 & 0.64 & 13.12.08 & 1.03 & 1440 & 1.5 & 9.3 \\ 
325 & Mrk 3 & 0.0135 & 1.5 & 0.42 & 13.12.08 & 1.29 & 1440 & 1.9 & 9.1 \\ 
766 & NGC 5899 & 0.0086 & 1.5 & 0.27 & 07.07.09 & 1.2 & 1440 & 2 & 8.7 \\ 
772 & MCG -01-40-001 & 0.0227 & 1.5 & 0.70 & 07.07.09 & 1.37 & 1440 & 1.9 & 10.4 \\ 
804 & CGCG 367-009 & 0.0274 & 1.5 & 0.85 & 06.07.09 & 1.6 & 1440 & 2 & 11.3 \\ 
1040 & 2MASX J19301380+3410495 & 0.0629 & 1.5 & 1.93 & 06.07.09 & 1.46 & 1440 & 1.5 & - \\ 
1177 & UGC 12282 & 0.017 & 1.5 & 0.53 & 07.07.09 & 1.03 & 1440 & 2 & 9.1 \\ 
\hline
\end{tabular}
\label{tab:Flamingos}

\tablenotetext{a}{\swiftbat 70-month hard X-ray survey ID (http://swift.gsfc.nasa.gov/results/bs70mon/).}
\tablenotetext{b}{AGN classification following \citet{Osterbrock1981}.}
\tablenotetext{c}{H 20mag/sq." isophotal fiducial elliptical aperture magnitudes from the 2MASS extended source catalog.}
\end{table*}

% Archival observations

\begin{table*}
\centering
\caption{Archival Observations}
\begin{tabular}{rlcccccl}
\hline

\multicolumn{1}{l}{ID\tablenotemark{a}} & Counterpart Name  & \multicolumn{1}{l}{Redshift } & Slit size & Slit size & Instrument & Type\tablenotemark{b} & Reference \\ 
 &  &  & [''] & \multicolumn{1}{c}{[kpc]}  &  & & \\ \hline
6 & Mrk 335 & 0.0258 & 0.8 & 0.43 & SpeX & 1.2 & Landt+2008 \\ 
33 & NGC 262 & 0.015 & 0.8 & 0.25 & SpeX & 1.9 & Riffel+2006 \\ 
116 & Mrk 590 & 0.0264 & 0.8 & 0.44 & SpeX & 1.5 & Landt+2008 \\ 
130 & Mrk 1044 & 0.0165 & 0.8 & 0.27 & SpeX & 1 & Rodr{\'\i}guez-Ardila+2002 \\ 
140 & NGC 1052  & 0.005 & 0.3 & 0.03 & GNIRS & 1.9 & Mason+2015 \\ 
157 & NGC 1144 & 0.0289 & 0.8 & 0.48 & SpeX & 2 & Riffel+2006 \\ 
173 & NGC 1275 & 0.0176 & 0.8 & 0.29 & SpeX & 1.5 & Riffel+2006 \\ 
226 & 3C 120 & 0.033 & 0.3 & 0.20 & GNIRS & 1.5 & Landt+2013 \\ 
266 & Ark 120 & 0.0323 & 0.8 & 0.53 & SpeX & 1 & Landt+2008 \\ 
269 & MCG-5-13-17 & 0.0125 & 0.8 & 0.21 & SpeX & 1.5 & Riffel+2006 \\ 
308 & NGC 2110 & 0.0078 & 0.8 & 0.13 & SpeX & 2 & Riffel+2006 \\ 
382 & Mrk 79 & 0.0222 & 0.8 & 0.37 & SpeX & 1.5 & Landt+2008 \\ 
404 & Mrk 1210 & 0.0135 & 0.8 & 0.22 & SpeX & 1.9 & Riffel+2006 \\ 
436 & NGC 2655  & 0.0047 & 0.3 & 0.03 & GNIRS & 2 & Mason+2015 \\ 
458 & Mrk 110 & 0.0353 & 0.8 & 0.58 & SpeX & 1.5 & Landt+2008 \\ 
477 & NGC 3031  & -0.0001 & 0.3 & 0.01 & GNIRS & 1.9 & Mason+2015 \\ 
484 & NGC 3079  & 0.0037 & 0.3 & 0.02 & GNIRS & 2 & Mason+2015 \\ 
497 & NGC 3227 & 0.0039 & 0.8 & 0.06 & SpeX & 1.5 & Landt+2008 \\ 
530 & NGC 3516  & 0.0088 & 0.675 & 0.12 & GNIRS & 1.2 & Landt+2013 \\ 
567 & H1143-192 & 0.0329 & 0.8 & 0.54 & SpeX & 1.2 & Riffel+2006 \\ 
579 & NGC 3998  & 0.0035 & 0.3 & 0.02 & GNIRS & 1.9 & Mason+2015 \\ 
585 & NGC 4051 & 0.0023 & 0.8 & 0.04 & SpeX & 1.5 & Riffel+2006 \\ 
595 & NGC 4151 & 0.0033 & 0.8 & 0.05 & SpeX & 1.5 & Landt+2008 \\ 
607 & NGC 4235 & 0.008 & 0.3 & 0.05 & GNIRS & 1.2 & Mason+2015 \\ 
608 & Mrk 766 & 0.0129 & 0.8 & 0.21 & SpeX & 1.5 & Riffel+2006 \\ 
609 & NGC 4258 & 0.00149 & 0.3 & 0.01 & GNIRS & 1.9 & Mason+2015 \\ 
615 & NGC 4388  & 0.0084 & 0.3 & 0.05 & GNIRS & 2 & Mason+2015 \\ 
616 & NGC 4395  & 0.0011 & 0.3 & 0.01 & GNIRS & 2 & Mason+2015 \\ 
631 & NGC 4593 & 0.009 & 0.8 & 0.15 & SpeX & 1 & Landt+2008 \\ 
641 & NGC 4748 & 0.0146 & 0.8 & 0.24 & SpeX & 1.5 & Riffel+2006 \\ 
665 & NGC 5033  & 0.0029 & 0.3 & 0.02 & GNIRS & 1.9 & Mason+2015 \\ 
697 & Mrk 279 & 0.0304 & 0.8 & 0.50 & SpeX & 1.5 & Riffel+2006 \\ 
717 & NGC 5548 & 0.0172 & 0.8 & 0.28 & SpeX & 1.5 & Landt+2008 \\ 
730 & Mrk 684 & 0.0461 & 0.8 & 0.76 & SpeX & 1 & Riffel+2006 \\ 
735 & Mrk 817 & 0.0314 & 0.8 & 0.52 & SpeX & 1.2 & Landt+2008 \\ 
739 & NGC 5728 & 0.0093 & 0.8 & 0.15 & SpeX & 1.9 & Riffel+2006 \\ 
774 & Mrk 290 & 0.0308 & 0.8 & 0.51 & SpeX & 1.5 & Landt+2008 \\ 
876 & Arp 102B & 0.0242 & 0.8 & 0.40 & SpeX & 1.9 & Riffel+2006 \\ 
994 & 3C 390.3  & 0.0561 & 0.675 & 0.78 & GNIRS & 1.5 & Landt+2013 \\ 
1090 & Mrk 509 & 0.0344 & 0.8 & 0.57 & SpeX & 1.2 & Landt+2008 \\ 
1180 & NGC 7465 & 0.0066 & 0.8 & 0.11 & SpeX & 1.9 & Riffel+2006 \\ 
1182 & NGC 7469 & 0.0163 & 0.8 & 0.27 & SpeX & 1.5 & Landt+2008 \\ 
1198 & NGC 7682 & 0.0171 & 0.8 & 0.28 & SpeX & 2 & Riffel+2013 \\ 
1287 & NGC 2273  & 0.006138 & 0.3 & 0.04 & GNIRS & 2 & Mason+2015 \\ 
1322 & PG 0844+349  & 0.064 & 0.8 & 1.05 & SpeX & 1 & Landt+2008 \\ 
1348 & NGC 3147  & 0.009346 & 0.3 & 0.06 & GNIRS & 2 & Mason+2015 \\ 
1387 & NGC 4579 & 0.00506 & 0.3 & 0.03 & GNIRS & 2 & Mason+2015 \\ 
\hline

\end{tabular}
\label{tab:arch_obs}
\tablenotetext{a}{\swiftbat 70-month hard X-ray survey ID (http://swift.gsfc.nasa.gov/results/bs70mon/).}
\tablenotetext{b}{AGN classification following \citet{Osterbrock1981}.}
\end{table*}

\subsection{NIR Spectral Data}
  While the data were taken from a number of observational campaigns, we maintain a uniform approach to data reduction and analysis. The observations where taken in the period 2010-2012. A summary of all the observational setups can be found in Table \ref{tab:observations_summary}. All programs used standard A0V stars with similar air masses to provide a benchmark for telluric correction. The majority of observations were done with the cross-dispersed mode of SpeX on the IRTF using a 0.8'' x 15'' slit \citep{Rayner2003} covering a wavelength range from 0.8 to 2.4 $\mu$m.  The galaxies were observed in two positions along the slit, denoted position A and position B, in an ABBA sequence by moving the telescope. This provided pairs of spectra that could be subtracted to remove the sky emission and detector artifacts. For 16 objects we have duplicate observations. In these cases, we chose the spectrum with the higher Signal-to-Noise ratio (S/N) in the continuum.
    
\subsubsection{Targeted Spectroscopic Observations}
\label{subsec:new_data}
For new observations the data reduction was performed using standard techniques with \spextool, a software package developed especially for IRTF SpeX observations \citep{Cushing2004}. The spectra were tellurically corrected using the method described by \cite{Vacca2002} and the IDL routine {\tt Xtellcor}, using a Vega model modified by deconvolution with the spectral resolution.   The routine {\tt Xcleanspec} was used to remove regions of the spectrum that were completely absorbed by the atmosphere. The spectra were then smoothed using a Savitzky-Golay routine, which preserves the average resolving power. We used the smoothed spectra for the measurements of the emission lines, whereas for the absorption lines analysis we used the unsmoothed spectra.  Further details are provided in \citet{Smith2014} and in the Appendix.

We also have seven spectra observed with the FLAMINGOS spectrograph at Kitt Peak over the wavelength range 0.9-2.3 $\mu$m using two setups, one with the JH grism and the other with the K grism both using a 1.5\arcsec\ slit.  The spectra were first flat-fielded, wavelength calibrated, extracted, and combined using IRAF routines.  Then telluric corrections and flux calibration were done in the same way as the IRTF observations using {\tt Xtellcorgeneral} from \spextool. The FLAMINGOS spectrograph has limitations in its cooling system and this induced thermal gradients and lower S/N particularly in the K-band, making this region unusable for analysis. 
Information about these observations are given in Tables \ref{tab:Table_IRTF} and \ref{tab:Flamingos}.

\subsubsection{Archival Observations}
The archival observations from the IRTF were reduced using \spextool\ in the same way as the targeted observations. Additional archival NIR spectra were from GNIRS at the Gemini North observatory observed with the cross-dispersed (XD) mode covering the 0.85 - 2.5 $\mu$m wavelength range processed using Gemini IRAF and the {\tt XDG-NIRS} task \citep{Mason2015}.
Information about the archival observations are given in Table \ref{tab:arch_obs} in the Appendix.

\begin{table*}
\centering
\caption{Summary of Instrumental Setups}
\begin{tabular}{ l l l c c c c c}
\hline
Reference & Telescope & Instrument & $N_{\rm spectra}$ & Slit size & Grating & \multicolumn{1}{c}{Resolving} & Wavelength  \\ 

 &  &  &   &  ['']& [l/mm] & \multicolumn{1}{c}{power} & range [$\mu$m]\\ \hline
This work & IRTF & SpeX & \NIRTF\ & 0.8 & - & 800 & 0.8 - 2.4 \\  
This work & KPNO & FLAMINGOS & \NFlamingos\ & 1.5 & - & 1000 & 1.0 - 1.8 \\ 
Riffel et al. (2006) & IRTF & SpeX & \NRiffel\ & 0.8 & - & 800 & 0.8 - 2.4 \\ 
Landt et al. (2008) & IRTF & SpeX & \NLandtIRTF\ & 0.8 & - & 800 & 0.8 - 2.4 \\ 

Rodr\'{i}guez-Ardila et al. (2002) & IRTF & SpeX & \NRodriguez\ & 0.8 & - & 800 & 0.8 - 2.4 \\ 
Riffel et al. (2013) & IRTF & SpeX & 1 & 0.8 & - & 800 & 0.8 - 2.4 \\ 
Mason et al. (2015) & Gemini North & GNIRS & 6 & 0.3 & 31.7 & 1300 & 0.9 - 2.5 \\ 
 &  & & 7 & 0.3 & 31.7 & 1800 & 0.9 - 2.5 \\ 
Landt et al. (2013) & Gemini North & GNIRS & 2 & 0.675 & 31.7 & 750 & 0.9 - 2.5 \\ 
 &  &  & 1 & 0.3 & 31.7 & 1800 & 0.9 - 2.5 \\ \hline
\end{tabular}
\label{tab:observations_summary}
\end{table*}

\section{Spectroscopic measurements}
In this section we present the spectroscopic measurements. First we explain the emission lines fitting method we used to measure the emission line flux and FWHM (Section \ref{sec:em_fit}). Then we describe absorption lines fitting method that we apply to measure the stellar velocity dispersion (Section \ref{sec:abs_fit}).

\subsection{Emission lines measurements}
\label{sec:em_fit}
\begin{figure*}
\centering
\subfigure{\includegraphics[width=0.4\textwidth]{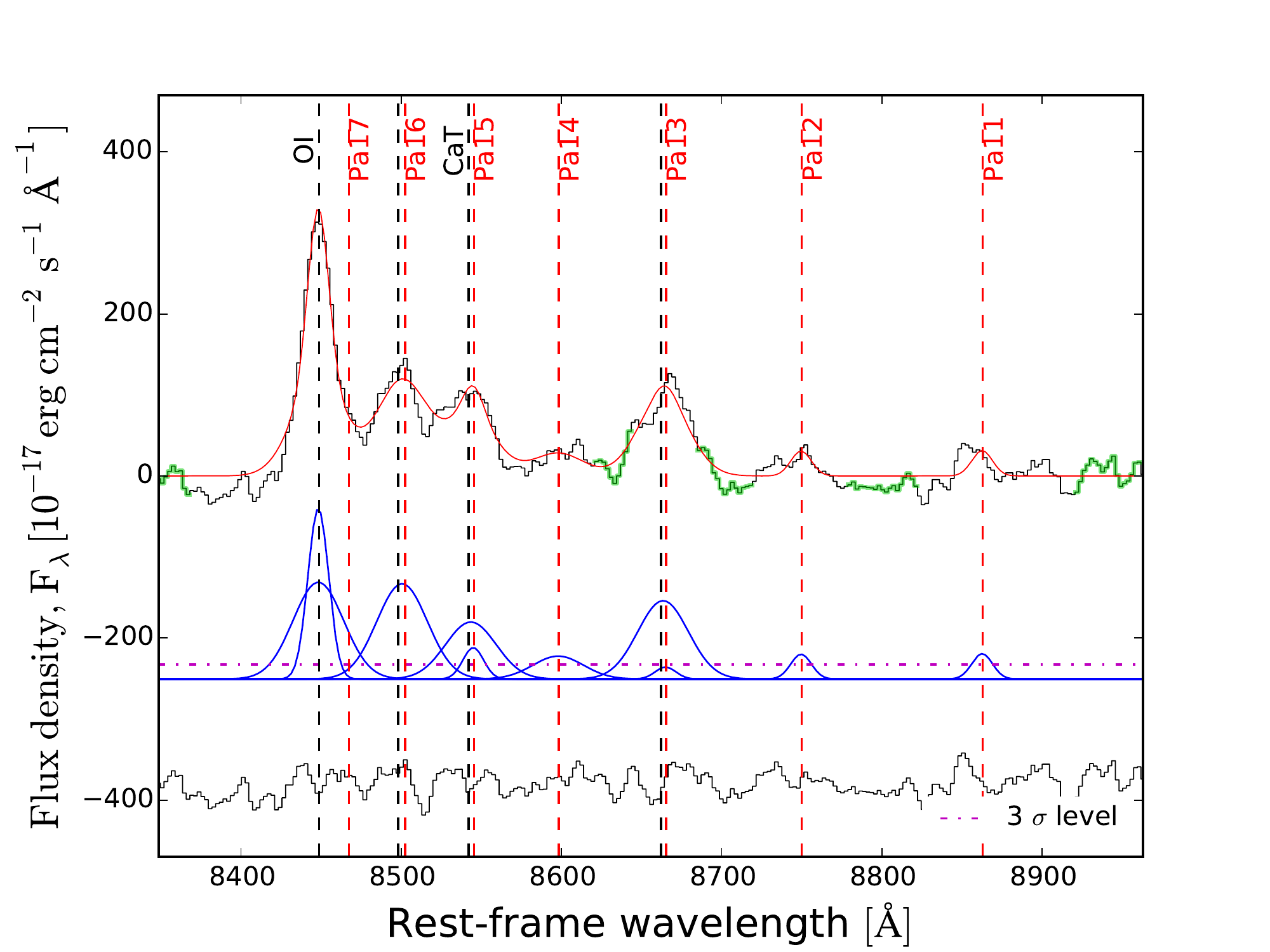}}
\subfigure{\includegraphics[width=0.4\textwidth]{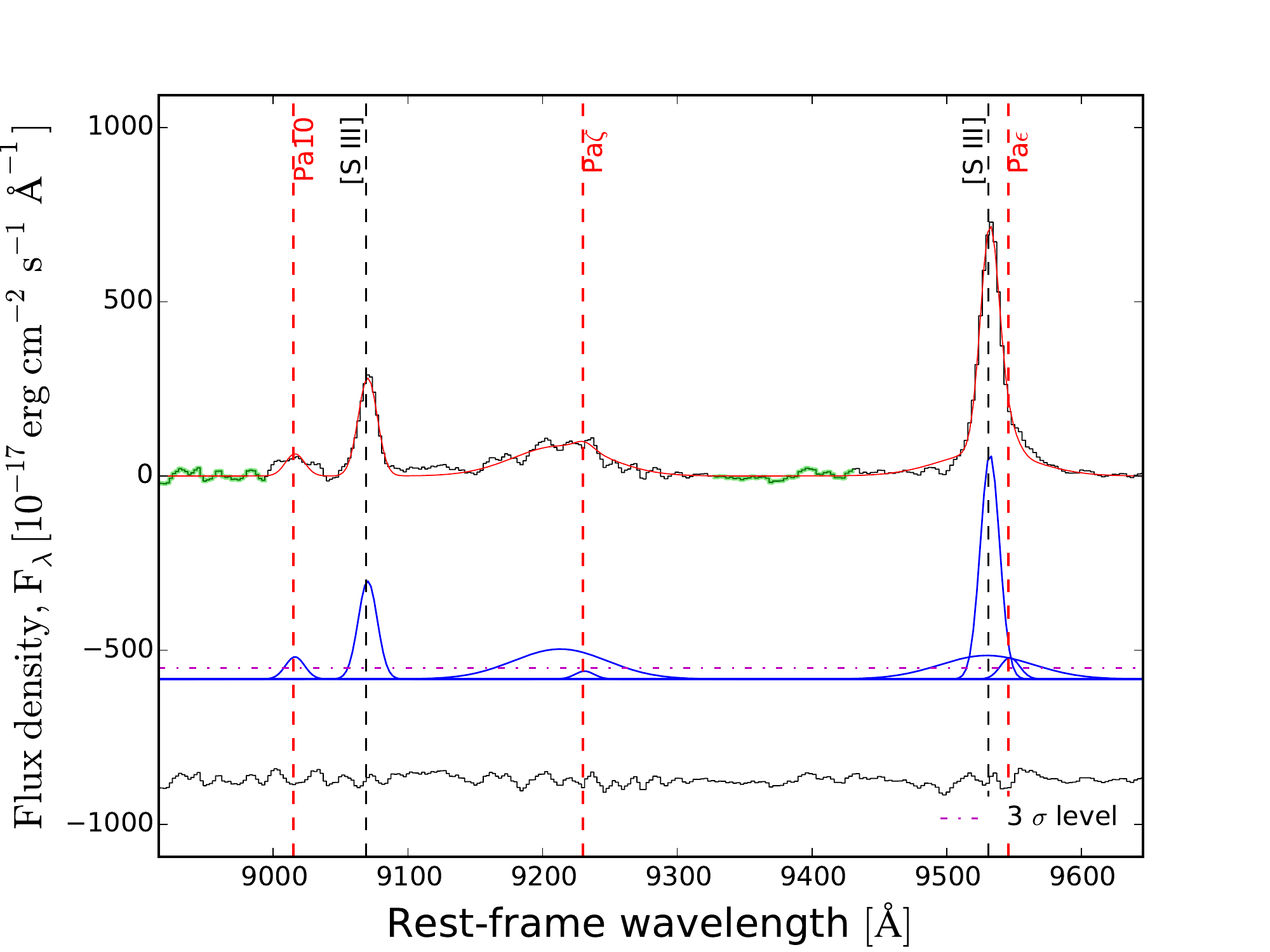}}
\vfill
\subfigure{\includegraphics[width=0.4\textwidth]
{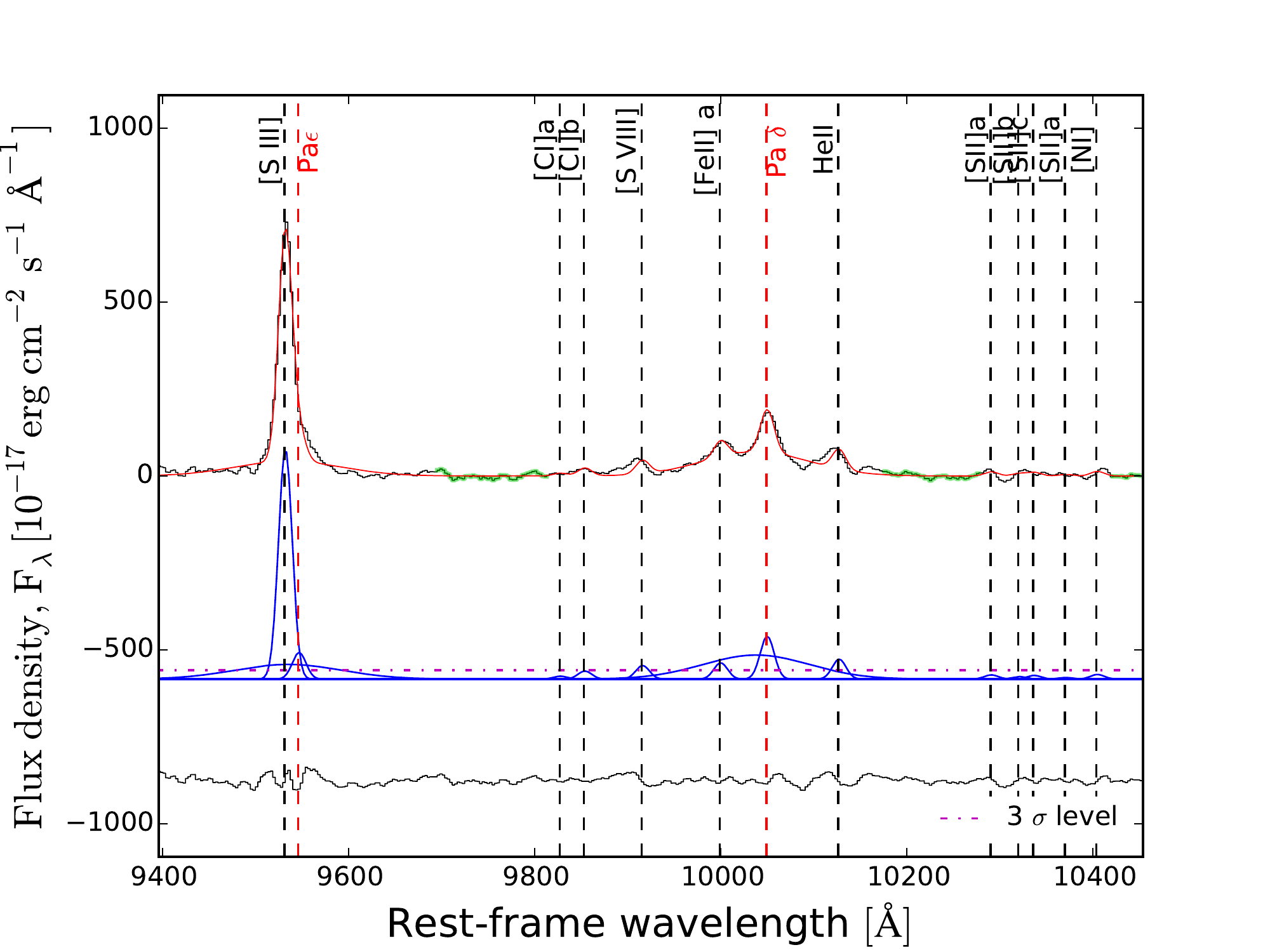}}
\subfigure{\includegraphics[width=0.4\textwidth]
{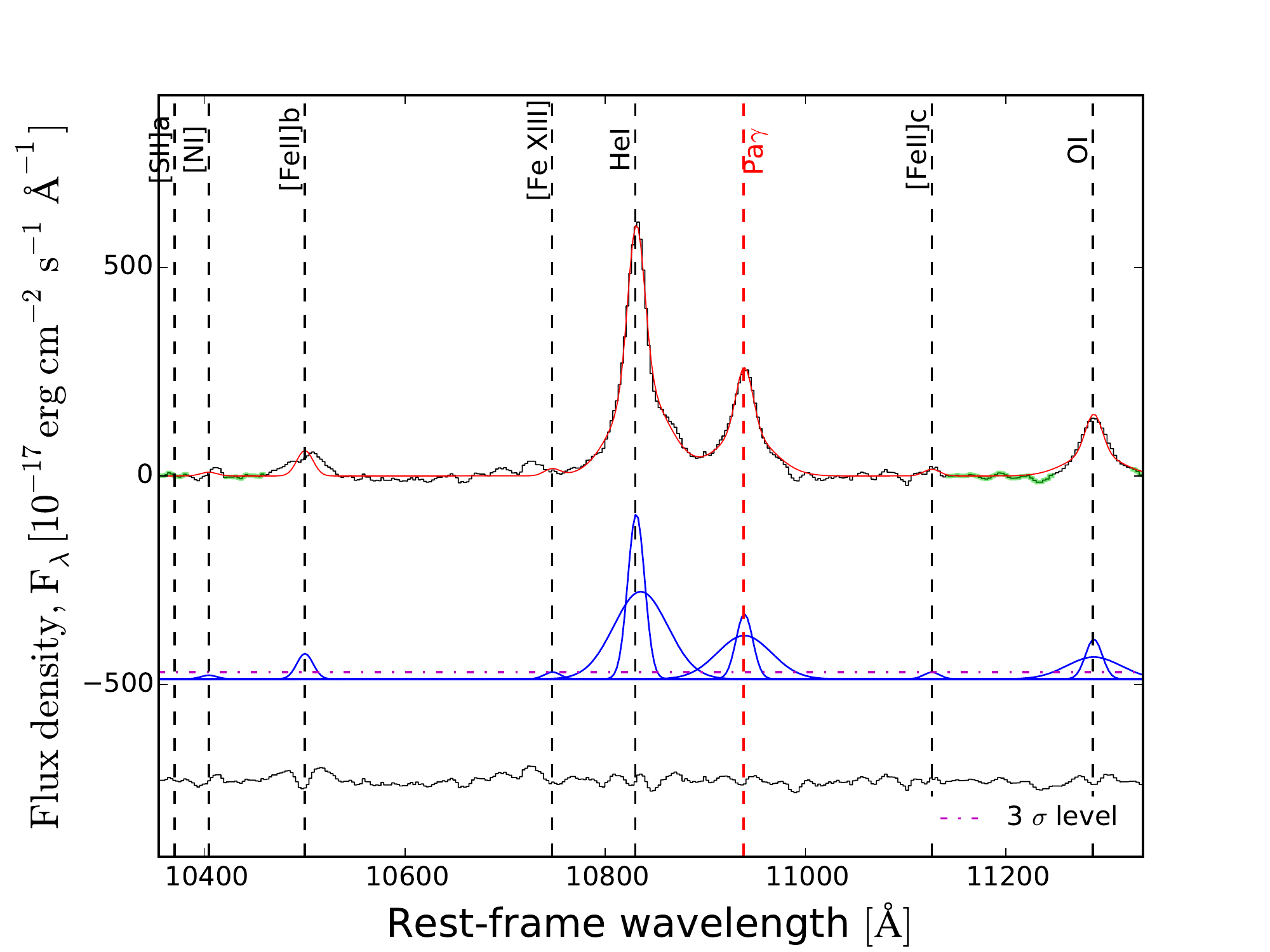}}

\subfigure{\includegraphics[width=0.4\textwidth]
{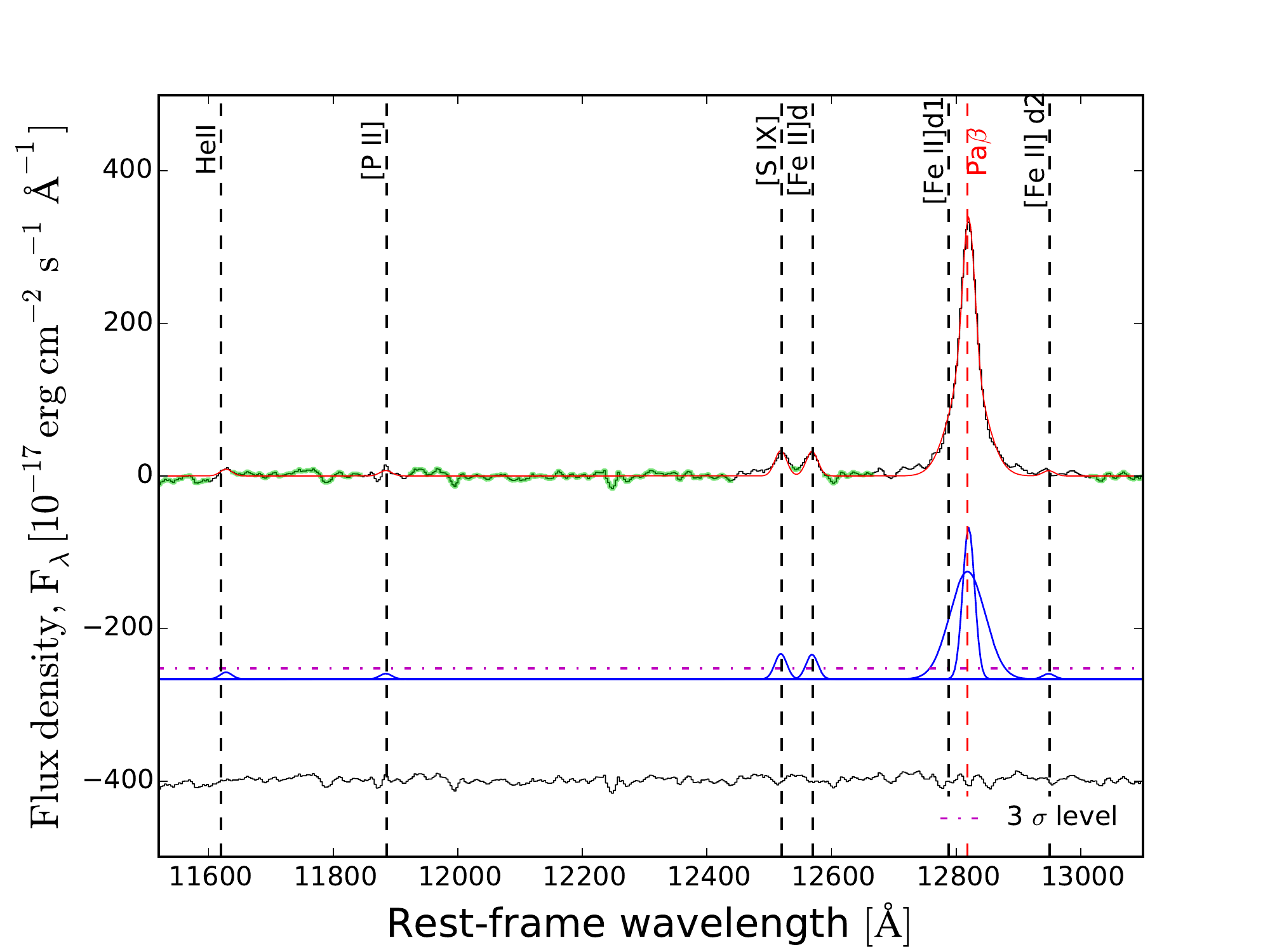}}
\subfigure{\includegraphics[width=0.4\textwidth]
{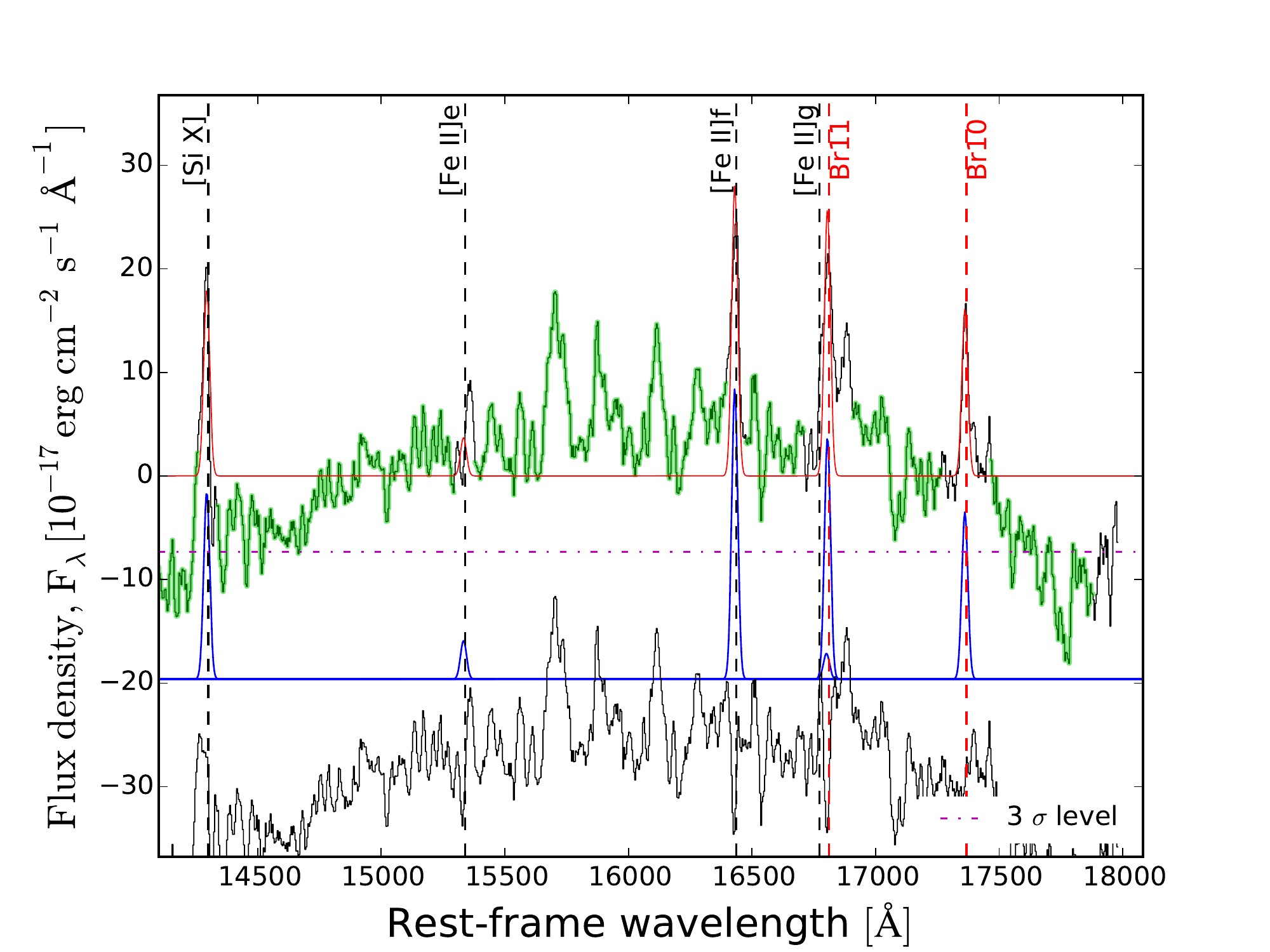}}
\vfill
\subfigure{\includegraphics[width=0.4\textwidth]
{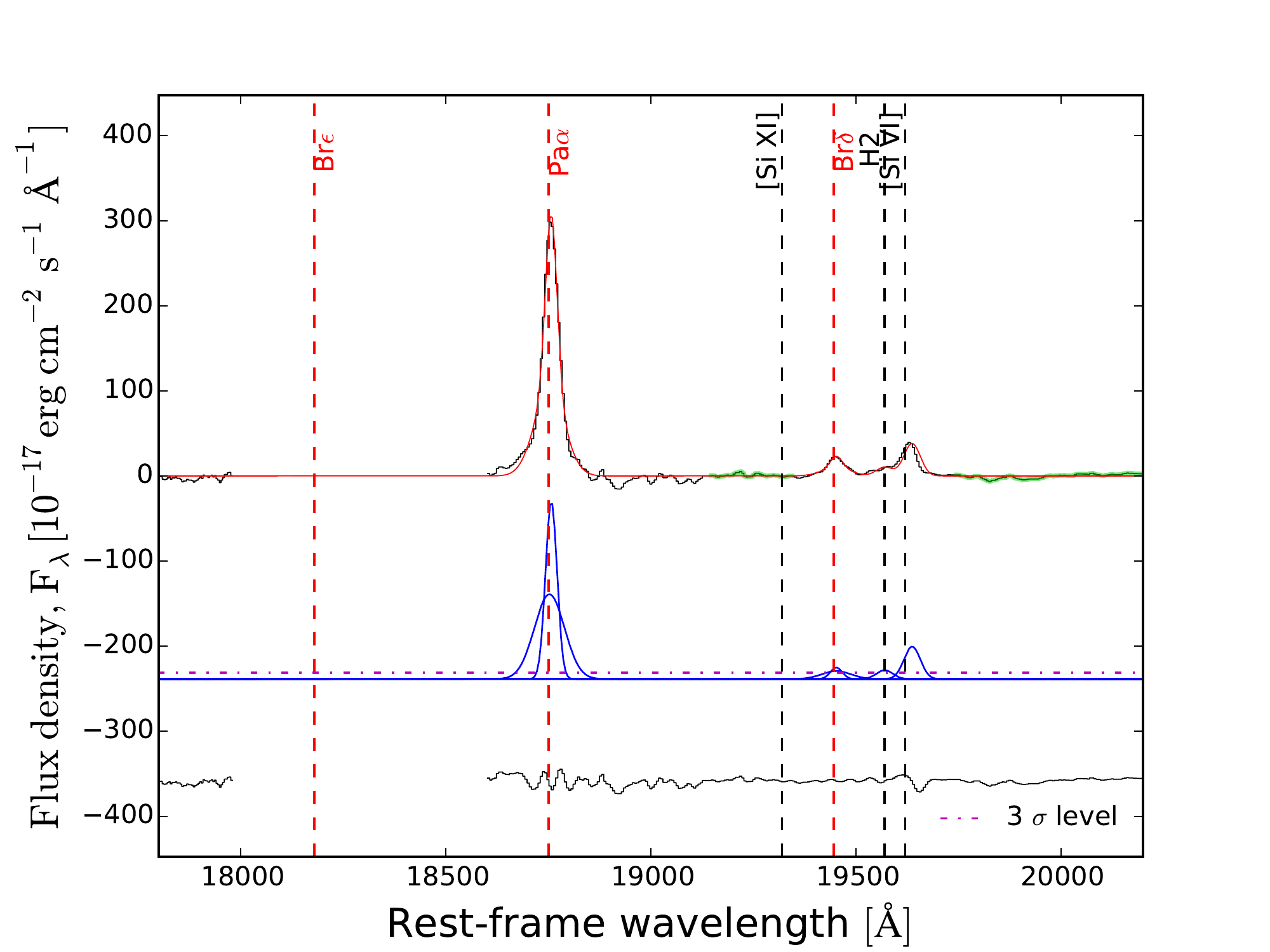}}
\subfigure{\includegraphics[width=0.4\textwidth]
{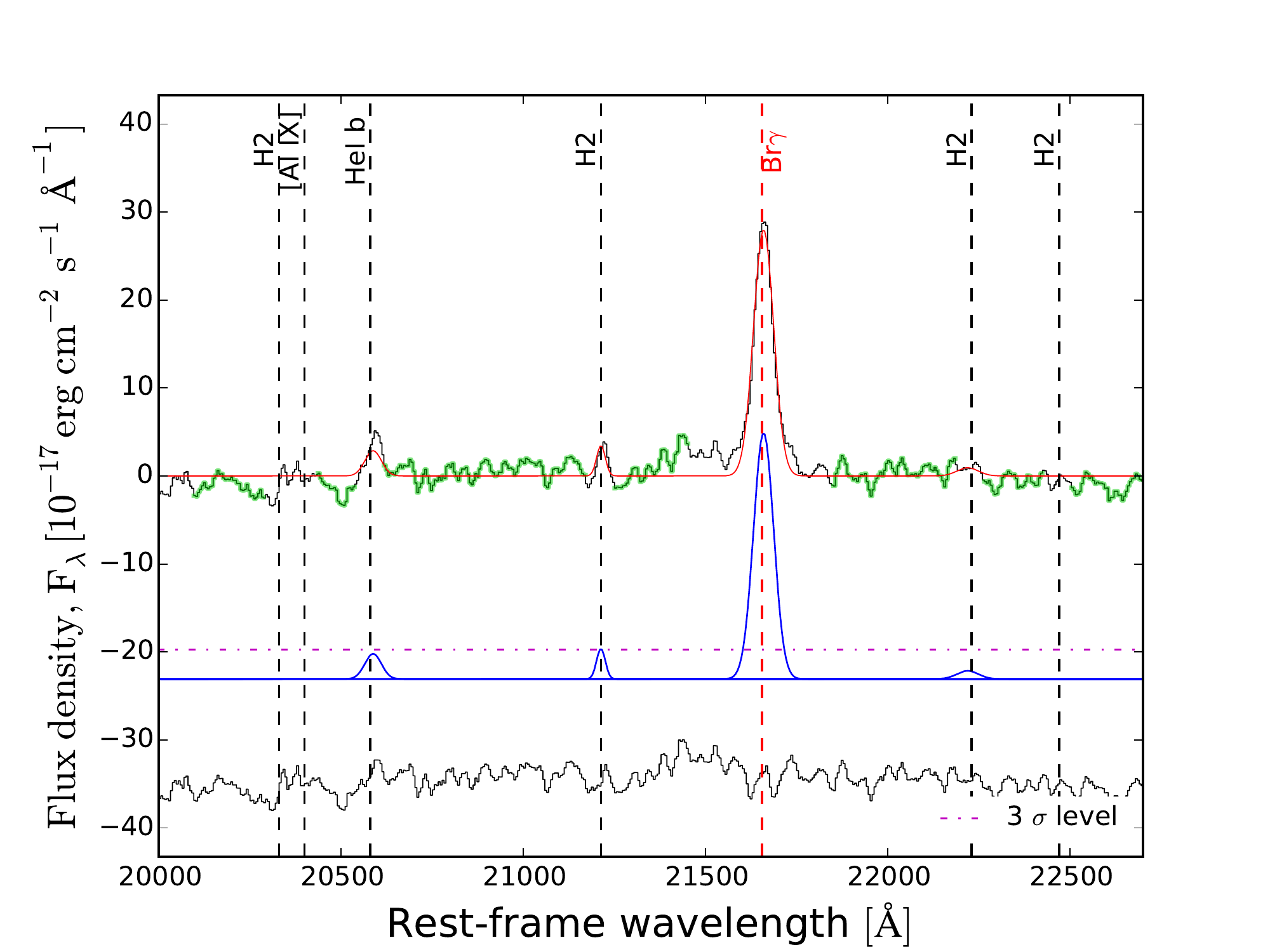}}

\caption{Example of emission lines fits for galaxy 2MASX J19373299-0613046 for which the spectrum was taken with IRTF. The 8 panels show the emission lines fits for the regions (from the upper left to the bottom right): Pa14, Pa$\zeta$, Pa$\delta$, Pa$\gamma$, Pa$\beta$, Br10, Pa$\alpha$ and Br$\gamma$. In the upper part of each figure, the spectrum is shown in black and the best fit in red. The regions where the continuum is measured are shown in green. In the middle part of the figure, the components of the fit are shown in blue and the magenta dashed line shows the threshold level for  detection (S/N $>\ $3). The lower part of the figure shows the residuals in black.
}
\label{em_fit}
\end{figure*}

We fit our sample of NIR spectra in order to  measure the emission line flux and FWHM.
We use {\tt PySpecKit}, an extensive spectroscopic analysis toolkit for astronomy, which uses a Levenberg-Marquardt algorithm for fitting \citep{GinsburgandMirocha2011}.
We fit separately the Pa14 ($\text{0.84-0.90}$ $\mu$m), Pa$\zeta$ (0.90-0.96 $\mu$m), Pa$\delta$ (0.96-1.04 $\mu$m), Pa$\gamma$ (1.04-1.15 $\mu$m),  Pa$\beta$ (1.15-1.30 $\mu$m), Br10 (1.30-1.80 $\mu$m), Pa$\alpha$ (1.80-2.00 $\mu$m) and Br$\gamma$ (2.00-2.40 $\mu$m) spectral regions.  An example fit is found in Figure \ref{em_fit}.
Before applying the fitting procedure, we de-redden the spectra using the galactic extinction value $E_{B-V}$ given by the IRSA Dust Extinction Service\footnote{http://irsa.ipac.caltech.edu/applications/DUST/docs/background.html} and a function from the {\tt PySpecKit} tool.

We employ a first order power-law fit to model the continuum.
For each spectral region, we estimated the continuum level from the entire wavelength range, except where the emission lines are located ($\pm$ 20 \AA\ for the narrow lines and $\pm$ 150 \AA\ for the broad lines).
For the modeling of the emission lines, we assume Gaussian profiles. The emission lines that we fit in each spectral region are listed in Table \ref{tab:emission_lines} in the Appendix. 
The position of the narrow lines are tied together. They are allowed to be shifted by a maximum of 8 \AA\ ($\simeq $ 160 \kmps) from the systemic redshift. The redshift  and widths of the narrow lines are also tied together. 
As an initial input value for the width of the narrow lines, we used the width of the \siiif $\lambda$0.9531 $\mu$m line, that is the strongest narrow emission line in the 0.8 - 2.4 $\mu$m wavelength range.
We set the maximum FWHM of the narrow lines to be 1200 \kmps.

For the Paschen, Brackett, and \hei $\lambda$1.083 $\mu$m lines, we allow the code to fit the line with a narrow component (FWHM $<\ $1200 km s$^{-1}$) and a broad component (FWHM $>\ $1200 km s$^{-1}$). We tied the width of the narrow component to the width of the other narrow lines present in the same spectral region. We also tied together the broad-line width of the Paschen lines that lie in the same fitting region. The broad components are allowed to be shifted by a maximum of 30 \AA\ ($\simeq$ 600 \kmps) from the theoretical position. This large velocity shift is motivated by the observations of a mean velocity shift of the broad H$\beta$ to the systemic redshift of 109 \kmps with a scatter of 400 \kmps  in a sample of 849 quasars \citep{Shen2016}. The largest velocity shifts are $\sim1000$ \kmps.
 We corrected the FWHM for the instrumental resolution, subtracting the instrumental FWHM from the observed FWHM in quadrature.\\

For line detection, we adopted a S/N of three compared to the surrounding continuum as the detection threshold. For each spectral region, we measured the noise level taking the dispersion of the continuum in a region without emission lines.
In a few cases near blended regions with strong stellar absorption features (e.g. \Silivi) a line detection seemed spurious by visual inspection even though it was above the detection limit.  We have noted those with a flag and quote upper limits for these sources.
If the S/N of the broad component of a particular line was below the threshold, we re-ran the fitting procedure for the spectral region of this line, without including a broad component. We inspected by eye all the spectra to verify that the broad component was not affected by residuals or other artifacts. After visual inspection, we decided to fit seven spectra with only a narrow component for Pa$\alpha$. 

To estimate the uncertainties on the line fluxes and widths, we performed a Monte Carlo simulation.  We repeated the fitting procedure 10 times, adding each time an amount of noise randomly drawn from a normal distribution with the deviation equal to the noise level.
Then we computed the median absolute deviation of the 10 measurements and we used this value as an estimate of the error at the one sigma confidence level.
We inspected visually all emission line fits to verify proper fitting and we assigned a quality flag to each spectral fit. 
We follow the classification nomenclature by visual inspection of the first BASS paper (Koss et al., submitted). Quality flag 1 refers to spectra that have small residuals and very good fit. Flag 2 means that the fit is not perfect, but it is still acceptable. Flag 3 is assigned to bad fit for high S/N source due to either the presence of broad line component or offset in emission lines. Flag 9 refers to spectra where no emission line is detected. Flag -1 means lack of spectral coverage.
The emission lines fluxes and FWHM are listed in Tables \ref{tab:Pa14}-\ref{tab:Brgamma}.% in the Appendix.

\subsection{Galaxy templates fitting}
\label{sec:abs_fit}
%\subsection{pPXF}
We used the Penalized Pixel-Fitting ({\tt pPXF}) code \citep{Cappellari2004} to extract the stellar kinematics from the absorption-line spectra. This method operates in the pixel space and uses a maximum penalized likelihood approach to derive the line-of-sight velocity distributions (LOSVD) from kinematical data \citep{Merritt1997}.
First the {\tt pPXF} code creates a model galaxy spectrum by convolving a template spectrum with a parametrized LOSVD. 
Then it determines  the best-fitting parameters of the LOSVD by minimizing the $\chi ^2$ value, which measures the agreement between the model and the observed galaxy spectrum over the set of good pixels used in the fitting process. 
 Finally, {\tt pPXF} uses the `best fit spectra' to calculate \sig\ from the absorption lines.
 
We used the same spectra as those used in section \ref{sec:em_fit}, de-redshifted to the rest-frame. Here we concentrate on three
narrow wavelength regions where strong stellar absorption features are present: CaT region ($0.846 - 0.870$ $\mu$m), CO band-heads in the H-band (1.570 - 1.720 $\mu$m), and CO band-heads in the K-band (2.250 - 2.400 $\mu$m). The absorption lines present in these three wavelength ranges are listed in Table \ref{abs_lines} in the Appendix. For the spectra from the FLAMINGOS spectrograph, we measured \sig\ only from the CO band heads in the H-band, due to limited wavelength coverage.

The {\tt pPXF} code uses a large set of stellar templates to fit the galaxy spectrum.
We used the templates from the Miles Indo-U.S. CaT Infrared (MIUSCATIR) library of stellar spectra \citep{Rock2015, Rock2016}. They are based on the MIUSCAT stellar populations models, which are an extension of the models of \cite{Vazdekis2010}, based on the Indo-U.S., MILES \citep{Sanchez-Blazquez2006}, and CaT \citep{Cenarro2001} empirical stellar libraries \citep{Vazdekis2012}. The IR models are based on 180 empirical stellar spectra  from the stellar IRTF library. This library contains the spectra of 210 cool stars in the IR wavelength range \citep{Cushing2005, Rayner2009}. The sample is composed of stars of spectral types F, G, K, M,  AGB-, carbon- and S-stars, of luminosity classes I-V. Some of the stars were discarded because of their unexpected strong variability, strong emission lines or not constant baseline.
The IR models cover the spectral range 0.815 - 5.000 $\mu$m at a resolving power R = 2000, which corresponds to a spectral resolution of FWHM $\sim$ 150 \kmps ($\sigma \sim$ 60 \kmps) equivalent to 10 \AA\  at 2.5 $\mu$m \citep{Rock2015}.  A comparison with Gemini Near-Infrared Late-type stellar (GNIRS) library \citep{Winge2009} in the range 2.15 - 2.42 $\mu$m at R = 5300 - 5900 (resolution of $\sim$ 3.2 \AA\ FWHM) is provided in the Appendix.  The spectral templates are convolved to the instrumental resolution of the observed spectra  under the assumption that the shapes of the instrumental spectral profiles are well approximated by Gaussians.

%\textbf{Mask for the emission lines}:
We applied a mask when fitting stellar templates around the following emission lines: Pa14 (in the CaT region), \feiif $\lambda$16436, \feiif $\lambda$16773, and Br11 $\lambda$16811 (in the H-band), and \Caviii $\lambda$23210 (in the K-band).
Since the \Caviii $\lambda$23210 emission line overlaps with the CO(3-1) $\lambda$23226 absorption line, we decided to mask the region around this line only if the emission line was detected.
Also for the Pa14 line we decided to mask the line only if the emission line was detected, because it is in the same position of the \caii $\lambda$8498 absorption line and therefore it is in a critical region for the measurement of the velocity dispersion.
We set the width of the emission lines mask to 1600 \kmps for the narrow lines and to 2000 \kmps for the Brackett  and Paschen lines (Br11 and Pa14).
The error on the velocity dispersion ($\Delta$ \sig)  is the formal error (1 $\sigma$) given by the {\tt pPXF} code. The error are in the range 1-20\% of the \sig values.

All absorption lines fits were inspected by eye to verify proper fitting.
We follow the nomenclature of fitting classification of the first BASS paper (Koss et al., submitted). We assigned the quality flag 1 to the spectra that have small residuals and very good fit of the absorption lines (average error $\langle \Delta \sigma_{*} \rangle$ = 6 \kms). Quality flag 2 refers to the spectra that have larger residuals and errors in the velocity dispersion value (average error $\langle \Delta \sigma_{*} \rangle$ = 12 \kms), but the absorption lines are well described by the fit. 
For spectra where the error of the velocity dispersion value is $>\ $50 \kmps and the fit is not good, we assigned the flag 9.

\section{Results}

We first compare the FWHM of broad Balmer and Paschen lines (Section \ref{sec:FWHM_comp}). Then we discuss AGN that show `hidden' broad lines in the NIR spectrum but not in the optical (Section \ref{sec:hidden_BLR}).
Next, we measure the velocity dispersion (Section \ref{sec:veldisp}) and black hole masses (Section \ref{sec:MBH}).
In Section \ref{sec:NIR_diagn}, we apply NIR emission line diagnostics to our sample. Then we discuss the presence of coronal lines in the AGN spectra (Section \ref{sec:coronal_lines}). Finally, we measure the correlation between coronal line and hard X-ray emission (Section \ref{sec:flux_comp}).

\subsection{Comparison of the line widths of the broad Balmer and Paschen lines}
\label{sec:FWHM_comp}

\begin{figure*}
\centering
\subfigure{\includegraphics[width=0.7\textwidth]{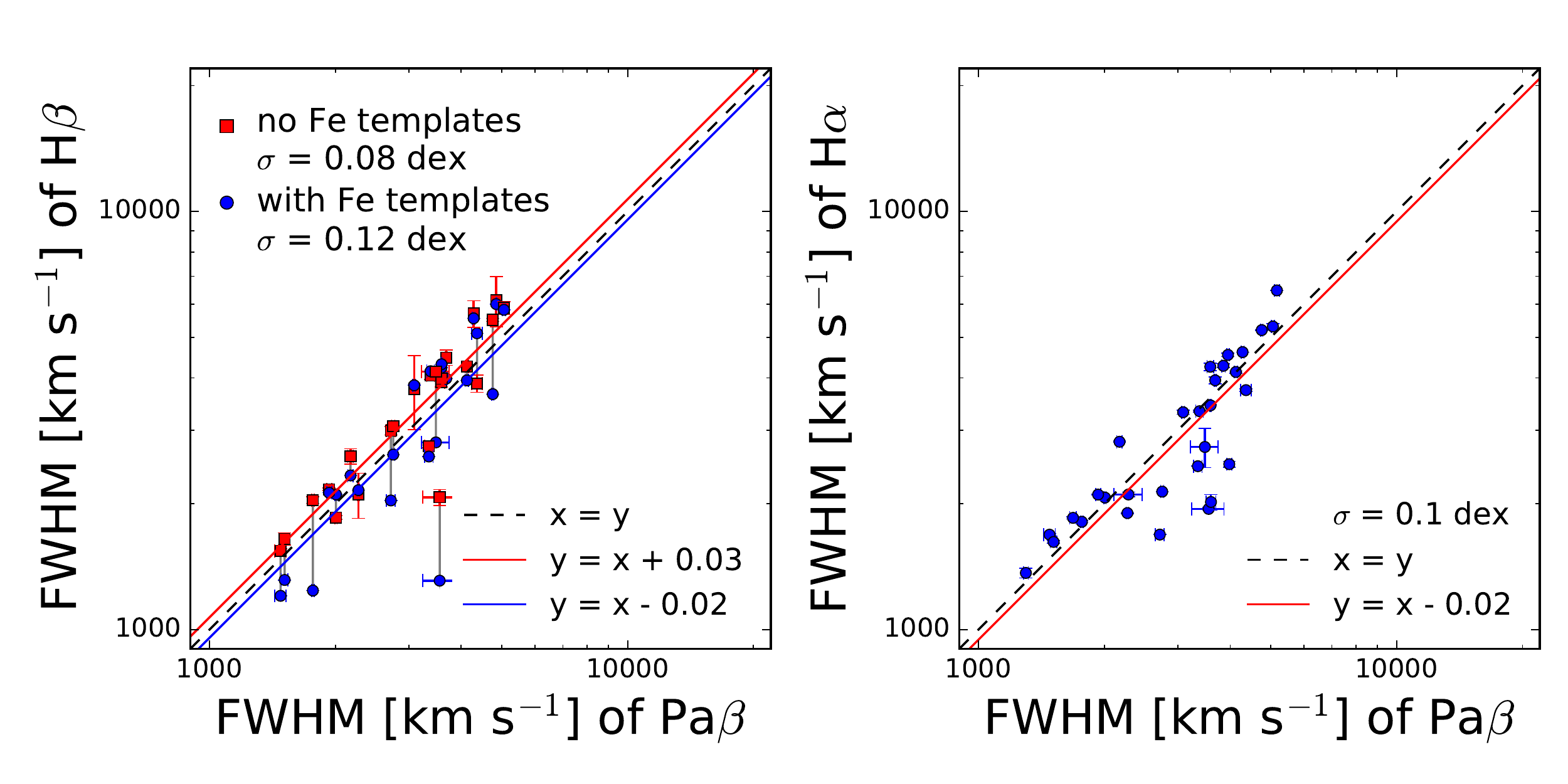} }
\subfigure{\includegraphics[width=1.\textwidth]{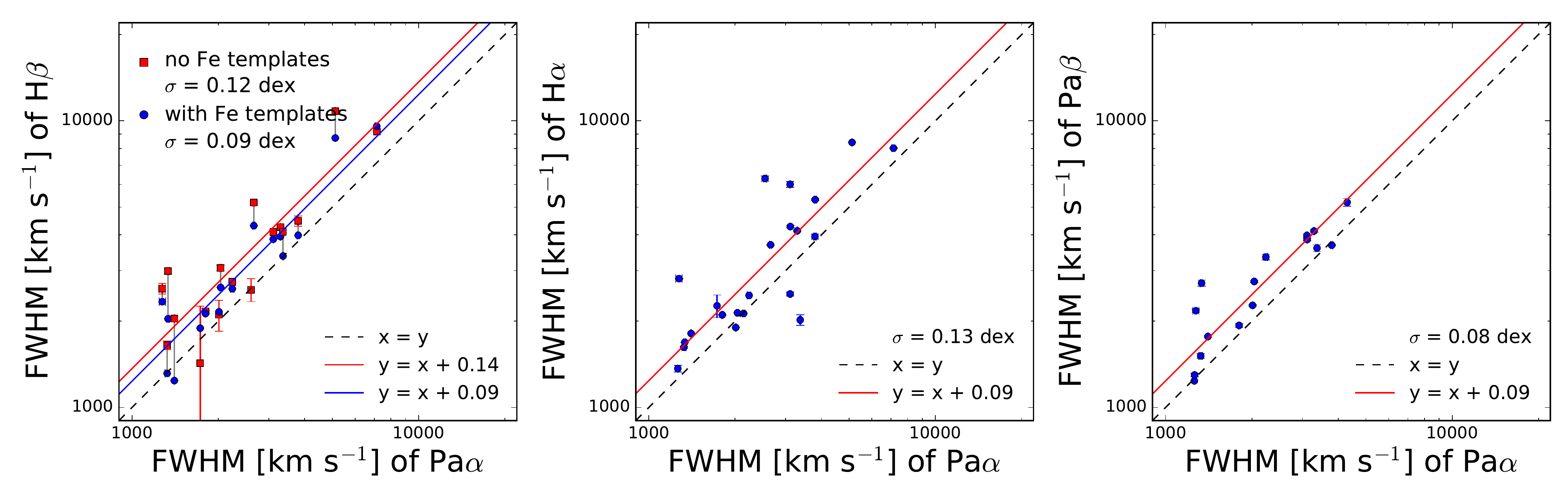} }

\caption{\textit{Upper panels}: Comparison between the FWHM of the broad components Pa$\beta$ and \Hbeta\ (left). The red points show the FWHM of H$\beta$ measured without taking into account the iron contamination and the instrumental resolution and the blue points the FWHM of H$\beta$ measured using iron templates and corrected for the instrumental resolution. The black dashed line shows the one-to-one relation, the red and blue lines show the the linear fit with slope 1.  The right panel shows the comparison between the FWHM of  the broad component of Pa$\beta$ and H$\alpha$.
\textit{Lower panels}: Comparison of the FWHM of the broad components of Pa$\alpha$ with H$\beta$ (left), H$\alpha$ (middle) and Pa$\beta$ (right). }
\label{Balmer_vs_Paschen}
\end{figure*}

We compared the line widths of the broad components of \Pab, \Paa, \Hbeta, and H$\alpha$.  Figure \ref{Balmer_vs_Paschen} shows the comparison between the FWHM of the broad \Pab\ and \Hbeta\ with or without fitting Fe templates to the \Hbeta\ region \citep[e.g.,][]{BorosonGreen1992,Trakhtenbrot2012}. We fitted the data using a linear relation with fixed slope of 1 and we searched for the best intercept value. We equally weight each data point in the line fit because the measured uncertainties from statistical noise are small ($<\ 5\%$). For the FWHM of the broad \Pab\ and \Hbeta\ with Fe fitting, we found the best fit to have an offset of $0.029\pm 0.003$ dex with a scatter of $\sigma$ = 0.08 dex. 
Without Fe fitting, we find a linear relation with an offset of $-0.019\pm 0.005$ dex and a larger scatter ($\sigma$ = 0.13 dex). 
After taking into account the effect of the iron contamination on H$\beta$, we did not find a significant difference between the FWHM of H$\beta$ and the FWHM of Pa$\beta$ (the p-value of the Kolomogorov-Smirnov test is 0.84). For the FHWM of Pa$\beta$ to H$\alpha$, we find an offset of $-0.023 \pm 0.03 $ dex and a scatter $\sigma$ = 0.1 dex with no significant difference between their distributions (Kolomogorov-Smirnov test p-value = 0.56).  

We observed a trend for the mean FWHM of Pa$\alpha$ to be smaller than the FWHM of the other lines. Comparing the FWHM of Pa$\alpha$ and \Hbeta\, we found the best fit to have an offset of $0.092\pm 0.005$ dex with a scatter $\sigma$ = 0.09 dex, while for the comparison between Pa$\alpha$ and H$\alpha$, the offset is $0.094\pm 0.006$ dex with a scatter $\sigma$ = 0.13 dex. 
For the comparison between Pa$\alpha$ and Pa$\beta$, the offset is $0.093 \pm 0.005$ dex with a scatter $\sigma$ = 0.08 dex. 
Considering the mean values, we found that that the mean value of the FWHM of Pa$\alpha$ ($2710 \pm 294$ \kms ) is smaller than the mean value of H$\alpha$ and H$\beta$ ($3495 \pm 436$ \kms  and  $3487 \pm 582$ \kms, respectively). For the sources  with measurements of the broad components of both Pa$\alpha$ and Pa$\beta$, the mean value of the FWHM of Pa$\alpha$ ($2309 \pm 250$ \kms) is also smaller than the mean value of Pa$\beta$ ($2841 \pm 283$ \kms). We note that none of these differences rises to the 3$\sigma$ level, so a larger sample would be required to study whether the FWHM of broad Pa$\alpha$ is indeed smaller than the other lines.

We used the Anderson-Darling and Kolomogorov-Smirnov to further test if the distribution of the FWHM of Pa$\alpha$ is significantly different from those of the other lines.
We find that both tests indicate that the populations are consistent with being drawn from the same intrinsic distribution at the greater than 20$\%$ level for all broad lines. For the comparison of Pa$\alpha$ with H$\beta$ the Kolomogorov-Smirnov test gives a p-value = 0.67, whereas for the comparison of Pa$\alpha$ with H$\alpha$ the p-value is 0.33 and for Pa$\alpha$ with Pa$\beta$ the p-value is 0.63 suggesting no significant difference in the distributions of the FWHM of the broad components.

\subsection{Hidden BLR}
\label{sec:hidden_BLR}
\begin{figure*}
\begin{center}
\subfigure{\includegraphics[width=0.44\textwidth]
{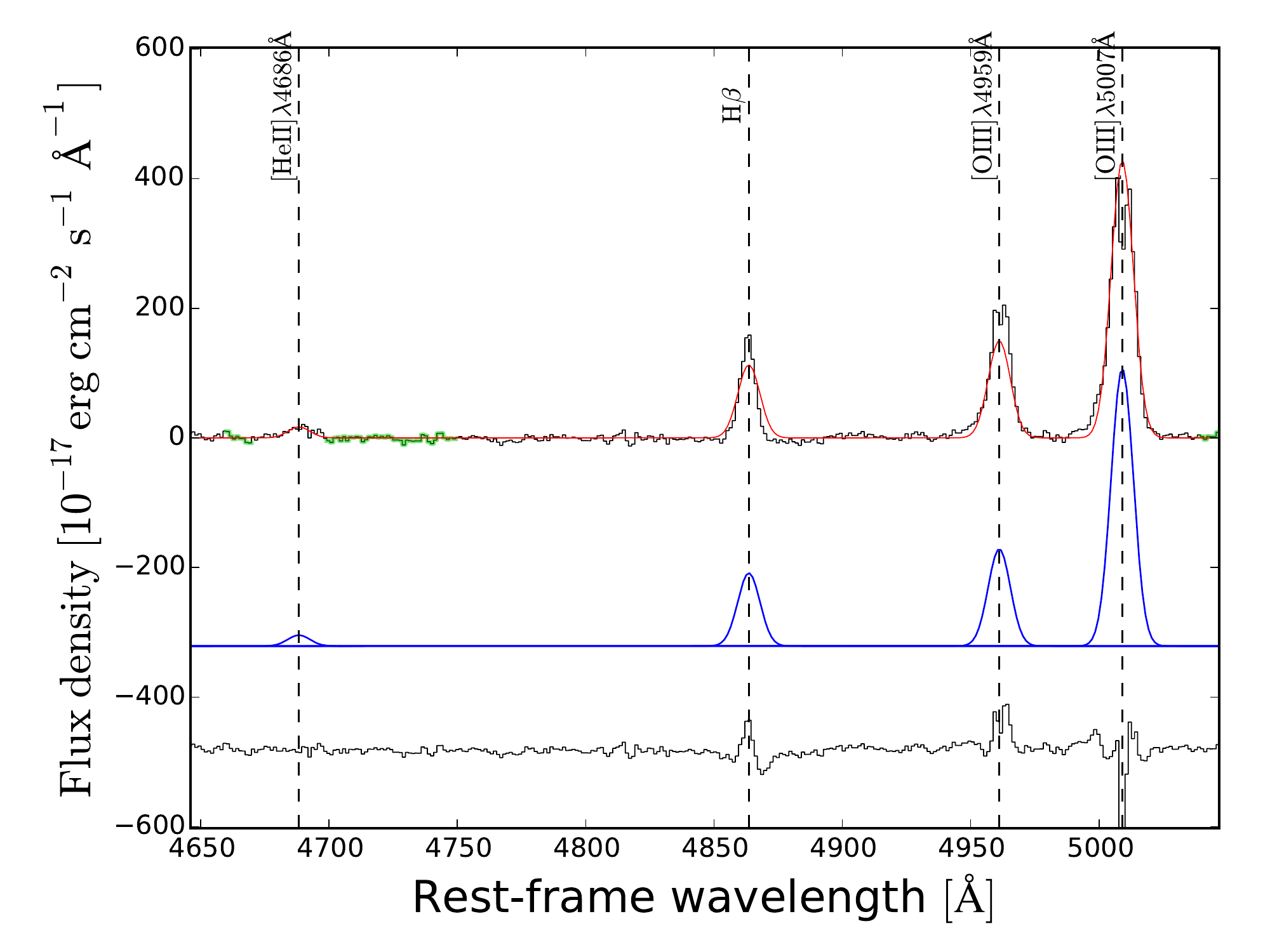} }
\subfigure{\includegraphics[width=0.44\textwidth]{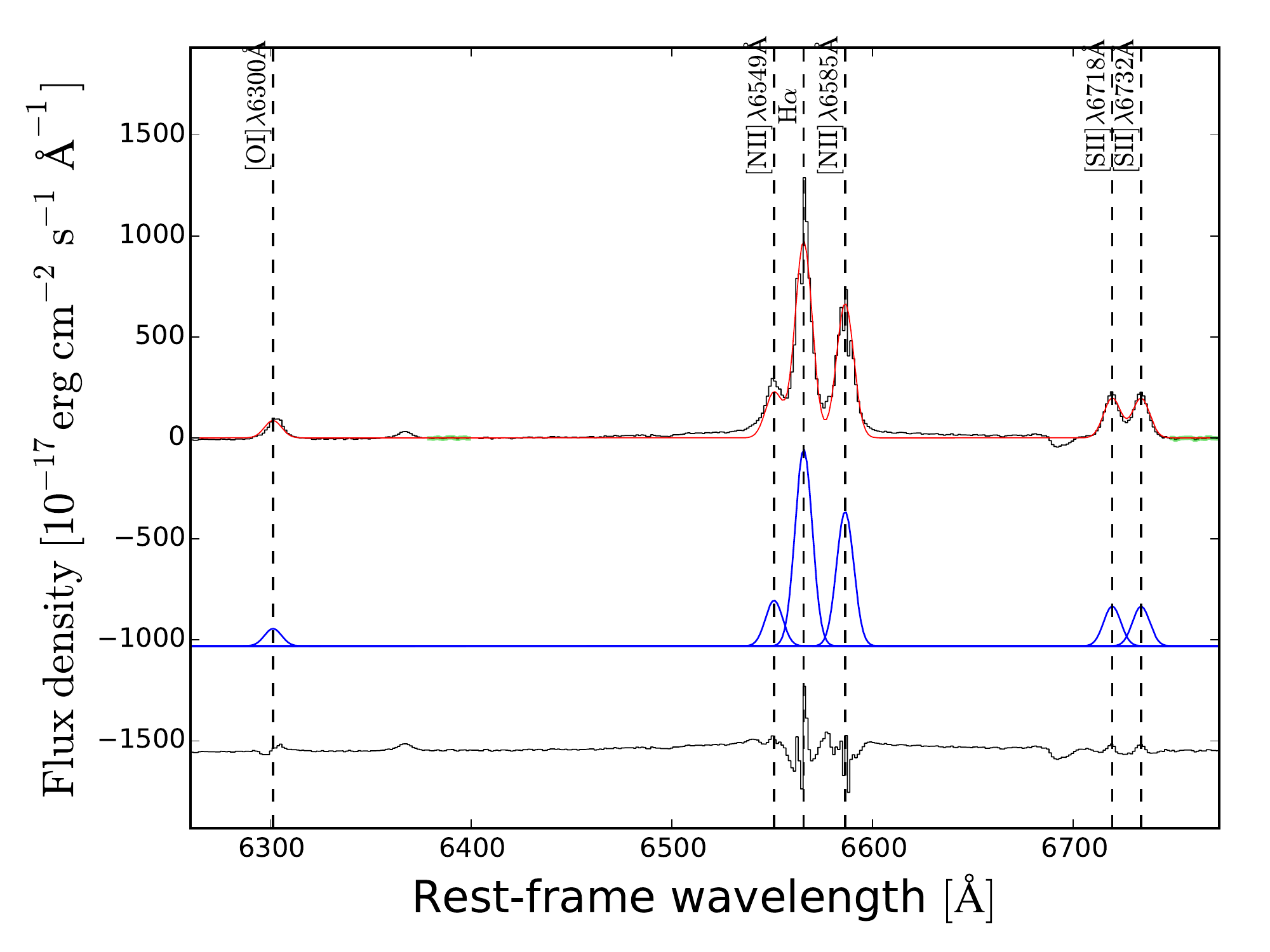} }
\vfill
\subfigure{\includegraphics[width=0.45\textwidth]{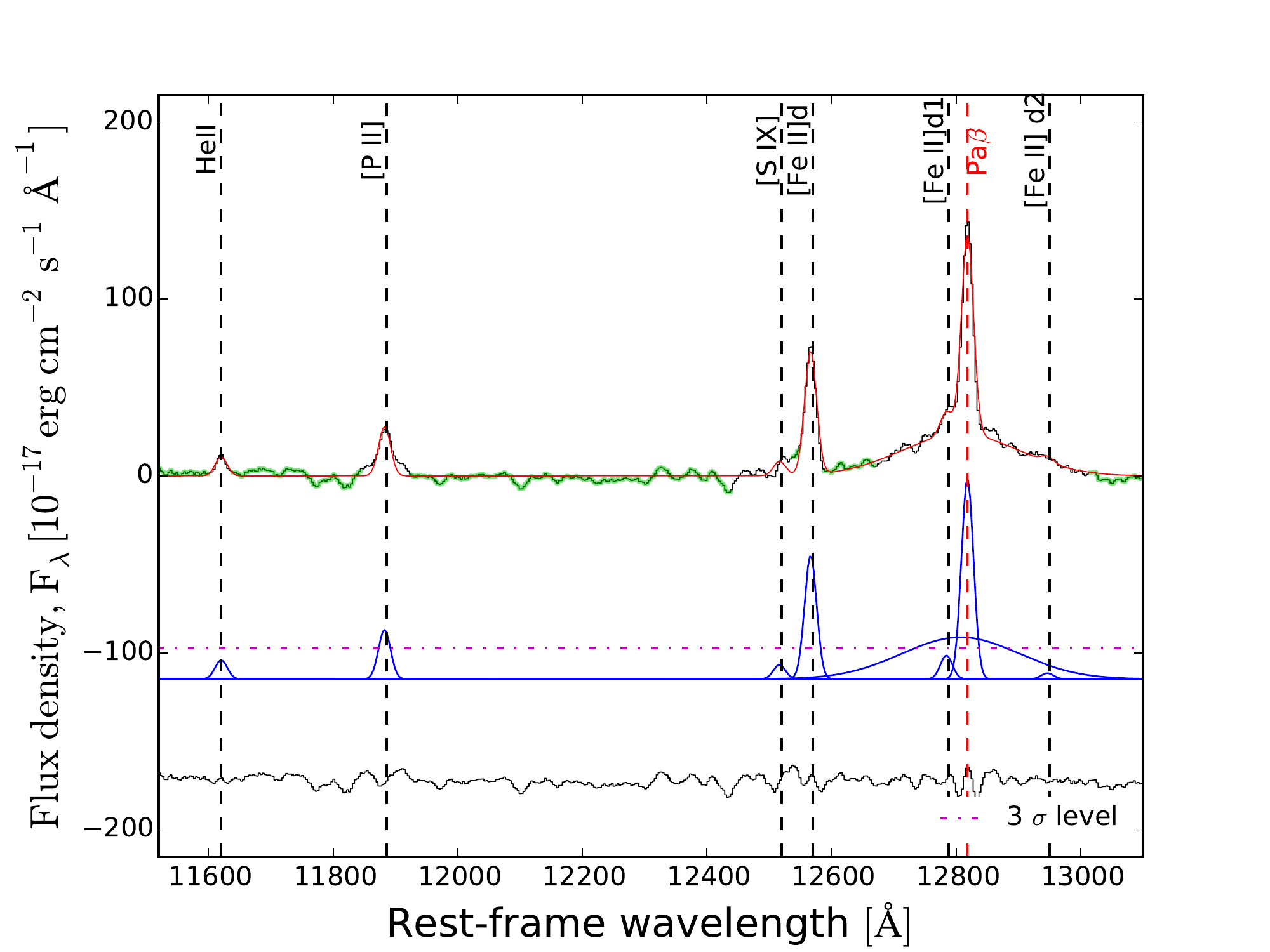} }
\subfigure{\includegraphics[width=0.45\textwidth]{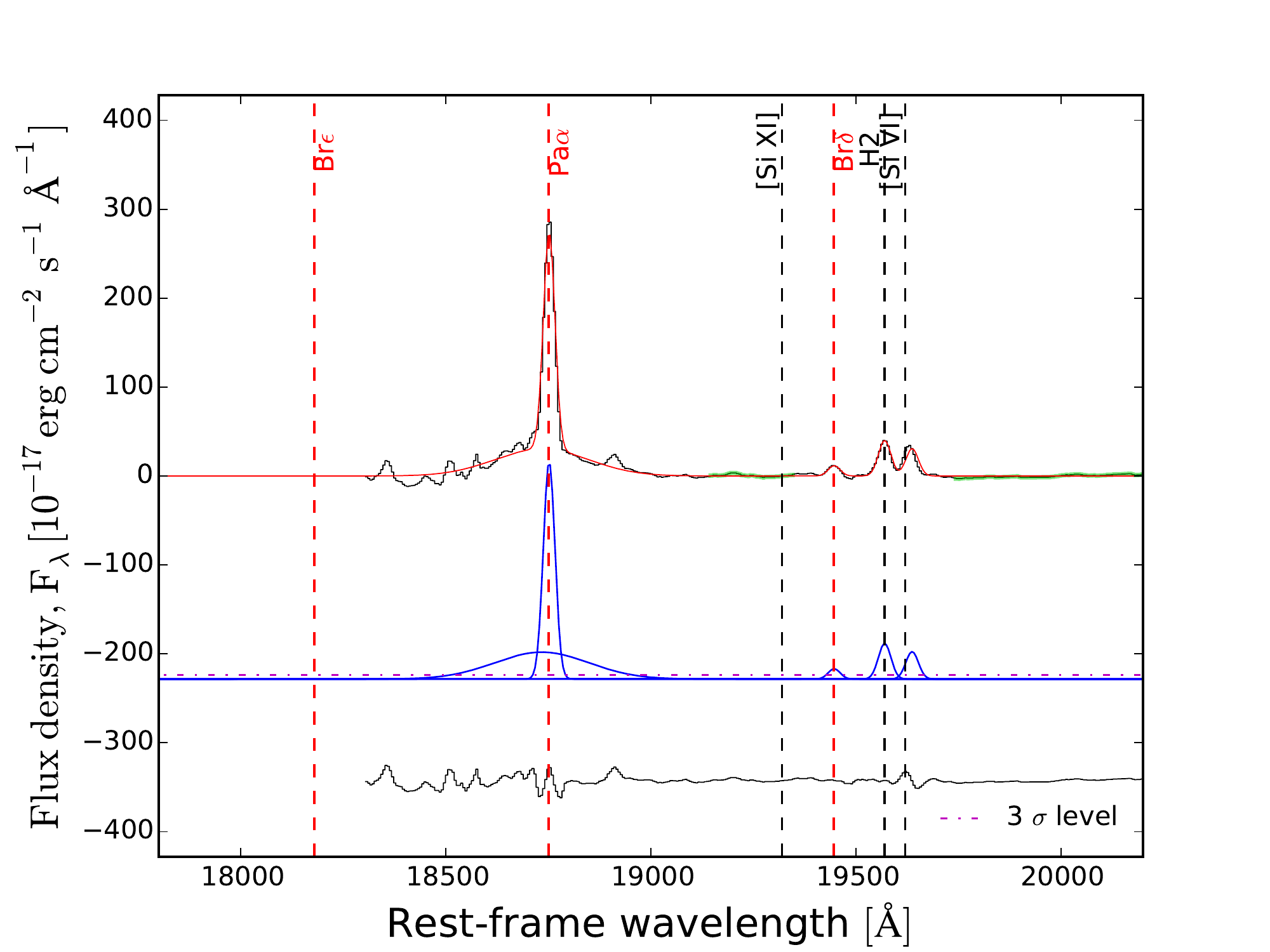} }
\caption{Optical and NIR spectra of Mrk 520, which is an example of Seyfert 2 galaxy displaying a "hidden" BLR in the NIR. \textit{Upper panels}: optical spectrum of the H$\beta$ and H$\alpha$ region. The best fit is in red, the model in blue, and the residuals in black.  \textit{Lower panels:} NIR spectrum of the Pa$\beta$ and Pa$\alpha$ region. In addition to the components explained above, the magenta dashed line in the middle part of the figures shows the detection threshold (S/N$>\ $3) with respect to the fitting continuum (blue). }
\label{Hidden_BLR_Mrk520}
\end{center}
\end{figure*}

In our sample there are \NSytwo\ AGN classified as Seyfert 2 based on the lack of broad Balmer lines in the optical spectral regime. We found three of these (9$\%$) to show broad components (FWHM $>\ $ 1200 \kms) in either Pa$\alpha$ and/or Pa$\beta$. An example is shown in Figure \ref{Hidden_BLR_Mrk520}. For Mrk 520 and NGC 5252 we detected the broad component of both Pa$\alpha$ and Pa$\beta$ and also of \hei\ 1.083 $\mu$m.  
For NGC 5231, we detected broad Pa$\alpha$, but the broad Pa$\beta$ component is undetected.
For all three galaxies, we did not observe a broad component in the other Paschen lines.
We do not detect a broad component in the Paschen lines for 16/\NSytwo\ (49$\%$) Seyfert 2 galaxies, in agreement with their optical measurements. For the remaining 14/\NSytwo\ (42$\%$) sources, we could not detect the broad components because the line is in a region with significant sky features.

\begin{figure}
\centering
\subfigure{\includegraphics[width=0.49\textwidth]{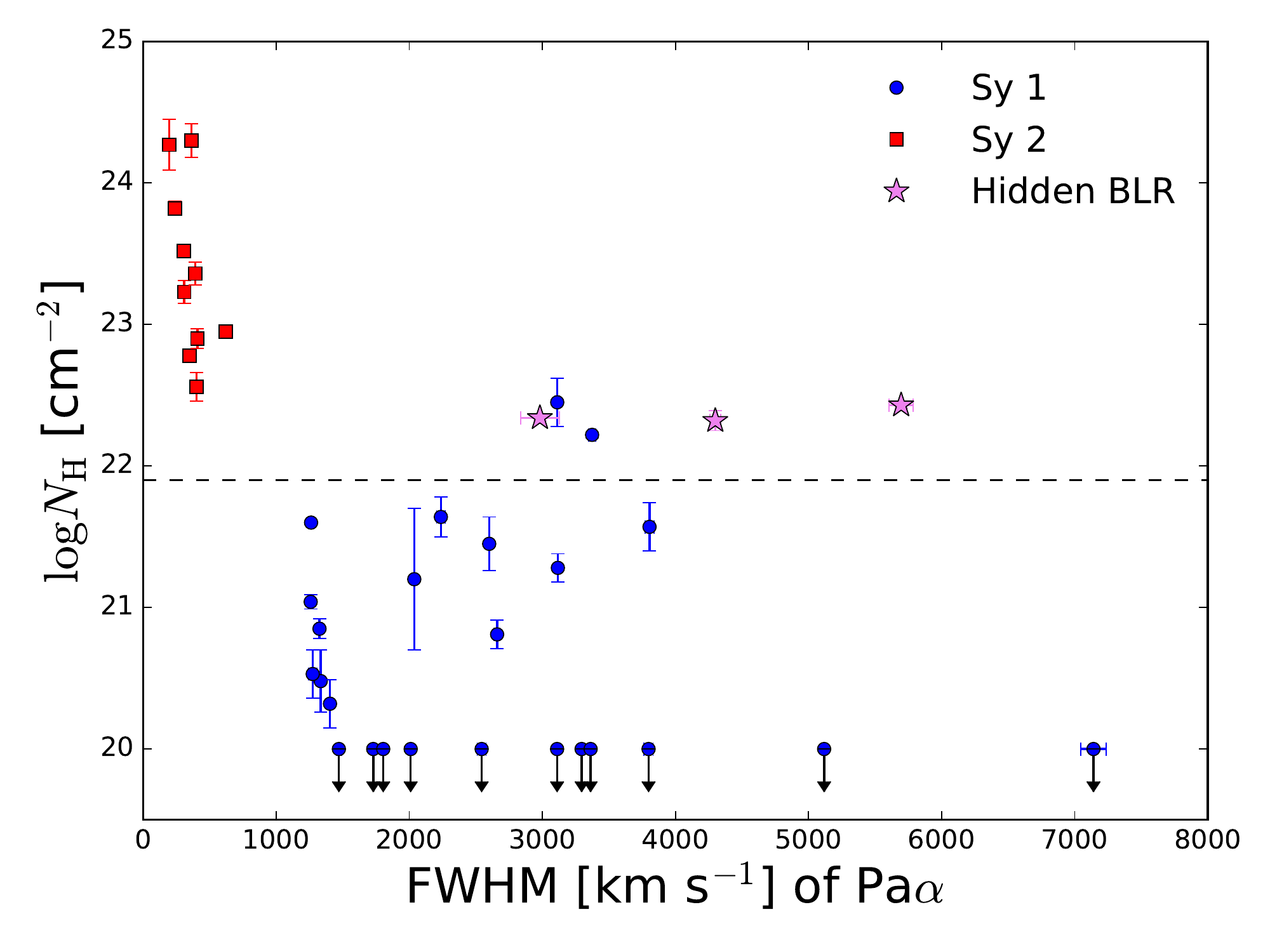}}
\caption{Distribution of column density $N_{\rm H}$ as a function of the FWHM of Pa$\alpha$. Blue points are Seyfert 1 which show a BLR in \Halpha, red squares are Seyfert 2  and violet stars are the Seyfert 2 with narrow lines in \Halpha\ that show a hidden BLR in the Pa$\alpha$. The dashed line shows the threshold $N_{\rm H}$ value that separates optical Seyfert 1 and Seyfert 2 using the \Hbeta\ line (Koss et al., submitted).}

\label{NH_vs_FWHM}
\end{figure}

We considered the values of $N_{\rm H}$ measured by \cite{Ricci2015} and Ricci et al., submitted. The Seyfert 2s in our sample have column densities in the range $\log N_{\rm H} = 21.3-25.1$ \nhunit\ with a median of $\log N_{\rm H} = 23.3$ \nhunit.
The three AGN showing a `hidden' BLR in the NIR belong to the bottom 11 percentile in $\log N_{\rm H}$ (Figure \ref{NH_vs_FWHM}) among the Seyfert 2s in our sample ($\log N_{\rm H} <\ 22.4$ \nhunit).

Next we considered the optically identified Seyfert 1 in our sample, and we investigated the presence of broad lines in their NIR spectra. In the Seyfert 1-1.5 in our sample, we did not find spectra that lack the broad Pa$\alpha$ or Pa$\beta$ component. 

For the optical Seyfert 1.9 in our sample, the NIR spectra in general show broad lines except for some objects with weak optical broad lines. We found broad Paschen lines in 10/23 (44$\%$) Seyfert 1.9. For 6/23 (26$\%$) of Seyfert 1.9  the spectra do not cover the Pa$\alpha$ and Pa$\beta$ regions. There are 7/23 Seyfert 1.9 (30$\%$) that do not show broad components in Pa$\alpha$ and Pa$\beta$. All these galaxies have $\log \NH >\ 21.5$ \nhunit. Four of them have a weak broad component of H$\alpha$ compared to the continuum (EQW[bH$\alpha] <\ 36$ \AA ). Further higher sensitivity studies are needed to test whether the  broad H$\alpha$ component is real or is a feature such as a blue wing. 

\
\subsection{Velocity dispersion}
\label{sec:veldisp}

We measured the stellar velocity dispersion (\sig) from the CaT region and from the CO band-heads in the H-band and in the K-band. The results are tabulated in Table \ref{tab:ppxf}. Example of \sig\ fitting are available in Appendix \ref{sec:comp_vel}. In total, we could measure \sig\ only for 10$\%$ (10/\Ntot) of objects using the CaT region.  The main reasons for this are:
lack of wavelength coverage of the CaT region, absorption lines too weak to be detected, and presence of strong Paschen emission lines (Pa10 to Pa16) in the same wavelength of the CaT. 
We have a good measurement of \sig\ for 31/\Ntot\ (30$\%$) objects from the CO band-heads in the K-band and for 54/\Ntot\ (53$\%$) from the CO band-heads in the H-band.  
We found that \sig\ measured from the CaT region and from the CO band-heads (H- and K-band) are in good agreement (median difference 0.03 dex). However, we compared these measurements with literature values of \sig\ measured in the optical and we found that the \sig\ measured from the NIR absorption lines are $\sim$ 30 \kms\ larger, on average, than the \sig\ measured in the optical range (median difference 0.09 dex).
Further details about the comparison of \sig\ measured in the different NIR spectral regions and in the optical is provided in Appendix \ref{sec:comp_vel}.

\subsection{Black hole masses}
\label{sec:MBH}

We derived the black holes masses from the velocity dispersion and from the broad Paschen lines. For the cases where we have both $\sigma_{*, \rm CO}$ and $\sigma_{*, \rm CaT}$, we use $\sigma_{*, \rm CO}$, since this method could be used for more sources with higher accuracy. We used the following relation from \citet{Kormendy2013} to estimate the \MBH\ from \sig :
\begin{equation}
\log{\left(\frac{M_{\rm BH}}{M_\odot}\right)}=4.38 \times \log{\left(\frac{\sigma_{*}}{200 \text{ \kmps}}\right)} + 8.49 \,\ .
\end{equation}
\\
This relation has an intrinsic scatter on $\log$ \MBH\ of 0.29 $\pm$ 0.03.
There are different studies that derived prescriptions to estimate \MBH\ from the broad Paschen lines (Pa$\alpha$ or \Pab). All these methods use the FWHM of the Paschen lines Pa$\alpha$ or Pa$\beta$. Since Pa$\alpha$ is near a region of atmospheric absorption, we have more measurements of the broad component of Pa$\beta$, therefore we prefer to use this line to derive \MBH. As an estimator of the radius of the BLR, previous studies used the luminosity of the 1 $\mu$m continuum \citep{Landt2011a}, the luminosity of  broad Pa$\alpha$ or Pa$\beta$ (\citealt{Kim2010}, \citealt{LaFranca2015}) or the hard X-ray luminosity \citep{LaFranca2015}. We use the luminosity of broad Pa$\beta$, since the continuum luminosity at 1 $\mu$m can be contaminated by emission from stars and hot dust and the hard X-ray luminosity is not observed simultaneously with the Paschen lines. We use the following formula from \cite{LaFranca2015} which was calibrated against reverberation measured masses assuming a virial factor $f=4.31$:

\begin{equation}
\begin{split}
\frac{M_{\rm BH}}{M_{\odot}} = 10^{7.83\pm0.03} \left(\frac{L_{\rm Pa\beta}}{10^{40} \text{ erg s$^{-1}$}} \right)^{0.436\pm0.02}\\ \times
\left(\frac{\text{FWHM}_{\rm Pa\beta}}{10^4 \text{ km s$^{-1}$}} \right)^{1.74\pm0.08} \,\ .
\end{split}
\label{eq:LaFranca}
\end{equation}
\\
This relation have an intrinsic scatter on $\log$ \MBH\ of 0.27.
The values of \MBH\ are listed in Table \ref{tab:MBH}. For 46/\Ntot\ (45$\%$) AGN we have the  \MBH\ from $\sigma_{*,\rm CO}$ and for 34/\Ntot\ (33$\%$) we have the \MBH\ from \Pab. Considering the cases where we have both measurements, we have \MBH\ measurements for 69/\Ntot\ (68$\%$) AGN in our sample. 
Using 11 objects that have both $\sigma_{*, \rm CO}$ and Pa$\beta$ measurements of \MBH, we found that \MBH\ derived from $\sigma_{*, \rm CO}$ are larger than \MBH\ derived from broad lines by up to 1 dex.
A detailed discussion of the results using different black hole mass estimators and comparison to mass measurements in the optical is provided in the Appendix Section \ref{sec:MBH_comp}.

\subsection{NIR emission line diagnostic diagram}
\label{sec:NIR_diagn}
\begin{figure}
\centering
\subfigure{\includegraphics[width=0.49\textwidth]{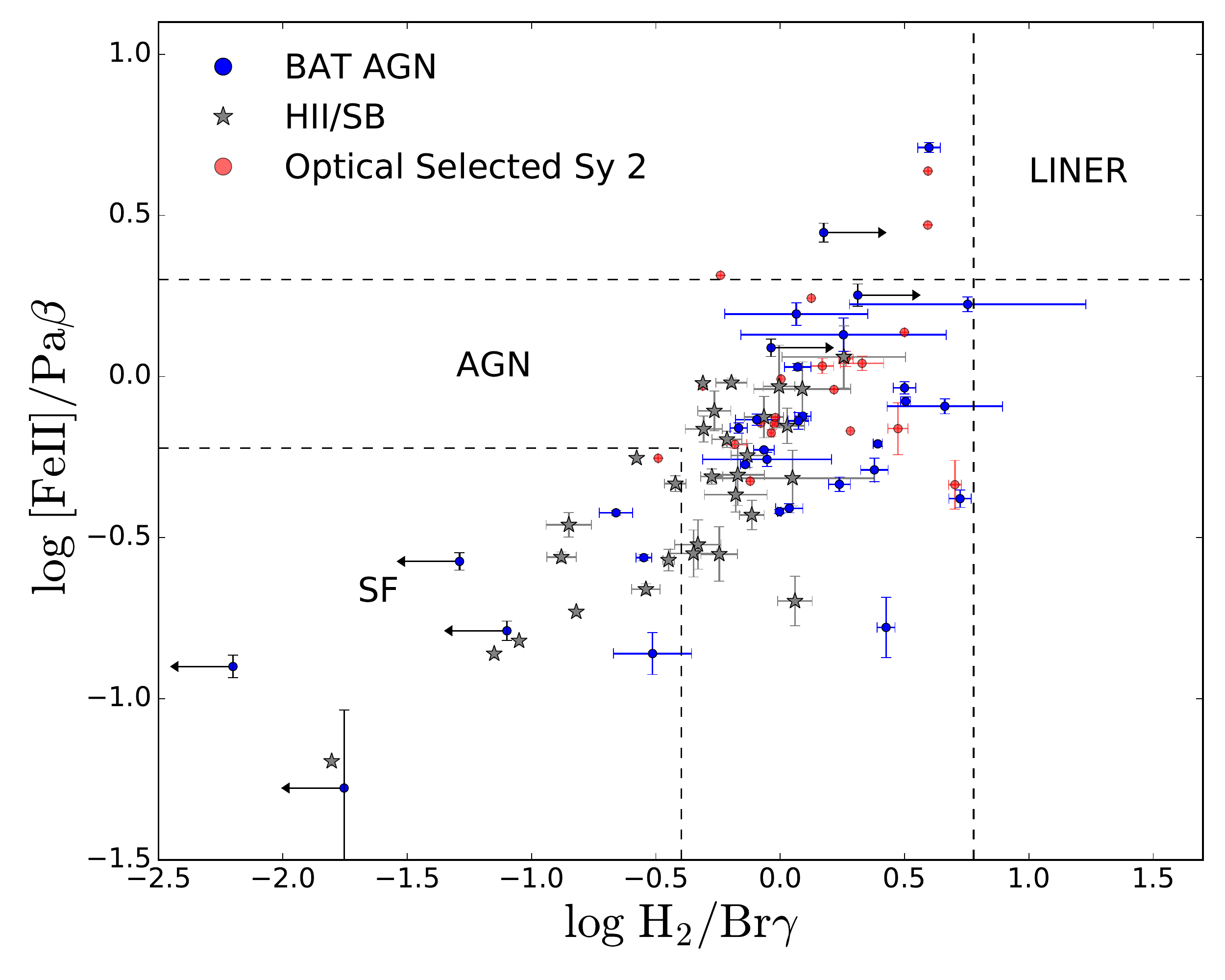} }
\caption{NIR emission lines diagnostic diagram from \citet{Riffel2013}. In blue are the AGN of our sample.
In grey are shown the sample of starburst galaxies (SBGs) from \citet{Riffel2006}; the sample of SBGs and HII region from \citet{Larkin1998}, \citet{Dale2004}, and \citet{Martins2013}. In red are shown the sample of Seyfert 2 from \citet{Riffel2006} and \citet{Veilleux1997}.}
\label{NIR_diag}
\end{figure}

NIR emission line diagnostics \citep[e.g.,][]{Riffel2013} use combinations of NIR lines (\feiif\ 1.257$\mu$m and Pa$\beta$ in the J-band, H$_2$ 2.12$\mu$m and Br$\gamma$ in the K-band) to identify AGN. We detected the four lines needed for the diagnostic in 25$\%$ (25/\Ntot) of AGN (Figure \ref{NIR_diag}).
Most of these objects (88$\%$, 22/25) are identified as AGN in the NIR diagnostic diagram, whereas three objects lie in the SF region. 

Moreover, star-forming galaxies from past studies (\citealt{Larkin1998, Dale2004, Martins2013}) overlap with AGN in the diagram. Considering the AGN in our sample, and SFGs and AGN line ratios from the literature, the fraction of SFGs in the AGN region is 20/30 ($67 \pm 13 \%$), whereas the fraction of AGN in the AGN region is 45/53 ($85 \pm 7 \%$).

For the AGN where we do not detect either H$_2$ or Br$\gamma$, we used upper limits.  All the three AGN that have upper limits on Br$\gamma$  are in the AGN region of the diagram, although if we consider the upper limit, they could also be in the LINER region. All the four objects with no detection of H$_2$ are in the star-forming region. We found that all the nine Seyfert 2 galaxies where emission lines were detected are selected as AGN, whereas the objects in the SF region are broad lines objects. 

The result of the Kolmogorov-Smirnov test is that the distributions of \feiif/Pa$\beta$ for AGN and SFGs are consistent with being drawn from the same distribution (p-value = 0.063), whereas distribution of  H2/Br$\gamma$ are significantly different (p-value = 0.00024).  Thus the SF line ratios \feiif/Pa$\beta$  are indistinguishable from AGN.

\subsection{Coronal lines}
\label{sec:coronal_lines}

\begin{figure}
\centering
\subfigure{\includegraphics[width=0.49\textwidth]{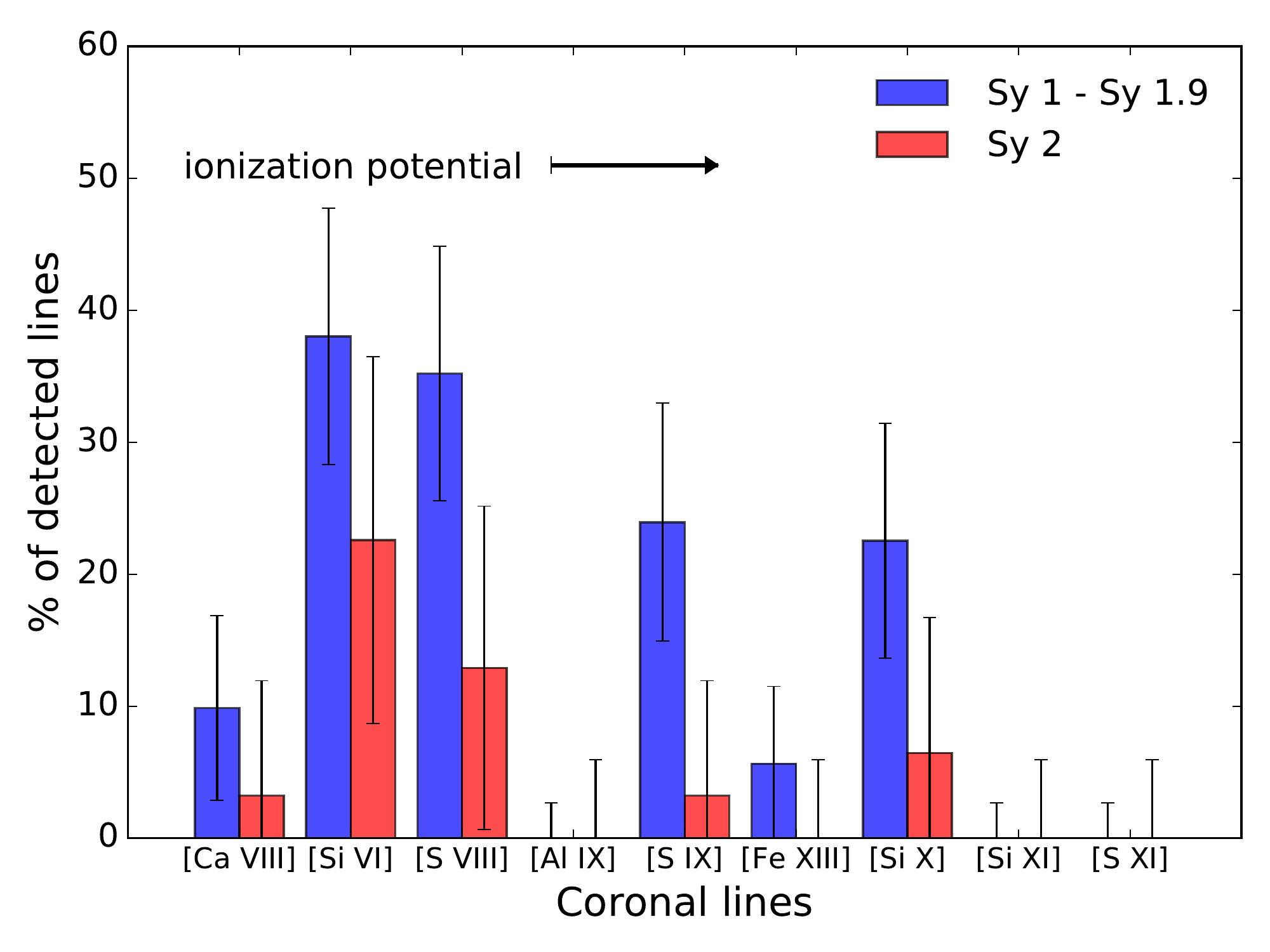} }
\caption{Percentage of coronal lines detections in Seyfert 1 and Seyfert 2. The coronal lines are ordered by increasing ionization potential (IP) from left to right (\Caviii: 127.7 eV, \Silivi: 166.8 eV, \Sviii: 280.9 eV, \Alix: 284.6 eV, \Six: 328.2 eV, \Fexiii: 330.8 eV, \Silix: 351.1 eV, \Silixi: 401.4 eV, \Sxi: 447.1 eV).}
\label{histo_coronal}
\end{figure}

In this section we report the number of CL detections in our sample. Table \ref{coronal_lines} in the Appendix provides a list of the fitted lines.  Figure \ref{histo_coronal} shows the percentage of spectra in which we detected each CL, divided by Seyfert 1 and Seyfert 2. 

We observe a trend for the number of CL detections to increase with decreasing ionization potential (IP). We found that the CL with the highest number of detections is \Silivi\ (34 detections, 33$\%$), followed by \Sviii\ (29 detections, 28$\%$). The \Caviii\ line does not follow this trend, since it is detected only in 8 objects (8$\%$). This line has the lowest IP (127.7 eV) among the CLs. There are three CLs (\Alix, \Silixi\ and \Sxi) that are not detected in any AGN spectra. \Silixi\ and \Sxi\ have the highest IPs among the CLs, while \Alix\ has an intermediate value.

We detect at least one CL in 44/\Ntot\ (43$\%$)  spectra in our sample, but only 18/\Ntot\ (18$\%$) have more than 2 CLs detected. Considering Seyfert 1 and Seyfert 2 separately, the percentage of objects with at least one CL detection is higher in Seyfert 1 ($53\pm 10 \%$) than in Seyfert 2 ($20 \pm 10 \%$).

\subsection{Coronal lines and X-ray emission}
\label{sec:flux_comp}

In order to test whether the strength of CL emission is stronger in Seyfert 1 than in Seyfert 2 for the same intrinsic AGN bolometric luminosity measured from the X-rays, we applied a survival analysis to take into account the fact that our data contains a number of upper limits 41\% (42/102) as well as emission line regions that were excluded because of atmospheric absorption 12\% (12/102).
We used the ASURV package \citep{Feigelson1985}  which applies the principles of survival analysis.
Specifically, we measure the ratio of \Silivi\ emission divided by the 14-195 keV X-ray emission which is a proxy for the AGN bolometric luminosity \cite[e.g.,][]{Vasudevan2009}.  We then compare the distributions of ratios for Seyfert 1 and Seyfert 2 using the ASURV Two Sample tests.  We find that the ratio of CL emission to X-ray emission in Sy 1 is significantly different than in Sy 2 at the less than 1$\%$ level, in the various survival analysis measures (e.g., Gehan's Generalized Wilcoxon Test, Logrank Test etc.). This means that for the same X-ray luminosity, CL emission is stronger in Sy 1 than in Sy 2, consistent with the higher detection fraction of CL in Sy 1.

\begin{figure*}
\centering
\subfigure{\includegraphics[width=0.9\textwidth]{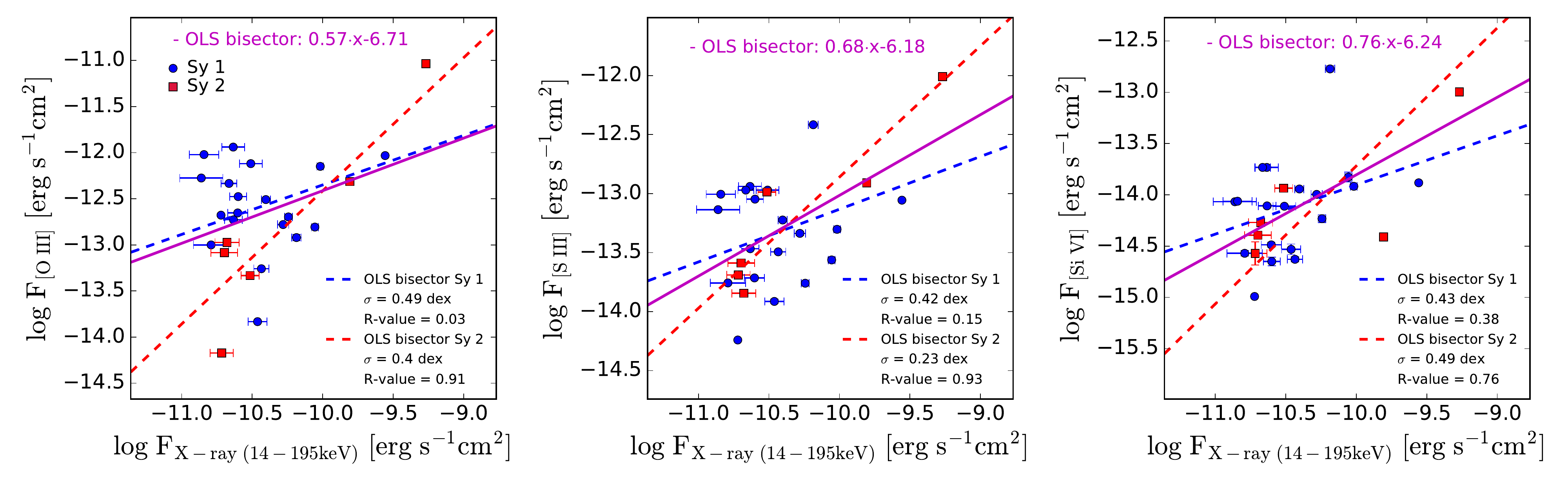} }

\subfigure{\includegraphics[width=0.9\textwidth]{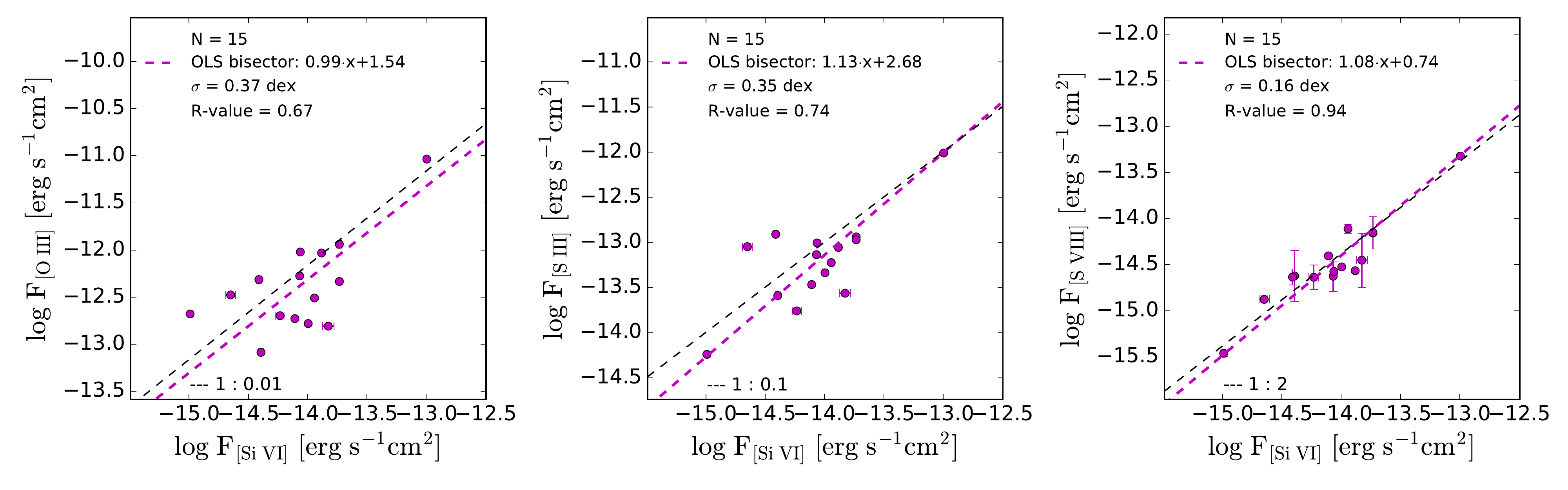} }

\subfigure{\includegraphics[width=0.9\textwidth]{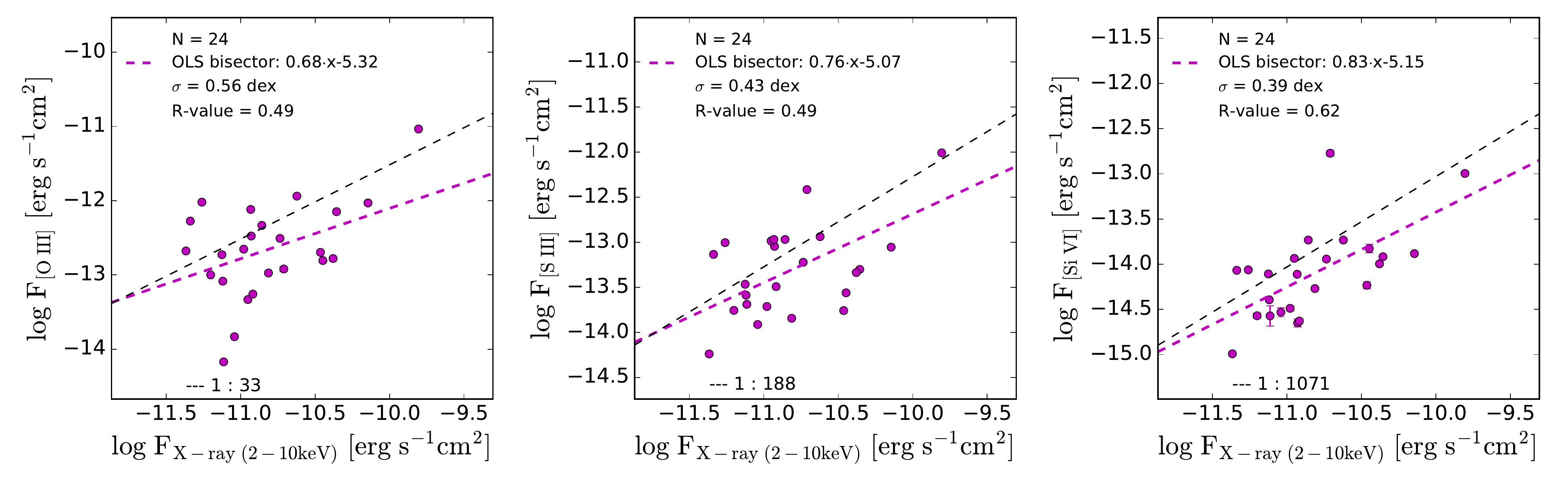} }

\subfigure{\includegraphics[width=0.9\textwidth]{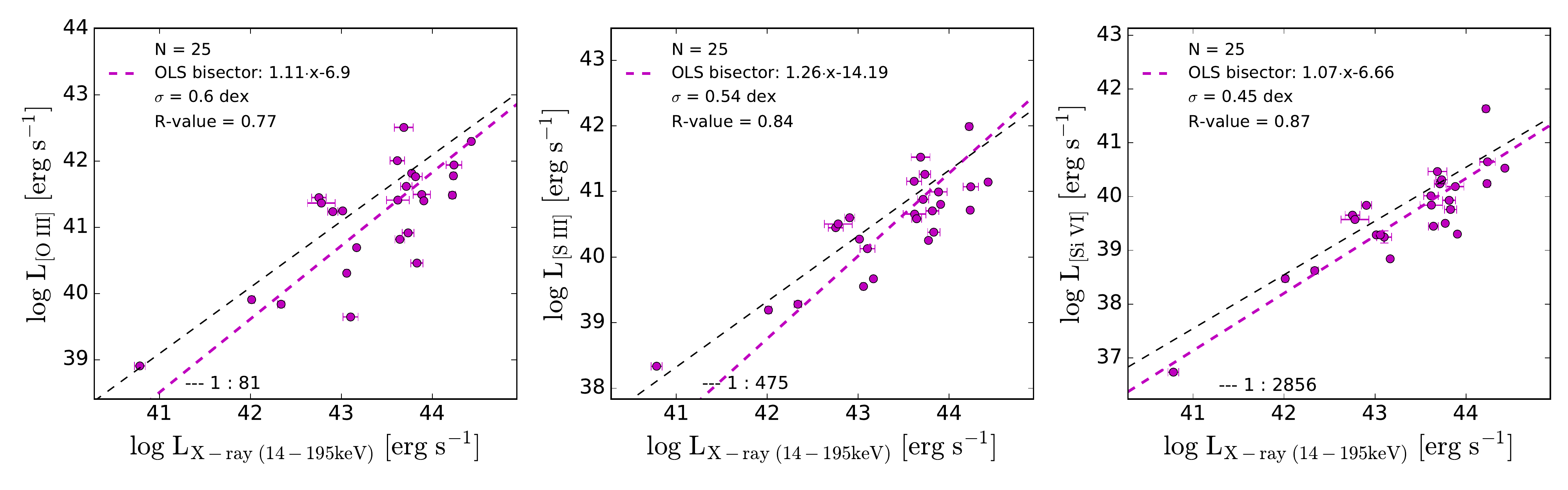} }

\caption{Plots comparing the scatter for a sample of AGN with line detections in each line to compare the overall scatter.  Blue points are Seyfert 1 and red squares are Seyfert 2.  The black dashed lines have slope 1 and are scaled according to the mean flux ratio. The magenta line is the ordinary least squares (OLS) bisector fit for the entire sample. \textit{Upper panels:} Relation of the hard X-ray emission (14-195 keV) with \oiii\ (left), \siiif\ $\lambda$ 0.9531 $\mu$m (middle), and \Silivi\ (right).  
\textit{Second row panels:} Relation of the \Silivi\ flux with \oiii\ (left), \siiif\ (middle) and \Sviii\ (right) for the same sample of AGN. 
\textit{Third row panels:} Relation of the hard X-ray flux (2-10 keV) with \oiii\ (left), \siiif\ (middle), and \Silivi\ (right) for the same sample of AGN. 
\textit{Bottom row panels:} Relation of the hard X-ray luminosity (14-195 keV) with \oiii\ (left), \siiif\ (middle), and with \Silivi\ (right) for the same sample of AGN.}

\label{Comp_SiVI_Xray}
\end{figure*}

We study the correlation between the coronal line and the hard X-ray continuum (14-195 keV) emission (Figure \ref{Comp_SiVI_Xray}). We focus on the \Silivi\ 1.962 $\mu$m  emission line (IP = 166.8 eV), because it is the coronal line which is detected in the largest number of spectra.
We fit the data using the ordinary least squares (OLS) bisector fit method as recommended when there are uncertainties in both the X and Y data \citep{Isobe1990}. The OLS fitting method takes into account the uncertainties in both fluxes. The relation between the \Silivi\ and the X-ray emission (14-195 keV) shows a scatter of $\sigma$ = 0.39 dex and a Pearson correlation coefficient $R_{\rm pear}$ = 0.57 ($p_{\rm Pear}=0.0005$). The relation between the \oiii\ flux and the X-ray flux shows a scatter of $\sigma$ = 0.53 dex and a Pearson correlation coefficient $R_{\rm pear}$ = 0.45 ($p_{\rm Pear}$ $= 0.009$). We ran a Fischer Z-test to investigate whether the intrinsic scatter in the correlation between \Silivi\ and X-ray emission is significantly better than the one between \oiii\ and X-ray emission and found a p-value = 0.18, suggesting the two correlations are not significantly different. Using the 2-10 keV intrinsic emission, the correlation with \Silivi\ is similar as at 14-195 keV ($\sigma = 0.40$, $R_{\rm Pear}$ = 0.58, $p_{\rm Pear}$ = 0.0004). We note that the uncertainties on the 2-10 keV flux are not provided, but are dominated by the difficulties in measuring the column density. Therefore for the fit we took into account only the the uncertainties on the line fluxes.

We compared also the flux of \Silivi\ and \oiii.
We found a scatter of $\sigma = 0.60$ that is slightly larger than the scatter in the \Silivi\ vs. hard X-ray ($\sigma = 0.51$). The correlation coefficient is $R_{\rm Pear}$ = 0.50 ($p_{\rm Pear}$ =  0.003). We ran the Z-test and we found p-value = 0.76. Thus, \Silivi\ correlates with \oiii\ almost at the same level as with the hard X-ray.  We found that the correlation between coronal lines is stronger than the one between \Silivi\ and \oiii. For example the correlation of \Silivi\ with \Sviii\ has a correlation coefficient $R_{\rm Pear}$ = 0.88 ($p_{\rm Pear}$ = $2.1\cdot 10^{-7}$), that is significantly stronger than the one with \oiii\ (Z-test $p$-value = $8.3 \cdot 10^{-5}$).

We also test the correlations in Seyfert 1 and Seyfert 2 separately. For the six Seyfert 2 with good flux measurements of \Silivi\ and \oiii, the relation between the \Silivi\ flux and the BAT X-ray flux shows a scatter of $\sigma$ = 0.27 dex and a Pearson correlation coefficient $R_{\rm Pear}$ = 0.59 ($p_{\rm Pear}$ = 0.22), whereas the relation between the  \oiii\ and X-ray emission shows a scatter of $\sigma$ = 0.65 dex and a Pearson correlation coefficient $R_{\rm Pear}$ = 0.53 ($p_{\rm Pear}$ = 0.27). A Fischer Z-test gives a p-value = 0.84, meaning that the two correlations are not significantly different. We have 27 Seyfert 1 galaxies with good flux measurements of \Silivi\ and \oiii. For these sources, we found the Pearson correlation coefficient $R_{\rm Pear} = 0.58$ ($p_{\rm Pear}$ = 0.001) in the comparison between \Silivi\ and hard X-ray and $R_{\rm Pear} = 0.43$ ($p_{\rm Pear}$ = 0.02) in the comparison between \oiii\ and hard X-ray. We ran the Z-test and we found a $p$-value = 0.16, meaning that the two correlations are not significantly different.

Next, we considered the correlation between \oiii, \Silivi\ and hard X-ray luminosity (14-195 keV).
The correlation coefficient of X-ray luminosity with the \Silivi\ luminosity ($R_{\rm Pear}$ = 0.86) is higher than the correlation coefficient with the \oiii\ luminosity ($R_{\rm Pear}$ = 0.74). The Fischer Z-test gives a p-value = 0.006, meaning that the two correlations are significantly different.  This suggests that the coronal line luminosity does show a significantly reduced scatter than the \oiii\ luminosity when compared to the X-ray emission.

 We considered also \siiif $\lambda$ 0.9531 $\mu$m, which is a NIR emission line with lower ionization potential (IP = 23.3 eV).  
The correlation of the hard X-ray with \Silivi\ luminosity ($R_{\rm Pear}$ = 0.87) is similar to the correlation with \siiif\ luminosity ($R_{\rm Pear}$ = 0.85). For the 26 sources which have both the \Silivi\ and \siiif\ line detections, the Fischer Z-test gives a $p$-value = 0.5, meaning that the two correlations are not significantly different.

For the 24 sources which also have 2-10 keV luminosity measurement,  the correlation of the 2-10 keV luminosity with \Silivi\ ($R_{\rm Pear}$ = 0.90, $\sigma$ = 0.39 dex) is stronger than the correlation with \siiif\ luminosity ($R_{\rm Pear}$ = 0.84, $\sigma$ = 0.52 dex). However, the Fischer Z-test gives a $p$-value = 0.08, suggesting that the \Silivi\ luminosity does not have a significantly reduced scatter than the \siiif\ luminosity when compared to the X-ray emission.

\begin{table*}
\caption{Relations of coronal lines (\Silivi\ and \Sviii) and \oiii\ and \siiif\ emission with the 14-195 keV emission and 2-10 keV emission.}
\begin{tabular}{|l|l|l|c|c|c|c|c}
\hline
Flux 1 & Flux 2 & Sample & \multicolumn{1}{c|}{N} & $\sigma$ [dex]& $R_{\rm Pear}$ & $p_{\rm Pear}$ & $p$-value \\
&  & (1) &(2) & (3) & (4) & (5) & (6) \\ \hline \hline

  &    &  Sy1  & 62 & 0.53 & 0.57 &  $1.1\cdot 10 ^{-6}$ & - \\  
{\oiii} $\lambda$0.5007 $\mu$m &  X-ray 14-195 keV  &  Sy2  & 26 & 0.75 & 0.41 & 0.04 & - \\  
  &    &  all  & 88 & 0.67 & 0.49 &  $1.3\cdot 10 ^{-6}$ & - \\ \hline
 &  & Sy1 & 26 & 0.41 & 0.58 & 0.001&0.16\\ 
{\Silivi} $\lambda$1.962 $\mu$m & X-ray 14-195 keV & Sy2 & 7 & 0.26 & 0.54 & 0.21 & 0.84\\ 
 &  & all & 33 & 0.39 & 0.57 & 0.0004 &0.18\\ \hline
 &  & Sy1 & 25 & 0.48 & 0.57 & 0.003&0.88\\ 
{\Sviii} $\lambda$ 0.9915 $\mu$m & X-ray 14-195 keV & Sy2 & 4 & 0.02 & 0.98 & 0.02& (sample too small) \\ 
 &  & all & 29 & 0.48 & 0.53 & 0.003& 0.85 \\\hline
   &    &  Sy1  & 48 & 0.42 & 0.64 & $1.0\cdot 10 ^{-6}$ &  0.65\\ 
{\siiif} $\lambda$ 0.9531 $\mu$m  &  X-ray 14-195 keV  &  Sy2  & 17 & 0.41 & 0.44 & 0.07 &  0.47 \\
  &    &  all  & 65 & 0.44 & 0.59 & \multicolumn{1}{l}{ $1.8\cdot 10 ^{-7}$} & 0.42
 \\ \hline

 &  &  & \multicolumn{1}{l|}{} & \multicolumn{1}{l|}{} & \multicolumn{1}{l|}{} \\ \hline
  &    &  Sy1  & 58 & 0.53 & 0.59 &  $1.2\cdot 10 ^{-6}$ & - \\ 
{\oiii} $\lambda$0.5007 $\mu$m &  X-ray 2-10 keV  &  Sy2  & 25 & 0.57 & 0.65 & 0.0004 & -  \\ 
  &    &  all  & 83 & 0.64 & 0.58 &  $1.1\cdot 10 ^{-8}$ &  -\\ \hline
 &  & Sy1 & 26 & 0.38 & 0.66 & 0.0002&0.04\\
{\Silivi} $\lambda$1.962 $\mu$m & X-ray 2-10 keV & Sy2 & 6 & 0.36 & 0.11 & 0.83 &0.01\\ 
 &  & all & 32 & 0.40 & 0.58 & 0.0004 &0.47 \\ \hline
 &  & Sy1 & 24 & 0.44 & 0.69 & 0.0002 &0.20\\ 
{\Sviii} $\lambda$ 0.9915 $\mu$m & X-ray 2-10 keV & Sy2 & 3 & 0.121 & 0.33& 0.79 &(sample too small)\\ 
 &  & all & 27 & 0.47 & 0.61 & 0.0007&0.87\\ \hline
 
   &    &  Sy1  & 42 & 0.43 & 0.55 & 0.0001 & 0.85 \\ 
{\siiif} $\lambda$ 0.9531 $\mu$m  &  X-ray 2-10 keV  &  Sy2  & 15 & 0.39 & 0.70 & 0.003 & 0.96 \\ 
  &    &  all  & 57 & 0.42 & 0.6 &  $1.2\cdot 10 ^{-6}$ & 0.38\\ 
 \hline
 
 &  &  & \multicolumn{1}{l|}{} & \multicolumn{1}{l|}{} & \multicolumn{1}{l|}{} \\ \hline
{\oiii} $\lambda$0.5007 $\mu$m  &  {\Silivi} $\lambda$1.962 $\mu$m &  all  & 33 & 0.60 & 0.5 & 0.003 \\ 
{\Sviii} $\lambda$ 0.9915 $\mu$m  &  {\Silivi} $\lambda$1.962 $\mu$m  &  all  & 22 & 0.20 & 0.88 &  $9.0\cdot 10 ^{-8}$ \\ 
{\siiif} $\lambda$ 0.9531 $\mu$m  & {\Silivi} $\lambda$1.962 $\mu$m & all & 26 & 0.33 & 0.80 & $1.1\cdot 10 ^{-6}$\\ \hline

\end{tabular}
\tablecomments{(1) AGN type; (2) size of the common sample in which Flux 1, Flux 2  and \oiii\ flux are available; (3) standard deviation and (4) Pearson R coefficient of the $\log$ Flux1- $\log$ Flux 2 relation; (5) Pearson $p$-value coefficient of the $\log$ Flux1- $\log$ Flux 2 relation; (6) $p$-value of the null hypothesis that this correlation coefficient and the \oiii correlation coefficient obtained from independent parent samples are equal. All coefficients are measured for the sample which have both coronal line emission and \oiii\ observations.)}
\label{tab:correlation_results}
\end{table*}

\begin{table*}
\caption{Relations of coronal lines (\Silivi\ and \Sviii), \siiif\ and \oiii\ luminosities with the 14-195 keV emission and 2-10 keV emission.}
\begin{tabular}{|l|l|l|c|c|c|c|c}
\hline
Luminosity 1 & Luminosity 2 &  &N & $\sigma$ [dex]& $R_{\rm Pear}$ & $p_{\rm Pear}$ & $p$-value \\
&  &  &(1) &(2) & (3) & (4) & (5) \\ \hline \hline

 {\oiii} $\lambda$0.5007 $\mu$m  &  X-ray 14-195 keV  &  all  & 88 & 0.67 & 0.73 &  $8.7\cdot 10 ^{-16}$  & - \\ \hline
{\Silivi} $\lambda$1.962 $\mu$m  &  X-ray 14-195 keV  &  all  & 34 & 0.41 & 0.86 & $5.1\cdot 10 ^{-11}$  &  0.006\\ \hline
{\Sviii} $\lambda$ 0.9915 $\mu$m  &  X-ray 14-195 keV  &  all  & 29 & 0.49 & 0.83 & $2.9\cdot 10 ^{-8}$  & 0.916 \\ \hline 
{\siiif} $\lambda$ 0.9531 $\mu$m   &  X-ray 14-195 keV  &  all  & 65 & 0.48 & 0.83 & $2.5\cdot 10 ^{-17}$  & 0.002\\ \hline

 &  &  & \multicolumn{1}{l|}{} & \multicolumn{1}{l|}{} & \multicolumn{1}{l|}{} \\ \hline

{\oiii} $\lambda$0.5007 $\mu$m  &  X-ray 2-10 keV  &  all  & 83 & 0.64 & 0.75 & $1.9\cdot 10 ^{-16}$   &  -\\ \hline 
{\Silivi} $\lambda$1.962 $\mu$m  &  X-ray 2-10 keV  &  all  & 32 & 0.41 & 0.87 & $1.6\cdot 10^{-10}$  &  0.019\\ \hline 
{\Sviii} $\lambda$ 0.9915 $\mu$m  &  X-ray 2-10 keV  &  all  & 27 & 0.46 & 0.86 & $1.1\cdot 10 ^{-8}$  &  0.744\\ \hline
{\siiif} $\lambda$ 0.9531 $\mu$m   &  X-ray 2-10 keV  &  all  & 57 & 0.49 & 0.84 & $2.1\cdot 10 ^{-16}$  &  0.002\\ \hline

\end{tabular}
\tablecomments{(1) size of the common sample in which both Luminosity 1 and Luminosity 2 are available; (2) standard deviation and (3) Pearson R coefficient of the $\log$ Flux1- $\log$ Flux 2 relation; (4) Pearson $p$-value coefficient of the $\log$ Flux1- $\log$ Flux 2 relation; (5) $p$-value of the null hypothesis that this correlation coefficient and the \oiii\ correlation coefficient obtained from independent parent samples are equal.}
\label{tab:lum_correlation_results}
\end{table*}

\begin{figure*}
\centering
\subfigure{\includegraphics[width=1\textwidth]{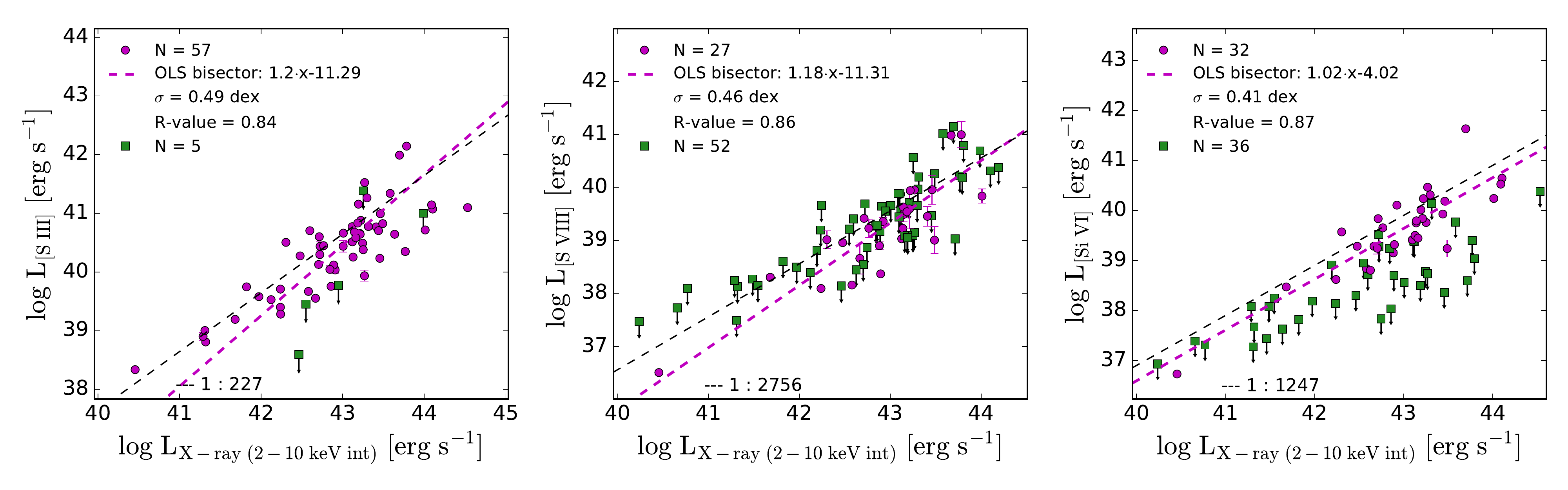} }
\subfigure{\includegraphics[width=1\textwidth]{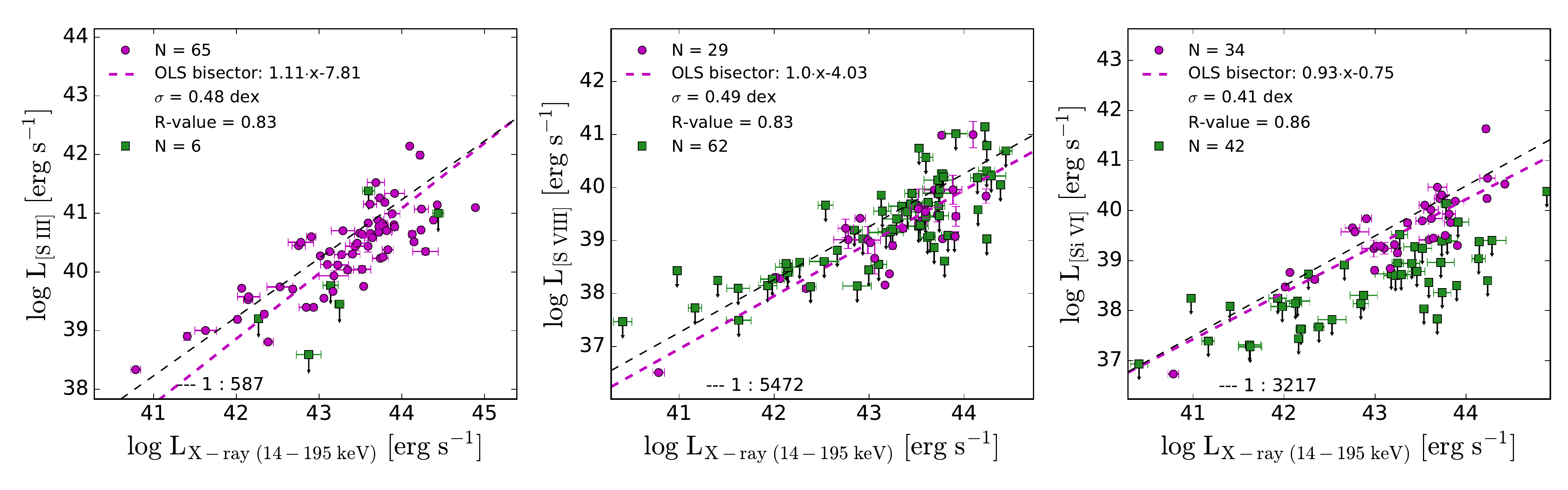} }

\caption{\textit{Upper panels:} Relation of the 2-10 keV X-ray luminosity with \siiif\ (left), \Sviii\ (middle), and \Silivi\ (right). The magenta dashed lines show the OLS bisector fit. The black dashed lines have slope 1 and are scaled according to the mean flux ratio. Upper limits (green squares) are not included in the measurements of the OLS bisector fit and mean flux ratio.
\textit{Lower panels:} Relation of the 14-195 keV X-ray luminosity (14-195 keV) with \siiif\ (left), \Sviii\ (middle), and \Silivi\ (right).}
\label{Comp_SiVI_Xray_lum}
\end{figure*}

\begin{table*}
\caption{Relations of coronal lines and \siiif\ luminosity with the 14-195 keV emission and 2-10 keV emission.}
\begin{tabular}{|l|c|c|c|c|c|c|c|c|c|c}
\hline
Line &  & &2-10 keV &  &  & & 14-195 keV &  &  \\
 &N$_{\text{undet}}$ & N$_{\text{det}}$&  X-ray to line& slope & intercept &N$_{\text{det}}$ &X-ray to line& slope & intercept \\
  & & &  ratio&  & & & ratio & &  \\
&  (1) &(2) & (3) & (4) & (5) & (6) & (7) & (8) & (9)\\ \hline \hline

{\siiif} $\lambda$0.9531 $\mu$m &6 &57  &227 &	1.20$\pm$0.11& 	-11.30$\pm$4.78
	& 65 & 587	&1.11$\pm$0.10 & -7.81$\pm$4.43
\\ \hline
{\Caviii}  $\lambda$2.3210 $\mu$m & 76 & 8& 7913 & 1.00$\pm$0.53 & -4.04$\pm$22.7& 8&	16448&	-0.67$\pm$0.37&	67.74$\pm$16.01
\\ \hline
{\Silivi} $\lambda$1.962 $\mu$m  &42&	32& 1247&	1.02$\pm$0.13& -4.02$\pm$5.92& 34&  3217& 0.83$\pm$0.13	& -0.75$\pm$5.75 \\ \hline
{\Sviii} $\lambda$0.9915 $\mu$m & 62& 27& 2756&	1.18$\pm$0.15& -11.31$\pm$6.52& 29&  5472& 1.00$\pm$0.15&	-4.03$\pm$6.35\\\hline
{\Six} $\lambda$1.2520 $\mu$m &76& 18& 4177& 1.00$\pm$0.22&	-3.92$\pm$9.45&	18&  9239& 0.91$\pm$0.17& -0.16$\pm$7.53\\ \hline
{\Fexiii} $\lambda$1.0747 $\mu$m & 84& 4& 3754& 0.44$\pm$0.70&	20.43$\pm$30.08& 4& 8886& 0.54$\pm$1.01& 15.6$\pm$43.37\\ \hline
{\Silix} $\lambda$1.4300 $\mu$m & 79& 17 & 3481 & 0.96$\pm$0.17& -1.77$\pm$7.42& 18&   8877& 1.02$\pm$0.23&	-5.14$\pm$9.9\\ \hline

\end{tabular}
\tablecomments{(1) number of sources with no line detection in clean atmospheric regions for which the upper limits are measured; (2) size of the sample in which both line luminosity and 2-10 keV intrinsic X-ray luminosity are measured;  (3) mean 2-10 keV X-ray luminosity to line luminosity ratio for the sources with line detection (upper limits are not considered); (4) slope $a$ and (5) intercept $b$ of the ordinary least squares (OLS) bisector fit $\log$ L$_{\rm Line} = a \times \log$ L$_{\rm 2-10\ keV} + b$ for the sources with line detection; (6) size of the sample in which both line luminosity and 14-195 keV X-ray luminosity are available; (7)  mean 14-195 keV X-ray luminosity to line luminosity ratio for the sources with line detection (upper limits are not considered); (8) slope $a$ and (9) intercept $b$ of the OLS bisector fit $\log$ L$_{\rm Line} = a \times \log$ L$_{\rm 14-195\ keV} + b$ for the sources with line detection.}
\label{tab:CL_Xray_ratio}
\end{table*}

\section{Discussion}

\subsection{Comparison of the FWHM of the broad Balmer and Paschen lines}

\cite{Landt2008} observed a trend for the FWHM of H$\beta$ to be larger than the FWHM of Pa$\beta$. They claimed that it might be due to the effect of the  H$\beta$ `red shelf', caused by the emission from the Fe II multiplets. 
This emission on the red side of H$\beta$ can contaminate the broad component of H$\beta$, so that the measured FWHM appear broader.
We compared the FWHM of Pa$\beta$ with the FWHM of H$\beta$, measured taking into account the emission from the \feii\ multiplets and we found that the FWHM of H$\beta$ is not systematically different from the FWHM of Pa$\beta$, suggesting that these two lines are produced in the same region.
Using a larger sample of broad line sources (N=22), we do not find a statistically significant difference in the intrinsic distributions of the FWHM of Pa$\alpha$ with respect to the other broad lines, as \cite{Landt2008} found with a smaller number of broad line AGN (N = 18). We note that the estimates of line FWHM performed in \cite{Landt2008} were done using a different method, where the narrow lines were subtracted before fitting broad lines as opposed to the simultaneous fit in this study.

A similar result is presented in recent paper by \cite{Ricci2016}. They analysed a sample of 39 AGN, and found a good correlation of the FWHM of the broad H$\alpha$, H$\beta$, Pa$\alpha$, Pa$\beta$, and \hei\ $\lambda$1.083 $\mu$m lines, with small offsets. They found less scatter around the relation, and this can be due to the fact that they have more `quasi-simultaneous' observations.

\subsection{Hidden BLR}
We detected broad lines in 3/\NSytwo\ (9$\%$) Seyfert 2 galaxies in our sample.
If we consider also intermediate class AGN (Seyfert 1.8 and Seyfert 1.9) together with Seyfert 2, we found broad lines in 19/62 AGN (31$\%$). This result is similar to the fraction (32$\%$) found by \cite{Onori2014,Onori2016}. They detected  broad emission line components in Pa$\alpha$, Pa$\beta$ or \hei\ in 13/41 obscured nearby AGN (Seyfert 1.8, Seyfert 1.9 and Seyfert 2). There are 8 objects in common between our sample and their sample. We detected broad lines in 2/8 objects, while they detected broad lines in two additional objects. For these two AGN, we did not detect broad lines because the spectra do not have enough S/N or have telluric features in the Pa$\alpha$ and Pa$\beta$ regions.
Also a previous study by \cite{Veilleux1997} found hidden BLR in 9/33 Seyfert 2 galaxies (27$\%$).
Our result suggests that for at least 10$\%$ of Seyfert 2 galaxies, we should expect to detect broad lines in the NIR, that can be used to derive a virial measure of \MBH.
We observed that the detection of broad lines in the NIR for Seyfert 2 galaxies is related to lower column densities. All the AGN with hidden BLRs are part of the bottom 11 percentile in $\log N_{\rm H}$.
This result suggests that the value of \NH\ can provide an indication of the probability to detect broad lines in the NIR.
Two of the AGN with hidden BLR, Mrk 520 and NGC 5231, are in mergers \citep{Koss2011, Koss2012}, while the third one (NGC 5252) shows signs of tidal material related to a merger \citep{Keel2015}. This suggests that the broad emission lines in the optical are obscured by host galaxy dust and not by the nuclear torus.

We also found that 6/22 (27$\%$) Seyfert 1.9 do not show broad lines in the NIR. In most of these AGN the broad H$\alpha$ component is weak as compared to the continuum (EQW[bH$\alpha] <\ 36$ \AA ), but this is unlikely to be related to dust obscuration as the broad lines are not detected in the NIR. We note that for weak broad lines the NIR has less sensitivity than the optical observations around H$\alpha$, unless the lines are significantly obscured.

\subsection{AGN diagnostic}
We tested the NIR diagnostic diagram on our sample of hard X-ray selected AGN.  Previous studies tested this diagnostic method on samples of AGN (Seyfert 2) and SFGs. They found that Seyfert 2 galaxies are clearly identified as AGN \citep{Rodriguez2004,Rodriguez2005,Riffel2013}, whereas the sample of SFGs lies both in the SF and in AGN regions of the diagram \citep{Dale2004,Martins2013}.  
Only 32/\Ntot\ (31$\%$) spectra show the emission lines necessary to apply the diagnostic method. We found that 25/\Ntot\ (25$\%$) objects are identified as AGN and 7/\Ntot\ (7$\%$) are identified as SFGs.  Thus, while it is true that all 9 Seyfert 2 AGN with all detected emission lines in our sample were in the AGN region, the NIR diagnostic diagram is not effective for finding AGN in typical surveys because of contamination by SFR galaxies and the difficulty detecting all the emission lines needed for the NIR AGN diagnostic.

\subsection{Coronal Line Emission}

We observed that in general the number of CLs detections increase with decreasing IP. However, there are some lines that do not follow this trend.
 For instance, the \Caviii\ $\lambda$ 2.321 $\mu$m line has the lowest IP among the CLs, but we consider it difficult to detect for two reasons. First, \Caviii\  is located near the red end of the K-band, and therefore for redshift z $ >\ $ 0.035, it falls outside the wavelength range covered by our spectra. Moreover, the \Caviii\ line is in the same wavelength position of the CO (3-1) absorption line. Therefore the CO (3-1) absorption line can attenuate the strength of the \Caviii\ line, making more difficult to detect it.  Additionally, the \Caviii\ line can be affected by metallicity effects.
There are three CLs (\Alix, \Silixi\ and \Sxi) that are not detected in our spectra. The \Alix\ emission line is affected by metallicity effects and depletion onto dust \citep{Rodriguez2011}.

\cite{Rodriguez2011} found that the non-detections of CL is associated with either loss of spatial resolution or increasing object distance. CL are emitted in the nuclear region and they loose contrast with the continuum stellar light in nearby sources. On the other hand, as the redshift increases the CL emission can be diluted by the strong AGN continuum. However, \cite{Rodriguez2011} claimed that in some AGN the lack of CL may be genuine: it can be due to a very hard AGN ionizing continuum, without photons with energy below a few keV. Using a survival analysis to compare the coronal line emission of Seyfert 1 and Seyfert 2 as compared to the bolometric emission, we do find that Seyfert 1 have more coronal line emission which may be a consequence of the torus obscuring some of the coronal line emission.

\subsection{Coronal lines and X-ray emission}

We found that the scatter in the comparison between the CL flux and hard X-ray flux (14-195 keV or 2-10 keV) is almost the same as the scatter between \oiii\ flux and hard X-ray flux.  However, the relation of the hard X-ray luminosity with \Silivi\  luminosity ($R_{\rm pear} = 0.86$) is stronger than the relation with \oiii\ ($R_{\rm pear} = 0.73$), based on a Z-test.  

In a past study, \cite{Rodriguez2011} also found a correlation between the CL emission and X-ray emission from ROSAT. They found that the large scatter in the correlation is introduced mostly by the Seyfert 2, while Seyfert 1 follow a narrower trend, with a strong correlation between the absorption corrected 2-10 keV luminosity and coronal line luminosity (correlation index of 0.97). We do not observe a stronger correlation between 14-195 keV X-ray flux and CL flux in Seyfert 1 than in  Seyfert 2, and this suggests that the weak correlation between CL emission and X-ray is not caused by obscuration.

The condition of the gas in the galaxy can also affect the CL emission. Factors that can influence the strength of the CLs are the gas electron density (N$_{\rm e}$), its temperature and its ionization state. Moreover, the covering factor of the CL region (CLR), which is defined by the spatial distribution of the gas and by the angular distribution of the ionizing radiation, can also affect the CLs emission \citep{Baskin2005}. 
\cite{Rodriguez2011} estimated the density of the CLR to be between BLR and NLR. \cite{Landt2015} instead found that the CL gas has a relatively low density of N$_{\rm e} = 10^3$ cm$^{-3}$.  A detailed study of the condition of the gas in the CLR is therefore crucial to quantify the impact of all these factors on the CL emission.

A possible explanation for this scatter remains AGN variability.  If CLs are emitted between the BLR and the NLR, we should expect them to be more correlated with the X-ray than \oiii. But if they are located in a region that extends also beyond the inner boundary of the NLR, they can be influenced by variability in a similar way as the NLR.  Until now, the exact location of the CLR is controversial.
\cite{Portilla2008} analysed the NIR spectra of a sample of 54 AGN and found that Seyfert 1 show more CLs than Seyfert 2. Their result suggested that CLs are emitted in an extended region and not on the inner surface of the torus. This is in agreement with the study of \cite{Rodriguez2006}, who found that the size of the CL region extends up to a few hundreds of parsecs.  \cite{Landt2015}, analysing the CL region in NGC 4151, found instead that extension of the CL gas is beyond the inner face of the torus.

\subsection{Outlook for \textit{James Webb Space Telescope} AGN Surveys}

Our survey provides a comprehensive census and legacy database of the nearest X-ray detected AGN with spectroscopic measurements in the NIR. 
The AGN in the BASS survey serve as a useful low redshift template for studies of AGN physics, as they have luminosities that are comparable to those of higher-redshift ($z>1$) AGN, detected in pencil-beam, deep X-ray surveys (Koss et al., submitted).

The Near-Infrared Spectrograph (NIRSpec) is a NIR multi-object dispersive spectrograph on board the \textit{JWST}, capable of simultaneously observing more than 100 slits over the wavelength range of $1 - 5$ \micron. 
The medium resolution mode is expected to provide $R=1000$, which is similar to the resolution of the spectra presented in this study, and thus sufficient for resolving the NIR coronal lines. 
An exposure of 100,000 s is expected to reach a limiting line flux of $\sim 5.7\times10^{-19}\,\ergcms$ at 2.0 \micron\ with $S/N=10$, or $\sim 2\times10^{-19}\,\ergcms$ with $S/N=3$.  This is a factor of $\sim$10,000 times more sensitive than our IRTF program for the 3$\sigma$ upper limits of line detection of the \Silivi\ line ($\sim 2\times10^{-15}\,\ergcms$).
Based on the typical scaling factor we found between the \Silivi\ and the 2-10 keV emission, of about 1270, this corresponds to an X-ray flux of roughly $2\times10^{-16}\,\ergcms$. 
This X-ray flux limit, in turn, corresponds to a completeness rate of about 20\% (6/30) for the Seyfert 2 sources in our survey. 
For comparison, the deepest X-ray data in Chandra Deep Field South (of 4 Ms) have very similar sensitivities in the hard band (2-8 keV), of $\sim2\times10^{-16}\,\ergcms$ at the 20\% completeness level \citep{Xue2011}. 
Thus, deep coronal line surveys of galaxies using {\it JWST}/NIRSpec could be as sensitive to AGN as the deepest X-ray surveys. 
This would be particularly important for detecting heavily obscured (Compton-thick) AGN that are missed by X-ray surveys, and/or nearby low luminosity AGN for which source confusion is difficult in the X-ray.

\section{Conclusions}

The goal of this work was to study the NIR spectroscopic properties of a large sample of nearby hard X-ray selected AGN.  We found:

\begin{itemize}

\item The FWHM of Pa$\beta$ is similar to the FWHM of H$\alpha$ and H$\beta$, if we take into account the emission from Fe multiplets. The FWHM of Pa$\alpha$ is smaller on average than the FWHM of the other hydrogen lines, but the difference is not statistically significant.

\item  AGN with a "hidden BLR" not observed in optical Balmer lines are the Seyfert 2 with the lowest values of \NH\ in our sample (bottom 11 percentile in $N_{\rm H}$) and show signs of ongoing mergers or tidal features suggesting the obscuration is related to the merger event.

\item We measured \MBH\ for the 68$\%$ of the AGN in our sample, either from the broad Paschen lines or from the velocity dispersion in the CO band-head. 

\item Overall, we find that the NIR region is significantly less effective at identifying X-ray selected AGN compared to emission line diagnostics in the optical. Only 25$\%$ (25/102) of our sample is identified as AGN, whereas 7$\%$ (7/102) are classified as star-forming galaxies, with the remaining majority of sources having too faint emission lines for classification.  While much deeper studies may identify many more spectra in the AGN region, the contamination with star forming regions is a significant problem. 

\item We found that the relation between \Silivi\ and hard X-ray flux is weak (scatter $\sigma = 0.43$ dex, correlation coefficient $R_{\rm pear} = 0.58$) and not significantly better than the one between \oiii\ and hard X-ray flux (scatter $\sigma = 0.59$ dex, correlation coefficient $R_{\rm pear} = 0.50$), based on a Z-test.   However, the relation of the hard X-ray luminosity with \Silivi\  luminosity ($R_{\rm pear} = 0.86$) is stronger than the relation with \oiii\ ($R_{\rm pear} = 0.74$), based on a Z-test.

\end{itemize}

\section*{Acknowledgements}

%\acknowledgements

M.\,K. acknowledges support from the Swiss National Science Foundation (SNSF) through the Ambizione fellowship grant PZ00P2\textunderscore154799/1.   M.\,K. was a visiting astronomer at the Infrared Telescope Facility, which is operated by the University of Hawaii under contract NNH14CK55B with the National Aeronautics and Space Administration (IRTF programs 2010A-059, 2011A-077, 2011B-104, 2012A-083). The authors wish to recognize and acknowledge the very significant cultural role and reverence that the summit of Mauna Kea has always had within the indigenous Hawaiian community. We are most fortunate to have the opportunity to conduct observations from this mountain. We acknowledge the work that the \swiftbat team has done to make this work possible.  The Kitt Peak National Observatory observations were obtained using MD-TAC time as part of the thesis of M.\,K. (2008B-0426, 2009A-0287, 2009B-0569) and also through NOAO time in program 2010A-0447 (PI M. Koss).  Kitt Peak National Observatory, National Optical Astronomy Observatory, is operated by the Association of Universities for Research in Astronomy (AURA), Inc., under cooperative agreement with the National Science Foundation. Data from Gemini programs (GN-2011B-Q-111,GN-2012A-Q-23, GN-2012B-Q-80, GN-2013A-Q-16, GN-2013A-Q-120) were used in this publication. M.\,K. would like to thank Dick Joyce at the NOAO for teaching him how use the Flamingos spectrograph on his first NIR spectroscopy run, the help of John Rayner with SPEX, and Michael Cushing for help with \spextool.  
K.\,S. gratefully acknowledges support from Swiss National Science Foundation Grant PP00P2\_138979/1.
C.\,R. acknowledges financial support from the CONICYT-Chile ''EMBIGGEN" Anillo (grant ACT1101), FONDECYT 1141218 and Basal-CATA PFB--06/2007. A.\,R.\,A acknowledges the Conselho Nacional de Desenvolvimento Cient\'ifico e Tecnol\'ogico (CNPq) for partial support to this work (grant 311935/2015-0). R.\,R. acknowledges support from CNPq and FAPERGS. 

This research made use of {\tt Astropy}, a community-developed core Python package for Astronomy (Astropy Collaboration, 2013).  This research has made use of the NASA/IPAC Extragalactic Database (NED) which is operated by the Jet Propulsion Laboratory, California Institute of Technology, under contract with the National Aeronautics and Space Administration.

%%%%%%%%%%%%%%%%%%%%%%%%%%%%%%%%%%%%%%%%%%%%%%%%%%



\bibliographystyle{mnras}

%\bibliographystyle{mn2e}
\bibliography{Bibliography/NIR_paper_biblio.bib}

%\begin{thebibliography}{99}
%\end{thebibliography}

%%%%%%%%%%%%%%%%%%%%%%%%%%%%%%%%%%%%%%%%%%%%%%%%%%
% TABLES

% Velocity dispersion table
\begin{table*}
\centering
\caption{Stellar velocity dispersion measurements}
\begin{tabular}{rcccccc}
\hline

ID\tablenotemark{a} &  $\sigma_{*, \rm CaT}$ & flag$_{\rm CaT}$ & $\sigma_{*, \rm CO}$ K-band &  flag$_{\rm CO}$ K-band  & $\sigma_{*, \rm CO}$ H-band & flag$_{\rm CO}$ H-band  \\   
  & [km/s]  & & [km/s] & &[km/s] &  \\ \hline 
6& ... & 9 & ... & 9  &... & 9 \\
33& ... & 9 & ... & 9  &... & 9 \\
116& ... & 9 & 207.0 $\pm$ 11.9 & 2 & 241.4 $\pm$ 7.8 & 1\\
130& ... & 9 & ... & 9  &... & 9 \\
140& ... & 9 & 211.8 $\pm$ 3.6 & 1 & 265.1 $\pm$ 12.3 & 2\\
157& ... & 9 & ... & 9  &275.0 $\pm$ 16.7 & 2\\
173& ... & 9 & ... & 9  &... & 9 \\
226& ... & 9 & ... & 9  &... & 9 \\
266& ... & 9 & ... & 9  &... & 9 \\
269& ... & 9 & 180.2 $\pm$ 7.9 & 2 & 187.6 $\pm$ 7.9 & 1\\
\hline

\end{tabular}
\label{tab:ppxf}
\tablecomments{(This table is available in its entirety in a machine-readable form in the online journal. A portion is shown here for guidance regarding its form and content.)}
\tablenotetext{a}{\swiftbat 70-month hard X-ray survey ID (http://swift.gsfc.nasa.gov/results/bs70mon/).}
\tablenotetext{b}{Quality flags: 1 = excellent fit with small error ($\langle \Delta \sigma_{*} \rangle$ = 6 \kms), 2 = larger errors than flag 1 ($\langle \Delta \sigma_{*} \rangle$ = 12 \kms), but acceptable fit, 3 = bad fit with high S/N, 9=bad fit}\\ 
\end{table*}

% M_BH table
\begin{table*}
\centering
\caption{Black hole mass measurements}
\begin{tabular}{rcccc}
\hline
  
ID\tablenotemark{a}& $\log M_{\rm BH}/M_\odot$ from Pa$\beta$ & flag$_{Pa\beta}$ &$\log M_{BH}/M_\odot$ from $\sigma_{CO}$ &  flag$_{CO}$ \\   
 %   & [$M_\odot$] &  & [$M_\odot$] &  \\
 \hline 

6& 7.39 $\pm$ 0.07 & 2& ... & 9 \\
33& ... & 9 & ... & 9 \\
116& ... & 9 & 8.85 $\pm$ 0.3 & 1\\
130& 6.5 $\pm$ 0.12 & 2& ... & 9 \\
140& ... & 9 & 9.03 $\pm$ 0.3 & 2\\
157& ... & 9 & 9.1 $\pm$ 0.3 & 2\\
173& ... & 9 & ... & 9 \\
226& ... & 9 & ... & 9 \\
266& 8.19 $\pm$ 0.04 & 2& ... & 9 \\
269& 7.31 $\pm$ 0.07 & 2& 8.37 $\pm$ 0.3 & 1\\

\hline
\end{tabular}
\label{tab:MBH}
\tablecomments{(This table is available in its entirety in a machine-readable form in the online journal. A portion is shown here for guidance regarding its form and content.)}
\tablenotetext{a}{\swiftbat-BAT 70-month hard X-ray survey ID (http://swift.gsfc.nasa.gov/results/bs70mon/).}
\end{table*}

% Flux tables
\begin{landscape}
% Flux table Pa14 (part 1)
\begin{table*}
%\begin{center}
\centering
\scriptsize
\begin{minipage}{\textwidth}
	\caption{Emission lines flux measurements in the 	 Pa14 region ($\text{0.84-0.99}$ $\mu$m) (part 1)}
	\begin{tabular}{rccccccccccc}
	\hline
  
ID\tablenotemark{a}& Flag\tablenotemark{b}  &  nFWHM\tablenotemark{c} & bFWHM\tablenotemark{d} & $\Delta v$\tablenotemark{e}& OI\tablenotemark{f}  & b OI    & Pa16   & b Pa16   & Pa15   & b Pa15    \\
& & [\kms]& [\kms]& [\kms]& $\lambda$0.8449$\mu$m & & $\lambda$0.8503$\mu$m & &$\lambda$0.8545$\mu$m &   \\
 \hline 

6 & 2 & ...& ...& ...& $<\ $ 10.2 & $<\ $ 10.2 & $<\ $ 18.6 & $<\ $ 18.6 & $<\ $ 8.0 & $<\ $ 8.0\\
33 & 9 & ...&...& ...& $<\ $ 8.2 & $<\ $ 8.2 & $<\ $ 1.1 & $<\ $ 1.1 & $<\ $ 2.3 & $<\ $ 2.3\\
116 & 9 & ...& ...& ...& $<\ $ 3.5 & $<\ $ 3.5 & $<\ $ 7.9 & $<\ $ 7.9 & $<\ $ 6.4 & $<\ $ 6.4\\
130 & 2 & 792 $\pm$ 388 & ...& ...& $<\ $ 0.6 & $<\ $ 0.6 & $<\ $ 0.3 & 0.7 $\pm$ 0.2 & 0.2 $\pm$ 0.3 & $<\ $ 0.9\\
140 & 9 & ...& ...& ...& ... & ... & ... & ... & ... & ...\\
157 & 9 & ...& ...& ...& ... & ... & ... & ... & ... & ...\\
173 & 9 & ...& ...& ...& ... & ... & ... & ... & ... & ...\\
226 & -1 & ...& ...& ...& ... & ... & ... & ... & ... & ...\\
266 & 9 & ...& 3299 $\pm$ 999 & ...& 67.9 $\pm$ 22.1 & 67.9 $\pm$ 22.1 & 8.4 $\pm$ 1.1 & 67.9 $\pm$ 19.9 & $<\ $ 28.4 & 60.7 $\pm$ 15.3\\
269 & 9 & ...& ...& ...& ... & ... & ... & ... & ... & ...\\

\hline
\end{tabular}
\medskip
\label{tab:Pa14}
\tablecomments{(This table is available in its entirety in a machine-readable form in the online journal. A portion is shown here for guidance regarding its form and content.)}
\tablenotetext{a}{\textit{Swift}-BAT 70-month hard X-ray survey ID (http://swift.gsfc.nasa.gov/results/bs70mon/).}
\tablenotetext{b}{Spectral fitting quality flags: 1 = good fit, 2 = acceptable fit, 9 = bad fit, -1 = lack of spectral coverage.}
\tablenotetext{c, d}{ nFWHM: FWHM of the narrow lines. bFWHM: FWHM of the broad lines. The FWHM are not corrected for the instrumental resolution.} 
\tablenotetext{e}{ Velocity offset of the broad components with respect to the systemic redshift.}
\tablenotetext{f}{Emission lines fluxes are in units of $10^{-15}$ erg s$^{-1}$ cm$^{-2}$. Negative numbers indicate upper limits.} 
\end{minipage}
%\end{center}
\end{table*}

% Flux table Pa14 region (part 2)
\begin{table*}
\begin{center}
%\centering
\scriptsize
\begin{minipage}{\textwidth}
\caption{Emission lines flux measurements in the Pa14 region ($\text{0.84-0.99}$ $\mu$m) (part 2)}
\begin{tabular}{rcccccccccc}
\hline
  
ID\tablenotemark{a}& Flag\tablenotemark{b}  & Pa14\tablenotemark{c}  & b Pa14  & Pa13 & b Pa13   & Pa12   & bPa12  & Pa11  & bPa11  \\
& & $\lambda$0.8600$\mu$m &  & $\lambda$0.8665$\mu$m & & $\lambda$0.8750$\mu$m& & $\lambda$0.8863$\mu$m& \\ 
 \hline

6 & 2 & $<\ $ 5.0 & $<\ $ 5.0 & $<\ $ 5.8 & $<\ $ 5.8 & $<\ $ 6.8 & $<\ $ 6.8 & 8.8 $\pm$ 0.9 & $<\ $ 8.7\\
33 & 9 & $<\ $ 4.8 & $<\ $ 4.8 & $<\ $ 4.4 & $<\ $ 4.4 & $<\ $ 2.1 & $<\ $ 2.1 & 2.0 $\pm$ 0.9 & $<\ $ 2.0\\
116 & 9 & $<\ $ 5.8 & $<\ $ 5.8 & $<\ $ 5.4 & $<\ $ 5.4 & $<\ $ 5.2 & $<\ $ 5.2 & $<\ $ 6.0 & $<\ $ 6.0\\
130 & 2 & $<\ $ 0.3 & $<\ $ 0.3 & 0.3 $\pm$ 0.0 & $<\ $ 0.2 & $<\ $ 0.3 & $<\ $ 0.3 & $<\ $ 0.3 & $<\ $ 0.3\\
140 & 9 & ... & ... & ... & ... & ... & ... & ... & ...\\
157 & 9 & ... & ... & ... & ... & ... & ... & ... & ...\\
173 & 9 & ... & ... & ... & ... & ... & ... & ... & ...\\
226 & -1 & ... & ... & ... & ... & ... & ... & ... & ...\\
266 & 9 & $<\ $ 18.1 & 31.0 $\pm$ 4.2 & $<\ $ 11.5 & $<\ $ 11.5 & $<\ $ 12.1 & $<\ $ 12.1 & $<\ $ 6.2 & $<\ $ 6.2\\
269 & 9 & ... & ... & ... & ... & ... & ... & ... & ...\\
\hline

\end{tabular}
\label{tab:Pa14_2}

\tablecomments{(This table is available in its entirety in a machine-readable form in the online journal. A portion is shown here for guidance regarding its form and content.)}
\tablenotetext{a}{\swiftbat 70-month hard X-ray survey ID (http://swift.gsfc.nasa.gov/results/bs70mon/).}
\tablenotetext{b}{Spectral fitting quality flags: 1 = good fit, 2 = acceptable fit, 9 = bad fit, -1 = lack of spectral coverage.}
\tablenotetext{c}{Emission lines fluxes are in units of $10^{-15}$ erg s$^{-1}$ cm$^{-2}$. Negative numbers indicate upper limits.} 

\end{minipage}
\end{center}
\end{table*}

% Flux table Pazeta region 
\begin{table*}

\begin{center}
%\centering
\scriptsize
\begin{minipage}{\textwidth}
\caption{Emission lines flux measurements in the Pa$\zeta$ region (0.90-0.96 $\mu$m)}
\begin{tabular}{rcccccccccccc}
\hline
ID\tablenotemark{a}& Flag\tablenotemark{b}  &  nFWHM\tablenotemark{c} & bFWHM\tablenotemark{d} & $\Delta v$\tablenotemark{e}& \siiif\ \tablenotemark{f}  & \siiif\  & Pa10   & b Pa10  & Pa$\zeta$  & b Pa$\zeta$ & Pa$\epsilon$ & b Pa$\epsilon$   \\
& & [\kms ]& [\kms ]& [\kms ] & $\lambda$0.9069$\mu$m & $\lambda$0.9531$\mu$m & $\lambda$0.9016$\mu$m & & $\lambda$0.9229$\mu$m& & $\lambda$0.9546$\mu$m& \\ 
 \hline 

6 & 2 & 605 $\pm$ 40 & 3153 $\pm$ 62 & -457 $\pm$ 60 & 9.8 $\pm$ 0.4 & 19.7 $\pm$ 1.0 & 8.5 $\pm$ 1.0 & 15.0 $\pm$ 13.2 & $<\ $ 18.7 & 75.6 $\pm$ 1.0 & 7.5 $\pm$ 0.7 & 53.5 $\pm$ 2.0\\
33 & 2 & 641 $\pm$ 5 & ...& ...& 48.7 $\pm$ 0.7 & 123.0 $\pm$ 1.3 & $<\ $ 6.8 & ... & $<\ $ 5.3 & ... & $<\ $ 47.8 & ...\\
116 & 2 & 654 $\pm$ 42 & ...& ...& 7.5 $\pm$ 0.3 & 16.6 $\pm$ 1.8 & $<\ $ 6.1 & ... & $<\ $ 6.0 & ... & $<\ $ 5.8 & ...\\
130 & 9 & ...& ...& ...& 0.5 $\pm$ 0.1 & $<\ $ 0.7 & $<\ $ 0.5 & $<\ $ 2.7 & 0.4 $\pm$ 0.1 & $<\ $ 2.0 & 0.7 $\pm$ 0.1 & $<\ $ 1.9\\
140 & 9 & 676 $\pm$ 32 & 4683 $\pm$ 208 & ...  & 28.9 $\pm$ 1.1 & 39.1 $\pm$ 1.4 & $<\ $ 21.6 & ... & $<\ $ 42.8 & 90.5 $\pm$ 2.8 & $<\ $ 50.8 & 66.3 $\pm$ 2.5\\
157 & 9 & 821 $\pm$ 77 & ...& ...& 3.0 $\pm$ 0.3 & 8.2 $\pm$ 0.9 & $<\ $ 0.8 & ... & $<\ $ 3.1 & ... & $<\ $ 1.9 & ...\\
173 & 2 & 1257 $\pm$ 1 & ...& ...& 72.1 $\pm$ 1.7 & 180.0 $\pm$ 11.8 & $<\ $ 25.8 & ... & $<\ $ 2.5 & ... & $<\ $ 86.4 & ...\\
226 & 2 & 314 $\pm$ 12 & 3146 $\pm$ 95 & 114 $\pm$ 70 & $<\ $ 4.0 & 30.0 $\pm$ 1.2 & $<\ $ 3.7 & $<\ $ 50.9 & $<\ $ 7.9 & 106.0 $\pm$ 2.5 & $<\ $ 5.5 & 98.5 $\pm$ 4.9\\
266 & 9 & 524 $\pm$ 40 & 5166 $\pm$ 80 & ...  & 12.0 $\pm$ 1.0 & 22.9 $\pm$ 1.5 & $<\ $ 7.5 & $<\ $ 79.0 & $<\ $ 5.3 & 228.0 $\pm$ 4.2 & $<\ $ 8.4 & 149.0 $\pm$ 3.2\\
269 & 9 & 534 $\pm$ 11 & ...& ...& 16.5 $\pm$ 1.1 & 34.2 $\pm$ 1.0 & $<\ $ 1.8 & $<\ $ 19.2 & $<\ $ 8.3 & $<\ $ 88.3 & $<\ $ 16.3 & 28.2 $\pm$ 1.5\\
\hline

\end{tabular}
\label{tab:Pazeta}
\tablecomments{(This table is available in its entirety in a machine-readable form in the online journal. A portion is shown here for guidance regarding its form and content.)}
\tablenotetext{a}{\swiftbat 70-month hard X-ray survey ID (http://swift.gsfc.nasa.gov/results/bs70mon/).}
\tablenotetext{b}{Spectral fitting quality flags: 1 = good fit, 2 = acceptable fit, 9 = bad fit, -1 = lack of spectral coverage.}
\tablenotetext{c, d}{ nFWHM: FWHM of the narrow lines. bFWHM: FWHM of the broad lines. The FWHM are not corrected for the instrumental resolution.} 
\tablenotetext{e}{ Velocity offset of the broad components with respect to the systemic redshift.} 
\tablenotetext{f}{Emission lines fluxes are in units of $10^{-15}$ erg s$^{-1}$ cm$^{-2}$. Negative numbers indicate upper limits.}
\end{minipage}
\end{center}
\end{table*}

% Flux table Padelta region 
\begin{table*}
\begin{center}
%\centering
\scriptsize
\begin{minipage}{\textwidth}
\caption{Emission lines flux measurements in the Pa$\delta$ region (0.96-1.04 $\mu$m) (part 1)}
\begin{tabular}{rccccccccccc}
\hline

ID\tablenotemark{a}& Flag\tablenotemark{b}  &  nFWHM\tablenotemark{c} & bFWHM\tablenotemark{d}  &$\Delta v$\tablenotemark{e} & \CI \tablenotemark{f}  & \CI & \Sviii  & \feiif  & \heii & \sii  
\\
& & [\kms ]& [\kms ]& [\kms ] & $\lambda$0.9827$\mu$m & $\lambda$0.9853$\mu$m & $\lambda$0.9915$\mu$m &  $\lambda$0.9999$\mu$m & $\lambda$1.0126$\mu$m & $\lambda$1.0290$\mu$m\\ 
 \hline

6 & 3 & 780 $\pm$ 598 & 4245 $\pm$ 532 & ...  & $<\ $ 3.6 & $<\ $ 2.9 & 3.4 $\pm$ 8.5 & $<\ $ 23.0 & 3.1 $\pm$ 7.6 & $<\ $ 4.0\\
33 & 3 & 660 $\pm$ 41 & ...& ...& 1.5 $\pm$ 22.5 & 3.9 $\pm$ 21.5 & 2.3 $\pm$ 20.0 & $<\ $ 0.8 & 3.6 $\pm$ 22.0 & 5.0 $\pm$ 15.1\\
116 & 3 & 596 $\pm$ 22 & ...& ...& $<\ $ 4.5 & 1.9 $\pm$ 1.8 & $<\ $ 3.0 & $<\ $ 1.9 & 2.1 $\pm$ 2.2 & $<\ $ 3.5\\
130 & 2 & 5158 $\pm$ 1 & 4899 $\pm$ 4 & -573 $\pm$ 30 & $<\ $ 0.1 & $<\ $ 0.1 & $<\ $ 0.2 & 1.6 $\pm$ 0.4 & $<\ $ 0.2 & $<\ $ 0.1\\
140 & 3 & 800 $\pm$ 51 & ...& ...& $<\ $ 26.8 & 11.3 $\pm$ 3.1 & $<\ $ 8.0 & $<\ $ 7.7 & $<\ $ 11.5 & 11.5 $\pm$ 7.3\\
157 & 2 & 804 $\pm$ 106 & ...& ...& $<\ $ 2.2 & 1.6 $\pm$ 1.2 & $<\ $ 0.5 & $<\ $ 0.6 & $<\ $ 1.7 & $<\ $ 1.4\\
173 & 3 & 1372 $\pm$ 154 & ...& ...& $<\ $ 28.7 & 21.6 $\pm$ 28.8 & $<\ $ 0.5 & $<\ $ 4.0 & $<\ $ 4.7 & 24.6 $\pm$ 34.1\\
226 & 2 & 335 $\pm$ 67 & 2837 $\pm$ 832 & -35 $\pm$ 6 & $<\ $ 4.0 & $<\ $ 7.8 & $<\ $ 4.4 & $<\ $ 23.8 & 4.4 $\pm$ 5.8 & $<\ $ 4.8\\
266 & 2 & 590 $\pm$ 84 & 5528 $\pm$ 121 & -85 $\pm$ 15 & $<\ $ 4.8 & $<\ $ 12.6 & $<\ $ 25.0 & $<\ $ 22.3 & 9.8 $\pm$ 12.1 & $<\ $ 7.7\\
269 & 3 & 624 $\pm$ 26 & 5910 $\pm$ 54 & ...  & $<\ $ 5.4 & 2.0 $\pm$ 2.3 & 2.2 $\pm$ 1.9 & $<\ $ 0.7 & 3.7 $\pm$ 1.2 & $<\ $ 3.7\\
\hline

\end{tabular}
\label{tab:Padelta}
\tablecomments{(This table is available in its entirety in a machine-readable form in the online journal. A portion is shown here for guidance regarding its form and content.)}
\tablenotetext{a}{\swiftbat 70-month hard X-ray survey ID (http://swift.gsfc.nasa.gov/results/bs70mon/).}
\tablenotetext{b}{Spectral fitting quality flags: 1 = good fit, 2 = acceptable fit, 9 = bad fit, -1 = lack of spectral coverage.}
\tablenotetext{c, d}{ nFWHM: FWHM of the narrow lines. bFWHM: FWHM of the broad lines. The FWHM are not corrected for the instrumental resolution.} 
\tablenotetext{e}{ Velocity offset of the broad components with respect to the systemic redshift.} 
\tablenotetext{f}{Emission lines fluxes are in units of $10^{-15}$ erg s$^{-1}$ cm$^{-2}$. Negative numbers indicate upper limits.}
\end{minipage}
\end{center}
\end{table*}

% Flux table Padelta region (part 2)
\begin{table*}
%\begin{center}
\centering
\scriptsize
\begin{minipage}{\textwidth}
\caption{Emission lines flux measurements in the Pa$\delta$ region (0.96-1.04 $\mu$m) (part 2)}
\begin{tabular}{rccccccccccc}
\hline

ID\tablenotemark{a}& Flag\tablenotemark{b} & \sii \tablenotemark{c}  & \sii  & \sii  & \NI &  Pa$\delta$ & b Pa$\delta$\\
&  & $\lambda$1.0320$\mu$m & $\lambda$1.0336$\mu$m &$\lambda$1.0370$\mu$m  & $\lambda$1.0404$\mu$m &$\lambda$1.0049$\mu$m  &\\
\hline

6 & 3 & $<\ $ 1.4 & $<\ $ 2.9 & $<\ $ 2.8 & $<\ $ 3.5 & 13.8 $\pm$ 8.2 & 110.0 $\pm$ 63.7\\
33 & 3 & 6.3 $\pm$ 12.6 & 5.0 $\pm$ 15.6 & 1.5 $\pm$ 15.6 & 1.4 $\pm$ 21.4 & 2.4 $\pm$ 13.4 & ...\\
116 & 3 & 1.3 $\pm$ 1.8 & $<\ $ 2.3 & $<\ $ 4.2 & $<\ $ 2.3 & $<\ $ 2.8 & ...\\
130 & 2 & $<\ $ 0.1 & $<\ $ 0.1 & $<\ $ 0.1 & $<\ $ 0.1 & 0.4 $\pm$ 0.2 & 1.6 $\pm$ 1.1\\
140 & 3 & 12.1 $\pm$ 6.6 & 5.8 $\pm$ 6.8 & $<\ $ 9.2 & 7.3 $\pm$ 5.8 & $<\ $ 11.1 & ...\\
157 & 2 & $<\ $ 2.0 & $<\ $ 1.8 & $<\ $ 2.9 & $<\ $ 1.6 & $<\ $ 0.7 & ...\\
173 & 3 & 29.5 $\pm$ 13.6 & 32.5 $\pm$ 23.7 & 14.3 $\pm$ 27.5 & 26.2 $\pm$ 16.1 & 3.3 $\pm$ 22.5 & ...\\
226 & 2 & $<\ $ 6.1 & $<\ $ 6.7 & $<\ $ 6.9 & $<\ $ 11.5 & $<\ $ 3.7 & 102.0 $\pm$ 74.9\\
266 & 2 & $<\ $ 3.4 & $<\ $ 4.7 & $<\ $ 3.6 & $<\ $ 4.2 & $<\ $ 4.4 & 343.0 $\pm$ 8.3\\
269 & 3 & 1.8 $\pm$ 3.0 & $<\ $ 1.9 & $<\ $ 4.3 & $<\ $ 1.0 & $<\ $ 1.8 & 35.7 $\pm$ 29.4\\
\hline

\end{tabular}
\label{tab:Padelta2}
\tablecomments{(This table is available in its entirety in a machine-readable form in the online journal. A portion is shown here for guidance regarding its form and content.)}
\tablenotetext{a}{\swiftbat 70-month hard X-ray survey ID (http://swift.gsfc.nasa.gov/results/bs70mon/).}
\tablenotetext{b}{Spectral fitting quality flags: 1 = good fit, 2 = acceptable fit, 9 = bad fit, -1 = lack of spectral coverage.}
\tablenotetext{c}{Emission lines fluxes are in units of $10^{-15}$ erg s$^{-1}$ cm$^{-2}$. Negative numbers indicate upper limits.}
\end{minipage}
%\end{center}
\end{table*}

\newpage
% Flux table Pagamma region 
\begin{table*}
%\begin{center}
\centering
\scriptsize
\begin{minipage}{\textwidth}
\caption{Emission lines flux measurements in the Pa$\gamma$ region (1.04-1.15 $\mu$m)}
\begin{tabular}{rccccccccccccc}
\hline
 
ID\tablenotemark{a}& Flag\tablenotemark{b}  &  nFWHM\tablenotemark{c} & bFWHM\tablenotemark{d} & $\Delta v$\tablenotemark{e}& \feiif\tablenotemark{f}  & \Fexiii  & \feiif  & \hei & b \hei & Pa$\gamma$ & b Pa$\gamma$ &  {\rm O}\,\textsc{i} & b {\rm O}\,\textsc{i}  \\
& & [\kms ]& [\kms ]& [\kms ]&  $\lambda$1.0500$\mu$m & $\lambda$1.0747$\mu$m & $\lambda$1.0370$\mu$m & $\lambda$1.0404$\mu$m & &$\lambda$1.0938$\mu$m  & & $\lambda$1.1287$\mu$m & \\ 
\hline 

6 & 2 & ...& ...& ...  & 7.7 $\pm$ 0.2 & $<\ $ 22.4 & $<\ $ 3.4 & 138.0 $\pm$ 0.6 & 234.0 $\pm$ 1.1 & 43.6 $\pm$ 0.4 & $<\ $ 90.1 & 23.4 $\pm$ 0.6 & $<\ $ 14.2\\
33 & 9 & ...& ...& ...  & $<\ $ 3.0 & $<\ $ 0.7 & $<\ $ 1.2 & 50.3 $\pm$ 1.1 & 20.4 $\pm$ 1.2 & 4.3 $\pm$ 0.2 & ... & $<\ $ 0.4 & ...\\
116 & 2 & ...& ...& ...  & $<\ $ 1.8 & $<\ $ 8.4 & $<\ $ 2.0 & 13.0 $\pm$ 0.5 & 59.0 $\pm$ 1.1 & $<\ $ 10.1 & ... & $<\ $ 2.0 & ...\\
130 & 2 & 795 $\pm$ 1 & 2045 $\pm$ 42 & -63 $\pm$ 3 & 0.2 $\pm$ 0.1 & 0.2 $\pm$ 0.1 & $<\ $ 0.2 & 1.0 $\pm$ 0.1 & 2.5 $\pm$ 0.1 & 0.3 $\pm$ 0.1 & 1.0 $\pm$ 0.1 & 0.2 $\pm$ 0.1 & 0.4 $\pm$ 0.1\\
140 & 9 & ...& ...& ...  & $<\ $ 10.0 & $<\ $ 23.1 & $<\ $ 8.2 & 80.1 $\pm$ 23.3 & ... & $<\ $ 34.1 & ... & $<\ $ 11.9 & ...\\
157 & 9 & ...& ...& ...  & $<\ $ 0.4 & $<\ $ 0.5 & $<\ $ 1.1 & $<\ $ 1.0 & ... & $<\ $ 2.8 & ... & $<\ $ 2.6 & ...\\
173 & 9 & ...& 2684 $\pm$ 16 & ...  & $<\ $ 1.5 & $<\ $ 74.7 & $<\ $ 1.2 & 102.0 $\pm$ 2.2 & 474.0 $\pm$ 2.9 & 12.2 $\pm$ 1.1 & 54.3 $\pm$ 2.4 & $<\ $ 4.4 & ...\\
226 & 2 & ...& 4571 $\pm$ 232 & 63 $\pm$ 3 & 7.1 $\pm$ 0.7 & $<\ $ 36.7 & $<\ $ 2.8 & 169.0 $\pm$ 10.9 & 538.0 $\pm$ 7.6 & 15.9 $\pm$ 4.2 & 161.0 $\pm$ 10.9 & 13.8 $\pm$ 2.9 & $<\ $ 58.4\\
266 & 2 & ...& 5581 $\pm$ 221 & 136 $\pm$ 19 & $<\ $ 5.0 & $<\ $ 87.6 & $<\ $ 14.1 & 93.0 $\pm$ 9.6 & 1020.0 $\pm$ 64.7 & $<\ $ 72.7 & 279.0 $\pm$ 70.4 & $<\ $ 4.8 & ...\\
269 & 2 & ...& ...& ...  & $<\ $ 2.2 & $<\ $ 10.6 & $<\ $ 2.2 & 58.9 $\pm$ 0.6 & 143.0 $\pm$ 1.6 & $<\ $ 14.8 & ... & $<\ $ 0.4 & ...\\

\end{tabular}
\medskip
\label{tab:Pagamma}
\tablecomments{(This table is available in its entirety in a machine-readable form in the online journal. A portion is shown here for guidance regarding its form and content.)}
\tablenotetext{a}{\swiftbat 70-month hard X-ray survey ID (http://swift.gsfc.nasa.gov/results/bs70mon/).}
\tablenotetext{b}{Spectral fitting quality flags: 1 = good fit, 2 = acceptable fit, 9 = bad fit, -1 = lack of spectral coverage.}
\tablenotetext{c, d}{ nFWHM: FWHM of the narrow lines. bFWHM: FWHM of the broad lines. The FWHM are not corrected for the instrumental resolution.} 
\tablenotetext{e}{ Velocity offset of the broad components with respect to the systemic redshift.} 
\tablenotetext{f}{Emission lines fluxes are in units of $10^{-15}$ erg s$^{-1}$ cm$^{-2}$. Negative numbers indicate upper limits.}
\end{minipage}
%\end{center}
\end{table*}

% Flux table Pabeta region 
\begin{table*}
%\begin{center}
\centering
\scriptsize
\begin{minipage}{\textwidth}
\caption{Emission lines flux measurements in the Pa$\beta$ region (1.15-1.30 $\mu$m)}
\begin{tabular}{rccccccccccccc}
\hline

ID\tablenotemark{a}& Flag\tablenotemark{b}  &  nFWHM\tablenotemark{c} & bFWHM\tablenotemark{d}  &b $\Delta v$\tablenotemark{e}& \heii\tablenotemark{f}  & \pii  & \Six & \feiif & \feiif & \feiif & \feiif & Pa$\beta$ & b Pa$\beta$\\
& & [\kms ]& [\kms ]& [\kms ] &$\lambda$1.1620$\mu$m & $\lambda$1.1886$\mu$m & $\lambda$1.2520$\mu$m &
$\lambda$1.2570$\mu$m & $\lambda$1.2788$\mu$m & $\lambda$1.2950$\mu$m & $\lambda$1.3201$\mu$m  & $\lambda$1.2818$\mu$m&\\ 
 \hline 

6 & 2 & ...& 2741 $\pm$ 67 & -133 $\pm$ 9 & $<\ $ 0.8 & $<\ $ 1.2 & 3.7 $\pm$ 0.2 & 1.9 $\pm$ 1.0 & 17.8 $\pm$ 0.7 & 1.7 $\pm$ 1.7 & $<\ $ 8.9 & 36.7 $\pm$ 0.9 & 117.0 $\pm$ 1.2\\
33 & 2 & 644 $\pm$ 285 & ...& 42 $\pm$ 6 & $<\ $ 0.7 & 3.0 $\pm$ 0.6 & $<\ $ 2.3 & 12.1 $\pm$ 0.1 & $<\ $ 10.6 & $<\ $ 3.0 & 2.6 $\pm$ 0.8 & 11.3 $\pm$ 0.2 & ...\\
116 & 9 & 635 $\pm$ 246 & ...& ...  & $<\ $ 1.9 & $<\ $ 2.1 & $<\ $ 1.5 & $<\ $ 0.9 & 3.4 $\pm$ 0.3 & $<\ $ 2.5 & 2.7 $\pm$ 0.6 & $<\ $ 1.8 & ...\\
130 & 2 & ...& 3572 $\pm$ 200 & -375 $\pm$ 28 & $<\ $ 0.1 & $<\ $ 0.1 & $<\ $ 0.2 & $<\ $ 0.1 & $<\ $ 1.8 & $<\ $ 0.1 & $<\ $ 0.2 & 1.0 $\pm$ 0.1 & 0.9 $\pm$ 0.1\\
140 & 3 & 788 $\pm$ 381 & ...& ...  & $<\ $ 6.8 & $<\ $ 4.1 & $<\ $ 21.2 & 31.1 $\pm$ 0.9 & 10.7 $\pm$ 2.0 & $<\ $ 6.2 & 10.0 $\pm$ 2.8 & $<\ $ 20.6 & ...\\
157 & 9 & ...& ...& ...  & $<\ $ 0.9 & $<\ $ 0.4 & $<\ $ 0.2 & $<\ $ 0.3 & $<\ $ 0.4 & $<\ $ 0.8 & $<\ $ 0.1 & $<\ $ 1.3 & ...\\
173 & 3 & 518 $\pm$ 1 & 518 $\pm$ 1 & ...  & $<\ $ 1.2 & 11.4 $\pm$ 0.2 & $<\ $ 15.8 & 50.6 $\pm$ 0.2 & $<\ $ 25.6 & $<\ $ 0.9 & 11.6 $\pm$ 0.4 & 12.6 $\pm$ 7.7 & 51.2 $\pm$ 25.8\\
226 & 3 & ...& 2842 $\pm$ 29 & ...  & $<\ $ 1.4 & $<\ $ 0.3 & 6.0 $\pm$ 0.4 & 3.3 $\pm$ 0.8 & $<\ $ 34.8 & 3.8 $\pm$ 2.5 & $<\ $ 62.5 & $<\ $ 9.3 & 239.0 $\pm$ 2.7\\
266 & 2 & 507 $\pm$ 1 & 4864 $\pm$ 15 & 362 $\pm$ 6 & $<\ $ 2.4 & $<\ $ 2.6 & $<\ $ 3.3 & 4.0 $\pm$ 0.7 & $<\ $ 22.6 & $<\ $ 40.5 & 26.2 $\pm$ 0.9 & $<\ $ 8.6 & 496.0 $\pm$ 1.3\\
269 & 2 & ...& 4379 $\pm$ 99 & -111 $\pm$ 36 & $<\ $ 0.6 & $<\ $ 0.6 & $<\ $ 1.4 & 4.6 $\pm$ 0.2 & $<\ $ 7.4 & $<\ $ 3.9 & $<\ $ 1.3 & 3.4 $\pm$ 0.3 & 48.9 $\pm$ 0.4\\

\hline

\end{tabular}
\label{tab:Pabeta}
\tablecomments{(This table is available in its entirety in a machine-readable form in the online journal. A portion is shown here for guidance regarding its form and content.)}
\tablenotetext{a}{\swiftbat 70-month hard X-ray survey ID (http://swift.gsfc.nasa.gov/results/bs70mon/).}
\tablenotetext{b}{Spectral fitting quality flags: 1 = good fit, 2 = acceptable fit, 9 = bad fit, -1 = lack of spectral coverage.}
\tablenotetext{c, d}{ nFWHM: FWHM of the narrow lines. bFWHM: FWHM of the broad lines. The FWHM are not corrected for the instrumental resolution.} 
\tablenotetext{e}{ Velocity offset of the broad components with respect to the systemic redshift.} 
\tablenotetext{f}{Emission lines fluxes are in units of $10^{-15}$ erg s$^{-1}$ cm$^{-2}$. Negative numbers indicate upper limits.}
\end{minipage}
%\end{center}
\end{table*}

% Flux table Br10 region 
\begin{table*}
%\begin{center}
\centering
\scriptsize
\begin{minipage}{\textwidth}
\caption{Emission lines flux measurements in the Br10 region (1.30-1.80 $\mu$m)}
\begin{tabular}{rcccccccccccc}
\hline

ID\tablenotemark{a}& Flag\tablenotemark{b}  &  nFWHM\tablenotemark{c} & bFWHM\tablenotemark{d} & b $\Delta v$\tablenotemark{e}& \Silix\tablenotemark{f}   & \feiif & \feiif & \feiif & Br10 & b Br10 & Br11 & b Br11 \\
& & [\kms ]& [\kms ] & [\kms ] & $\lambda$1.4300$\mu$m & $\lambda$1.5339$\mu$m & $\lambda$1.6436$\mu$m & $\lambda$1.6807 $\mu$m & $\lambda$1.7367 $\mu$m & & $\lambda$1.6811 $\mu$m &  \\
 \hline

6 & 9 & ...& ...& ...  & $<\ $ 5.1 & $<\ $ 2.8 & 4.3 $\pm$ 0.4 & $<\ $ 1.2 & $<\ $ 1.2 & $<\ $ 1.2 & 4.2 $\pm$ 0.7 & $<\ $ 1.1\\
33 & 1 & 763 $\pm$ 1 & ...& ...  & 2.0 $\pm$ 0.2 & 2.0 $\pm$ 0.2 & 9.4 $\pm$ 0.1 & $<\ $ 0.9 & $<\ $ 1.1 & $<\ $ 1.1 & $<\ $ 1.1 & $<\ $ 1.1\\
116 & 9 & ...& ...& ...  & $<\ $ 2.0 & $<\ $ 1.9 & $<\ $ 2.4 & $<\ $ 1.5 & $<\ $ 1.8 & $<\ $ 1.8 & $<\ $ 0.5 & $<\ $ 0.5\\
130 & 9 & ...& ...& ...  & $<\ $ 0.1 & $<\ $ 0.1 & $<\ $ 0.1 & $<\ $ 0.1 & $<\ $ 0.1 & $<\ $ 0.1 & $<\ $ 0.1 & $<\ $ 0.1\\
140 & 9 & ...& ...& ...  & $<\ $ 8.4 & $<\ $ 7.3 & 27.0 $\pm$ 1.6 & $<\ $ 10.2 & $<\ $ 11.8 & $<\ $ 11.8 & $<\ $ 12.3 & $<\ $ 12.3\\
157 & 9 & ...& ...& ...  & $<\ $ 0.5 & $<\ $ 0.5 & $<\ $ 1.5 & $<\ $ 0.4 & $<\ $ 0.5 & $<\ $ 0.5 & $<\ $ 0.5 & $<\ $ 0.5\\
173 & 2 & ...& ...& ...  & $<\ $ 1.0 & 6.9 $\pm$ 0.2 & 49.2 $\pm$ 0.4 & $<\ $ 5.0 & $<\ $ 5.4 & $<\ $ 5.4 & $<\ $ 1.5 & $<\ $ 1.5\\
226 & 9 & ...& ...& ...  & $<\ $ 3.8 & $<\ $ 1.7 & $<\ $ 0.8 & $<\ $ 2.5 & $<\ $ 2.2 & $<\ $ 2.2 & $<\ $ 2.3 & $<\ $ 2.3\\
266 & 9 & ...& ...& ...  & $<\ $ 8.2 & $<\ $ 5.6 & $<\ $ 1.3 & $<\ $ 2.4 & $<\ $ 2.7 & $<\ $ 2.7 & $<\ $ 2.8 & $<\ $ 2.8\\
269 & 9 & ...& ...& ...  & $<\ $ 2.6 & $<\ $ 0.5 & $<\ $ 4.6 & $<\ $ 3.7 & $<\ $ 4.0 & $<\ $ 4.0 & $<\ $ 1.0 & $<\ $ 1.0\\
\hline

\end{tabular}
\label{tab:Br10}
\tablecomments{(This table is available in its entirety in a machine-readable form in the online journal. A portion is shown here for guidance regarding its form and content.)}
\tablenotetext{a}{\swiftbat 70-month hard X-ray survey ID (http://swift.gsfc.nasa.gov/results/bs70mon/).}
\tablenotetext{b}{Spectral fitting quality flags: 1 = good fit, 2 = acceptable fit, 9 = bad fit, -1 = lack of spectral coverage.}
\tablenotetext{c, d}{ nFWHM: FWHM of the narrow lines. bFWHM: FWHM of the broad lines. The FWHM are not corrected for the instrumental resolution.} 
\tablenotetext{e}{ Velocity offset of the broad components with respect to the systemic redshift.} 
\tablenotetext{f}{Emission lines fluxes are in units of $10^{-15}$ erg s$^{-1}$ cm$^{-2}$. Negative numbers indicate upper limits.}
\end{minipage}
%\end{center}
\end{table*}

% Flux table Paalpha region 
\begin{table*}
%\begin{center}
\centering
\scriptsize
\begin{minipage}{\textwidth}
\caption{Emission lines flux measurements in the Pa$\alpha$ region (1.80-2.00 $\mu$m)}
\begin{tabular}{rccccccccccc}
\hline

ID\tablenotemark{a}& Flag\tablenotemark{b}  &  nFWHM\tablenotemark{c} & bFWHM\tablenotemark{d} & b $\Delta v$\tablenotemark{e}& \Silixi\tablenotemark{f}   & H$_2$ & \Silivi & Pa$\alpha$ & b Pa$\alpha$ & Br$\delta$ & b Br$\delta$ \\
& & [\kms ]& [\kms ] & [\kms ] & $\lambda$1.9320$\mu$m & $\lambda$1.9564$\mu$m & $\lambda$1.9620$\mu$m &  $\lambda$1.8751$\mu$m &  & $\lambda$1.9446$\mu$m  &\\
 \hline 

6 & 2 & 374 $\pm$ 1 & 1386 $\pm$ 16 & -145 $\pm$ 8 & $<\ $ 2.1 & $<\ $ 7.8 & $<\ $ 0.4 & 27.3 $\pm$ 1.3 & 127.0 $\pm$ 1.7 & 8.6 $\pm$ 1.1 & $<\ $ 5.6\\
33 & 3 & ...& ...& ...  & $<\ $ 0.1 & $<\ $ 3.0 & 4.1 $\pm$ 0.2 & 42.6 $\pm$ 0.2 & $<\ $ 33.5 & $<\ $ 0.3 & $<\ $ 0.3\\
116 & 9 & ...& 4659 $\pm$ 290 & ...  & $<\ $ 0.5 & ... & $<\ $ 1.2 & $<\ $ 1.4 & 30.1 $\pm$ 1.4 & $<\ $ 0.2 & $<\ $ 0.2\\
130 & 9 & 374 $\pm$ 13 & ...& ...  & $<\ $ 0.2 & $<\ $ 0.1 & $<\ $ 0.3 & 0.7 $\pm$ 0.1 & $<\ $ 0.2 & $<\ $ 0.1 & $<\ $ 0.1\\
140 & 9 & ...& ...& ...  & $<\ $ 4.7 & ... & $<\ $ 3.0 & 24.7 $\pm$ 1.0 & $<\ $ 11.8 & $<\ $ 8.5 & $<\ $ 8.5\\
157 & 9 & ...& ...& ...  & $<\ $ 0.7 & ... & $<\ $ 0.2 & $<\ $ 1.6 & $<\ $ 1.6 & $<\ $ 0.4 & $<\ $ 0.4\\
173 & 3 & ...& 2799 $\pm$ 43 & ...  & $<\ $ 1.2 & ... & $<\ $ 3.2 & 44.9 $\pm$ 0.5 & 146.0 $\pm$ 0.9 & $<\ $ 0.8 & 66.6 $\pm$ 24.1\\
226 & 2 & ...& 2606 $\pm$ 16 & 372 $\pm$ 8 & $<\ $ 1.6 & $<\ $ 1.0 & 13.2 $\pm$ 2.0 & 12.5 $\pm$ 0.7 & 336.0 $\pm$ 2.0 & $<\ $ 5.2 & $<\ $ 5.2\\
266 & 3 & 953 $\pm$ 59 & 4633 $\pm$ 18 & ...  & $<\ $ 7.0 & 24.8 $\pm$ 2.0 & 20.2 $\pm$ 1.4 & $<\ $ 12.7 & 678.0 $\pm$ 2.4 & $<\ $ 7.9 & 69.4 $\pm$ 2.0\\
269 & 2 & 432 $\pm$ 82 & 3954 $\pm$ 76 & 139 $\pm$ 30 & $<\ $ 0.3 & 2.6 $\pm$ 0.5 & 4.3 $\pm$ 0.4 & 20.7 $\pm$ 0.3 & 58.3 $\pm$ 1.0 & $<\ $ 0.1 & $<\ $ 0.1\\
\hline

\end{tabular}
\label{tab:Paalpha}
\tablecomments{(This table is available in its entirety in a machine-readable form in the online journal. A portion is shown here for guidance regarding its form and content.)}
\tablenotetext{a}{\swiftbat 70-month hard X-ray survey ID (http://swift.gsfc.nasa.gov/results/bs70mon/).}
\tablenotetext{b}{Spectral fitting quality flags: 1 = good fit, 2 = acceptable fit, 9 = bad fit, -1 = lack of spectral coverage.}
\tablenotetext{c, d}{ nFWHM: FWHM of the narrow lines. bFWHM: FWHM of the broad lines. The FWHM are not corrected for the instrumental resolution.} 
\tablenotetext{e}{ Velocity offset of the broad component with respect to the systemic redshift.} 
\tablenotetext{f}{Emission lines fluxes are in units of $10^{-15}$ erg s$^{-1}$ cm$^{-2}$. Negative numbers indicate upper limits.}
\end{minipage}
%\end{center}
\end{table*}

% Flux table Brgamma region 
\begin{table*}
\begin{center}
\scriptsize
\begin{minipage}{\textwidth}
\caption{Emission lines flux measurements in the Br$\gamma$ region (2.00-2.40 $\mu$m)}
\begin{tabular}{rccccccccccccc}
\hline

ID\tablenotemark{a}& Flag\tablenotemark{b}  &  nFWHM\tablenotemark{c} & bFWHM\tablenotemark{d} & b $\Delta v$\tablenotemark{e}& H$_2$\tablenotemark{f} & \Alix  & \hei &H$_2$ & H$_2$ & \Caviii & H$_2$  & Br$\gamma$ & b Br$\gamma$  \\
& & [\kms ]& [\kms ]& [\kms ] &  $\lambda$2.0330$\mu$m & $\lambda$2.0400$\mu$m &  $\lambda$2.0580$\mu$m & $\lambda$2.1213$\mu$m & $\lambda$2.2227$\mu$m   & $\lambda$2.3210$\mu$m & $\lambda$2.2467$\mu$m & $\lambda$2.1655$\mu$m \\
\hline

6 & 2 & ...& 5592 $\pm$ 647 & -110 $\pm$ 33 & $<\ $ 0.5 & $<\ $ 1.0 & 3.8 $\pm$ 0.9 & $<\ $ 0.3 & $<\ $ 0.9 & $<\ $ 2.0 & $<\ $ 0.6 & 16.9 $\pm$ 1.4 & 16.4 $\pm$ 7.4\\
33 & 2 & 442 $\pm$ 33 & ...& ...  & 0.8 $\pm$ 0.3 & $<\ $ 0.5 & $<\ $ 0.7 & 1.8 $\pm$ 0.1 & $<\ $ 0.4 & $<\ $ 0.5 & $<\ $ 0.8 & 1.5 $\pm$ 0.2 & $<\ $ 0.2\\
116 & 9 & ...& ...& ...  & $<\ $ 0.7 & $<\ $ 0.3 & $<\ $ 0.1 & $<\ $ 0.7 & $<\ $ 0.3 & $<\ $ 1.9 & $<\ $ 0.2 & $<\ $ 2.0 & $<\ $ 0.1\\
130 & 9 & ...& ...& ...  & $<\ $ 0.1 & $<\ $ 0.1 & $<\ $ 0.1 & $<\ $ 0.1 & $<\ $ 0.1 & $<\ $ 0.1 & $<\ $ 0.1 & 0.3 $\pm$ 0.1 & $<\ $ 0.1\\
140 & 2 & ...& ...& ...  & $<\ $ 2.9 & $<\ $ 0.9 & $<\ $ 1.4 & 3.6 $\pm$ 0.2 & $<\ $ 1.8 & $<\ $ 12.1 & $<\ $ 0.6 & $<\ $ 12.2 & $<\ $ 2.6\\
157 & 9 & 515 $\pm$ 38 & ...& ...  & 0.5 $\pm$ 0.3 & $<\ $ 0.1 & $<\ $ 0.1 & 0.8 $\pm$ 0.1 & $<\ $ 0.2 & $<\ $ 0.8 & $<\ $ 0.1 & $<\ $ 0.8 & $<\ $ 0.2\\
173 & 2 & 534 $\pm$ 7 & ...& ...  & 14.8 $\pm$ 0.6 & $<\ $ 3.1 & 3.6 $\pm$ 1.2 & 42.9 $\pm$ 0.7 & 10.9 $\pm$ 0.3 & $<\ $ 0.2 & 4.0 $\pm$ 1.7 & $<\ $ 0.2 & $<\ $ 0.1\\
226 & 9 & ...& ...& ...  & $<\ $ 0.7 & $<\ $ 1.1 & $<\ $ 0.2 & $<\ $ 0.4 & $<\ $ 1.4 & $<\ $ 1.4 & $<\ $ 2.7 & $<\ $ 1.4 & $<\ $ 4.6\\
266 & 2 & ...& 5171 $\pm$ 183 & 110 $\pm$ 1 & $<\ $ 3.1 & $<\ $ 1.4 & $<\ $ 1.4 & $<\ $ 2.3 & $<\ $ 1.8 & $<\ $ 2.0 & 3.8 $\pm$ 24.2 & $<\ $ 2.1 & 72.4 $\pm$ 1.8\\
269 & 2 & ...& 5758 $\pm$ 286 & 63 $\pm$ 18 & $<\ $ 0.8 & $<\ $ 0.3 & $<\ $ 0.7 & 1.4 $\pm$ 0.1 & $<\ $ 0.5 & $<\ $ 2.9 & $<\ $ 0.3 & 0.8 $\pm$ 0.7 & 11.6 $\pm$ 3.4\\

\hline

\end{tabular}
\label{tab:Brgamma}
\tablecomments{(This table is available in its entirety in a machine-readable form in the online journal. A portion is shown here for guidance regarding its form and content.)}
\tablenotetext{a}{\swiftbat 70-month hard X-ray survey ID (http://swift.gsfc.nasa.gov/results/bs70mon/).}
\tablenotetext{b}{Spectral fitting quality flags: 1 = good fit, 2 = acceptable fit, 9 = bad fit, -1 = lack of spectral coverage.}
\tablenotetext{c, d}{ nFWHM: FWHM of the narrow lines. bFWHM: FWHM of the broad lines. The FWHM are not corrected for the instrumental resolution.} 
\tablenotetext{e}{ Velocity offset of the broad component with respect to the systemic redshift.} 
\tablenotetext{f}{Emission lines fluxes are in units of $10^{-15}$ erg s$^{-1}$ cm$^{-2}$. Negative numbers indicate upper limits.}
\end{minipage}
\end{center}
\end{table*}

\end{landscape}

%%%%%%%%%%%%%%%%%%%%%%%%%%%%%%%%%%%%%%%%%%%%%%%%%%

%%%%%%%%%%%%%%%%% APPENDICES %%%%%%%%%%%%%%%%%%%%%

\appendix


\section{}
Here we provide additional information about the spectra and fitting.  Plots of each of the new spectra obtained in the observing programs on SpeX and FLAMINGOS are found in Figures A1-A3.  In Section \ref{lines_tables} we reported a list of the emission and absorption lines  present in the NIR range (0.8 - 2.4 $\mu$m). In the next section we show an example of {\tt pPXF} absorption line fitting. In Section \ref{sec:comp_vel} we also performed a variety of checks comparing our velocity dispersion measurements obtained from different spectral region in the NIR and in the optical.
%, which are described in Section \ref{sec:comp_vel}. 
 Finally, in Section \ref{sec:MBH_comp} we show the comparison of our black hole mass measurements.

%% Plot of the new observed spectra
\begin{figure*}
\centering
\subfigure{\includegraphics[width=1.\textwidth]
{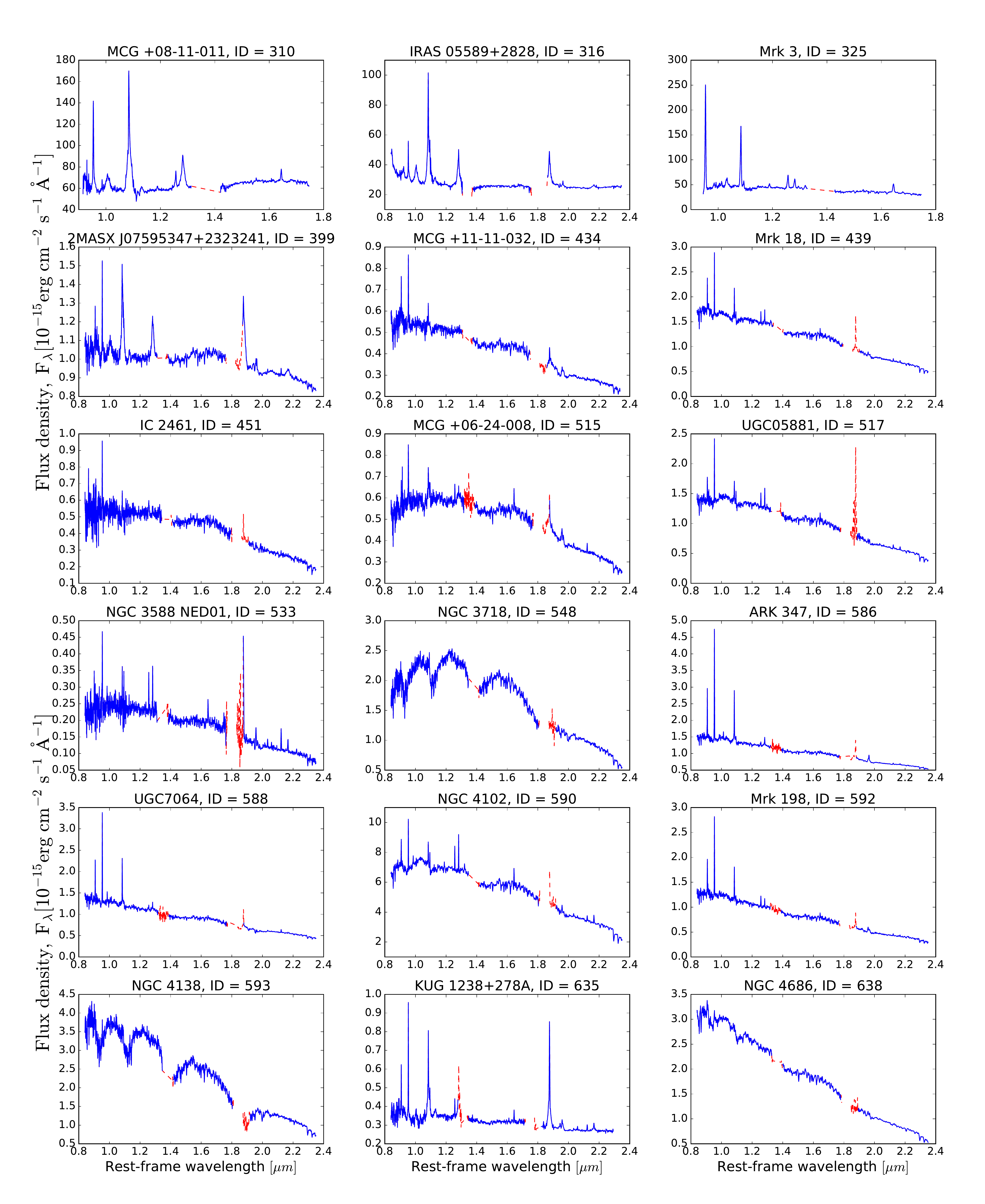}}
\caption{Galaxy spectra from new observations arranged in order of increasing ID (\swiftbat 70-month hard X-ray survey ID (http://swift.gsfc.nasa.gov/results/bs70mon/). The red color indicates indicate regions of atmospheric absorption. Spectra are taken from the SpeX instrument on the IRTF with a 0.8'' slit for all sources, except MCG +08-11-011 (ID = 310) and Mrk 3 (ID = 325), which were taken from the FLAMINGOS instrument on the Kitt Peak 4m telescope with a 1.5'' slit.}
\label{spectra1}
\end{figure*}

\begin{figure*}
\centering
\subfigure{\includegraphics[width=1.\textwidth]
{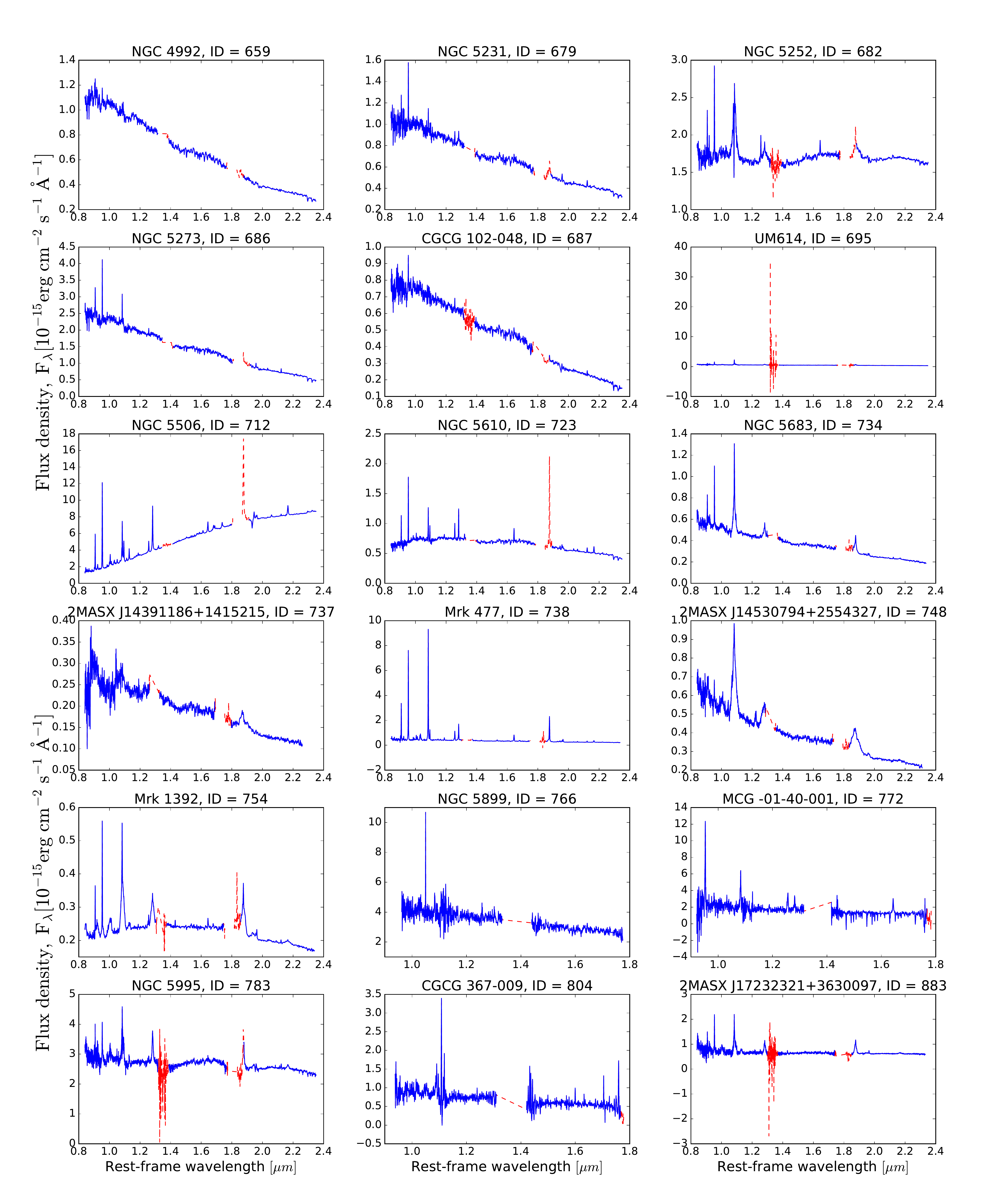}}
\caption{Galaxy spectra from new observations arranged in order of increasing ID (\swiftbat 70-month hard X-ray survey ID (http://swift.gsfc.nasa.gov/results/bs70mon/). The red color indicates indicate regions of atmospheric absorption. Spectra are taken from the SpeX instrument on the IRTF with a 0.8'' slit for all sources, except NGC 5899 (ID = 766), MCG -01-40-001 (ID = 772) and CGCG 367-009 (ID = 804), which were taken from the FLAMINGOS instrument on the Kitt Peak 4m telescope with a 1.5'' slit.}
\label{spectra2}
\end{figure*}

\begin{figure*}
\centering
\subfigure{\includegraphics[width=1.\textwidth]
{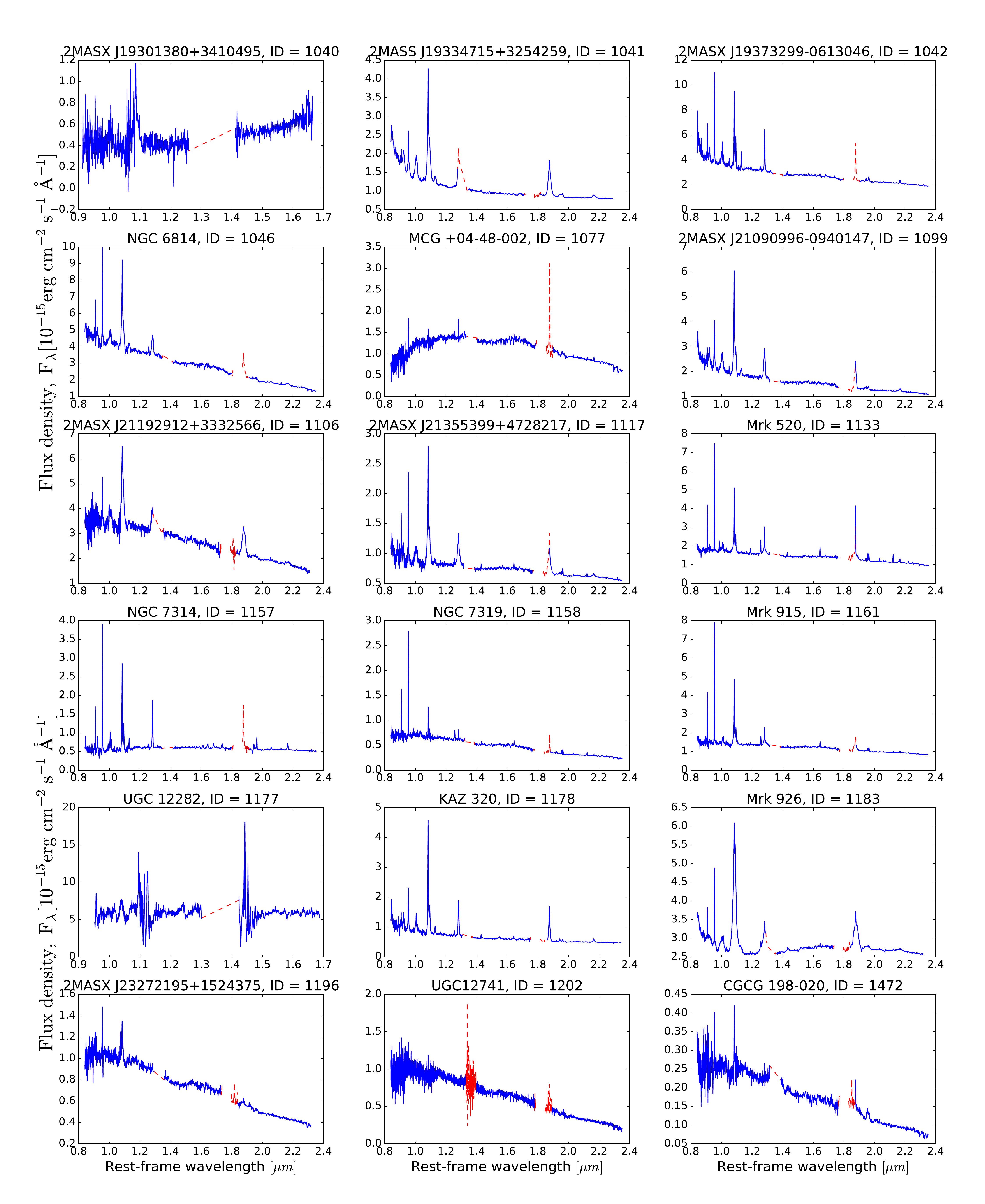}}
\caption{Galaxy spectra from new observations arranged in order of increasing ID (\swiftbat 70-month hard X-ray survey ID (http://swift.gsfc.nasa.gov/results/bs70mon/). The red color indicates indicate regions of atmospheric absorption. Spectra are taken from the SpeX instrument on the IRTF with a 0.8'' slit for all sources, except 2MASX J19301380+3410495 (ID = 1040) and UGC 12282 (ID = 1177), which were taken from the FLAMINGOS instrument on the Kitt Peak 4m telescope with a 1.5'' slit.}
\label{spectra3}
\end{figure*}

\begin{figure*}
\centering
\subfigure{\includegraphics[width=1.\textwidth]
{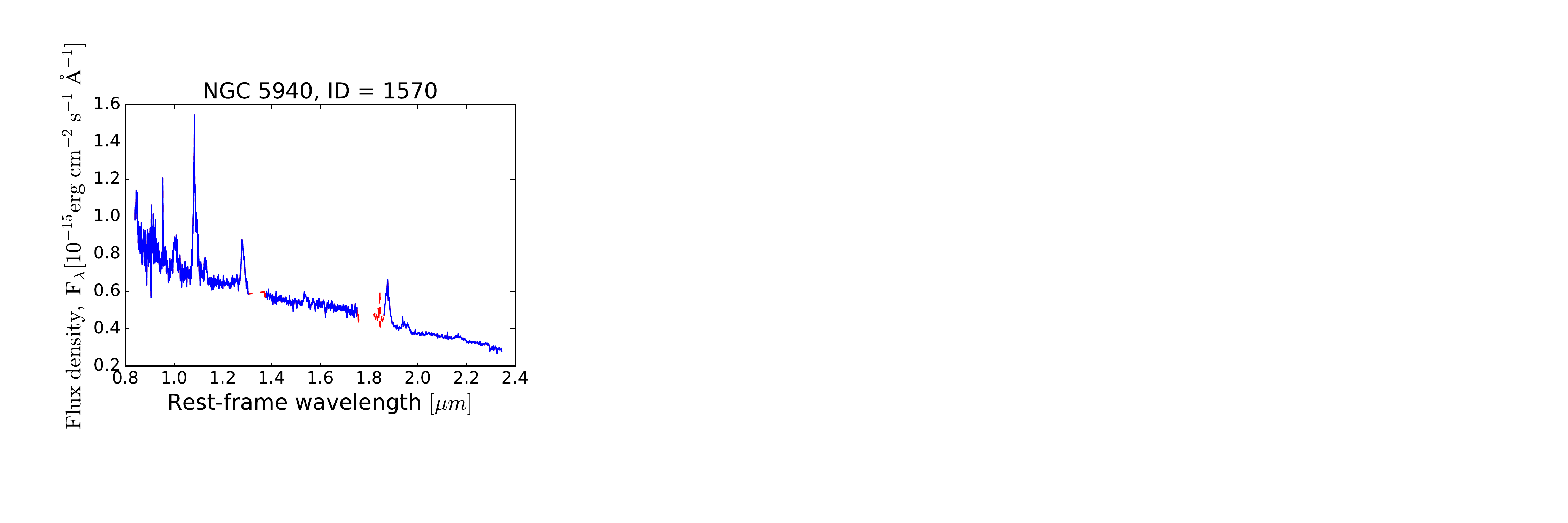}}
\caption{Galaxy spectra from new observations arranged in order of increasing ID (\swiftbat 70-month hard X-ray survey ID (http://swift.gsfc.nasa.gov/results/bs70mon/). The red color indicates indicate regions of atmospheric absorption. This spectrum is taken from the SpeX instrument on the IRTF with a 0.8'' slit.}
\label{spectra4}
\end{figure*}

\subsection{Emission and absorption lines tables}
\label{lines_tables}
In Table \ref{tab:emission_lines} are listed the emission lines present in the range 0.8-2.4 $\mu$m. Table \ref{abs_lines} shows the absorption lines  from which we measured the velocity dispersion using {\tt pPXF}, while Table \ref{emlinesmask} shows the emission lines that we masked in the {\tt pPXF} fitting procedure. 

% List of emission lines
\begin{table}
\caption{Emission lines}
\centering
\begin{tabular}{|l|c|c|}
\hline
Emission Line & Wavelength & Spectral region \\ 
 &  [$\mu$m]&  \\ \hline

{\rm O}\,\textsc{i} & 0.8449 & Pa14 \\ 
%Pa17  & 0.8467 &  \\ 
Pa16  & 0.8503 &  \\ 
Pa15  & 0.8545 &  \\ 
Pa14  & 0.8600 &  \\ 
Pa13  & 0.8665 &  \\ 
Pa12  & 0.8750 &  \\ 
Pa11  & 0.8863 &  \\ 
 & &  \\ 
Pa10  & 0.9016 & Pa$\zeta$  \\
\siiif\ & 0.9069 &  \\ 
\feiif\ & 0.9227 &  \\ 
Pa$\zeta$  & 0.9229 &  \\ 
\siiif\ & 0.9531 &  \\ 
Pa$\epsilon$ & 0.9546 &  \\ 
 & \multicolumn{1}{l|}{} &  \\ 
\CI & 0.9827 & Pa$\delta$  \\
\CI & 0.9853 &  \\
\Sviii  & 0.9915 &  \\
\feiif\  & 0.9999 &  \\
Pa$\delta$  & 1.0049 &  \\
\HeII & 1.0126 &  \\
\sii & 1.0290 &  \\
\sii & 1.0320 &  \\ 
\sii & 1.0336 &  \\ 
\sii & 1.0370 &  \\ 
\NI & 1.0404 &  \\ 
 & \multicolumn{1}{l|}{} &  \\ 
\feiif\ & 1.0500 & Pa$\gamma$ \\ 
\Fexiii & 1.0747 & \\
\hei & 1.0830 &  \\ 
Pa$\gamma$ & 1.0938 &  \\ 
\feiif\ & 1.1126 &  \\ 
{\rm O}\,\textsc{i}  & 1.1287 &  \\ 
& & \\
\HeII & 1.1620 & Pa$\beta$  \\ 
\pii & 1.1886 &  \\ 
\NiII &  1.1910 &  \\
\Six & 1.2520 &  \\ 
\feiif\  & 1.2570 &  \\ 
\feiif\ & 1.2788 &  \\ 
Pa$\beta$  & 1.2818 &  \\ 
\feiif\ & 1.2950 &  \\ 
\feiif\ & 1.3201 &  \\ 
&  &  \\ 
\Silix & 1.4300 & Br10 \\ 
\feiif\ & 1.5339 &  \\ 
\feiif\  & 1.6436 &  \\ 
\feiif\  &  1.6807 &  \\ 
Br11  & 1.6811 &  \\ 
Br10  & 1.7367 &  \\ 
 &  &  \\ 
%Br$\epsilon$  & 1.8179 & Pa$\alpha$  \\ 
Pa$\alpha$  & 1.8751 & Pa$\alpha$  \\ 
\Silixi & 1.9320 & \\ 
Br$\delta$  & 1.9446 &  \\ 
H$_2$ & 1.9564 &  \\ 
\Silivi & 1.9620 &  \\ 
&  &  \\ 
H$_2$ & 2.0330 &  Br$\gamma$\\ 
\Alix &2.0400 & \\ 
\hei & 2.0580 &  \\ 
H$_2$ & 2.1213 &  \\
Br$\gamma$ & 2.1655 &  \\ 
H$_2$ & 2.2227 &  \\ 
\Caviii & 2.3210 &  \\ 
H$_2$ & 2.2467 &  \\ \hline
\end{tabular}
\label{tab:emission_lines}
\end{table}

% Table of absorption lines
\begin{table}
\caption{Absorption lines used in {\tt pPXF} host galaxy fitting}
\centering
\begin{tabular}{l|r|l}
\hline
Absorption lines & \multicolumn{1}{l|}{Wavelength} & Spectral region \\ 
 & \multicolumn{1}{l|}{[$\mu$m]} &  \\ \hline 
\caii & 0.8498 & CaT \\ 
\caii & 0.8542 & CaT \\ 
\caii & 0.8662 & CaT \\ 
 &  &  \\
CO (4-1) & 1.5800 & H-band \\ 
CO (5-2) & 1.6000 & H-band \\
CO (6-3) & 1.6200 & H-band \\ 
CO (7-4) & 1.6400 & H-band \\ 
CO (8-5) & 1.6600 & H-band \\ 

%Si I & 1.59 & H-band \\ 
%Mg I  & 1.71 & H-band \\ \hline
 &  &  \\ 
$^{12}$CO (2-0) & 2.2935 & K-band \\ 
$^{12}$CO (3-1) & 2.3226 & K-band \\ 
$^{13}$CO (2-0) & 2.3448 & K-band \\ 
$^{12}$CO (4-2) & 2.3514 & K-band \\ 
$^{12}$CO (5-3) & 2.3834 & K-band \\ \hline
\end{tabular}
\label{abs_lines}
\end{table}

\begin{table}
\caption{Emission lines masked in {\tt pPXF} host galaxy fitting}
\centering
\begin{tabular}{l|r|l}
\hline
Emission line & \multicolumn{1}{l|}{Wavelength} & Spectral region \\ 
 & \multicolumn{1}{l|}{[$\mu$m]} &  \\ \hline
Pa14 & 0.8598 & CaT \\
\feiif & 1.6436 & CO H-band \\ %\hline
\feiif & 1.6773 & CO H-band \\ %\hline
Br11 & 1.6811 & CO H-band \\ %\hline
\Caviii & 2.3210 & CO K-band \\ \hline

\end{tabular}
\label{emlinesmask}
\end{table}

% Coronal lines
\begin{table}
\caption{Coronal lines in the wavelength range 0.8-2.4 $\mu$m}
\begin{center}
\begin{tabular}{l|c|c}
\hline
Coronal line & Wavelength & Ionization potential \\ 
 & [$\mu$m] & [eV] \\ \hline 
{\Caviii} & 2.3210 & 127.7 \\ 
{\Silivi} & 1.9620 & 166.8 \\ 
{\Sviii} & 0.9915 & 280.9 \\ 
{\Alix} & 2.0400 & 284.6 \\ 
{\Six} & 1.2520 & 328.2 \\ 
{\Fexiii} & 1.0747 & 330.8 \\ 
{\Silix} & 1.4300 & 351.1 \\ 
{\Silixi} & 1.9320 & 401.4 \\
{\Sxi} & 1.9196 & 447.1 \\
\hline
\end{tabular}
\label{coronal_lines}
\end{center}
\end{table}

\subsection{Galaxy template fitting and velocity dispersion comparison}
\label{sec:comp_vel}

\begin{figure*}
\centering
\subfigure{\includegraphics[width=0.49\textwidth]{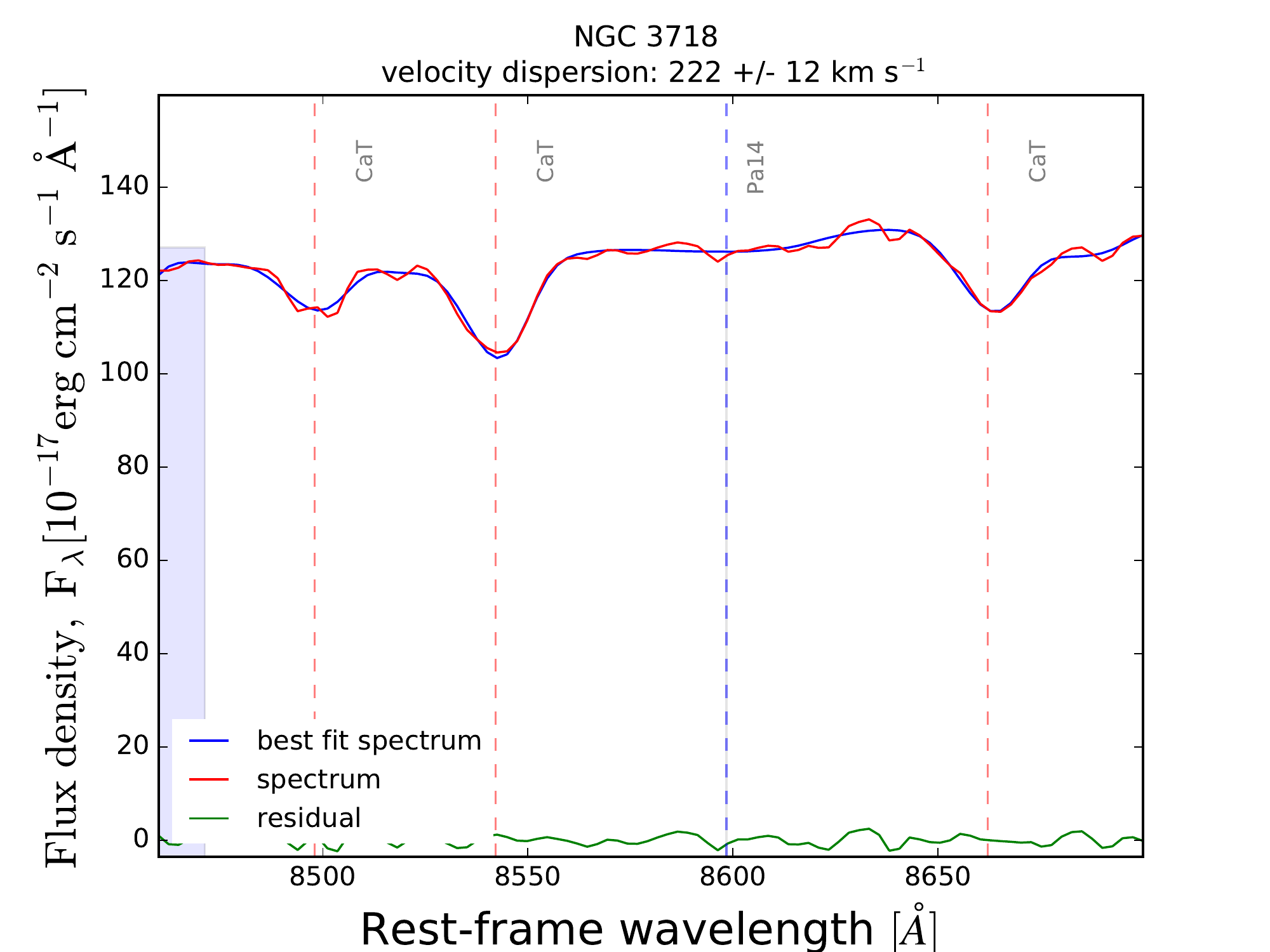}}
\subfigure{\includegraphics[width=0.49\textwidth]{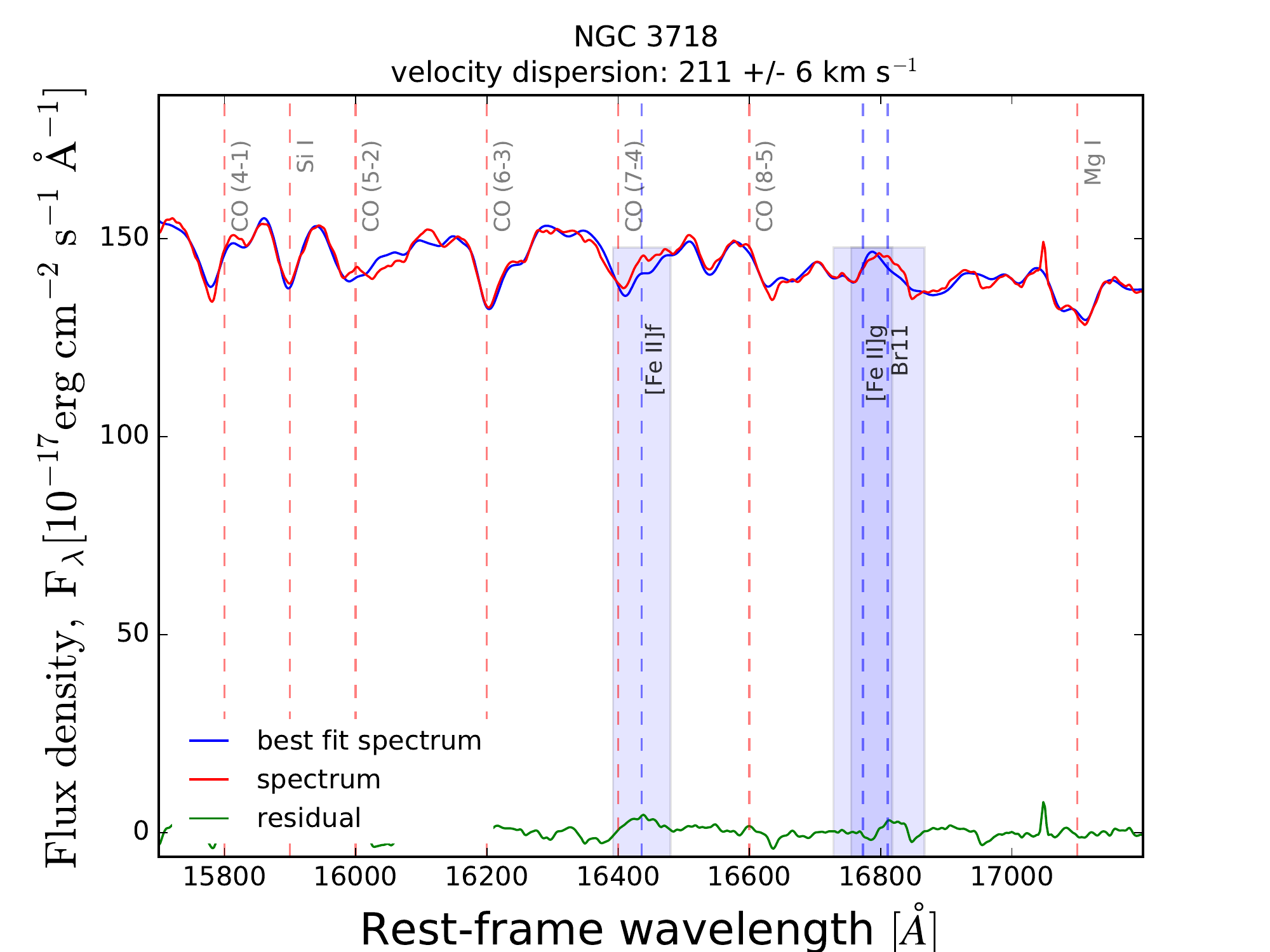}}
\subfigure{\includegraphics[width=0.49\textwidth]{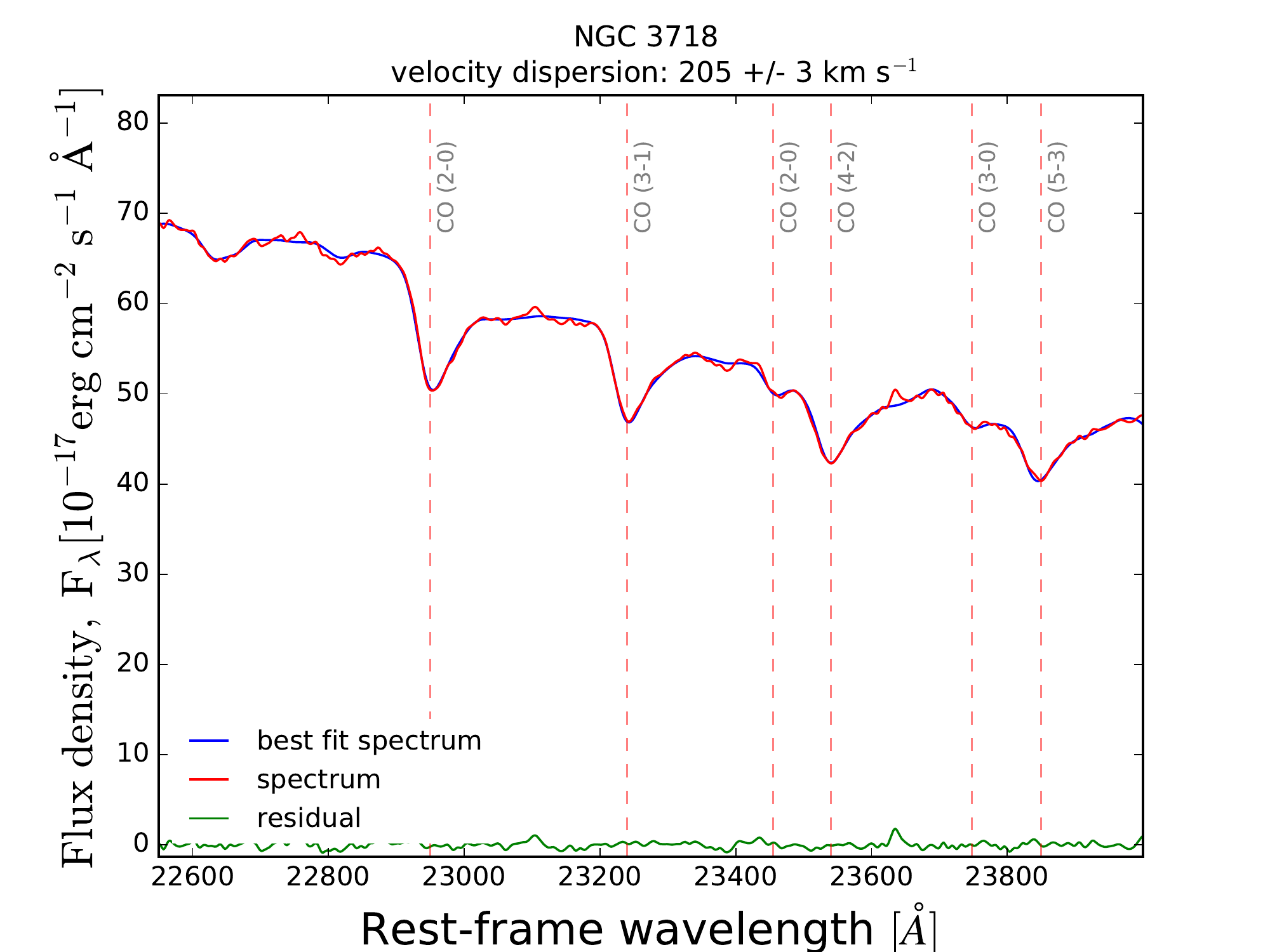}}
\caption{Example of {\tt pPXF} fit of the CaT region (0.846-0.870 $\mu$m), CO band-head in the H-band (1.570-1.720 $\mu$m) and in the K-band (2.255-2.400 $\mu$m).}
\label{ppxf_ex}
\end{figure*}

\begin{figure*}
\centering
\subfigure{\includegraphics[width=0.49\textwidth]{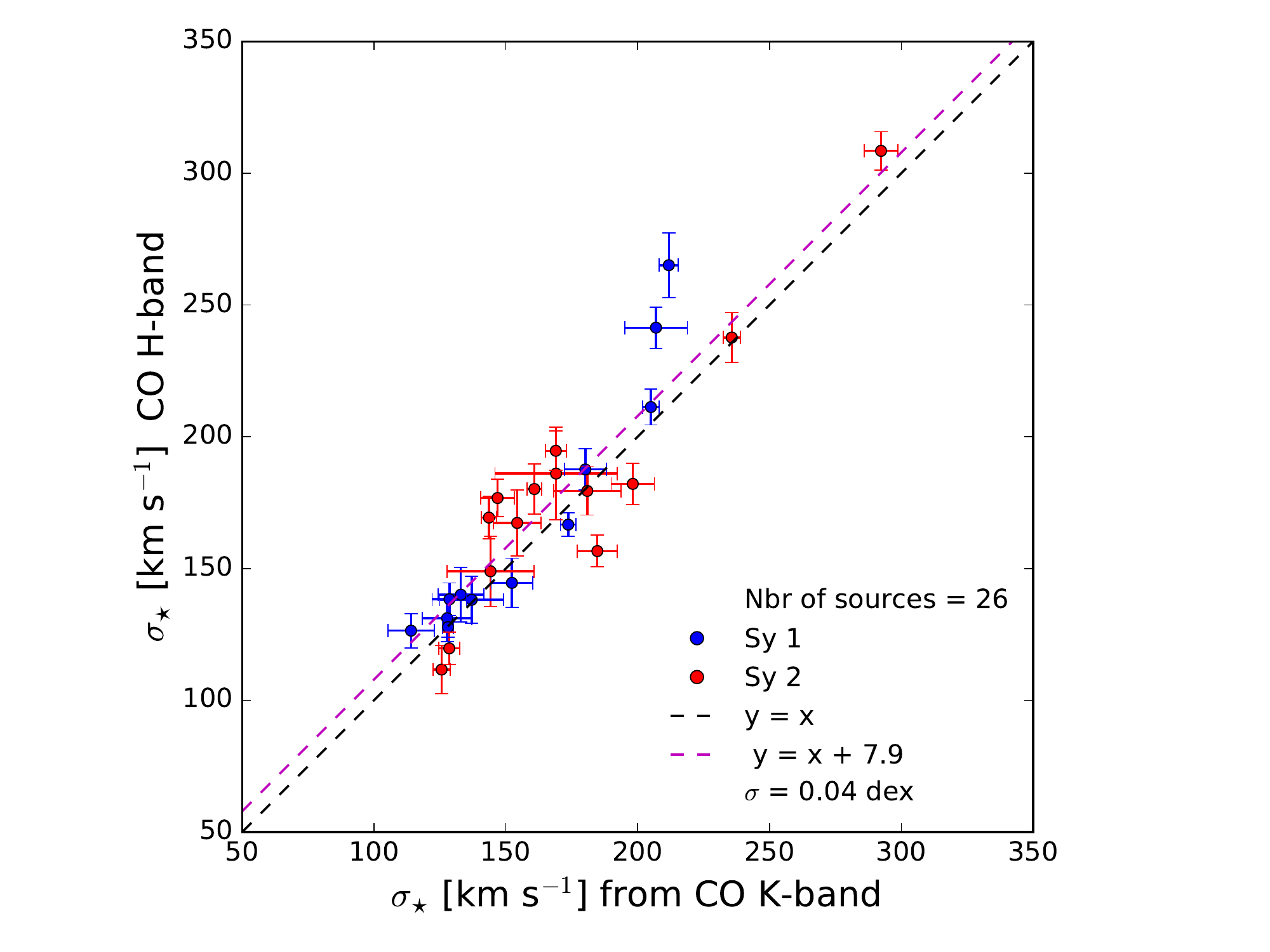} }
\subfigure{\includegraphics[width=0.49\textwidth]{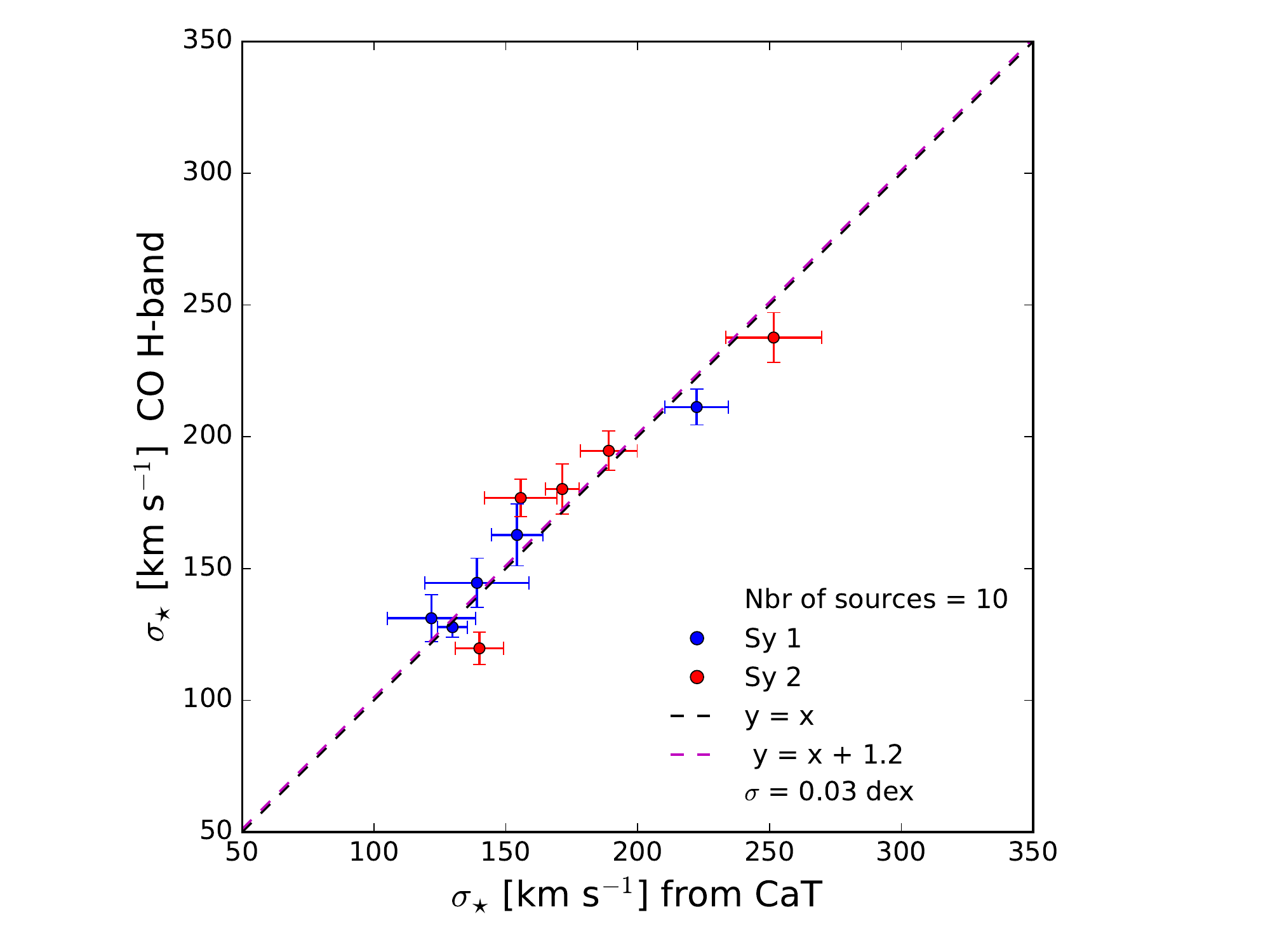} }
%\vfill
\subfigure{\includegraphics[width=0.49\textwidth]{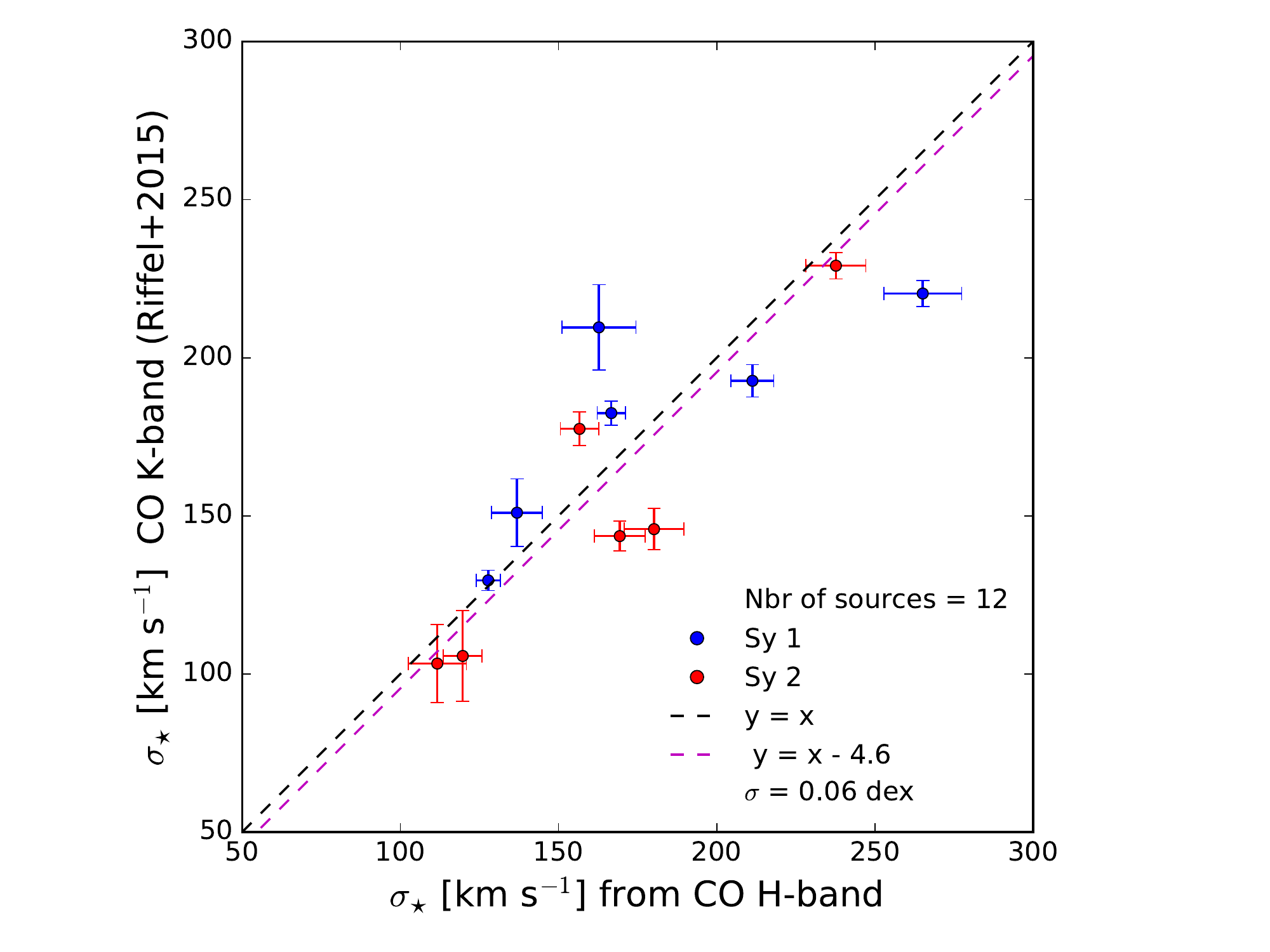} }
\subfigure{\includegraphics[width=0.49\textwidth]{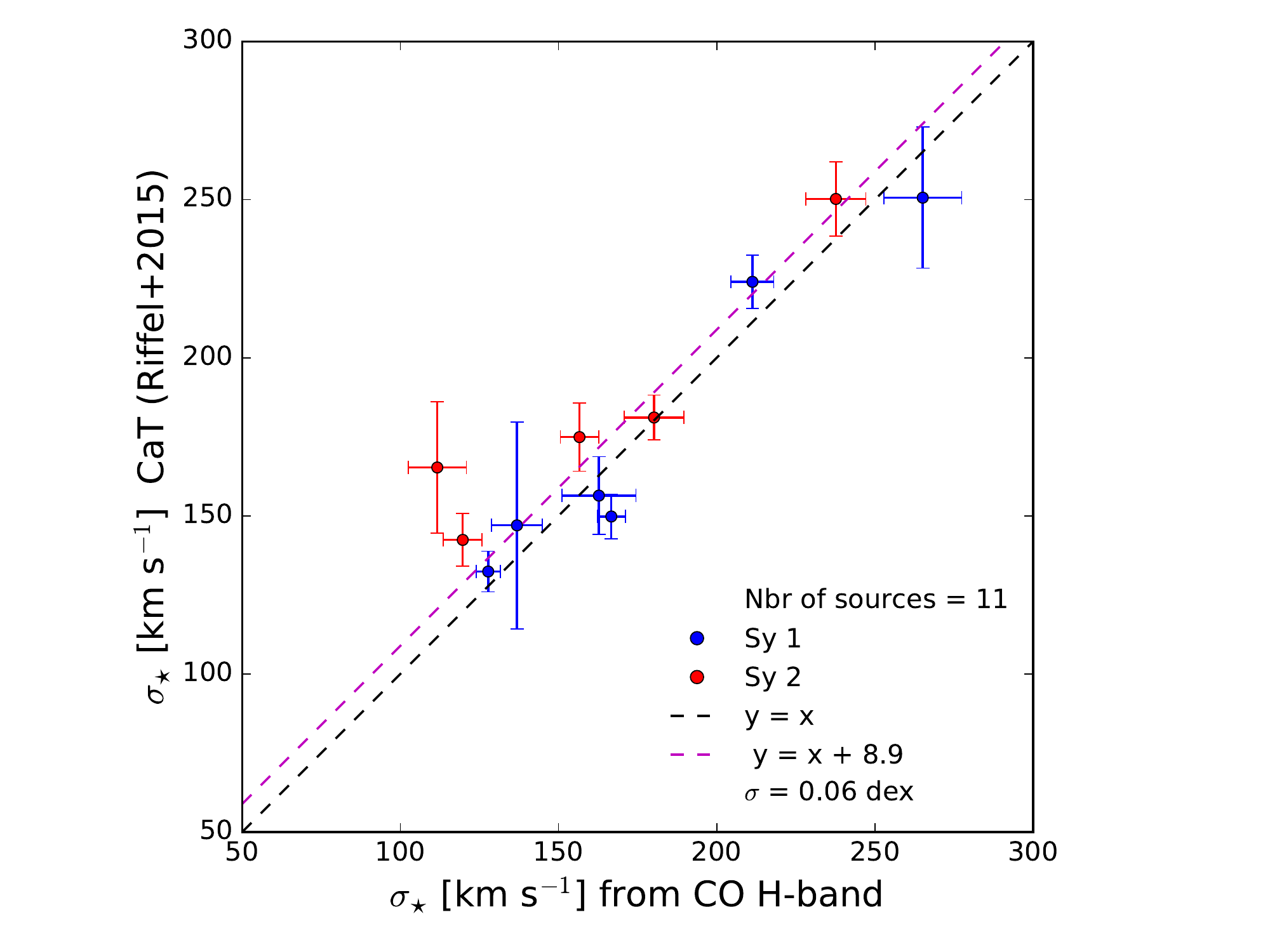} }
%\vfill

\subfigure{\includegraphics[width=0.49\textwidth]
{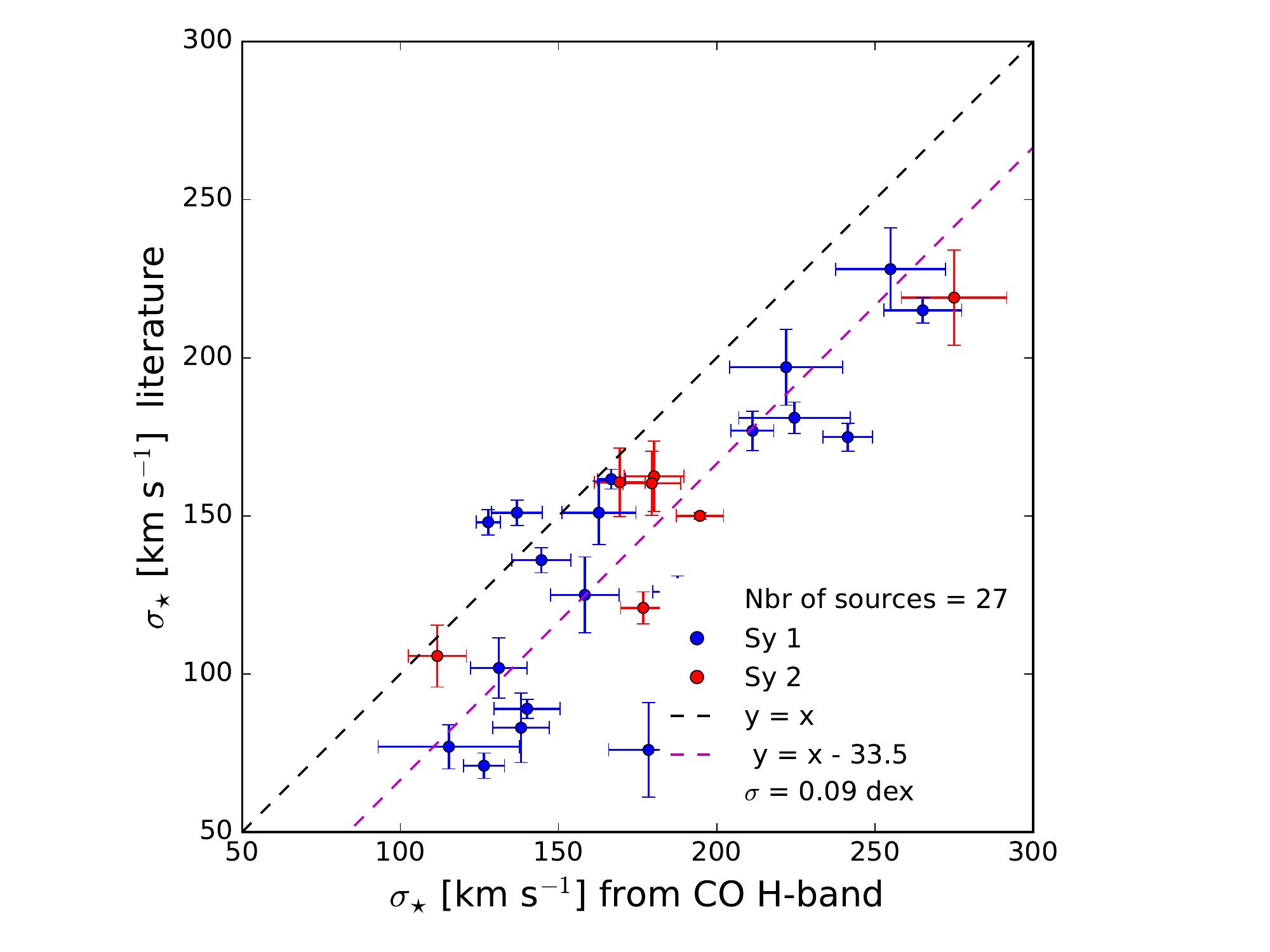} }
\subfigure{\includegraphics[width=0.49\textwidth]{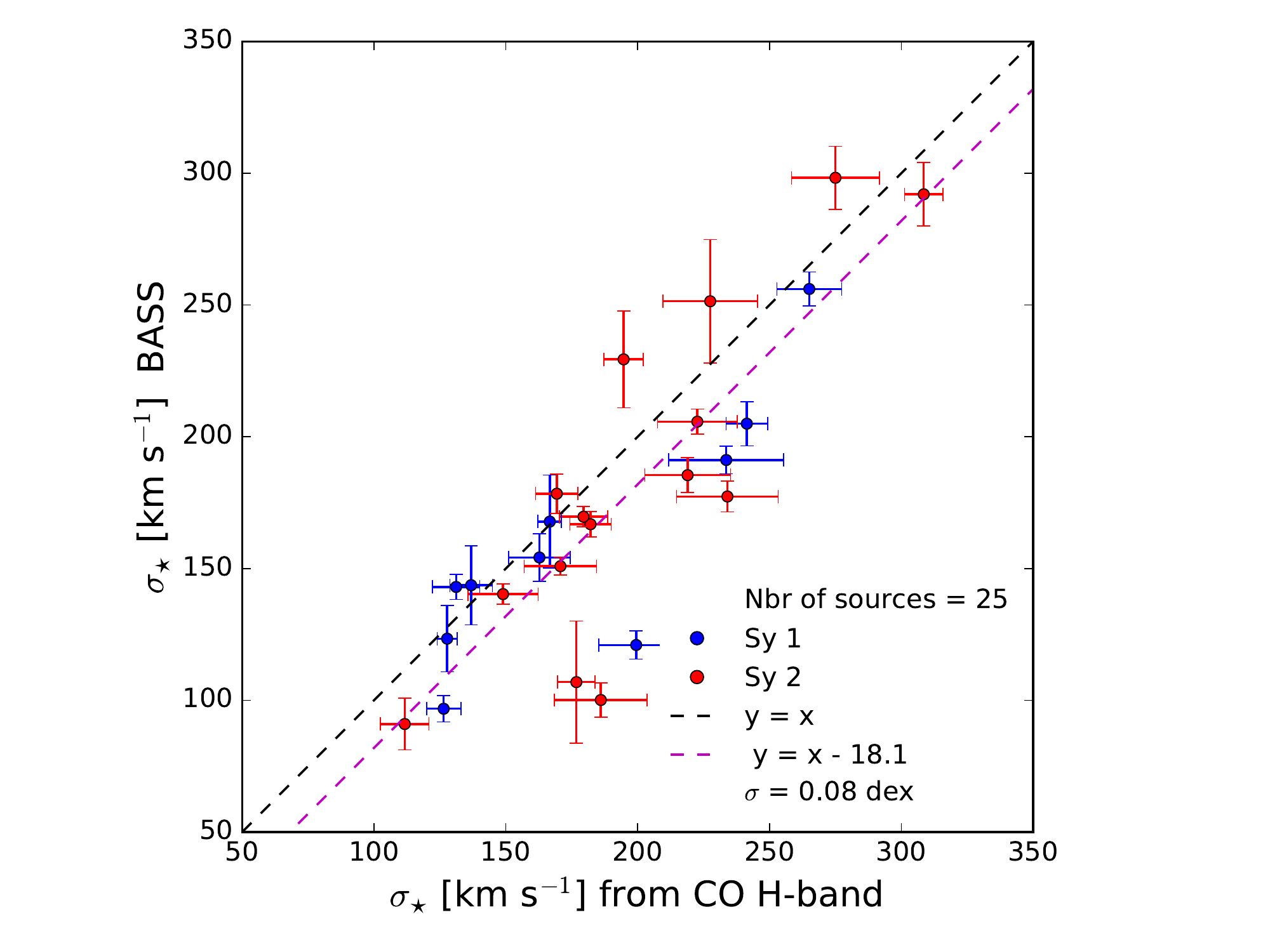} }

\caption{Comparison of velocity dispersion $\sigma_{*}$ measurements. The black dashed line shows the one-to-one relation ($x = y $) and the magenta dashed line shows the fit of a one-to-one relation allowing an offset.
\textit{Upper panels:} Comparison of $\sigma_{*,\rm CO}$ measured in the K band and in the H band (left) and comparison of $\sigma_{*,\rm CO}$ measured in the H band and $\sigma_{*,\rm CaT}$ (right).
%\label{Comp_CaT_CO}
\textit{Middle panles:} Comparison of our $\sigma_{*, \rm CO}$ measured in the H band and the $\sigma_{*}$ measured by \citet{Riffel2015} from the CO band-head in the K-band (left) and from the CaT (right).
%\label{Comp_Riffel}
\textit{Lower panels:} Comparison of $\sigma_{*,\rm CO}$ (H band) with literature values (left) and with the values of \sig\ from the CaH and K and the MgIb absorption lines from the BASS catalog (right).}

\label{Comp_CO}
\end{figure*}

In Figure \ref{ppxf_ex} we show an example of the stellar velocity dispersion fit in the three wavelength regions (CaT, CO bandhead in the H-band and K-band).
We performed some tests to compare our velocity dispersion measurements in the different NIR wavelength ranges and with literature values (Figure \ref{Comp_CO}).

\begin{figure}
\centering

\subfigure{\includegraphics[width=0.49\textwidth]{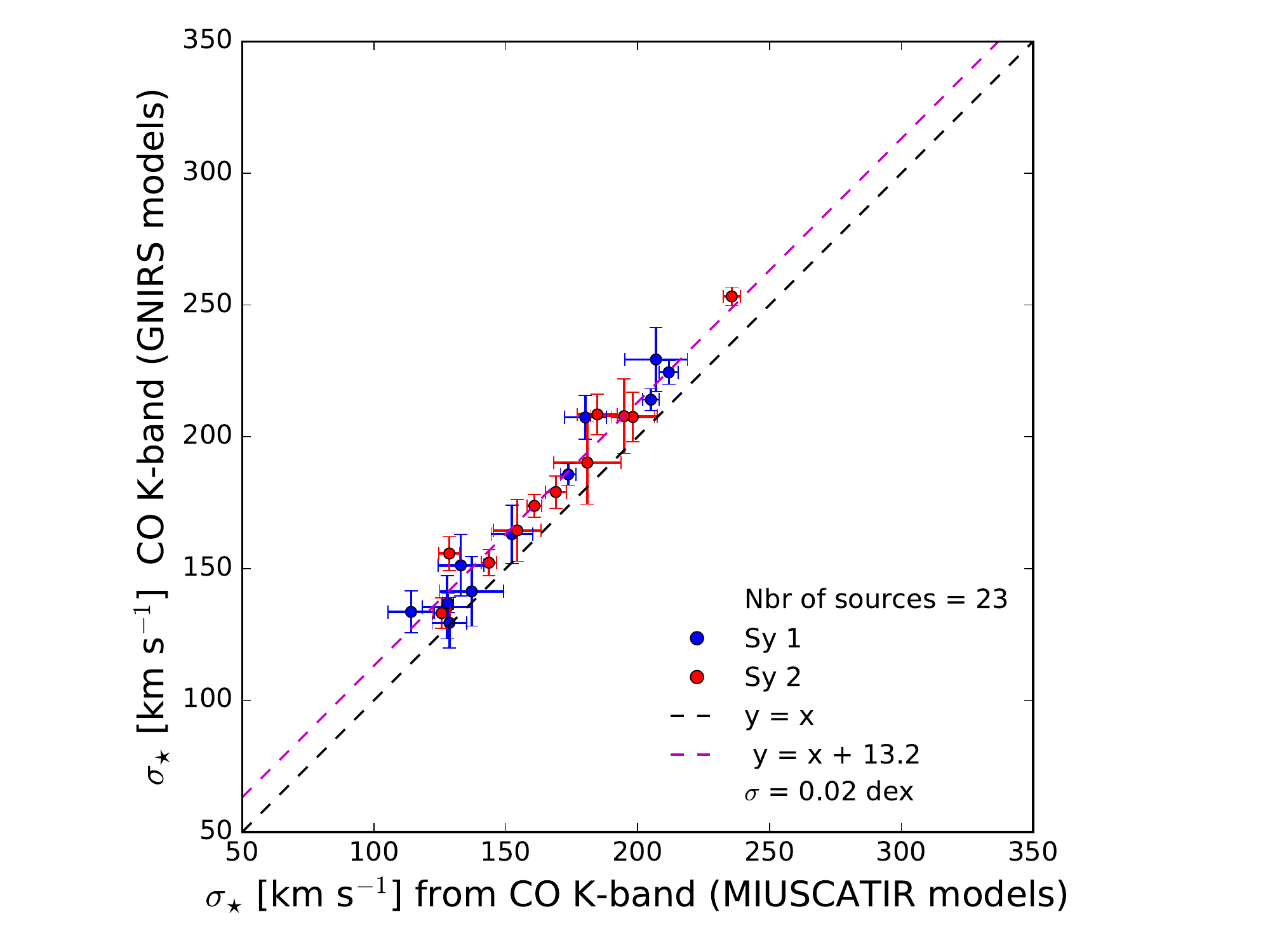}}
%\vfill
\caption{Comparison of $\sigma_{* \rm CO}$ measured in the K band region using the MIUSCATIR models and the GNIRS models. The black dashed line shows the one-to-one relation ($x = y $) and the magenta dashed line shows the fit of a one-to-one relation allowing an offset. }
\label{Comp_MIUSCAT_GNIRS}
\end{figure}

For the CO band-heads in the K-band (2.255-2.400 $\mu$m), we compared the results obtained using the MIUSCATIR models and the GNIRS models (Figure \ref{Comp_MIUSCAT_GNIRS}).
The values of \sig\ measured with the GNIRS models are systematically larger than the values measured with the MIUSCATIR models. However, the offset is quite small (13 \kmps) and lies within the error bars. Thus, we conclude that the difference between the two libraries is negligible.
We decided to use the MIUSCATIR models, since they have a larger wavelength coverage and they can be used also for the H-band.

We also compared the \sig\ measured from the CO band-heads in the K-band (2.255-2.400 $\mu$m) and in the H-band (1.57-1.72 $\mu$m). The values of $\sigma_{*,\rm CO}$ measured in the H and K bands are in good agreement. The offset (8 \kmps) is within the measurement errors.  We compared also the measurements from the CO band-heads and from the CaT. The results are in very good agreement, with a scatter around the one-to-one relation of 0.03 dex.

We compared our results with the \sig\ values measured in the NIR by \cite{Riffel2015}. They measured \sig\ from the CO band-head in the K-band and from the CaT for 48 galaxies, of which 13 are also in our sample. Our measurements of $\sigma_{*,\rm CO}$ in the H-band are not in perfect agreement with the $\sigma_{*,\rm CO}$ and $\sigma_{*,\rm CaT}$ measured by \cite{Riffel2015}. However, the scatter is quite small (0.06 dex) and the offset from the one-to-one relation ($<\ 10 \kmps$) is comparable to the size of the error bars.

We  compared our values of \sig\ with the literature values measured from the optical absorption lines. The \sig\ that we measured in the NIR is, on average, larger than the \sig\ from the optical ($\sigma_{*,\rm opt}$). Fitting a one-to-one relation we found an offset of 33 \kmps. 
We also compared the values of $\sigma_{*,\rm CO}$ with the values measured from the optical absorption lines (\Cahk \AA\ and \Mgb \AA) from the BASS catalog. The scatter is quite large and the values of $\sigma_{*,\rm CO}$ are larger on average than \sig\ from the optical lines (offset = 18 \kmps). 
These discrepancies can be due to different reasons. 
The size of the observed part of the galaxy from which the spectrum is extracted depends on the slit size and on the redshift on the galaxy. The \sig\ measured in the nuclear region of the galaxy is larger than the \sig\ measured in a larger region. The slit size used for our NIR observations is relatively small (0.8''-1.5'') and this can cause our \sig\ measurements to be larger than other measurements obtained using a larger slit.
Another effect that can affect the \sig\ measurements is the broadening of the absorption line due to the rotational motion of the stars. The absorption line width include components from both random motion and rotation and therefore the measured \sig\ need to be corrected for the rotation effect. 
Another possible explanation is that CO absorption lines probe the younger red supergiants, therefore $\sigma_{*,\rm CO}$ can be different from $\sigma_{*,\rm opt}$ if there is an age radial gradient in the stellar distribution \citep{Rothberg2010, Rothberg2013}.
 \cite{Kang2013} observed that $\sigma_{*,\rm CO}$ measured in the H-band is in good agreement with the $\sigma_{*,\rm opt}$, based on a sample of 31 nearby galaxies.
\cite{Riffel2015} instead found that $\sigma_{*,\rm CO}$ is on average smaller than $\sigma_{*,\rm CaT}$, whereas $\sigma_{*,\rm CaT}$ is in good agreement with $\sigma_{*,\rm opt}$.

\subsection{Black hole mass comparison}
\label{sec:MBH_comp}
In this section we compare our \MBH\ measurements from the broad Pa$\beta$ line  with \MBH\ measurements from the \sigs, and  with values taken from literature.
 To calculate \MBH\ from the broad Pa$\beta$ line, we used the  formula from \cite{LaFranca2015} (equation \ref{eq:LaFranca}) which is based on the FWHM and luminosity of Pa$\beta$.
We considered also the formula from \cite{Kim2010}
\begin{equation}
%\begin{split}
\frac{M_{BH}}{M_{\odot}} = 10^{7.40} \left(\frac{L_{Pa\beta}}{10^{42} \text{ erg s$^{-1}$}} \right)^{0.46\pm0.02} \left(\frac{\text{FWHM}_{Pa\beta}}{10^3 \text{ km s$^{-1}$}} \right)^{1.41\pm0.09} \,\ .
%(\pm 10^{0.27})
%\end{split}
\end{equation}

\begin{figure*}
\centering
\subfigure{\includegraphics[width=0.49\textwidth]{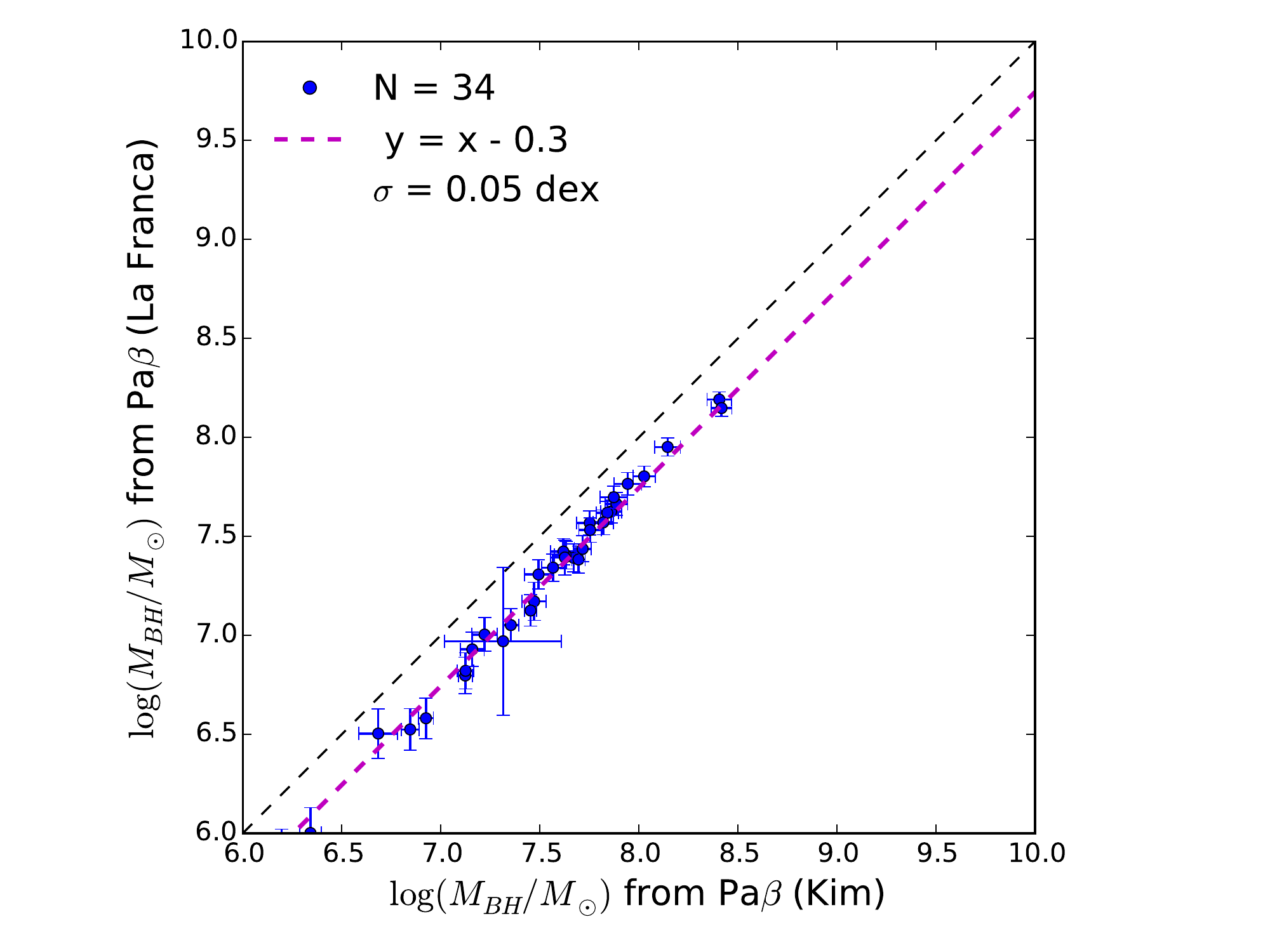} }
\subfigure{\includegraphics[width=0.49\textwidth]{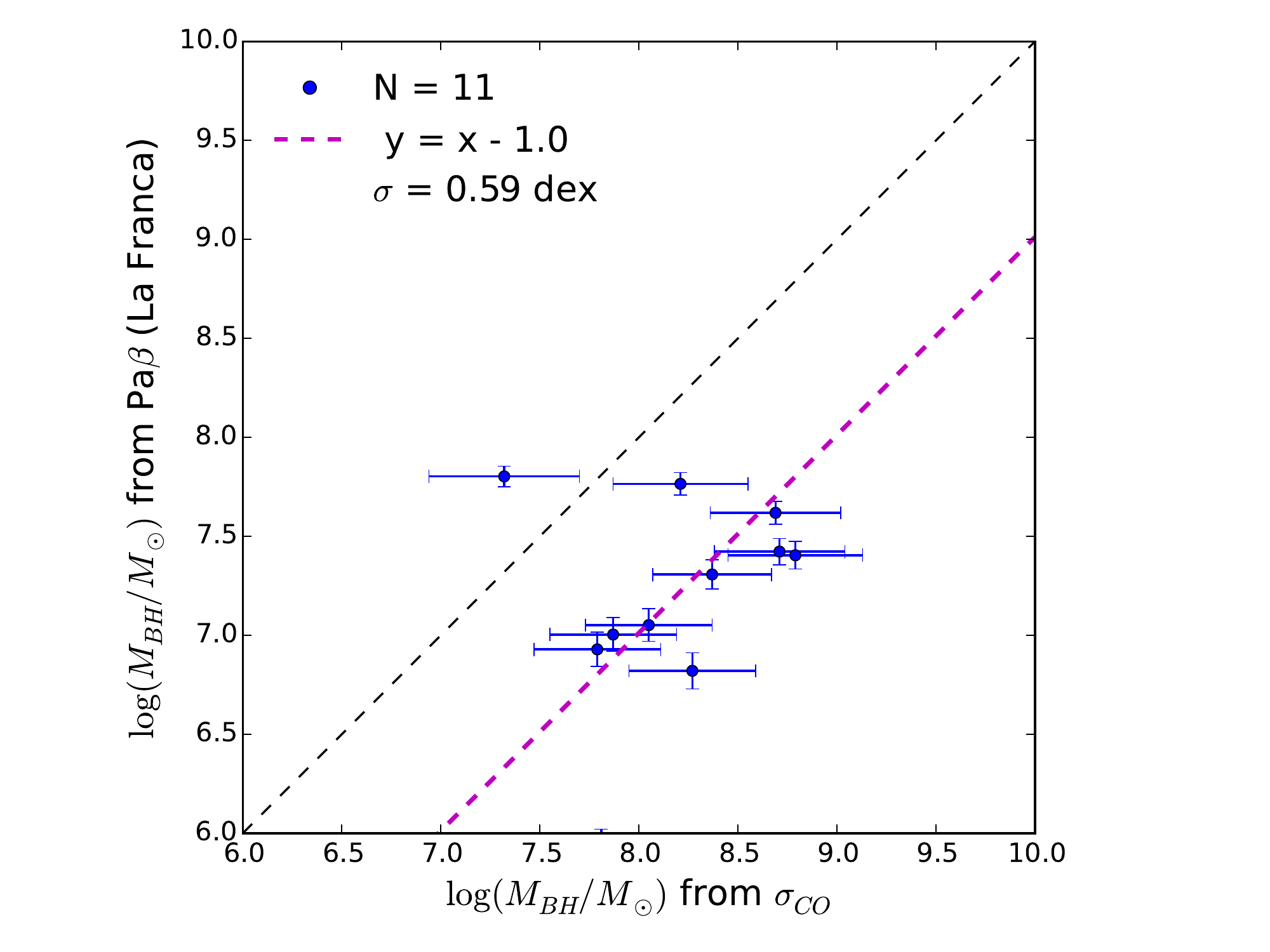} }

\subfigure{\includegraphics[width=0.49\textwidth]{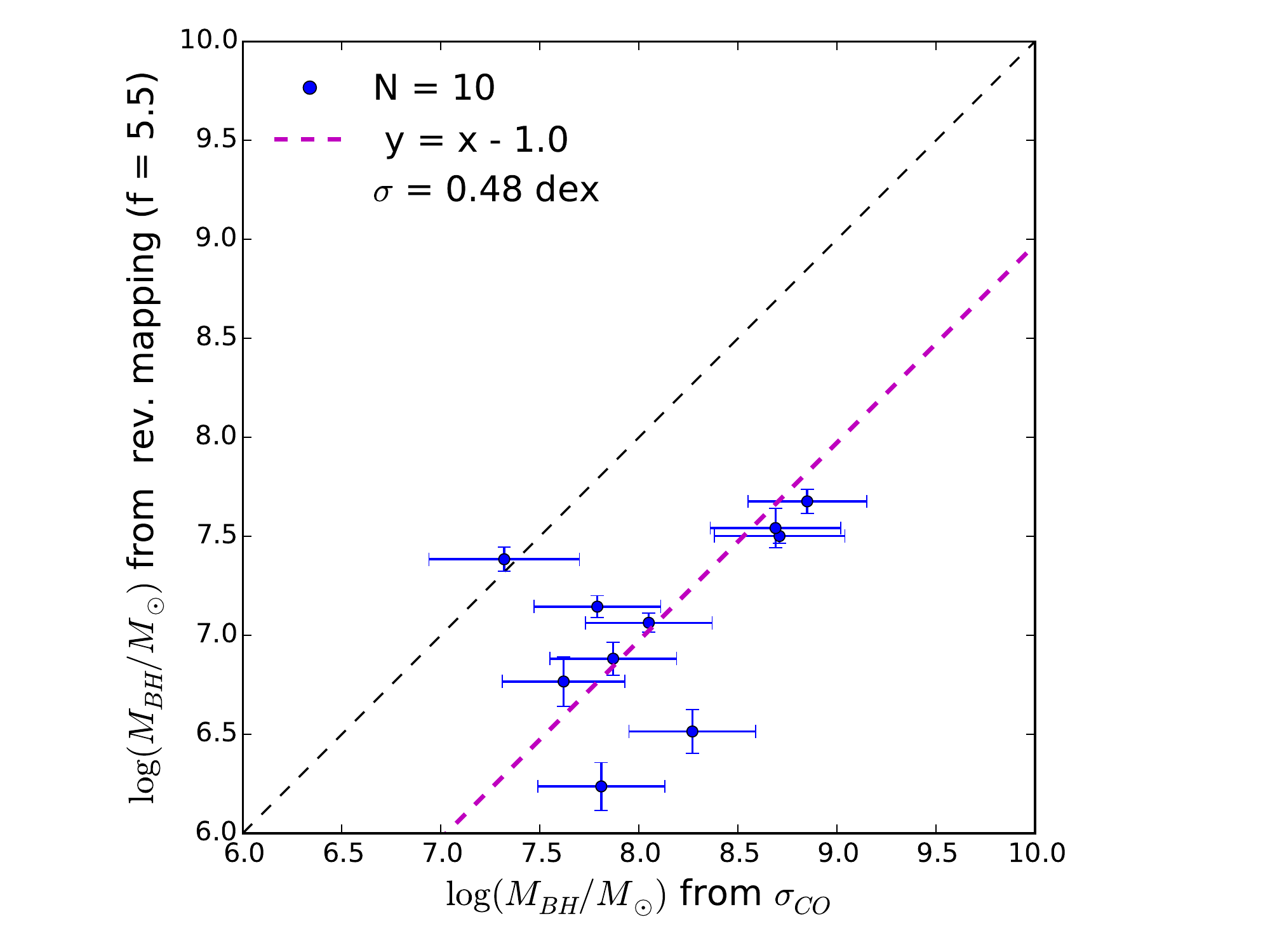} }
\subfigure{\includegraphics[width=0.49\textwidth]{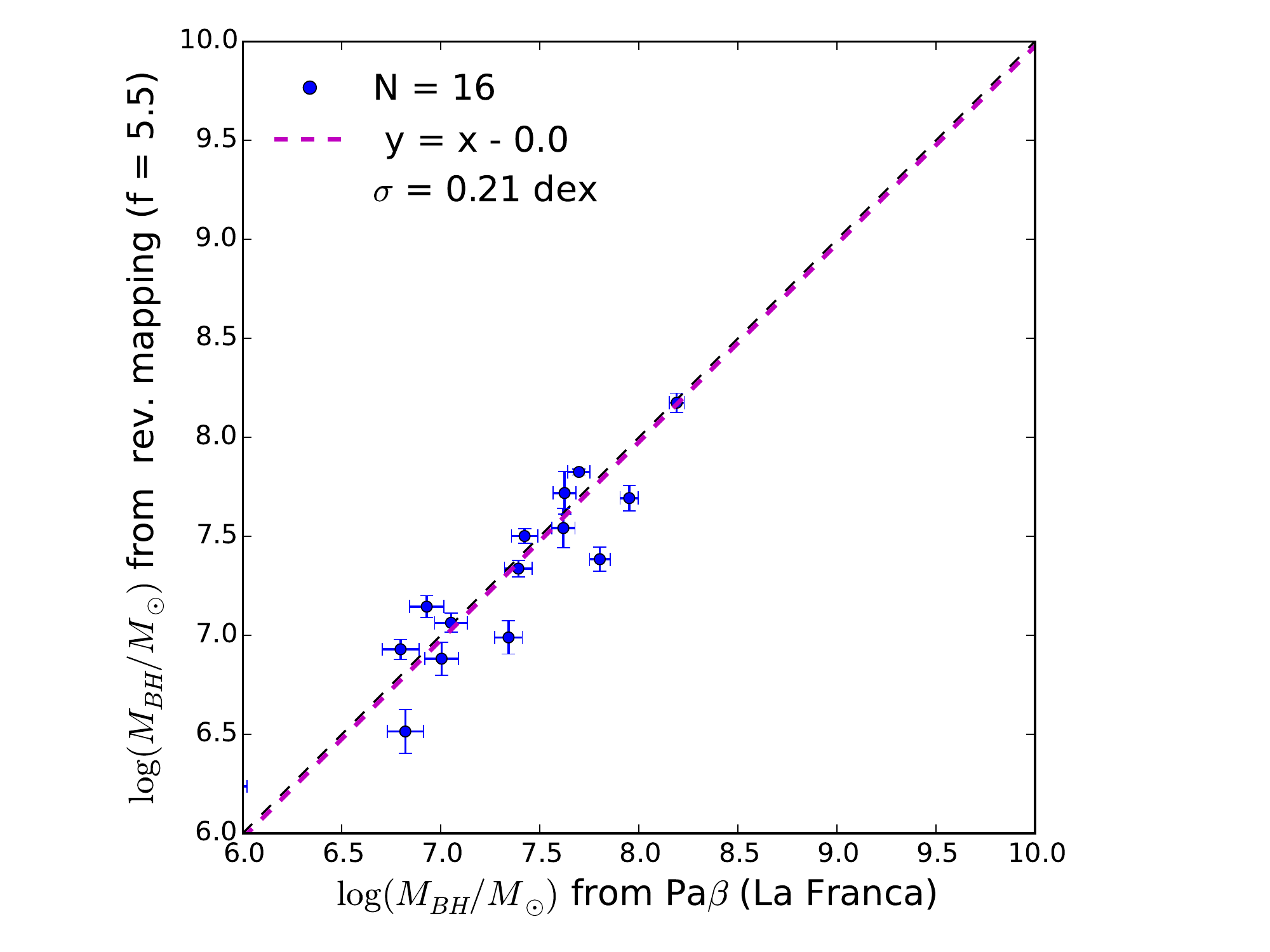} }

\subfigure{\includegraphics[width=0.49\textwidth]{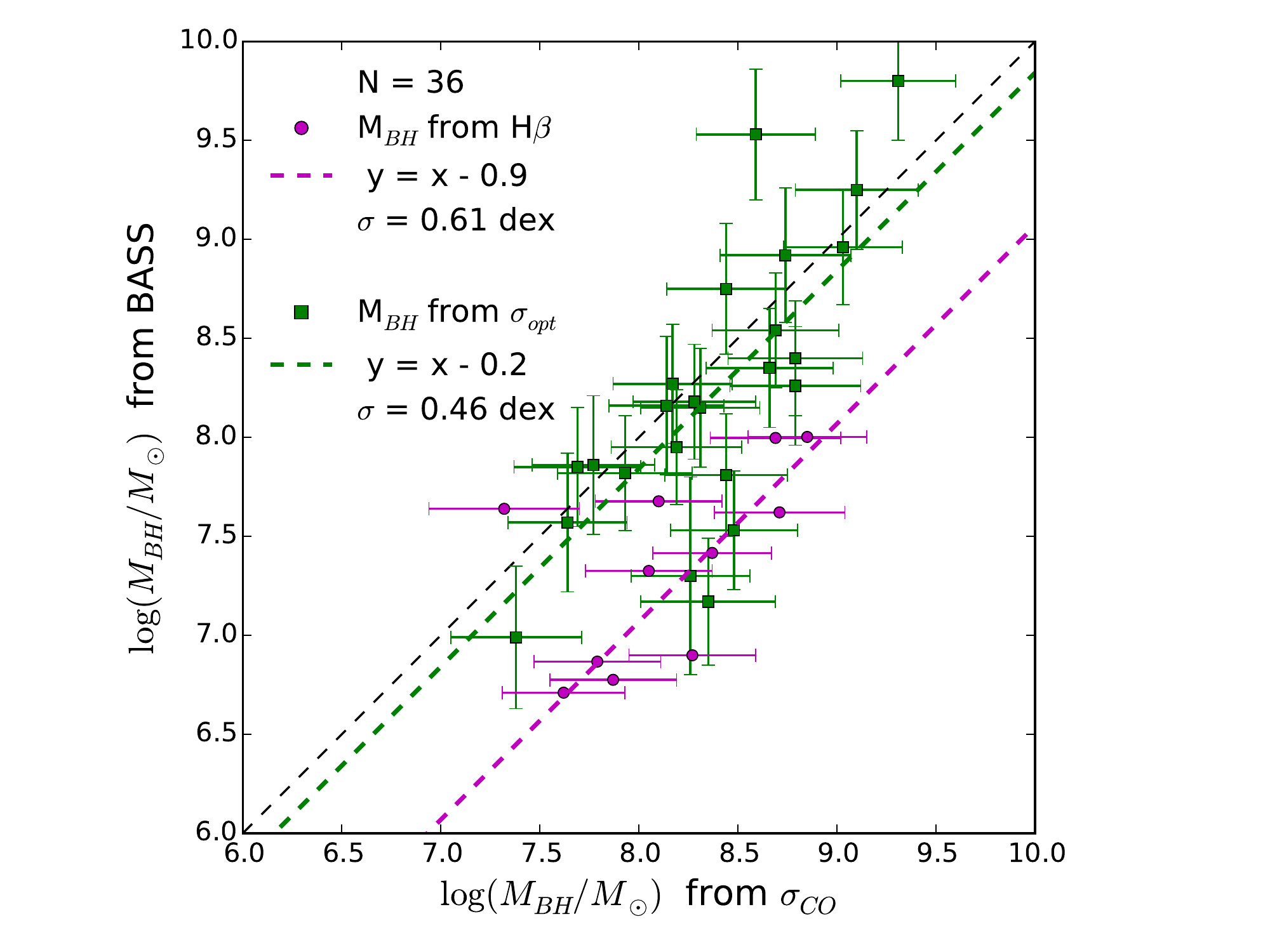} }
\subfigure{\includegraphics[width=0.49\textwidth]{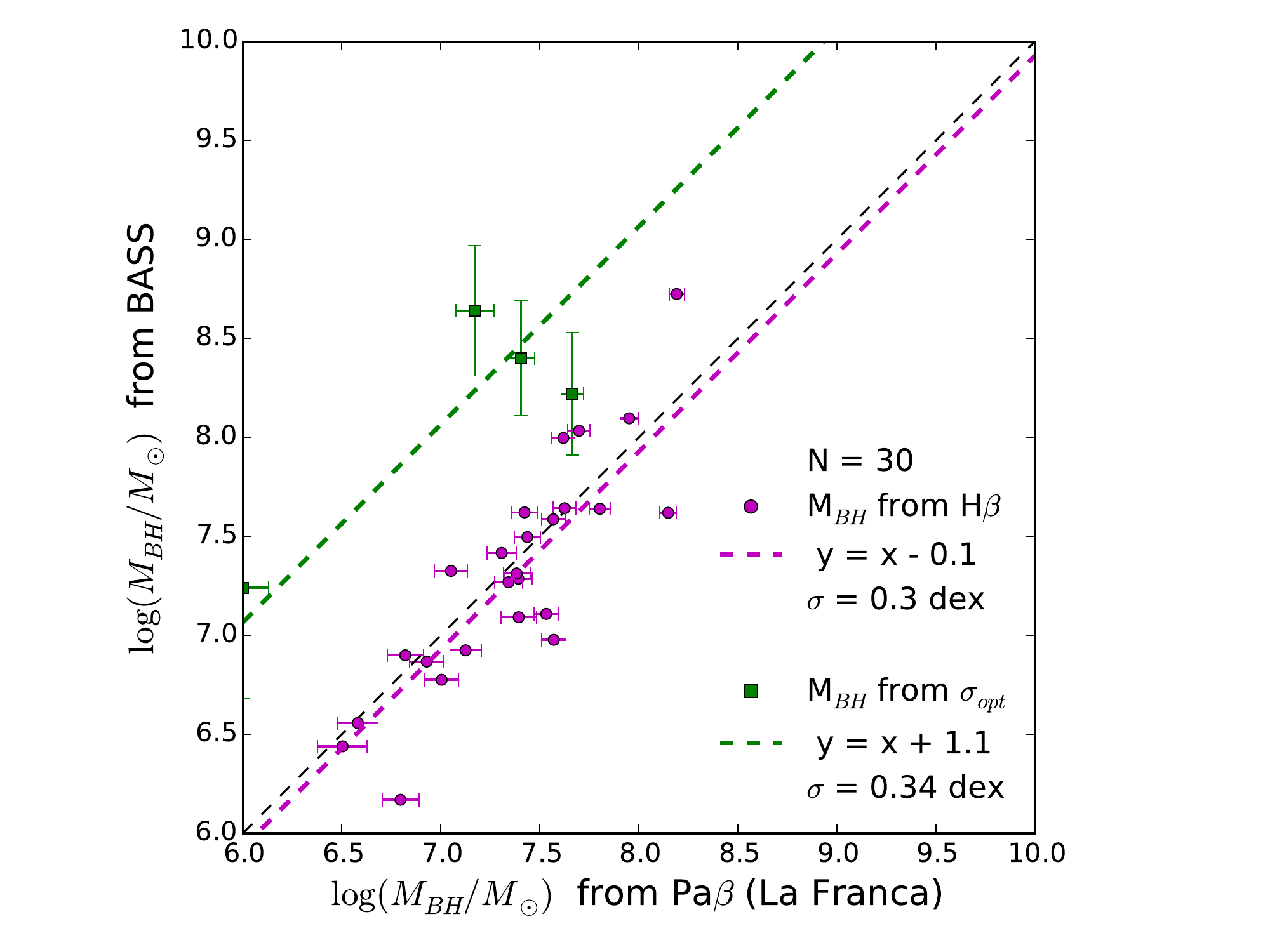} }

\caption{Comparison of \MBH\ measurements. The black dashed line shows the one-to-one relation and the magenta dashed line show the fit with a one-to-one relation allowing an offset. \textit{Upper panels:} Comparison of the \MBH\ estimated from broad Pa$\beta$ with the prescriptions from \citet{Kim2010} and from \citet{LaFranca2015} (right). Comparison of the \MBH\ estimated from $\sigma_{*, \rm CO}$ and from broad Pa$\beta$ with the prescriptions from \citet{LaFranca2015} (left).
%\label{Comp_MBH}
 \textit{Middle panels:} Comparison of the \MBH\ estimated from $\sigma_{*, CO}$ (left) and from broad Pa$\beta$ (right) with the \MBH\ values from reverberation mapping \citep{Bentz2015}. 
 \textit{Lower panels:} Comparison of the \MBH\ estimated from $\sigma_{3,\rm CO}$ (left) and from broad Pa$\beta$ (right) with the \MBH\ values from BASS. In green are the \MBH\ estimated from broad H$\beta$ and in magenta the \MBH\ estimated from \sig\ measured in the optical. The magenta and green dashed lines show the corresponding fit of a one-to-one relation allowing an offset.}
\label{Comp_MBH}
\end{figure*}

This relation have an intrinsic scatter on log MBH of 0.27.
Comparing the values obtained using these two relations (Figure \ref{Comp_MBH}),  we found an offset of 0.3 dex, that is small enough considering that the scatter of the relations is $\sim$ 0.2 dex. 
We compared also the \MBH\ derived from $\sigma_{CO}$ and from the broad Pa$\beta$. The \MBH\ estimated from $\sigma_{CO}$ are 1.1 dex larger than the values from Pa$\beta$. 

We compared our results with the M$_{BH}$ measured 
from reverberation mapping from `The AGN Black Hole Mass Database'\footnote{http://www.astro.gsu.edu/AGNmass/} assuming a virial factor $\left\langle f \right\rangle$= 5.5 \citep{Bentz2015}. We note that the relation from \cite{LaFranca2015} was calibrated assuming a virial factor $f = 4.31$. We found a good agreement between \MBH\ from Pa$\beta$ and the reverberation mapping values (average difference: $0.18 \pm 0.03$ dex). This is not surprising, since the relation from \cite{LaFranca2015} was calibrated against the reverberation mapping \MBH. For the \MBH\ from $\sigma_{\rm CO}$, we observe again that the \MBH\ measured from $\sigma_{\rm CO}$ seem to be overestimated.

For 43 AGN, we could compare our results with the \MBH\ values from the BASS catalog.
The \MBH\ from $\sigma_{*,\rm CO}$ are in agreement with the \MBH\ values from the optical \sig, even if there is a large scatter ($\sigma$ = 0.46 dex). There is an offset of 0.9 dex between the \MBH\ measured from $\sigma_{*,\rm CO}$ and \MBH\ measured from H$\beta$.
The \MBH\ from Pa$\beta$ are in good agreement with the values of \MBH\ measured from H$\beta$ (offset = 0.1 dex, scatter = 0.3 dex), but not with \MBH\ measured from $\sigma_{*, \rm opt}$ (offset = 1.1 dex, scatter = 0.3 dex).

The \MBH\ calculated from the M-\sig\ relation, using $\sigma_{*,\rm CO}$ or $\sigma_{*,\rm opt}$, agree with each other. In the same way, the measurements from broad lines, from single epoch H$\beta$ or Pa$\beta$ observations or from reverberation mapping, are in good agreement with each other. 

There is a significant offset of 0.9 dex between the values of \MBH\ derived using the M-\sig\ relation and \MBH\ measured from the broad emission lines. Using the \cite{Gultekin2009} M-\sig\ relation instead of \cite{Kormendy2013} the offset is still 0.7 dex.  We note that even the best virial mass estimators are known to suffer from systematic uncertainties of at least 0.3 dex \cite[see, e.g.,][and references therein]{Shen2013,Peterson2014} despite being tied to reproduce similar masses for systems where both are applicable \cite[e.g.]{Graham2011, Woo2013}.  Another possibility might be that sources having both a black hole mass measurement and velocity dispersion are more likely to suffer residual AGN contamination in the host galaxy template that may bias the velocity dispersion.  Further studies, such as using an integral field unit where the AGN and host galaxy emission can be separated and fit spacially as well as in wavelength space, are needed to fully understand this issue.

\begin{figure*}
\begin{center}
\subfigure{\includegraphics[width=0.3\textwidth]
{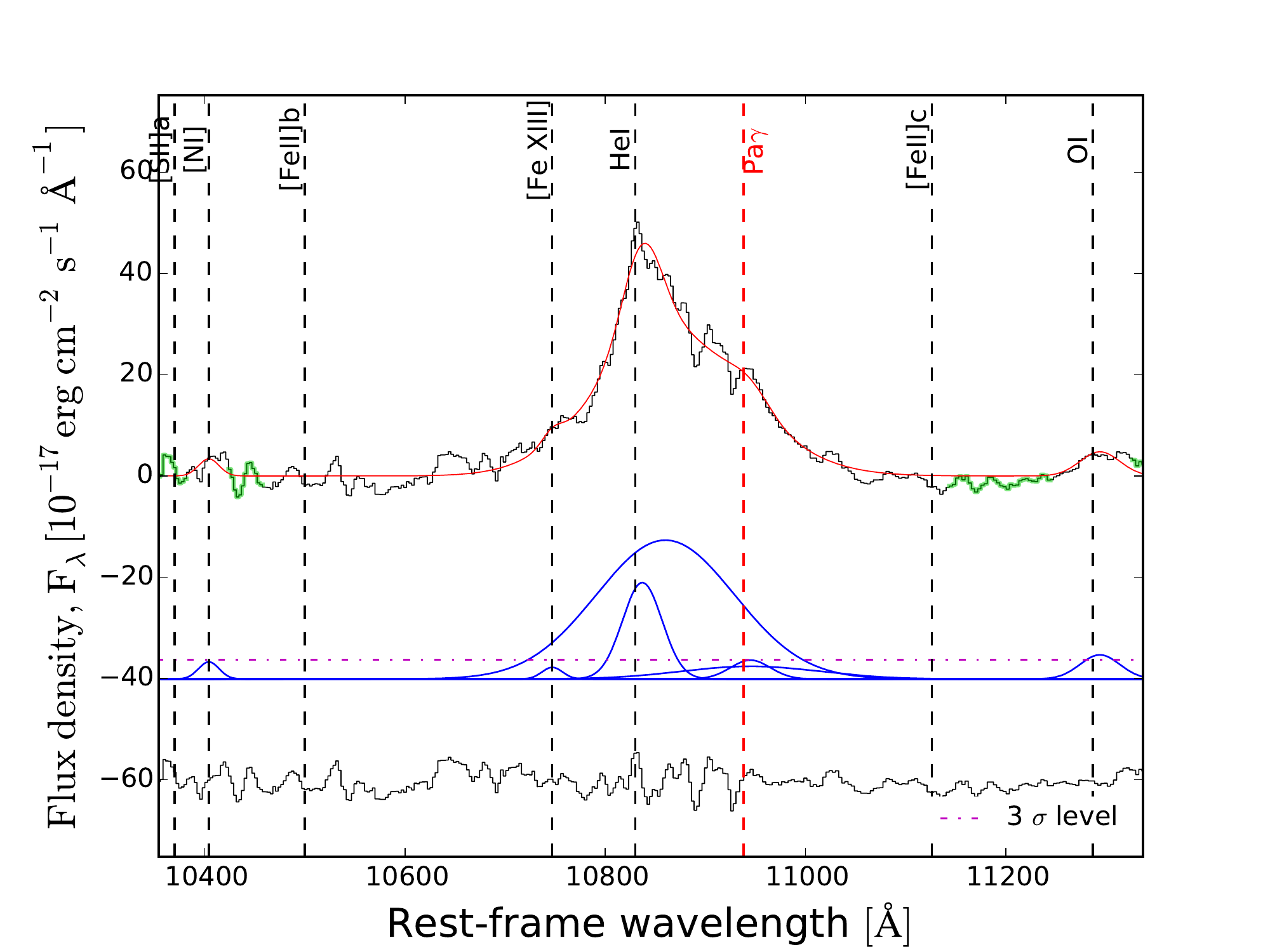} }
\subfigure{\includegraphics[width=0.3\textwidth]
{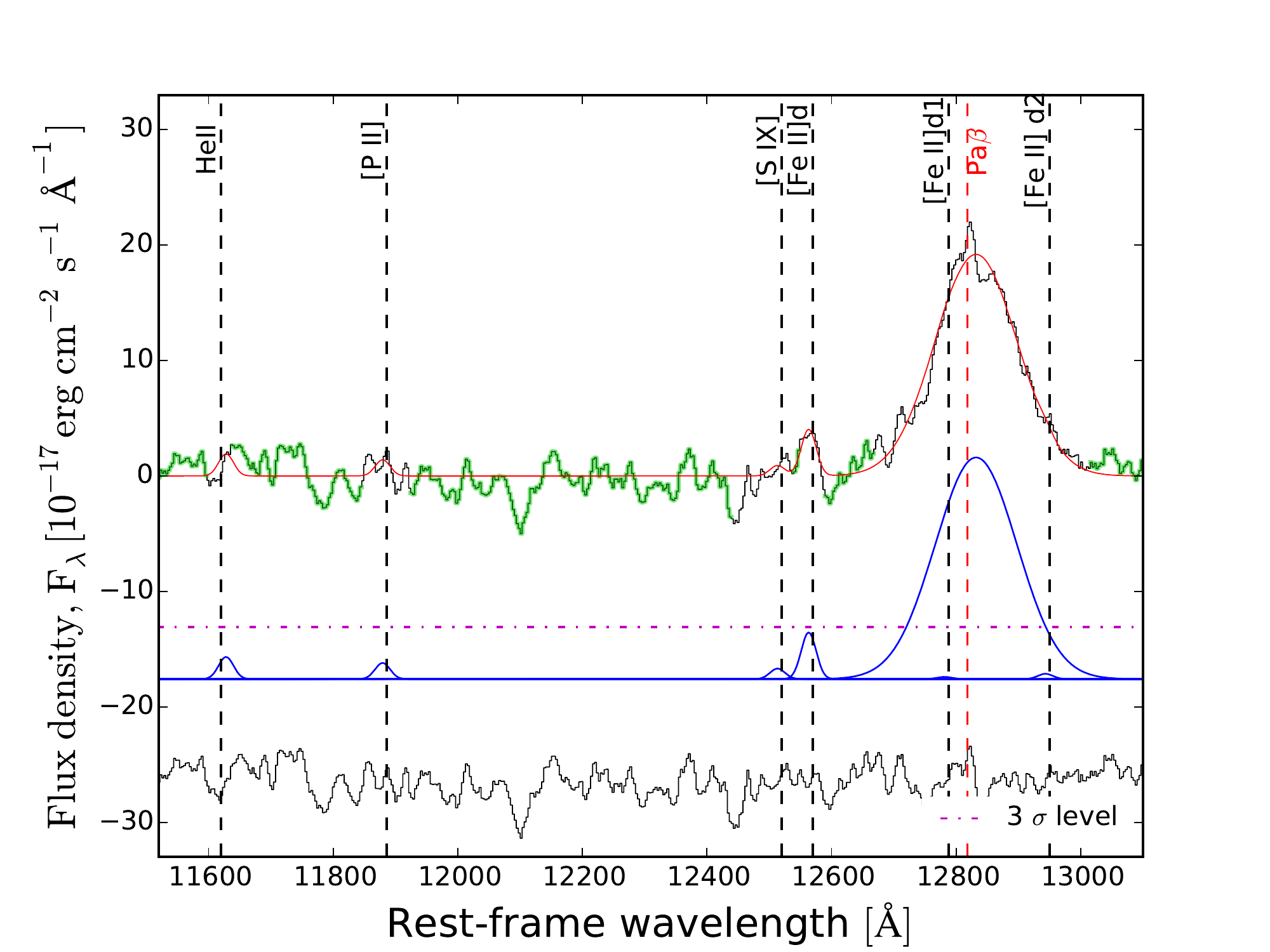} }
\subfigure{\includegraphics[width=0.3\textwidth]
{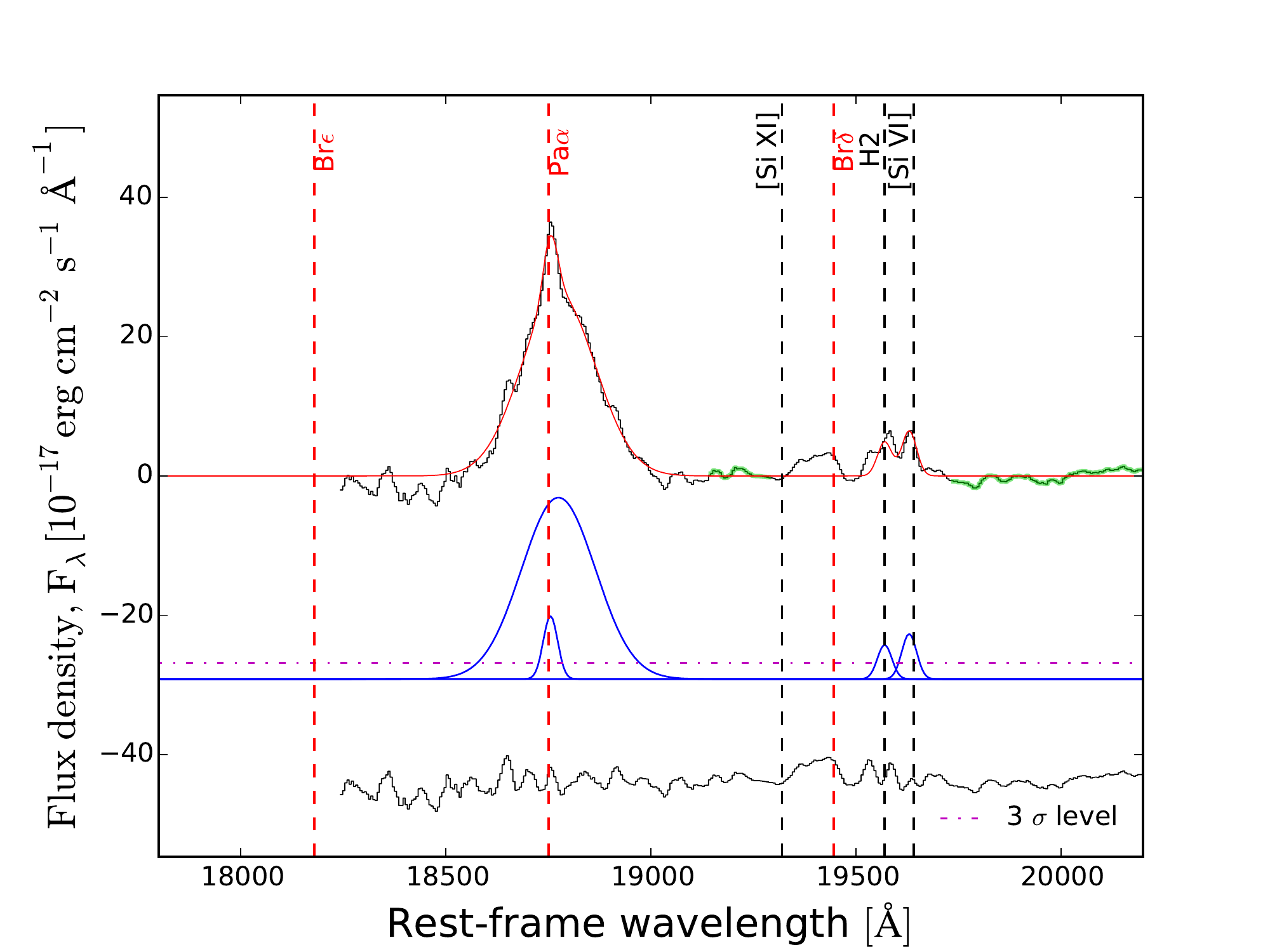} }

\subfigure{\includegraphics[width=0.3\textwidth]{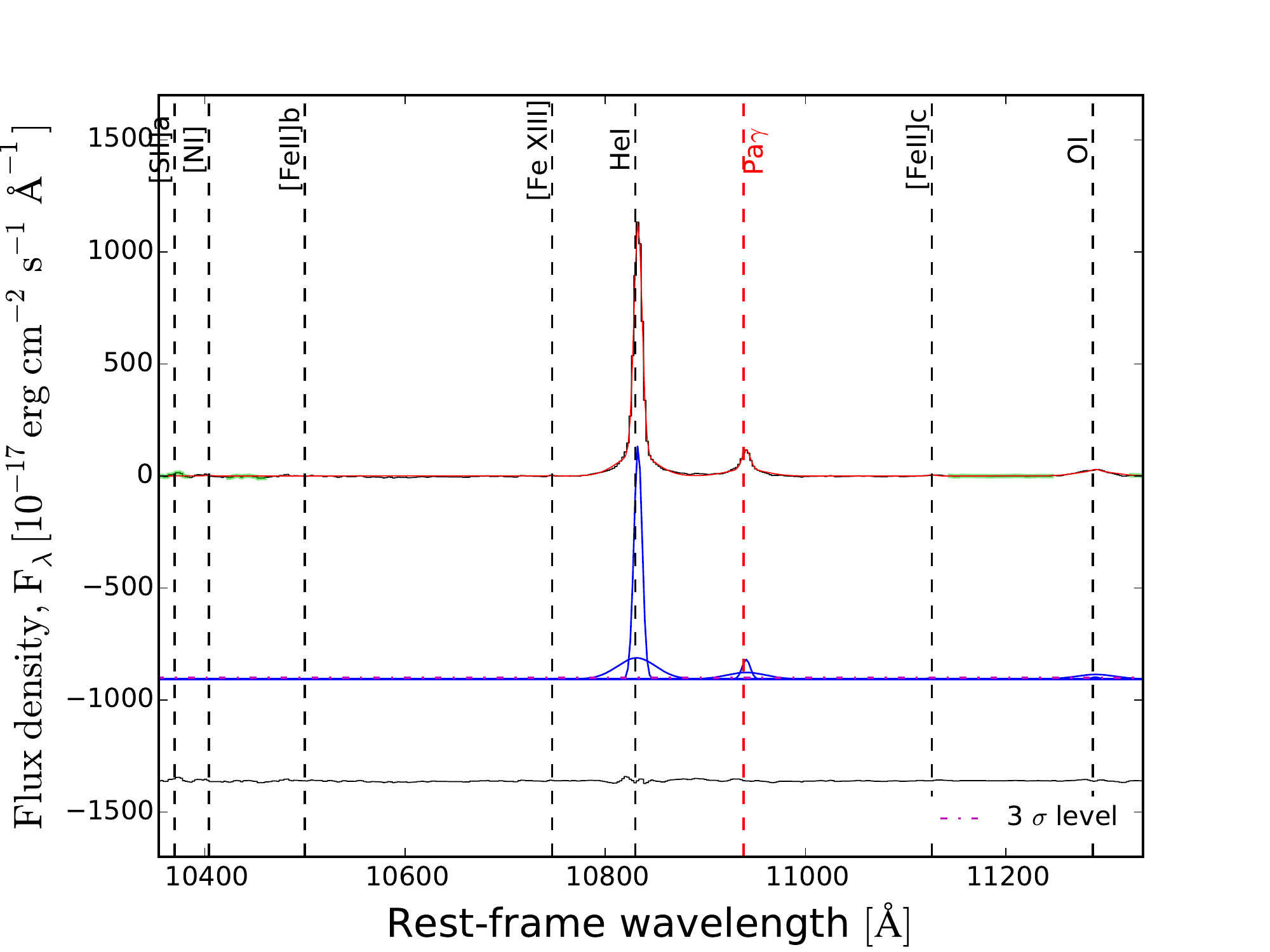} }
\subfigure{\includegraphics[width=0.3\textwidth]{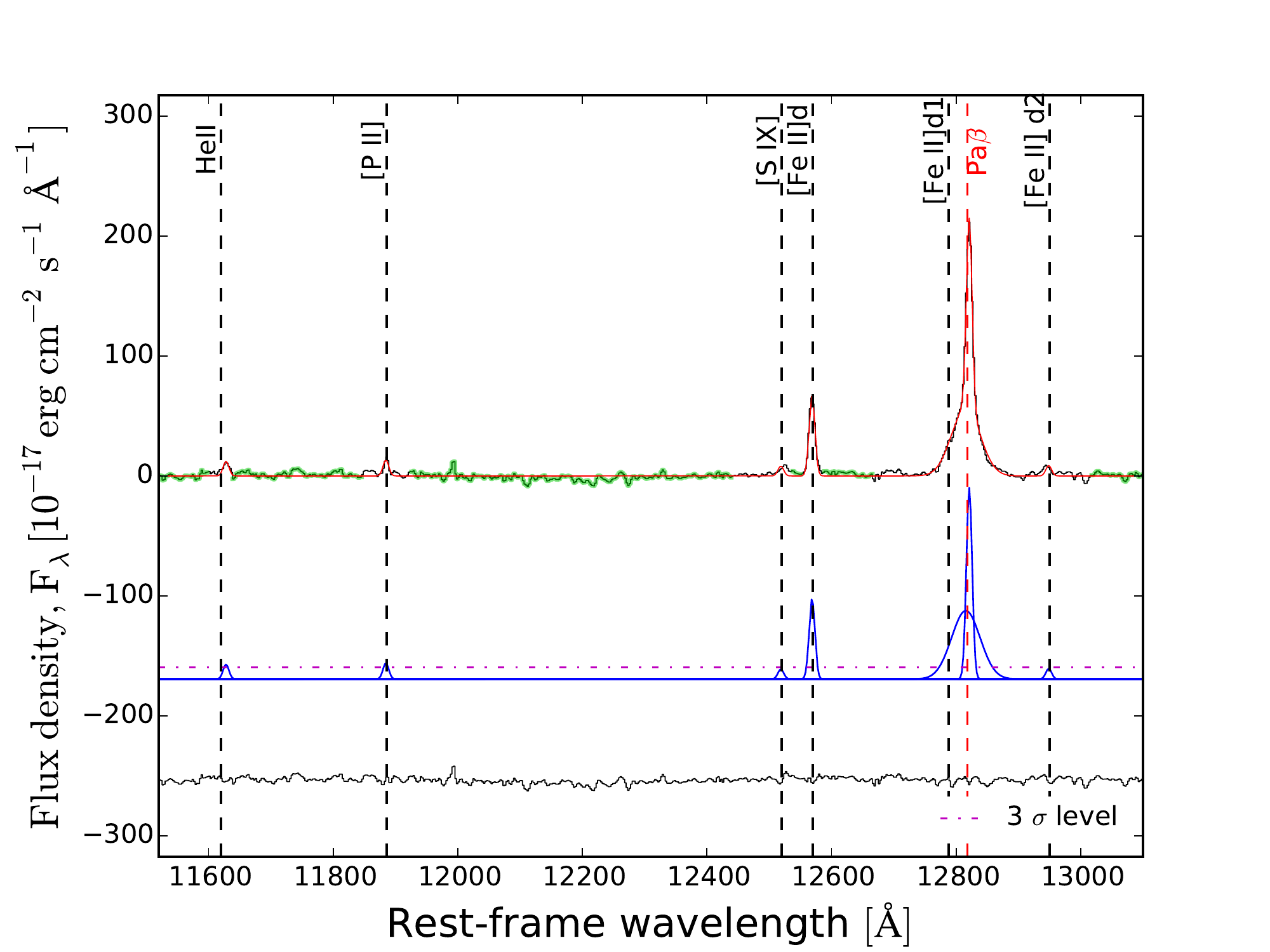} }
\subfigure{\includegraphics[width=0.3\textwidth]{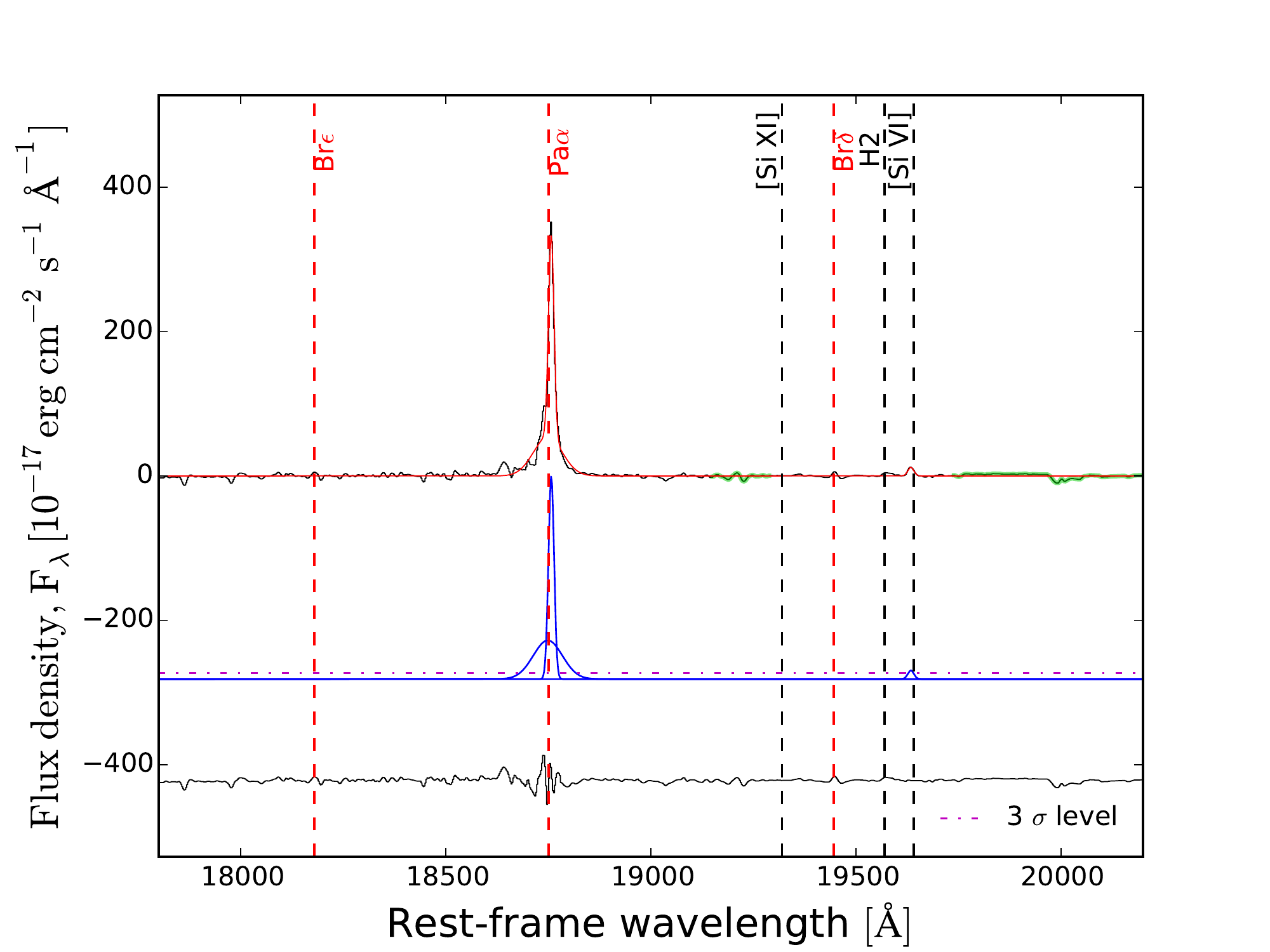} }

\subfigure{\includegraphics[width=0.3\textwidth]{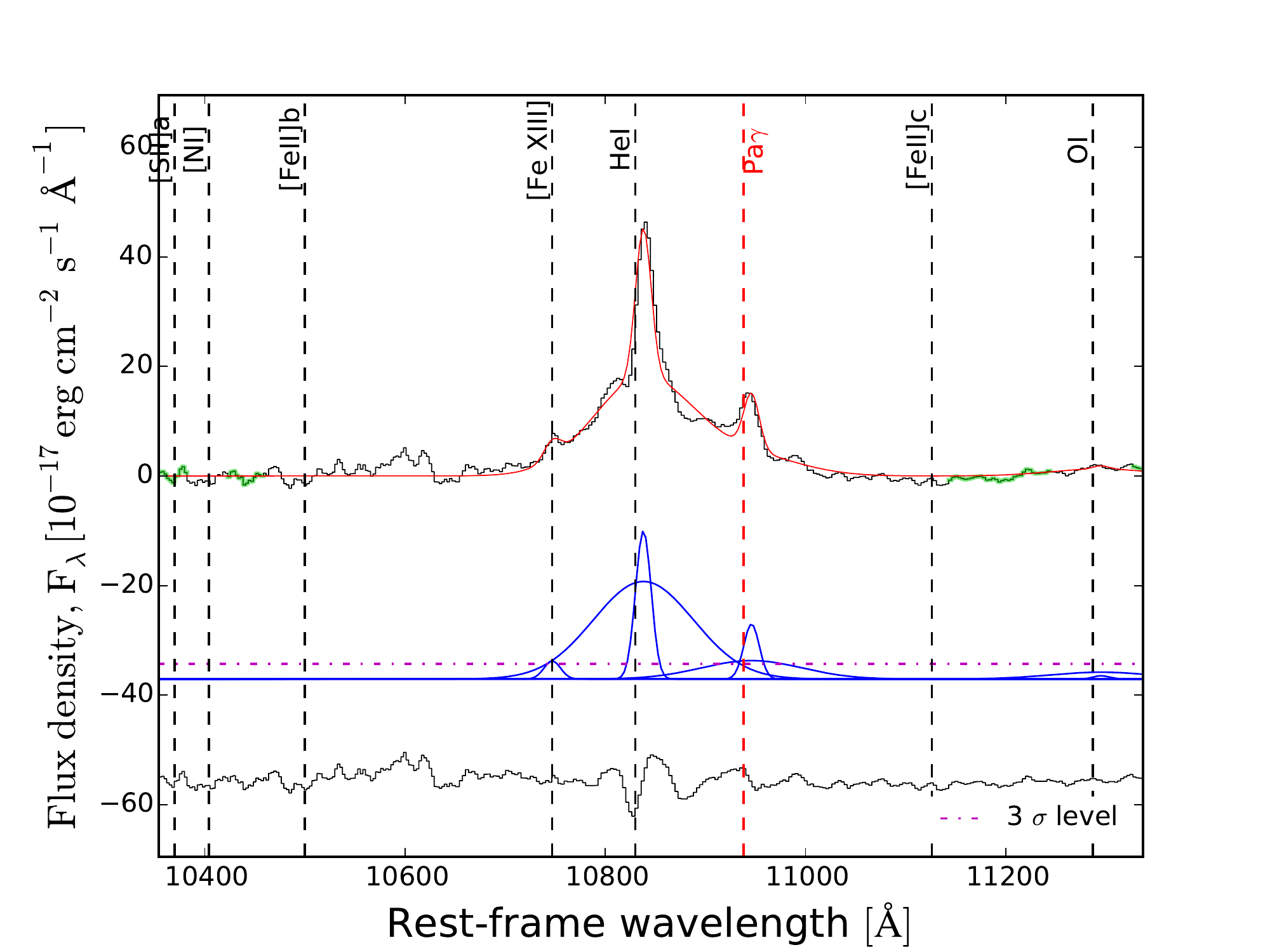} }
\subfigure{\includegraphics[width=0.3\textwidth]{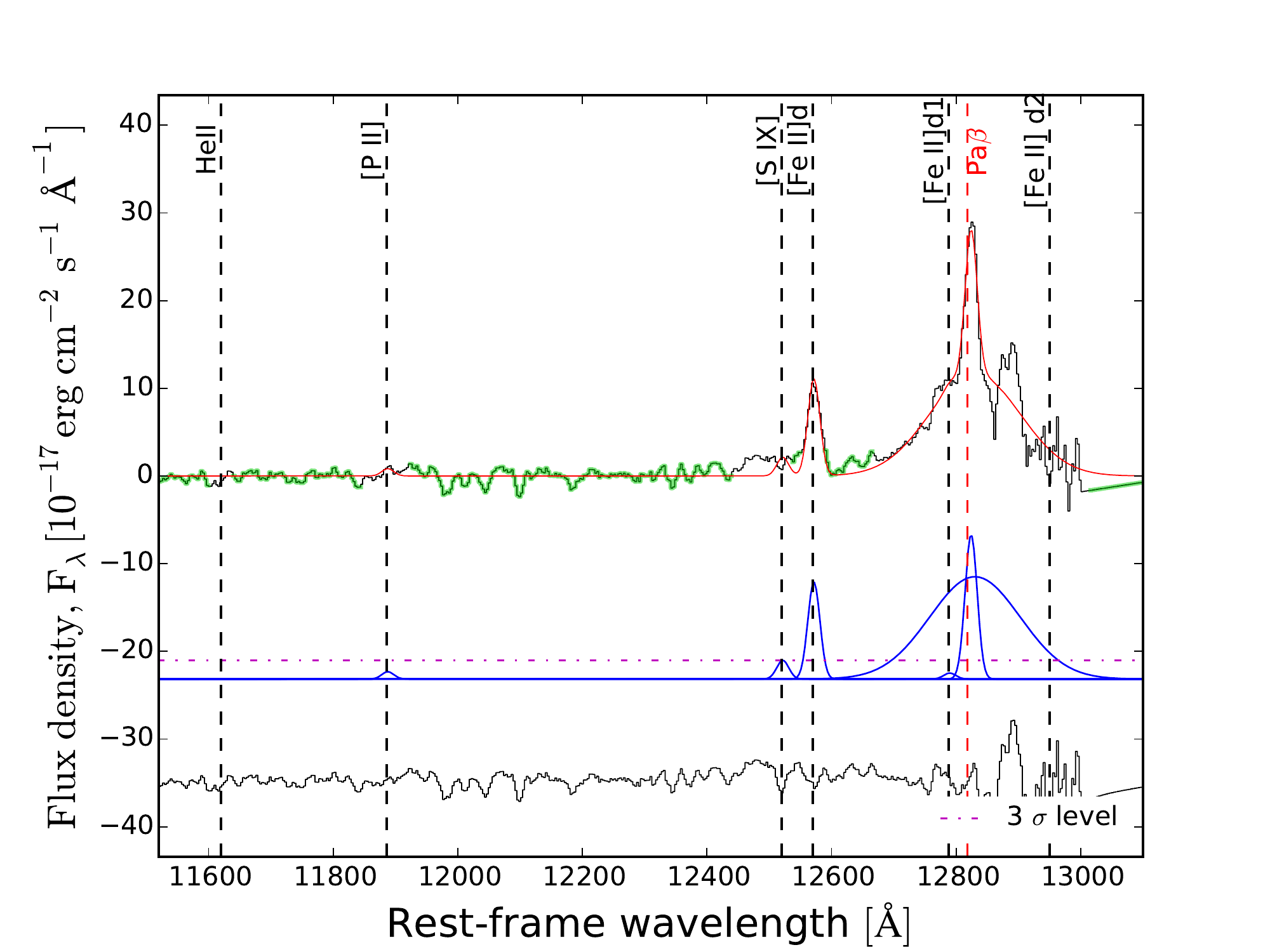} }
\subfigure{\includegraphics[width=0.3\textwidth]{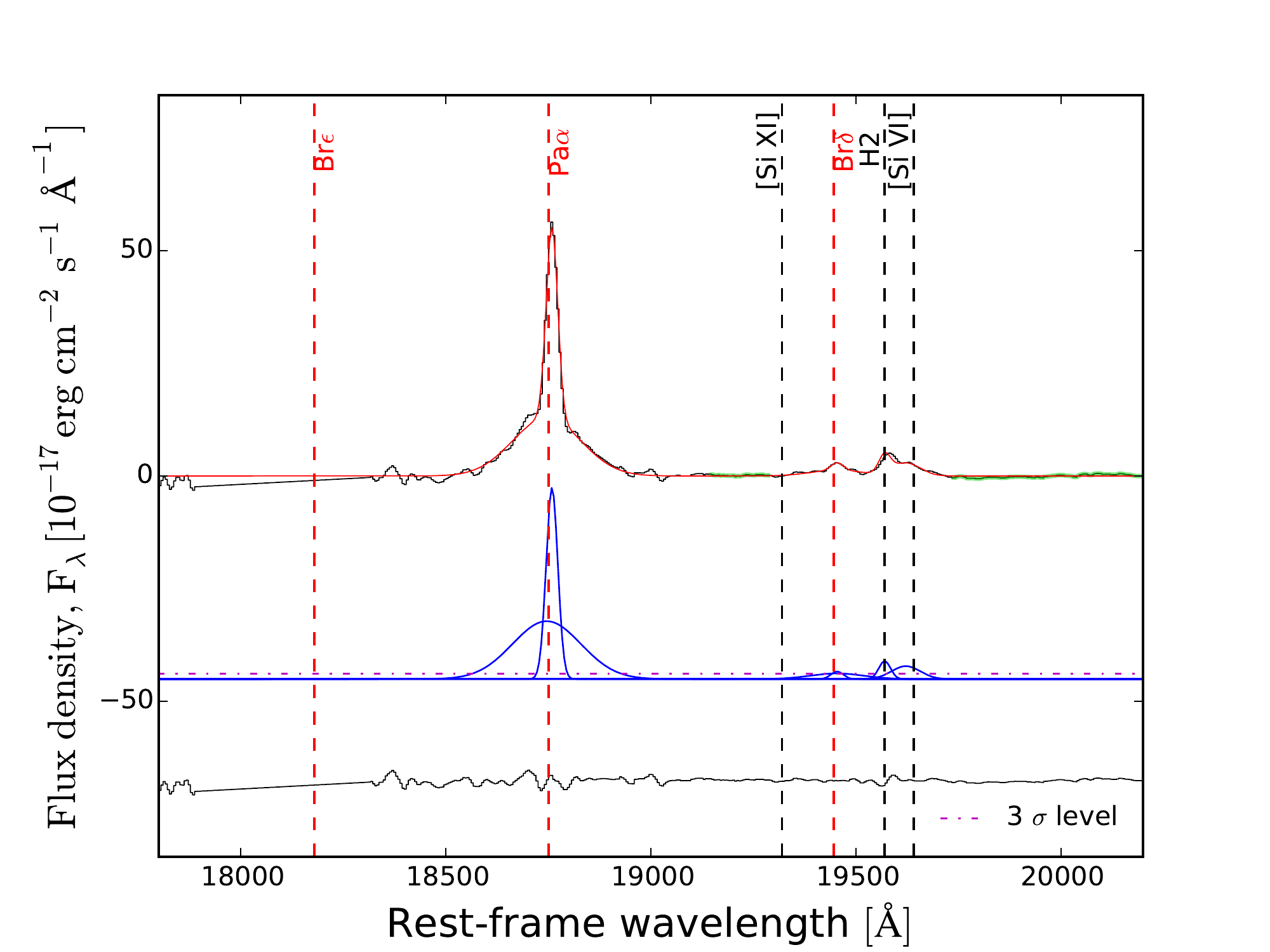} }

\subfigure{\includegraphics[width=0.3\textwidth]{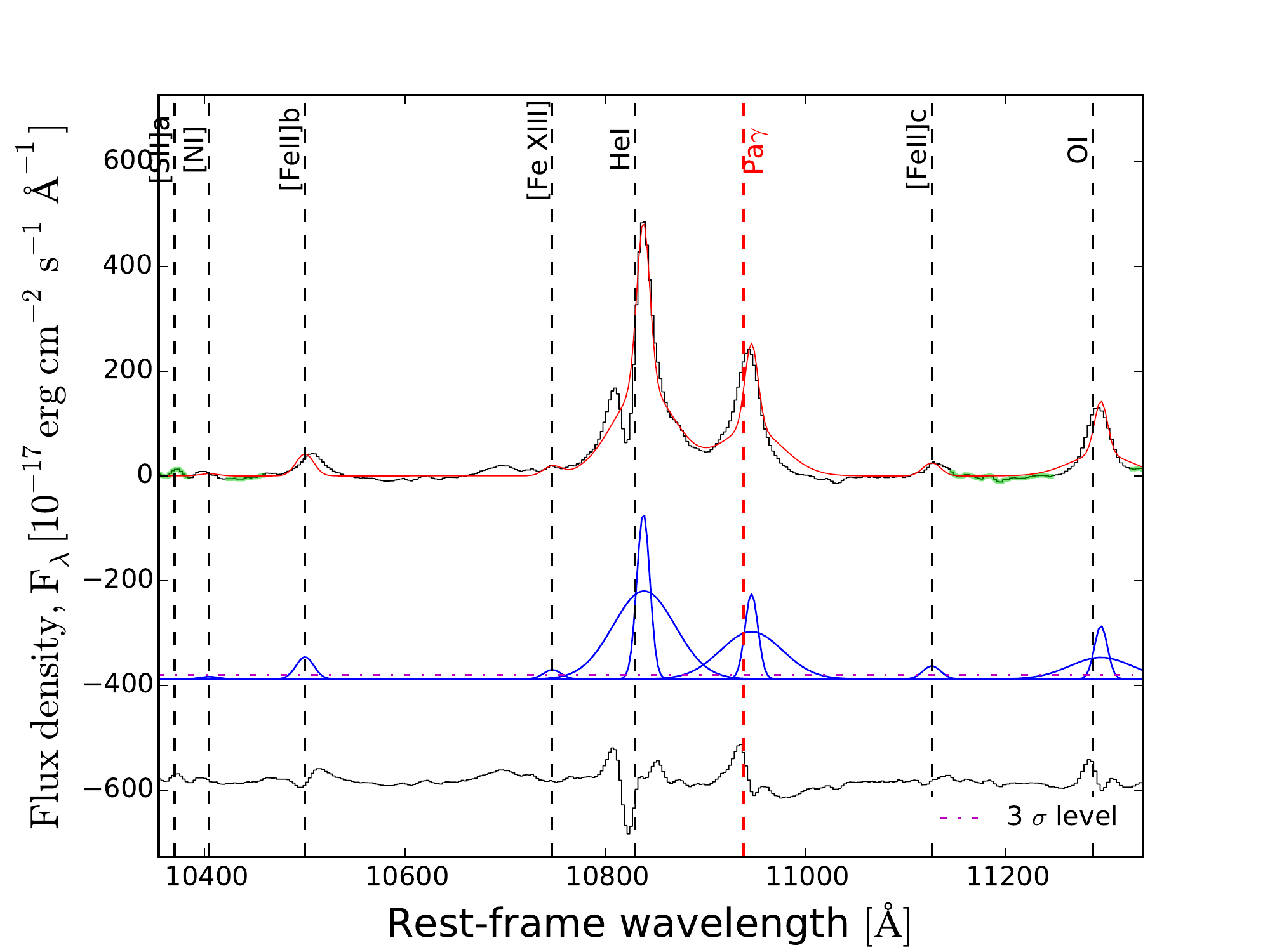} }
\subfigure{\includegraphics[width=0.3\textwidth]{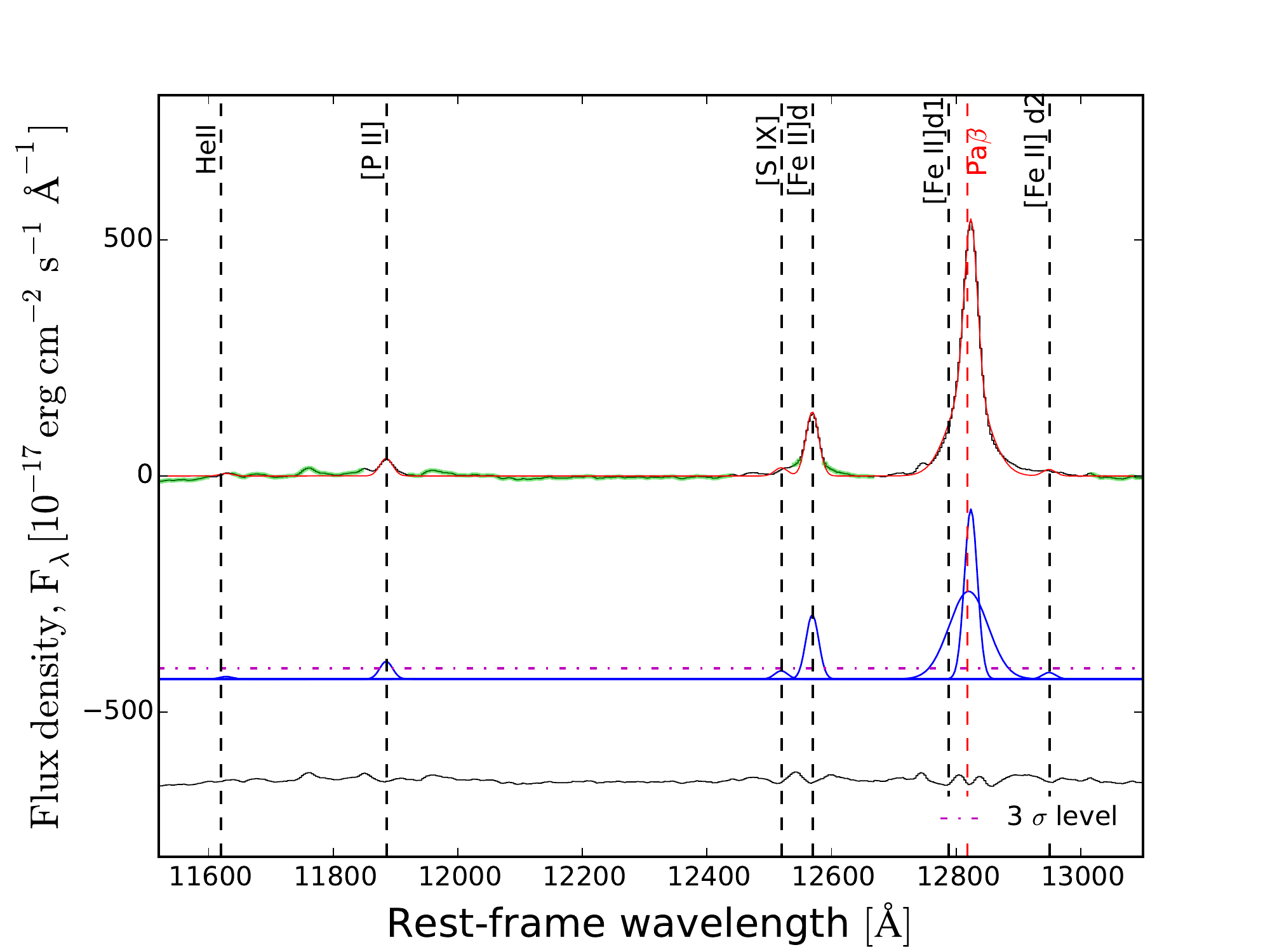}}
\subfigure{\includegraphics[width=0.3\textwidth]{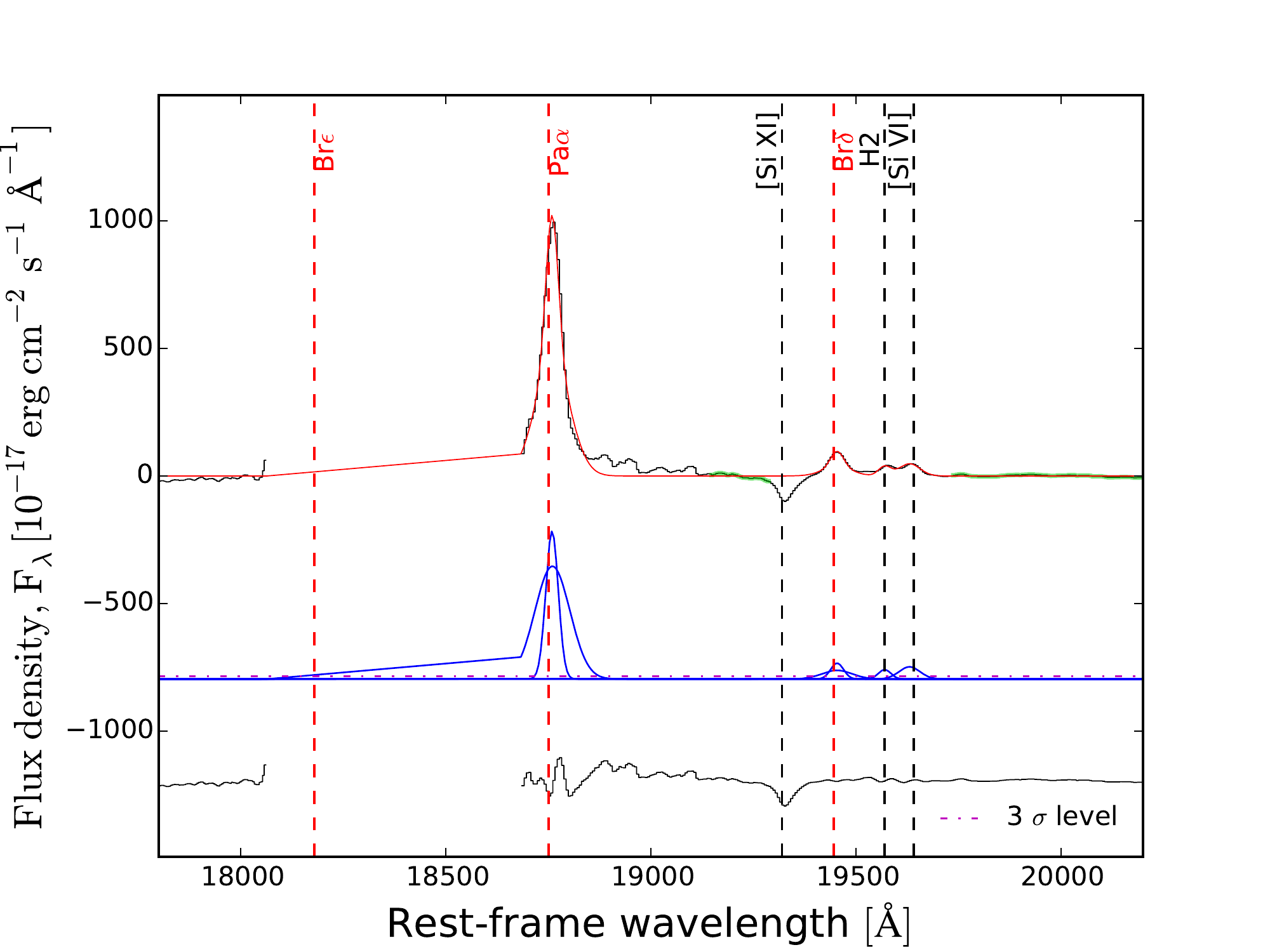} }

\subfigure{\includegraphics[width=0.3\textwidth]{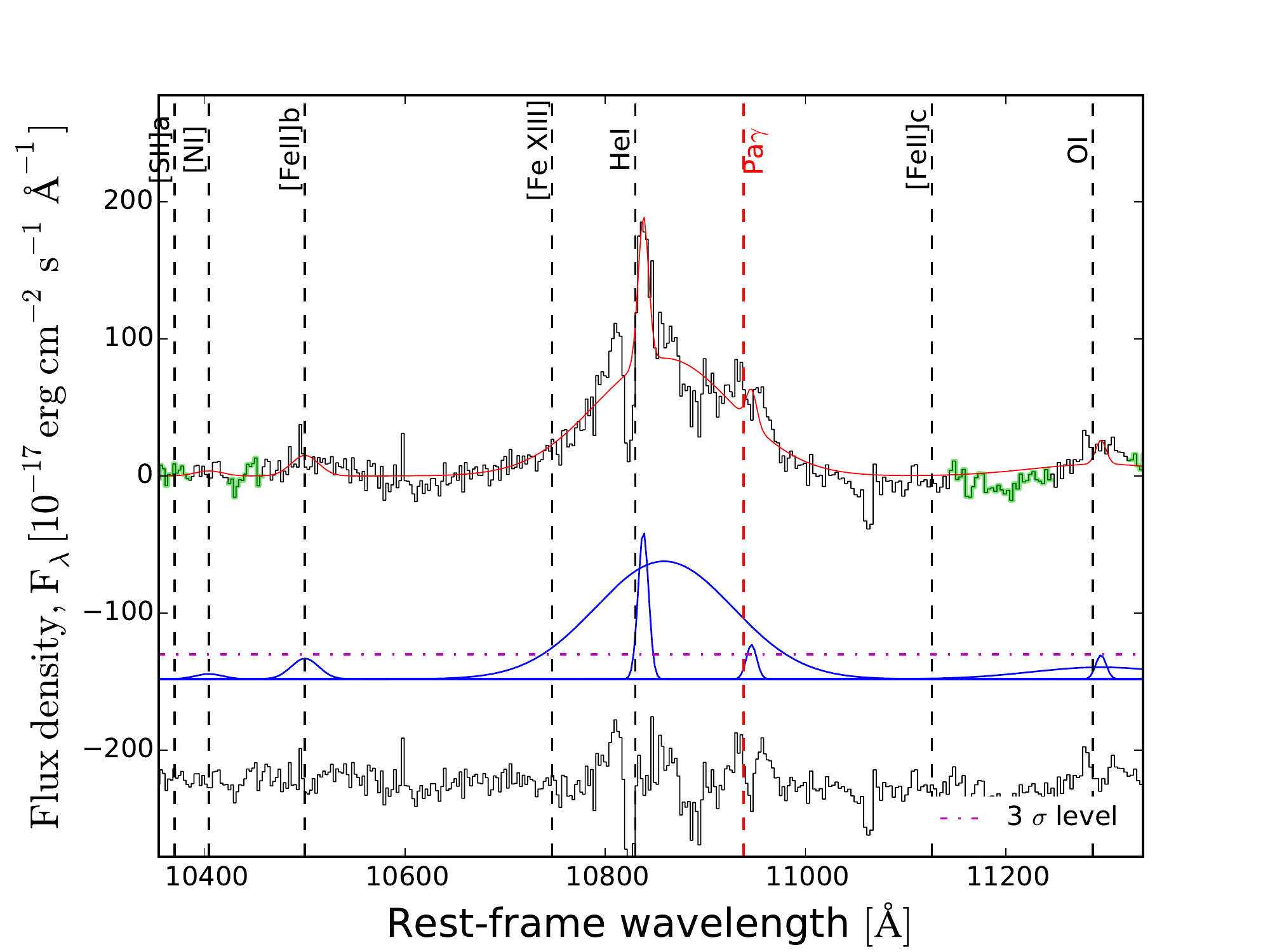} }
\subfigure{\includegraphics[width=0.3\textwidth]{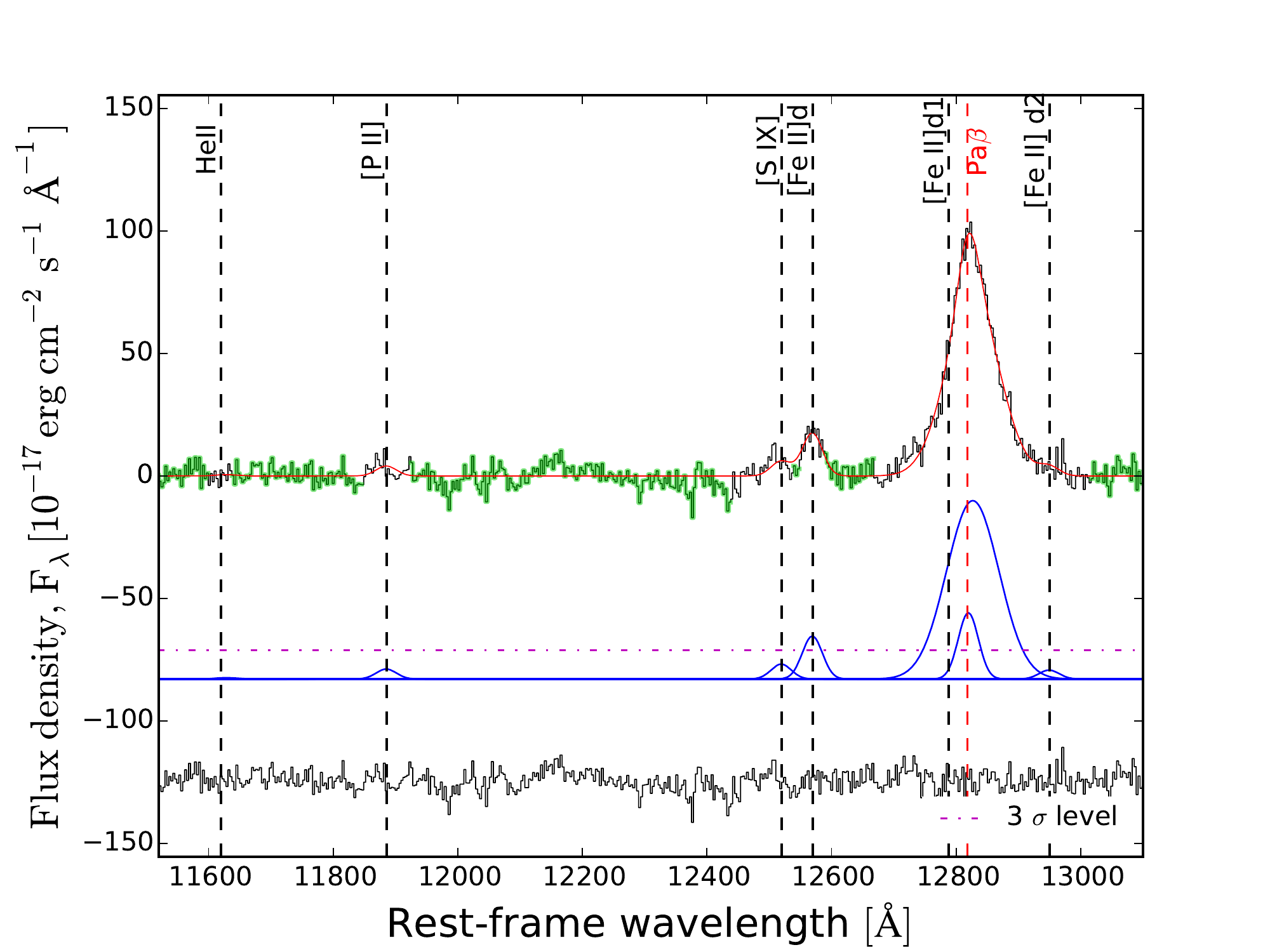} }
\subfigure{\includegraphics[width=0.3\textwidth]{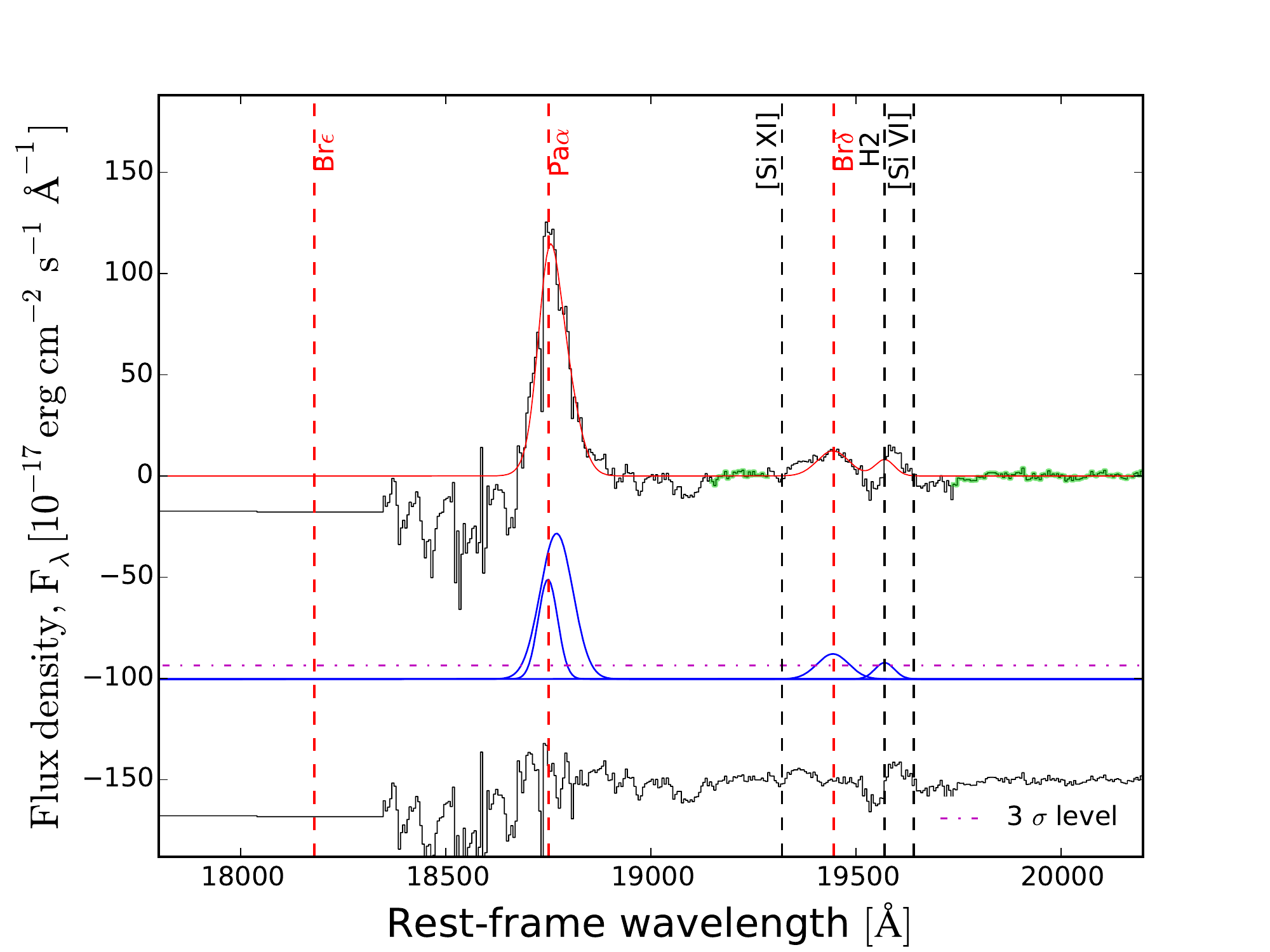} }

\caption{Fit of the HeI $\lambda 1.0830$ $\mu$m, Pa$\beta$ and Pa$\alpha$ regions for the Seyfert 1.9 and Seyfert 2 for which the broad Pa$\beta$ was used to measure \MBH. From the upper row: 2MASX J07595347+2323241, NGC 4395, KUG 1238+278A, NGC 5506, and NGC 5995. The best fit is in red, the model in blue, and the residuals in black. The magenta dashed line in the middle part of the figures shows the detection threshold (S/N$>\ $3) with respect to the fitting continuum (blue).  }
\label{Fit_Hidden_BLR}
\end{center}
\end{figure*}

\begin{figure*}
\begin{center}

\subfigure{\includegraphics[width=0.3\textwidth]{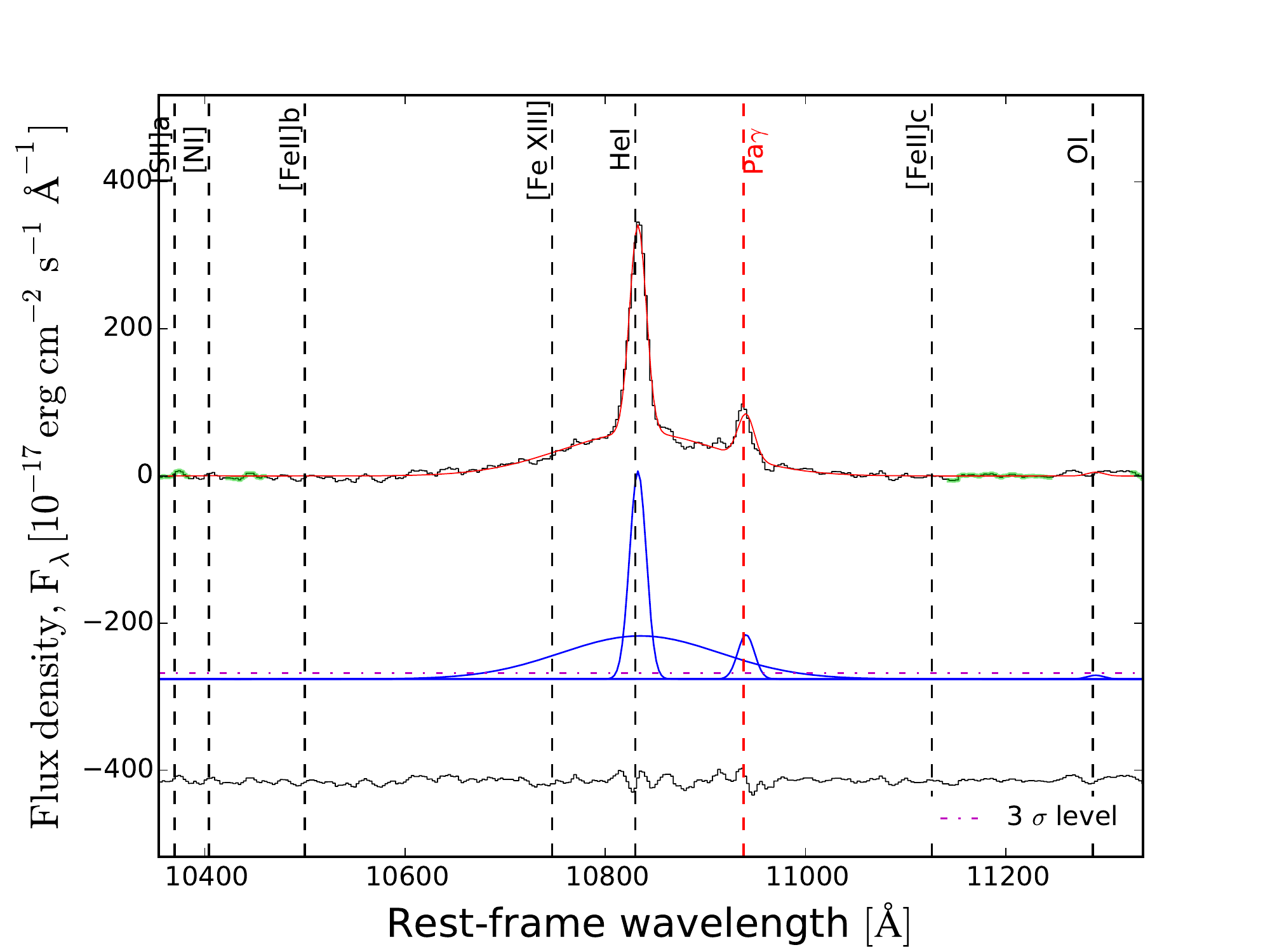} }
\subfigure{\includegraphics[width=0.3\textwidth]{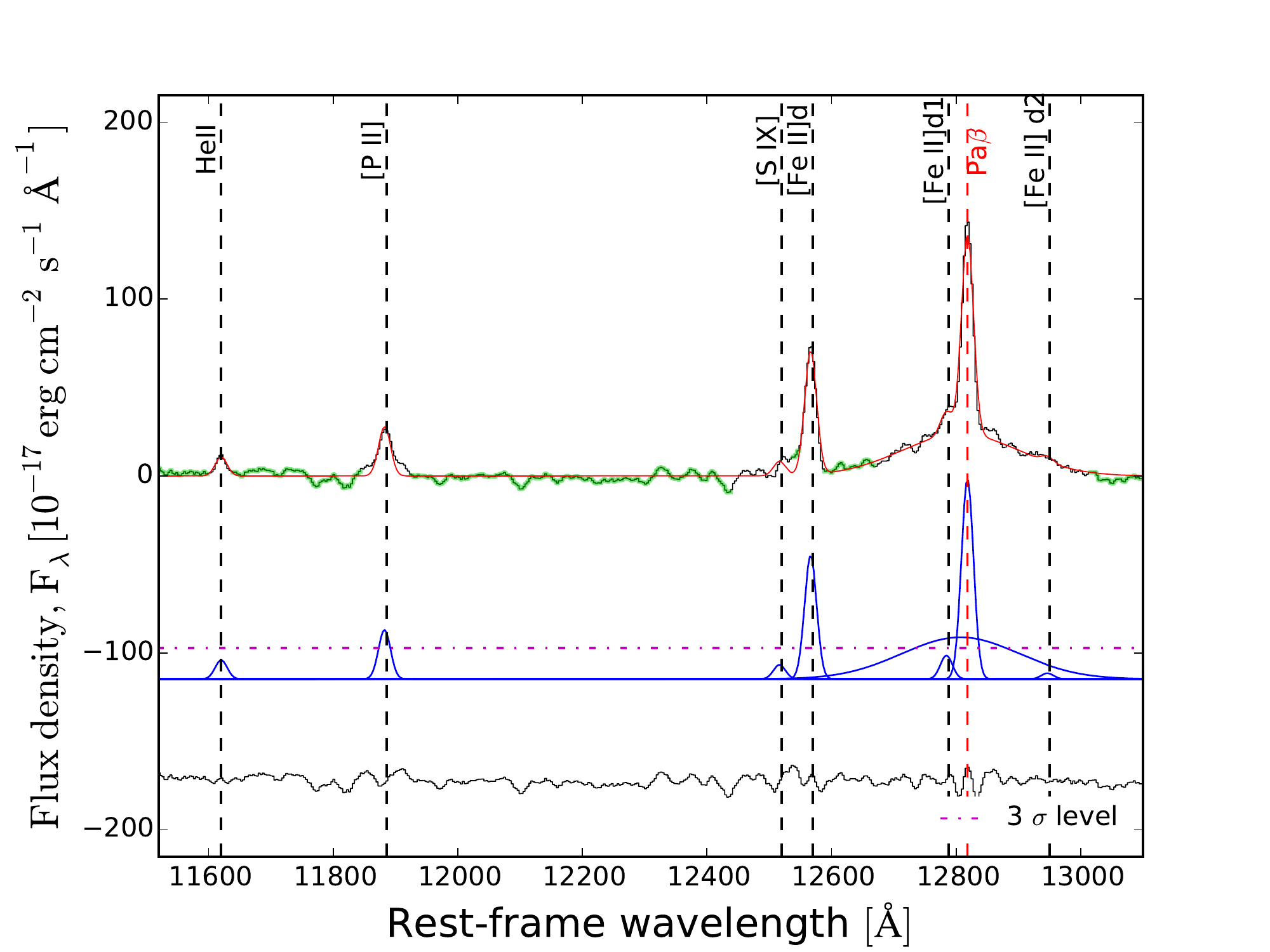} }
\subfigure{\includegraphics[width=0.3\textwidth]{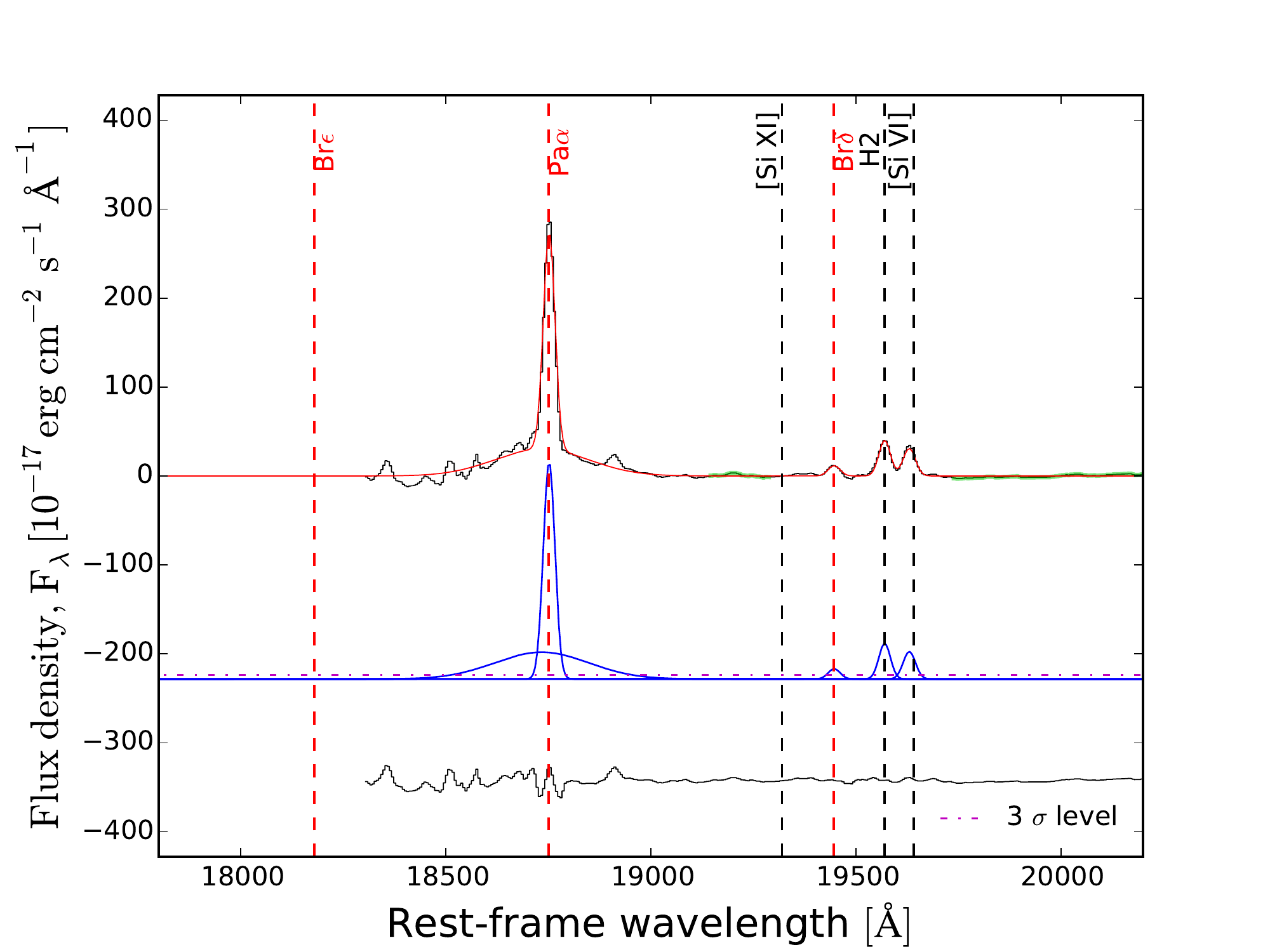} }

\subfigure{\includegraphics[width=0.3\textwidth]{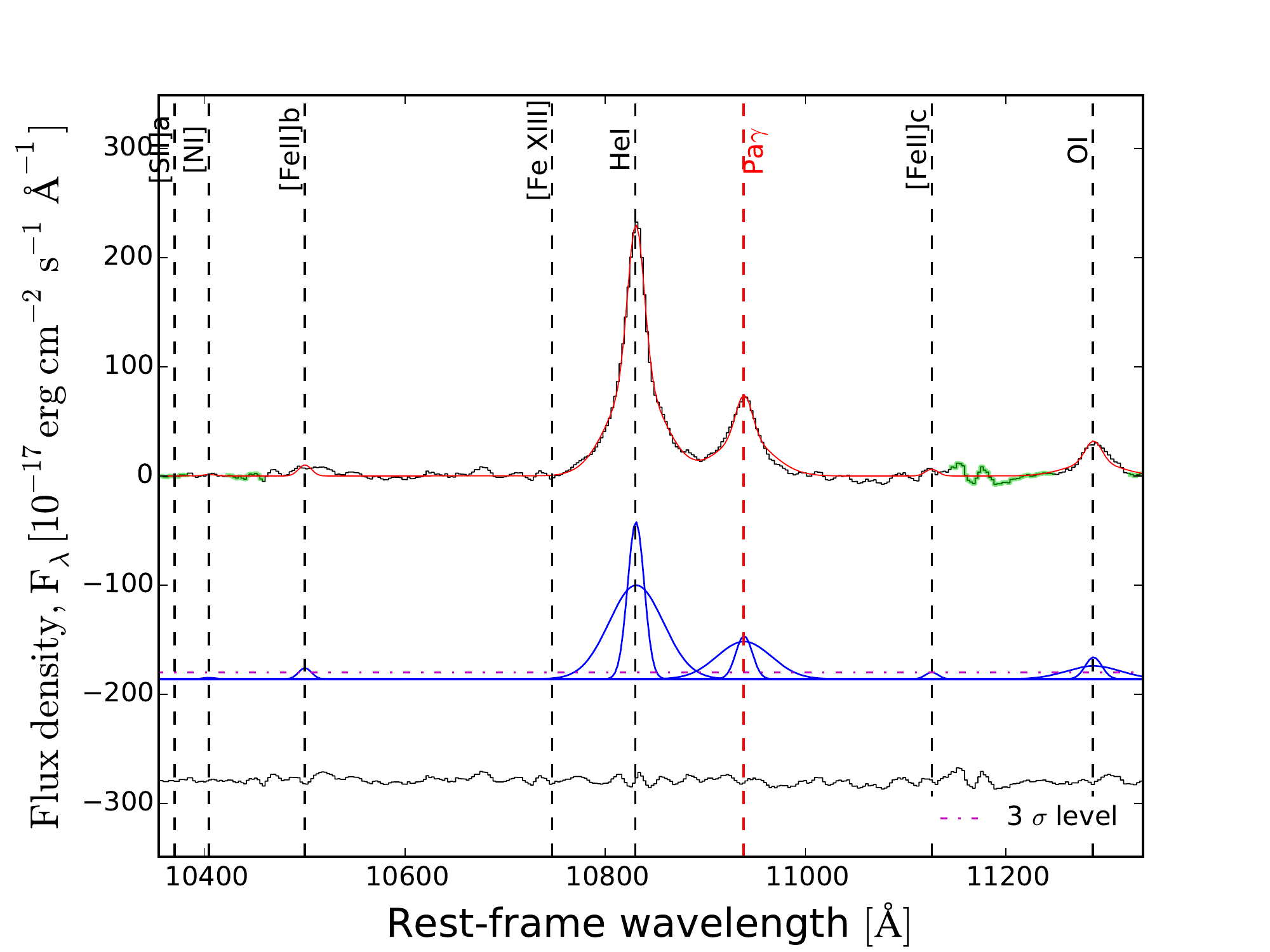} }
\subfigure{\includegraphics[width=0.3\textwidth]{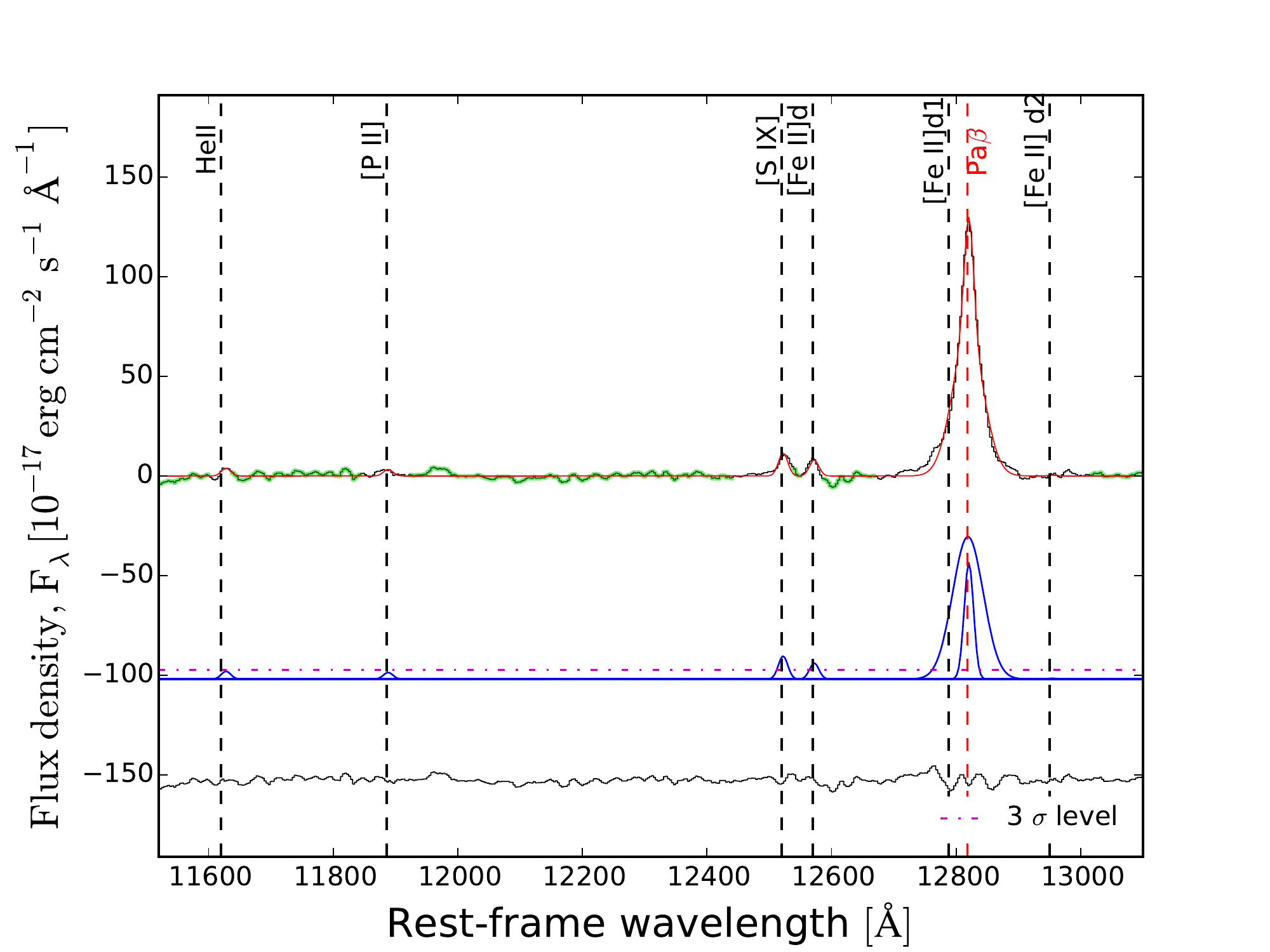} }
\subfigure{\includegraphics[width=0.3\textwidth]{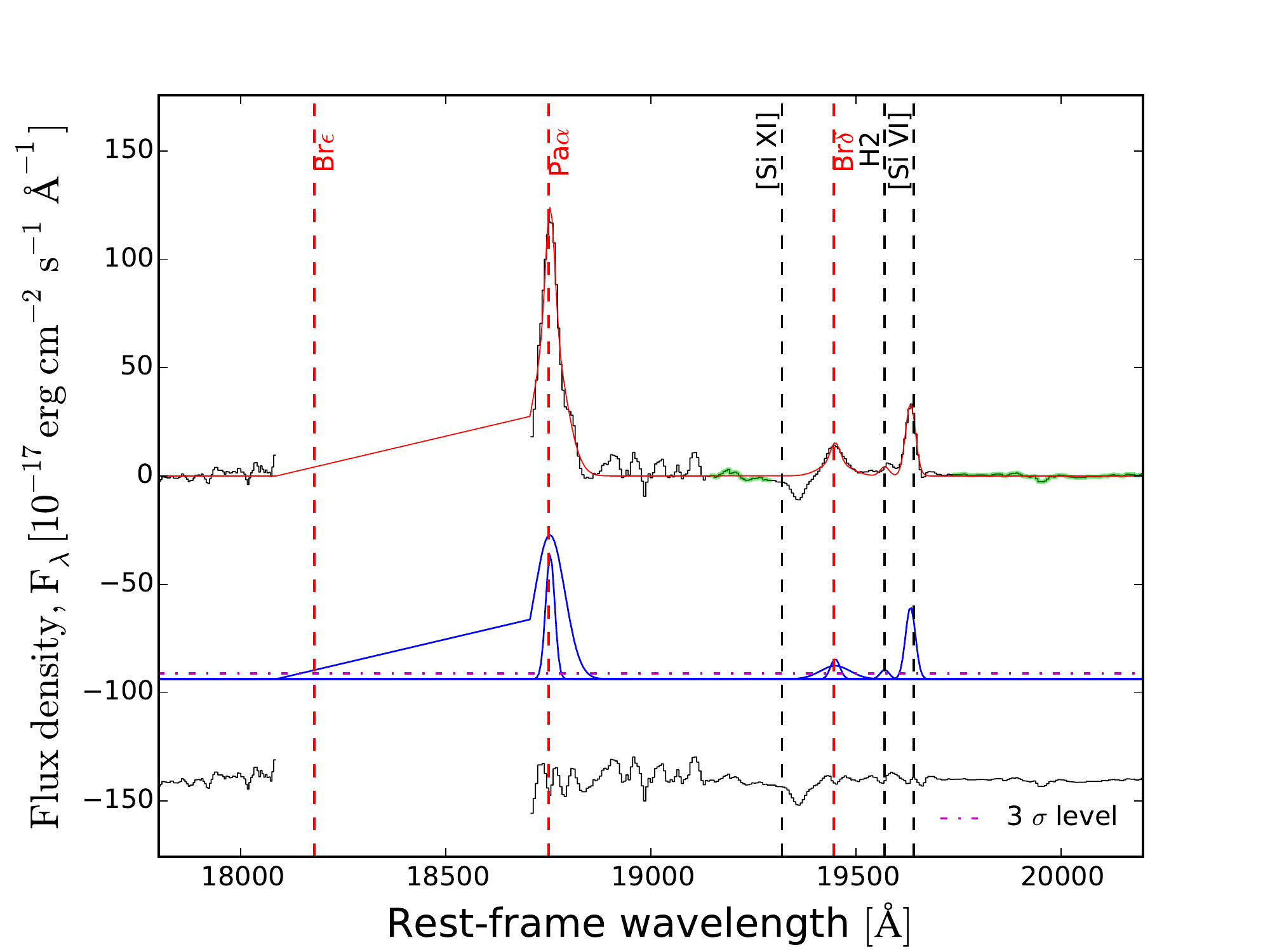} }

\caption{Fit of the HeI $\lambda 1.0830$ $\mu$m, Pa$\beta$ and Pa$\alpha$ regions for the Seyfert 1.9 and Seyfert 2 for which the broad Pa$\beta$ was used to measure \MBH. From the first row: Mrk 520 and NGC 7314. The best fit is in red, the model in blue, and the residuals in black.  In addition to the components explained above, the magenta dashed line in the middle part of the figures shows the detection threshold (S/N$>\ $3) with respect to the fitting continuum (blue). }
\label{Fit_Hidden_BLR_2}
\end{center}
\end{figure*}

%%%%%%%%%%%%%%%%%%%%%%%%%%%%%%%%%%%%%%%%%%%%%%%%%%

% Don't change these lines
\bsp	% typesetting comment
\label{lastpage}
\end{document}

% End of mnras_template.tex